\documentclass[12pt,a4paper,twoside]{book}

\usepackage{a4wide, amsmath, latexsym, epsfig, psfig,
  psfrag, graphicx, rotating, fancyhdr, tabularx}
\usepackage{feynmf}

\begin{document}
\frontmatter     

\begin{titlepage}
\begin{center}
{
\Large
\noindent Technische Universit\"at M\"unchen

\vspace{0.4cm}
\noindent Max-Planck-Institut f\"ur Physik\\
(Werner-Heisenberg-Institut)

\vspace{1.0cm}
\Huge
{\bfseries Precision Calculation for \\
 \mbox{Supersymmetric Particle Decays} 
}

\vspace{1.5cm} \Large
\noindent Qingjun~Xu

\vspace{1.5cm}
\normalsize
\noindent Vollst\"andiger Abdruck der von der Fakult\"at f\"ur Physik\\
\noindent der Technischen Universit\"at M\"unchen\\
\noindent zur Erlangung des akademischen Grades eines\\
\noindent {\bfseries Doktors der Naturwissenschaften (Dr.\ rer.\ nat.)}\\
\noindent genehmigten Dissertation.

\vspace{1.5cm}
\large
  \begin{tabular}{lll}
    Vorsitzender : & &  Univ.-Prof. Dr.\ L.\ Oberauer \\[16pt]
    Pr\"ufer der Dissertation : \\
                                          & 1. & Hon.-Prof. Dr.\ W.\ F.\ L.\ Hollik \\
                                          & 2. & Univ.-Prof. Dr.\ M.\ Lindner
                                         
  \end{tabular} \newline
\vspace{1.5cm}

\noindent Die Dissertation wurde am 23.\,03.\,2006 \\
\noindent bei der Technischen Universit\"at M\"unchen eingereicht \\
\noindent und durch die Fakult\"at f\"ur Physik am 30.\,05.\,2006 angenommen.
}
\end{center}
\end{titlepage}
\cleardoublepage

\tableofcontents
\mainmatter
\chapter{Introduction}
\hspace*{7mm}The Standard Model (SM) \cite{Glashow} is very successful in describing the known \mbox{phenomena} of \mbox{particle} physics.
However, it also has some drawbacks.
\mbox {Supersymmetry} (SUSY) \cite{susy, Hilles}, which is a symmetry which connects fermions with bosons,
is one of the best motivated extensions of the SM.

In the Minimal \mbox {Supersymmetric} Standard Model (MSSM) \cite{susy, Hilles},
we introduce a superpartner to every known particle, i.e. 
scalar superpartners (sfermions) to the SM fermions, 
fermionic superpartners (gauginos and higgsinos) to the gauge bosons and Higgs bosons.
After \mbox {electroweak} symmetry is broken, 
fields with same quantum numbers can mix.  
We therefore have sfermion mixing, and higgsino and gaugino mixing into charginos and \mbox {neutralinos}.
Moreover two Higgs doublets are necessary in the MSSM.
This leads to five physical Higgs bosons after the electroweak symmetry is 
\mbox {broken}.
If $R$-parity \mbox{ is conserved}, the lightest supersymmetric particle (LSP), which in many scenarios
is the lightest neutralino $\tilde{\chi}_1^0$, appears at the end of the decay
chain of each supersymmetric \mbox{particle}. The LSP \mbox{escapes} the detector, giving
the characteristic SUSY signature of missing energy.
If SUSY exists at the electroweak scale, experiments at future high energy colliders should be able to discover
the superpartners of the known particles, and to study their properties \cite{LHC, TESLA}. 
%

At the Large Hadron Collider (LHC), the total SUSY production-cross section is \mbox{expected} to be
dominated by the production of gluinos and squarks, which decay into lighter
charginos or neutralinos.
Of particular interest are decay chains leading to
the next-to-lightest neutralino $\tilde{\chi}_2^0$.  $\tilde \chi_2^0$ in
turn can always decay into the LSP $\tilde\chi_1^0$ and two fermions $f \bar f$, 
at least for light SM fermions. 
Depending on the neutralino, sfermion and Higgs boson masses, the possible decays
of $\tilde \chi_2^0$ are three-body decays $\tilde\chi_2^0\rightarrow \tilde \chi_1^0 f \bar f$,
cascade two-body decays $\tilde\chi_2^0 \rightarrow \tilde f \bar f \rightarrow \tilde \chi_1^0 f \bar f$ 
and/or $\tilde \chi_2^0 \rightarrow \tilde \chi_1^0 Z/ \tilde \chi_1^0\phi^0 \rightarrow
\tilde \chi_1^0 f \bar f$,
where $\phi^0$ stands for one of the neutral Higgs bosons
or the neutral Goldstone boson of the MSSM. 
The leptonic final states
are of particular interest, since they can be identified relatively easily
even at the LHC. 
Moreover, the dilepton invariant mass distribution can be measured accurately at the LHC. 
In particular, its endpoint is used in several analyses which aim to reconstruct the mass differences of 
the supersymmetric particles \cite{LHC, LHC/LC}.
Under favorable circumstances it has been shown
that this endpoint can be measured to an accuracy of $0.1\%$ at the LHC
\cite{LHC}. In order to match this accuracy in the theoretical prediction, at least one-loop \mbox{corrections} to
$\tilde \chi_2^0$ decays have to be included.
At the planned $e^+e^-$ linear collider ILC~\cite{TESLA},
the lighter supersymmetric particles can be produced directly.
The detailed analysis of $\tilde \chi_2^0$
decays can then yield information about the supersymmetric particles. 
Moreover, the masses of the supersymmetric particles are expected to be determined with
high precision at the ILC~\cite{TESLA}, again making the inclusion of quantum corrections
mandatory to match the experimental precision.

Leptonic two-body decays $\tilde{\chi}_2^0\rightarrow \tilde{l}^\pm_1 l^\mp
\rightarrow \tilde{\chi}_1^0 l^- l^+$ have been
investigated at tree level in Ref.\cite{LHC/LC}, where $\tilde
l_1$ stands for the lighter slepton of a given flavor. 
Three-body decays of
$\tilde{\chi}_2^0$ have also been studied at tree level in
Refs.\cite{treedonea,noya}.
In this thesis, we calculate leptonic $\tilde \chi_2^0$ decays at one-loop
level. 
Cases where two-body decays $\tilde{\chi}_2^0\rightarrow \tilde{l}^\pm_1 l^\mp
\rightarrow \tilde{\chi}_1^0 l^- l^+$ are kinematically alowed are 
treated both completely and in a single-pole approximation.
In the complete calculation one has to employ complex slepton masses in the relevant propagators and
one-loop integrals. The single-pole approximation in this case is performed in the way that
the $\tilde{\chi}_2^0$ decays are treated as the production and decay of the sleptons
$\tilde{l}_1$. 
We compare the results from the complete and approximate calculations 
and find a good agreement.
We also analyze a scenario where $\tilde \chi_2^0$ only has three-body decays. 
The calculations for these decays are similar to the complete calculation except that
we do not have to introduce complex masses. 
The differential decay width of $\tilde{\chi}_2^0$ as a function of the dilepton
invariant mass $M_{l^+l^-}$ is calculated. The shape of the $M_{l^+l^-}$ distribution is 
expected to be altered by real photon emission contributions, which  
must be added to the one-loop corrrections in order to cancel the infrared divergences
in the virtual contributions.
In order to obtain the total decay width of $\tilde\chi_2^0$ and hence the 
branching ratios of its leptonic decays,
the invisible decays 
$\tilde\chi_2^0 \rightarrow \tilde \chi_1^0 \nu_l \bar \nu_l$ and the hadronic decays
$\tilde\chi_2^0 \rightarrow \tilde \chi_1^0 q \bar q$ are also calculated.

The general MSSM has more than one hundred unknown free parameters. 
We assume CP-conserving MSSM with real parameters. For specific numerical evaluation we consider
the SPS1a parameter set \cite{SPS1aa, SPS1ab},
which is one of the standardized benchmark scenarios.
It gives rise to a particle spectrum where many
states are accessible both at the LHC and at ILC 
\cite{LHC/LC}. Note in particular that the
two-body decays $\tilde{\chi}_2^0\rightarrow \tilde{l}^\pm_1 l^\mp
\rightarrow \tilde{\chi}_1^0 l^- l^+$ are kinematically allowed in the SPS1a scenario. No
other two-body decay mode is open.
\newpage
The outline of this thesis is as follows:

\vspace*{5mm}
In Chapter \ref{MSSM} the basic ideas of SUSY are presented first,
where we introduce the SUSY algebra, the superfields and
the SUSY Lagrangian in detail. 
Then the MSSM Lagrangian and the soft SUSY-breaking Lagrangian are constructed.
At last all the physical fields of the MSSM are discussed and the expressions for their masses are presented.

\vspace*{3mm}
The MSSM is renormalized in Chapter \ref{ReMSSM} where we follow the strategy of Refs.\cite{SMDenner, ReNeu, ReSle}.
All relevant parameters in the MSSM are assumed to be real quantities. 
This amounts to the assumption that the soft SUSY-breaking terms conserve CP.
We introduce two renormalization schemes, the $\overline {\rm DR}$ and the on-shell scheme.
The SM sector, the chargino and neutralino sector and the sfermion sector 
are renormalized via the on-shell renormalization scheme,
while the Higgs sector is renormalized via a mixing of on-shell and $\overline {\rm DR}$ schemes.

\vspace*{3mm}
In Chapter \ref{calculations} the leptonic decays  $\tilde{\chi}_2^0\rightarrow \tilde{\chi}_1^0 l^- l^+$ 
are calculated at one-loop level. 
The tree-level calculations for these leptonic decays
are outlined in \mbox{Section} \ref{tree},
where we calculate the leptonic decays in both a complete and
an approximate way where $\tilde l_1$ can be on shell.
In Section \ref{complete} we discuss how to calculate these decays completely at one-loop level,
where the virtual corrections and the real photon bremsstrahlung are investigated in detail.
The approximate one-loop calculations for these decays where $\tilde l_1$ can be on shell are given in Section \ref{approx}.
We study the total decay width of $\tilde\chi_2^0$ and the branching ratios of the leptonic decays 
in Section \ref{brlepton}.
In Section \ref{numerical} the explicit values of the MSSM parameters and the spectrum of 
the supersymmetric particles at the SPS1a benchmark point are given first. Then 
the numerical results of the calculations are presented.

\chapter{The Minimal Supersymmetric Standard Model (MSSM)}\label{MSSM}
\section{The Standard Model}
The Standard Model (SM) of particle physics describes the electroweak and strong interactions with very good precision. 
It is based on the gauge group $SU(3)_C\times SU(2)_L\times U(1)_{Y}$\cite{Glashow}, where $SU(3)_C$ and $SU(2)_L\times U(1)_{Y}$ 
describe the strong and electroweak interactions, respectively.

The elementary particles include spin-$\frac{1}{2}$ fermions and spin-$1$ gauge bosons.
The interactions in the nature are described as the exchange of different gauge bosons.
The SM gauge bosons are eight gluons $G_{\mu}^k (8, 1, 0)$,
three weak bosons $W_{\mu}^a (1, 3, 0)$ and a hypercharge boson $B_{\mu} (1, 1, 0)$, where 
we have indicated their quantum numbers with respect to the gauge group $SU(3)_C\times SU(2)_L\times U(1)_{Y}$.
These gauge bosons are the mediators of the strong, weak and electromagnetic interactions.

The SM fermions are chiral because the left-handed and right-handed fermions transform as 
doublets and singlets of the gauge group $SU(2)_L$, respectively. 
The fermions, i.e. leptons and quarks, can be arranged in three generations as below,
\begin{table}[htb]
\begin{tabular}{ccccc}
1st generation & 2nd generation & 3th generation & group representation\\ 
$\left(\begin{array}{c}\nu_e\\ e \end{array} \right)_L$, &
$\left(\begin{array}{c}\nu_\mu\\ \mu  \end{array} \right)_L$, &
$\left(\begin {array}{c}\nu_\tau\\ \tau \end{array} \right)_L$,&
\ \ \ $ (\underline{1}, \underline{2},-1)$\\
$e_R$,&
$\mu_R$,& $\tau_R$,& \ \ \ $(\underline{1},\underline{1}, -2)$\\
 $\left(\begin{array}{c}u\\d \end{array}\right)_L$,
& $\left(\begin{array}{c}c\\s \end{array}\right)_L$,&
$\left(\begin{array}{c}t
\\b \end{array} \right)_L$,& \ \ \ $(\underline{3},\underline{2},1/3)$\\
 $u_{R}$,& $c_{R}$,& $t_{R}$,& \ \ \ 
$(\underline{3}^*, \underline{1}, 4/3)$\\ 
$d_{R}$,&
  $s_{R}$,& $b_{R}$,& \ \ \ $(\underline{3}^*, \underline{1}, -2/3)$\, .
 \end{tabular}
\end{table}

In the SM we introduce one scalar doublet $H$ with hypercharge $Y = + 1$,
which implements  
the spontaneous breaking of the electroweak $SU(2)_L \times U(1)_{Y}$ symmetry 
down to the electromagnetic $U(1)_{EM}$ symmetry,
\begin{eqnarray}
SU(3)_C\times SU(2)_L\times U(1)_{Y} & \rightarrow & SU(3)_C\times U(1)_{EM}\, .
\nonumber
\end{eqnarray}
This is called the Higgs mechanism\cite{Higgs}, which 
gives masses to the fermions and  
produces three massive vector bosons $W^{\pm}$ and $Z$, and a massless photon $\gamma$.  
The Higgs mechanism also
predicts a new particle: the Higgs boson.

\begin{figure}[b]
\begin{center}
\begin{tabular}[2]{ll}
\includegraphics[width=0.30\linewidth]{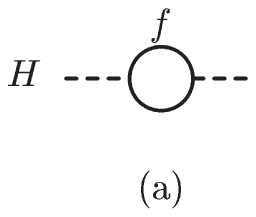} & \includegraphics[width=0.30\linewidth]{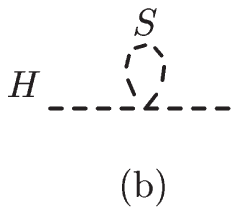}
\end{tabular}
\end{center}
\caption{Radiative corrections to the Higgs boson mass}
\end{figure}

The SM provides an extremely successful description of the known phenomena of particle physics.
All the particles that it predicts, except for the Higgs boson, have been discovered experimentally.
The mass of the Higgs boson is severely constrained from electroweak precision data\cite{test-smb}. 
Moreover, the present experimental data agree with the SM predictions very well\cite{test-smb, test-sma}.
However, the SM also has some drawbacks and unsolved problems. Here we focus on why we need supersymmetry. 

The SM is an "effective low-energy theory" for energy scales up to $100$ GeV.
More reasonably, we regard the SM as part of a larger fundamental theory which describes physics at arbitrarily high energies.
The Quantum gravitational effects become important at the Planck scale $M_p \simeq 2.4\times 10^{18}$GeV.
A Grand Unification Theory (GUT) may appear at a somewhat lower energy scale.
The surprising thing is that the ratio $\frac{M_P}{m_W}\approx 10^{16}$ is so huge. This is called
"hierarchy problem"\cite{hierarchy}. 

Moreover, radiative corrections must be included in the theoretical predictions in order to match 
the precision of experimental measurements.  
When we consider radiative corrections to the Higgs boson mass, 
$m_H^2$ is found to be quadratically divergent at one-loop level\cite{Hooft}. 
For example, consider the contributions to $m_H^2$ from a loop containing the fermion $f$ 
as shown in Figure~2.1(a) and suppose the Lagrangian term is $-\lambda_f Hf\bar{f}$. Then the correction is given by
\begin{equation}
\Delta m_H^2 \sim - \frac{\lambda_f^2}{8\pi^2}\Lambda^2 + {\rm logarithmically \ divergent \ terms}\, ,
\end{equation}
where $\Lambda$ is a momentum cutoff used to regulate the loop integral, which represents the energy scale at which 
new physics appears. 
If $\Lambda$ is of order of the Planck scale $M_p$,
the one-loop quantum correction $\Delta m_H^2$ is some 30 orders of magnitude larger than 
$m_H^2\leq (1TeV)^2$, which is needed to preserve unitary in the scattering of the longitudinal 
gauge bosons\cite{Higgslimit}. This is the technical aspect of the hierarchy problem. 
One needs extreme fine-tuning of the parameters to cancel the large quadratic contributions
against a counterterm,
leaving a resulting Higgs boson mass of about 1~TeV. Such a cancellation is unnatural and
in general not acceptable. 
Additionally, in every order of the perturbative theory, the parameters must be re-tuned.


The only known way to cancel these quadratic divergences is by introducing a new partner with spin differing by $\frac{1}{2}$ for
every known particle. The properties of the known particles and their new partners are related by a symmetry, which 
is known as supersymmetry. As shown in Figure~2.1(b), suppose there is a new partner $S$ for fermion $f$, 
which couples to the Higgs boson with a Lagrangian term $-\lambda_S H^2 S^2$, then the contributions are
\begin{equation}
\Delta m_H^2 \sim \frac{\lambda_S}{8\pi^2}\Lambda^2 + {\rm logarithmically \ divergent \ terms}\, .
\end{equation}
If each of the SM quarks and leptons has a new supersymmetric partner with $\lambda_S = \lambda_f^2$, 
then the quadratic divergences neatly cancel. 

\section{Supersymmetry}\label{SUSY}
In order to protect the Higgs boson mass from quadratically divergent radiative corrections,
we introduce supersymmetry(SUSY), which is a symmetry which connects fermions and bosons.
In this section we first introduce the SUSY algebra and the different superfields, and then
come to the construction of the SUSY Lagrangian. The breaking of SUSY is also discussed. 
The discussions in this section are based on Refs. \cite{susy, Hilles, sspace}.   
\subsection{SUSY Algebra}\label{algbra}
The generators of SUSY must turn a fermionic state into a bosonic state, and vice versa,
 $$ Q| boson\rangle = |fermion\rangle, \ \ \  Q| fermion\rangle = | boson\rangle .$$
This implies that the SUSY generators are fermionic, so they must satisfy the anticommutation relations below \cite{superalgbra}, 
\begin{eqnarray}
\left\{Q_{\alpha}\, , Q_{\beta}\right\} &= & \left\{\bar{Q}_{\dot{\alpha}}\,,\bar{Q}_{\dot{\beta}}\right\} =0\, ,\label{eqn:SUSY1a}\\
\left\{Q_{\alpha}\, , \bar{Q}_{\dot{\beta}}\right\} & = & 2\sigma_{\alpha\dot{\beta}}^{\mu}P_{\mu}\, ,\label{eqn:SUSY1b}\\
\bigl[Q_{\alpha}\, , P_{\mu}\bigr] &=& \bigl[\bar Q_{\dot\beta}\, , P_{\mu}\bigr] = 0\, ,\label{eqn:SUSY1c}\\
\bigl[Q_{\alpha}\, , M_{\mu\nu}\bigr] &=& i \left(\sigma^{\mu\nu}\right)_\alpha^\beta Q_\beta\, .
\label{eqn:SUSY1d}
\end{eqnarray}
Here the SUSY generators are 2-component (Weyl) spinors, their indices $\alpha, \beta$ and $\dot{\alpha}, \dot{\beta}$
equal $1$ or $2$, $P^{\mu}$ and $M^{\mu\nu}$ are the four-momentum and angular momentum operators,
and $\sigma^{\mu}$ and $\sigma^{\mu\nu}$ are defined in appendix~\ref{Notations}. 
We have used the simplest case which is called $N=1$ SUSY, 
where $N$ denotes the number of SUSY generators. 
\subsection{Superfields}\label{superfields}
In order to describe the SUSY transformations, we introduce the superspace \cite{sspace} which 
is different from the normal space by adding
two new "Grassmann" coordinates $\theta_{\alpha} (\alpha = 1, 2)$ and $\bar{\theta}_{\dot{\beta}} (\dot{\beta} = 1, 2)$.
They are anticommuting
\begin{eqnarray}
\left\{\theta_{\alpha}, \theta_{\beta}\right\} = \left\{\bar{\theta}_{\dot{\alpha}}, \bar{\theta}_{\dot{\beta}}\right\}=
 \left\{\theta_{\alpha}, \bar{\theta}_{\dot{\beta}}\right\}  =0\, .
\label{Grassmann}
\end{eqnarray}
A "finite" SUSY transformation in the superspace ($x_{\mu}, \theta, \bar{\theta}$)
can be defined as
 \begin{eqnarray}
G(x, \theta, \bar{\theta}) & = & \exp\bigl[i\left(\theta Q + \bar{Q}\bar{\theta}- x_{\mu}P^{\mu}\right)\bigr]\, .
\end{eqnarray}
Here the indices have been dropped. The superfields $\Phi$ should be functions of $\theta, \bar{\theta}$ and $x_{\mu}$
such that they transform under the SUSY transformations as follows,
\begin{eqnarray}
G(y, \eta, \bar{\eta}) \Phi(x, \theta, \bar{\theta}) & = & \Phi(x+y -i\eta\sigma\bar{\theta} + i
\theta\sigma\bar{\eta}, \theta + \eta, \bar{\theta} + \bar{\eta})\, ,
\end{eqnarray}
where $\eta$ and $\bar{\eta}$ are again "Grassmann" variables.
Considering infinitesimal SUSY transformations on the superfield
\begin{eqnarray}
\delta_S(\eta, \bar{\eta})\Phi(x, \theta, \bar{\theta})& =& i\left(\eta Q + \bar{Q}\bar{\eta}- x_{\mu}P^{\mu}\right)
\Phi(x, \theta, \bar{\theta})\, \nonumber \\
& = & -\Bigl[\eta \frac{\partial}{\partial\theta} + \bar{\eta} \frac{\partial}{\partial\bar{\theta}} - i
\left (\eta\sigma_{\mu}\bar{\theta}-\theta\sigma_{\mu}\bar{\eta}\right)\partial^{\mu}\Bigr]\Phi(x, \theta, \bar{\theta})\, ,
\label{eqn:SUSYtrans}
 \end{eqnarray}
one obtains the representation of the SUSY generators,
 \begin{eqnarray}
Q_{\alpha}& =& i\left (\frac{\partial}{\partial\theta_{\alpha}} - 
i\sigma^{\mu}_{\alpha\dot{\beta}}\bar{\theta}^{\dot{\beta}}\partial_{\mu}\right )\, , \\
\bar{Q}_{\dot{\alpha}} &= & i\left (\frac{\partial}{\partial\bar{\theta}_{\dot{\alpha}}} +
 i\theta^{\beta}\sigma^{\mu}_{\beta\dot{\alpha}}\bar{\theta}^{\dot{\beta}}\partial_{\mu}\right )\, .
\end{eqnarray}
The SUSY covariant derivatives anticommute with the SUSY transformations~(\ref{eqn:SUSYtrans}), i.e.
\begin{eqnarray}
D_{\alpha}\left(\delta_S\Phi\right) & = & -\delta_S\left(D_{\alpha}\Phi \right)\, .
\end{eqnarray}
This gives the expressions for the SUSY covariant derivatives,
\begin{eqnarray}
D_{\alpha}& =& \frac{\partial}{\partial\theta_{\alpha}} + i\sigma^{\mu}_{\alpha\dot{\beta}}\bar{\theta}^{\dot{\beta}}\partial_{\mu}
\, , \\
\bar{D}_{\dot{\alpha}} &= &\frac{\partial}{\partial\bar{\theta}_{\dot{\alpha}}} -
 i\theta^{\beta}\sigma^{\mu}_{\beta\dot{\alpha}}\bar{\theta}^{\dot{\beta}}\partial_{\mu}\, .
\end{eqnarray}

Because the SM fermions are chiral, the first superfield we need is a chiral superfield, 
which can describe the left- and right-handed fermions as well as their superpartners. 
The left- and right-handed chiral superfields are defined via
\begin{eqnarray}
\bar{D}\Phi_L &=&0\, ,
\label{SUSYchiralL}\\
D\Phi_R & =&0
\label{SUSYchiralR}\, ,
\end{eqnarray}
respectively. The definition of the
SUSY transformation, SUSY generators and SUSY covariant derivatives in the L(R) representation
are similar to the definitions above~\cite{Hilles}. 
The L(R) representation of the SUSY covariant derivatives are
\begin{eqnarray}
D_L &= &\frac{\partial}{\partial\theta} + 2i \sigma^{\mu}\bar{\theta}\partial_{\mu}\, , \hspace*{3mm}
\bar{D}_L =  \frac{\partial}{\partial\bar{\theta}}\, ,
\label{derivativeL}\\
D_R &=& \frac{\partial}{\partial\bar{\theta}} - 2i \theta\sigma^{\mu}\partial_{\mu}\, , \hspace*{3mm}
\bar{D}_R = \frac{\partial}{\partial\theta}\, .
\label{derivativeR}
\end{eqnarray}
From (\ref{SUSYchiralL}) we know that $\Phi_L$ is independent of
$\bar{\theta}$, therefore we can expand it as,
\begin{eqnarray}
\Phi_L(x, \theta) & = & \phi(x) +\sqrt 2\theta^{\alpha}\psi_{\alpha}(x) + \theta^{\alpha}\theta^{\beta}\varepsilon_{\alpha\beta}F(x)\, ,
\label{SUSYchiralLexp}
\end{eqnarray}
where $\varepsilon_{\alpha\beta}$ is the anti-symmetric tensor in two dimensions.
From (\ref{Grassmann}) we know that $\theta_{\alpha}^2 =0$, therefore 
the terms with three or more factors of $\theta$ vanish.
The field $\phi$ is a complex scalar field, and $\psi$ is a left-handed Weyl spinor. 
They are called superpartners to each other.
The field $F$ is an auxiliary field, which we can get rid of by using the equations of motion.

Under SUSY transformations the "component" fields $\phi, \psi, F$ transform as
 \begin{eqnarray}
\delta_S(\eta, \bar{\eta})\phi &= &\sqrt 2 \eta \psi\, ,
\label{SUSYphi}\\
\delta_S(\eta, \bar{\eta})\psi & = & \sqrt 2 F + i \sqrt 2 \sigma_{\mu}\bar{\eta}\partial^{\mu}\phi\, ,
\label{SUSYpsi}\\
\delta_S(\eta, \bar{\eta}) F & = & -i \sqrt 2 \partial^{\mu}\psi\sigma_{\mu}\bar{\eta}\, ,
\label{SUSYF}
 \end{eqnarray}
which shows that SUSY transforms fermions into bosons
and the variation of the $F$ field is a total derivative. 

The chiral superfield contains the spin-0 bosons and spin-$\frac{1}{2}$ fermions.
One also needs superfields which describe the spin-1 gauge bosons of the SM.
Hence we introduce a vector superfield $V(x, \theta, \bar{\theta})$ which satisfies
\begin{eqnarray}
V(x, \theta, \bar{\theta}) & = & V^{\dagger}(x, \theta, \bar{\theta})\, .
\end{eqnarray}  
In component form it can be written as,
\begin{eqnarray}
V(x, \theta, \bar \theta) & = & C(x) + i\theta \chi (x) -i\bar
\theta \bar \chi (x)  \nonumber \\
 & + & \frac{i}{2} \theta \theta [M(x) + iN(x)] - \frac{i}{2} \bar
 \theta \bar \theta [M(x) - iN(x)]  \nonumber \\
 & - & \theta \sigma^{\mu} \bar \theta v_{\mu}(x) + i \theta \theta
\bar \theta [\bar \lambda (x) + \frac{i}{2}\bar \sigma^{\mu} \partial
_{\mu} \chi (x)]  \nonumber \\
 & - & i\bar \theta \bar \theta \theta [\lambda + \frac{i}{2}
\sigma^{\mu} \partial_{\mu} \bar \chi (x)] + \frac{1}{2} \theta
\theta \bar \theta \bar \theta [D(x) + \frac{1}{2}\partial_{\mu}\partial^{\mu}C(x)]\, ,
 \label{p}
 \end{eqnarray}
where $C, M, N, D$ are real spin-0 fields, $\chi, \lambda$ are Weyl spinors and
$v_{\mu}$ is a spin-1 gauge field. 
Under the non-Abelian supersymmetric gauge transformation the vector superfield transforms as
\begin{eqnarray}
e^{2gV} &\longrightarrow & e^{-i 2g\Lambda^{\dagger}}e^{2gV}e^{i 2g\Lambda}\, ,
\label{nonabelian}
 \end{eqnarray}
where $\Lambda(x, \theta, \bar{\theta})$ is a chiral superfield, $g$ is the gauge coupling,
and $V = V^a T^a$, $\Lambda = \Lambda^a T^a$ where $T^a$ are the generators of the non-Abelian gauge group. 
For the Abelian case, the transformation simplifies to
\begin{eqnarray}
V &\longrightarrow & V +i \left(\Lambda - \Lambda^{\dagger}\right)\, .
 \end{eqnarray}
One can choose the Wess-Zumino gauge\cite{WZgauge}, where
\begin{eqnarray}
C(x) = \chi(x) = M(x) = N(x) \equiv 0\, ,
\end{eqnarray}
leaving the gauge field $v_{\mu}$, its superpartner $\lambda$ (gaugino) and the auxiliary field $D$. 
In this gauge,
the vector superfield is expressed as
\begin{eqnarray}
V(x, \theta, \bar{\theta}) & = & - \theta \sigma^{\mu} \bar \theta v_{\mu}(x) + i \theta
\theta \bar \theta \bar \lambda (x) -i\bar \theta \bar \theta
\theta \lambda (x) + \frac{1}{2} \theta \theta \bar \theta \bar
\theta D(x)\, .
 \end{eqnarray}
Similarly to the chiral superfield, one finds that the D component of the vector field transforms 
into a total derivative under SUSY transformations.
This is very important for the construction of the SUSY Lagrangian.
\subsection{Construction of the SUSY Lagrangian}\label{construction}
In this section SUSY Lagrangian is constructed in the notation of superfields. The principle of construction is
that the action should be invariant under SUSY transformations, i.e.
$\delta_S\int d^4 x\mathcal{L}(x)= 0$.
It is satisfied when $\mathcal{L}$ transforms into a total derivative under SUSY transformations.
From the discussions above we know that the highest components of chiral (F-term) and vector (D-term) superfields
satisfy this demand. They can be obtained with integration over the Grassmann variables
$\theta$ and $\bar \theta$ via the definition,
\begin{eqnarray}
\int d\theta_{\alpha} =0\, , \hspace*{3mm} \int \theta_{\alpha} d\theta_{\beta} = \delta_{\alpha\beta}\, .
\end{eqnarray}
Therefore, the general form of the SUSY invariant Lagrangian is,  
 \begin{eqnarray}
\mathcal{L} & =& \int d^2 \theta \mathcal{L}_F+ \int d^2\theta d^2\bar \theta \mathcal{L}_D\, ,
\label{totalLagrangian}
 \end{eqnarray}
where the Lagrangian densities $\mathcal{L}_F$ and $\mathcal{L}_D$ are chiral and vector superfields, respectively.
 
In order to obtain the explicit expressions for the  SUSY invariant Lagrangian, we consider
the product of left-chiral superfields. 
From the definition of the chiral superfield in (\ref{SUSYchiralL}),
it is easy to see that the product of left-handed superfields is always a left-handed superfield.
Therefore it can take the role of $\mathcal{L}_F$. 
The superpotential $W(\Phi)$ (corresponding to $\mathcal{L}_F$) is introduced as
\begin{eqnarray}
W(\Phi_i) & = &\sum_i\lambda_i \Phi_i + \frac{1}{2}\sum_{ij}m_{ij}\Phi_i \Phi_j + \frac{1}{3}\sum_{ijk}y_{ijk}
\Phi_i \Phi_j \Phi_k \, ,
 \end{eqnarray}
where $\Phi_i$ are left-chiral superfields, and $m_{ij}$ and $y_{ijk}$ are totally symmetric constants.  
Because of the renormalizability constraint the superpotential does not contain terms 
with four or more powers of superfields.
The F-component of the superpotential can be written as
\begin{eqnarray}
\int d^2 \theta W(\Phi_i) & =& \sum_i \lambda_i F_i + \sum_{ij}m_{ij}\left(\phi_i F_j -\frac{1}{2}\psi_i \psi_j\right)
+ \sum_{ijk}y_{ijk}\left(\phi_i \phi_j F_k -\psi_i \psi_j \phi_k\right)\nonumber \\
& = & \sum_j \frac{\partial W(\phi)}{\partial \phi_j}F_j -\frac{1}{2}\sum_{jk}
\frac{\partial^2 W(\phi)}{\partial\phi_{j}\partial \phi_k }\psi_j\psi_k\, .
\end{eqnarray}
This gives a fermion mass term and Yukawa interactions.

Considering the product of the left-chiral superfield and its conjugate, one finds that
$\Phi_i^+ \Phi_i$ is a vector superfield. Hence, it can take the role of $\mathcal{L}_D$ in (\ref{totalLagrangian}).
The D-component of $\Phi_i^+ \Phi_i$ is obtained by
\begin{eqnarray}
\int d^2\theta d^2\bar \theta \Phi_i^+ \Phi_i & = & F_iF_i^{\ast} + \partial_{\mu}\phi_i\partial^{\mu}\phi_i^{\ast}
+i\bar\psi_i\sigma_{\mu}\partial^{\mu}\psi_i\, ,
 \end{eqnarray}
 which contains the kinetic terms for the scalar component $\phi_i$ and the fermionic component $\psi_i$. 
Therefore the SUSY invariant Lagrangian can be written as
 \begin{eqnarray}
\mathcal{L} & = &\sum_i \int d^2\theta d^2\bar \theta ~\Phi_i^+ \Phi_i+\biggl[\int d^2\theta W(\Phi_i) + h.c.\biggr]\, .
 \end{eqnarray}

In the discussions above, only the SUSY invariant Lagrangian is considered. But this is not sufficient. 
In order to describe the SUSY theory, we need a SUSY Lagrangian
which is not only SUSY invariant, but also gauge invariant.  
Considering the non-Abelian supersymmetric gauge transformations, the vector superfield transforms as in (\ref{nonabelian}),
and the chiral superfield transforms as $\Phi\rightarrow e^{-i 2g\Lambda}\Phi$.
The kinetic terms for the chiral superfield can be written as 
 \begin{eqnarray}
\int d^2\theta d^2\bar \theta ~\Phi_i^+ \Phi_i &\rightarrow &
\int d^2\theta d^2\bar \theta \Phi_i^+ e^{2gV} \Phi_i\, \nonumber \\
& = & D_{\mu}\phi_i D^{\mu}\phi_i^{\ast}+
i \bar\psi_i \bar\sigma^{\mu}D_{\mu}\psi_i - \sqrt 2 g\left(\bar \psi_i \bar \lambda \phi_i +
\phi_i^{\ast}\lambda \psi_i\right) + \, \nonumber \\
&& g\phi_i^{\ast}D\phi_i + F_i^2\, ,
\label{eqn:minimalcoupling}
 \end{eqnarray}
where the gauge-covariant derivative $D_{\mu}$ is defined by 
 \begin{eqnarray}
D_{\mu}= \partial_{\mu} + i g v_{\mu}^a T^a\, ,
 \end{eqnarray}
and $T^a$ are the generators of the gauge group.
The first two terms of the second line in (\ref{eqn:minimalcoupling}) are the kinetic terms for scalars 
and fermions, and they also describe the interactions of matter fields (fermions and scalars) 
with the gauge fields.  
The couplings of fermions and scalars with gauginos can be found in the third term
of the second line, while the third line contains the auxiliary fields $D$ and $F$.

The kinetic term for the vector superfield is produced with the help of the field strength tensor,
which is defined by
\begin{eqnarray}
W_\alpha & = & - \frac{1}{4}\bar D\bar D\bigl[\exp(-2gV)D_\alpha \exp(2gV)\bigr]\, ,
\end{eqnarray}
where $D$ and $\bar D$ are SUSY covariant derivatives and $W_\alpha$ is a left-chiral superfield
because $\bar D_{\bar\alpha}W_{\alpha}=0$. It is easy to show that the product $W_{\alpha}W^{\alpha}$
is gauge invariant. Therefore its F-component can appear in the Lagrangian,
\begin{eqnarray}
\frac{1}{16g^2}\int d^2\theta \bigl[ W^a_\alpha W^{\alpha a} + h.c.\bigr]=
-\frac{1}{4}F_{\mu\nu}^a F^{\mu\nu a} + i \bar \lambda^a \bar\sigma^{\mu}\left(D_{\mu}\lambda\right)^a
+ \frac{1}{2}D^a D^a\, ,
\label{eqn:vectorkinetic}
\end{eqnarray}
where 
\begin{eqnarray}
F_{\mu\nu}^a = \partial_{\mu}v^a_\nu - \partial_{\nu}v^a_\mu +f^{abc}v_{\mu}^b v_{\nu}^c
\end{eqnarray}
and $f^{abc}$ are the group structure constants. The kinetic terms for the gauge fields and gauginos fields,
as well as the coupling of the gauginos to the gauge bosons are contained in (\ref{eqn:vectorkinetic}).  

The Lagrangian invariant under SUSY and gauge transformations can be written as
\begin{eqnarray}
\mathcal{L} & = & \sum_i \int d^2\theta d^2\bar \theta \Phi_i^+ e^{2gV} \Phi_i + 
\frac{1}{16g^2}\int d^2\theta \bigl[ W^a_\alpha W^{\alpha a} + h.c.\bigr]+\, \nonumber \\
&&\biggl[\int d^2\theta W(\Phi_i) + h.c.\biggr]\, .
\label{eqn:SUSYLagrangian}
\end{eqnarray}
The auxiliary fields can be integrated out with the help of their equations of motion,
 \begin{eqnarray}
\frac{\partial \mathcal{L}}{\partial F_j} =0 &\Rightarrow & F_j^{\ast} = - \frac{\partial W(\phi)}{\partial \phi_j}\, ,\label{eqn:auxiliarya} \\
\frac{\partial \mathcal{L}}{\partial D_a} =0 &\Rightarrow &D_a = -g \sum_{i}\phi_i^{\ast}T_a \phi_i\, .\label{eqn:auxiliaryb}  
\end{eqnarray}
Substituting these expressions for the F- and D-fields into the SUSY Lagrangian (\ref{eqn:SUSYLagrangian}),
the auxiliary fields in the Lagrangian disappear. The scalar potential can then be written as
 \begin{eqnarray}
V &= & V_F + V_D = \sum_i \biggl[ \frac{\partial W(\phi)}{\partial \phi_j}\biggr]^2 +
 \frac{g^2}{2} \biggl [\sum_{i}\phi_i^{\ast}T_a \phi_i\biggr]^2\, ,
\end{eqnarray}
where $V_F$ and $V_D$ are the F- and D-term of the scalar potential, respectively.
\subsection{Spontaneous Breaking of SUSY}\label{SUSYbreaking}
The superpartners have the same masses as the corresponding SM particles, but no superpartners have been discovered yet. 
Hence SUSY must be broken. From the relation of the SUSY algebra (\ref{eqn:SUSY1b}) one can derive 
\begin{eqnarray}
H &=& P^0 = \frac{1}{4}\left(\bar{Q}_1 Q_1 + Q_1 \bar{Q}_1 + \bar{Q}_2 Q_2 + Q_2\bar{Q}_2 \right)\geq 0\, ,
\end{eqnarray}
where $H$ is the Hamiltonian. Suppose the vacuum state $|0\rangle$ is supersymmetric, which means
\begin{eqnarray}
Q_{\alpha}|0\rangle &=& \bar{Q}_{\dot\alpha}|0\rangle=0\, , 
\end{eqnarray}
then the vacuum expectation value is
\begin{eqnarray}
E_{vac} &\equiv & \langle0|H|0\rangle =0\, . 
\end{eqnarray}
This implies that SUSY is spontaneously broken ($Q_{\alpha}|0\rangle \neq 0$) if the  vacuum expectation value 
is positive, i.e. $E_{vac} > 0$. 
It can be achieved if the scalar potential $V$ does not vanish for any field configuration. 
Therefore we can break SUSY spontaneously with $D_a \neq 0$ (D-term breaking) by the Fayet-Iliopoulos mechanism \cite{Fayet},
or $F_i \neq 0$ (F-term breaking) by the O'Raifeartaigh mechanism \cite{O'R}. Unfortunately, none of these mechanisms works
in the Minimal Supersymmetric Stand Model (MSSM). 
In the MSSM one introduces all the possible "soft-breaking" terms to the Lagrangian
instead of assuming an explicit SUSY-breaking mechanism~\cite{Girardello}.
After these terms are introduced the quadratic divergences in $\delta m_H^2$ are canceled. 
The details of the soft SUSY-breaking will be explained in Section (\ref{MSSMLagrangian}).

\section{The MSSM}\label{MSSMss}
\subsection{The Field Content of the MSSM}\label{fields}
The minimal supersymmetric extension of the SM, which is called the Minimal Supersymmetric Standard Model (MSSM)
\cite{MSSMa, MSSMb}, 
keeps as few interactions and particles as possible. 
It means that the MSSM is also based on the gauge group $SU(3)_C\times SU(2)_L\times U(1)_{Y}$.
It is not possible that one of the known SM particles is the superpartner of another one.
Hence there must be a new superpartner for each known particle of the SM \cite{MSSMb}.

The SM chiral fermions and their superpartners are described by chiral superfields.
The left-handed superfields can be arranged in $SU(2)$-doublets and the charged conjugates of the 
right-handed ones in $SU(2)$-singlets.
For every generation we have five chiral superfields: $\hat L$, $\hat E$, $\hat Q$, $\hat U$, and $\hat D$. 
$\hat L$ and $\hat Q$ are the $SU(2)$-doublet chiral superfields which contain lepton(slepton) and quark (squark) doublets,
respectively. 
$\hat E$ is the $SU(2)$-singlet chiral superfield containing lepton (slepton) singlets.
There is no right-handed neutrino in the SM, hence the corresponding superpartner does not exit.
The $SU(2)$-singlet chiral superfields $\hat U$ and $\hat D$ contain 
the up-quark (up-squark) and down-quark (down-squarks) singlets, respectively.   
In the SM gauge sector we introduce gluinos $\lambda_s^k ( k = 1, 2, \dots 8)$ as the superpartners of the gluons 
$G_\mu^k$, winos $\lambda^a ( a = 1, 2, 3)$ as the superpartners of the weak bosons $W_\mu^a$,
and a bino as the superpartner of the hypercharge boson $B_\mu$. They are described by the vector superfields 
$V^k_s$, $V^a $, and $v^\prime$, respectively.
\begin{table}[htb]
\begin{center}
\begin{tabular}{|l|l|l|ccc|}\hline
{\bf Superfield} &  {\bf \ \ \ \ \ \ \ \ Bosonic field} & {\bf \ Fermionic field} &
$SU_c(3)$& $SU_L(2)$ & $U_Y(1)$ \\ \hline \hline & \ \ \ \ \ \ \ \ \ sleptons & \ \ \ leptons & & & \\
\ \ \ \ \ $\hat L$ & \ \ \ \ \ \ \ 
$\tilde L = \left(\begin{array}{l}\tilde{\nu}_L \\ \tilde e_L\end{array}\right)$ & \ 
$L = \left(\begin{array}{l}\nu_L\\ e_L \end{array}\right)$ &
 1 & 2 & -1 \\ 
$\ \ \ \ \ \hat E$ & $\ \ \ \ \ \ \ \ \ \ \ \tilde e_R^\dagger$ & \ \ \ \ \ $e_R^c$ & 1 & 1 & 2\\ \hline
 & \ \ \ \ \ \ \ \ \ squarks & \ \ \ quarks & & & \\
$\ \ \ \ \ \hat Q$ & $\ \ \ \ \ \ \ \tilde Q = \left(\begin{array}{l}\tilde{u}_L\\ \tilde d_L \end{array}\right )$ & \ 
$Q = \left(\begin{array}{l} u_L \\ d_L \end{array}\right )$ &
$3$ & 2 & $1/3$ \\ 
$\ \ \ \ \ \hat U$ & $\ \ \ \ \ \ \ \ \ \ \ \tilde{u}_R^\dagger$ & $\ \ \ \ \ u_R^c$ & $3^\ast$ & 1 & $-4/3$ \\  
$\ \ \ \ \ \hat D$ & $\ \ \ \ \ \ \ \ \ \ \ \tilde{d}_R^\dagger$ & $\ \ \ \ \ d_R^c$ & $3^\ast$ & 1 & $2/3$ \\  
\hline $\ \ \ \ \ V_s^k$   & gluons $\ \ \ \ \ \ \ \ \ \ \ \ \ \ \ \ \ \ G_{\mu}^k$ &
gluinos $\ \ \ \lambda_s^k$ & 8 & 1 & 0 \\ \hline 
$\ \ \ \ \ V^a $ & weak bosons\ \ \ \ \ \ \ \ \ \ \ $W_{\mu}^a$ &
winos $\ \ \ \ \lambda^a$ & 1 & 3& 0 \\ \hline 
$\ \ \ \ \ v'$   & hypercharge boson \ \ $B_{\mu}$ & bino $\ \ \ \ \ \ \lambda'$ & 1 & 1& 0 \\
\hline 
& \ \ \ \ \ \ \ \ \ Higgses &\ \ higgsinos & & &\\
$\ \ \ \ \hat H_1$ & $\ \ \ \ \ \ H_1 = \left(\begin{array}{l} H_1^1 \\ H_1^2 \end{array}\right )$ & 
$\tilde H_1 = \left(\begin{array}{l} \tilde{H}_1^1 \\ \tilde{H}_1^2 \end{array}\right )$ &
1 &2 &-1\\
$\ \ \ \ \hat H_2$ & \ \ \ \ \ \ $H_2 = \left(\begin{array}{l} H_2^1 \\ H_2^2 \end{array}\right )$ & 
$\tilde H_2 = \left(\begin{array}{l} \tilde{H}_2^1 \\ \tilde{H}_2^2 \end{array}\right )$ &
1 &2 &1
\\ \hline
\end{tabular}
\end{center}
\caption{Field Content of the MSSM\label{particleMSSM}}
\end{table}

The Higgs scalar boson has spin $0$, hence we must describe it and its superpartner by a chiral superfield.
But it turns out that one chiral superfield is not enough. One reason is related to the chiral (triangle)
anomaly, which is proportional to $Tr\left[Y^3\right]$, where $Y$ denotes the weak hypercharge.
In the SM, the chiral anomaly is canceled by the known quarks and leptons.
However if one introduces only one chiral Higgs superfield, which contains chiral fermions 
with weak hypercharge $Y= 1$ or $Y= -1$, it leads to a nonvanishing contribution
to the chiral anomaly, hence spoils gauge invariance. One therefore has to add a second Higgs doublet
with opposite hypercharge. From Section \ref{construction} we know that the superpotential 
does not contain any products of left- and right-chiral superfields. 
Hence the Yukawa coupling term $HQu_R^c$ cannot be replaced by $\bar H Qu_R^c$ and 
the $HQd_R^c$ and $HLe_R^c$ terms cannot be replaced by $\bar H Qd_R^c$ and $\bar H Le_R^c$, respectively. 
Therefore it is impossible to give masses to both up and down 
quarks if we only have one Higgs doublet. This is another reason why we need two Higgs doublets. 
Starting from the SM, 
we get the MSSM by introducing a superpartner to each known particles and adding another Higgs doublet (with its superpartner).
The fields of the MSSM are summarized in Table \ref{particleMSSM} where the color and generation indices are suppressed.

\subsection{The MSSM Lagrangian}\label{MSSMLagrangian}
As discussed in Sections \ref{construction} and \ref{SUSYbreaking}, the MSSM Lagrangian can be written as
\begin{eqnarray}
\mathcal{L} & = \mathcal{L}_{SUSY} + \mathcal{L}_{soft}\, ,
\end{eqnarray}
where $\mathcal{L}_{SUSY}$ is the SUSY-invariant term and $\mathcal{L}_{soft}$ is the soft SUSY-breaking term.
All superfields of the MSSM were presented in Section~\ref{fields}. 
The coupling constants of the gauge groups $U(1)$, $SU(2)$ and $SU(3)$ are denoted by $g'$, $g$ and $g_s$, respectively.
We define the field strengths of these groups as 
\begin{eqnarray}
W_{\alpha}' & = & -\frac{1}{4}g'DD\bar{D}_{\bar\alpha}V'\, , \nonumber \\
W_{\alpha}^a & = &  - \frac{1}{4}\bar D\bar D\bigl[\exp(-2gV)D_\alpha \exp(2gV)\bigr]\, , \nonumber \\
W_{s\alpha}^k & = &  - \frac{1}{4}\bar D\bar D\bigl[\exp(-2g_sV_s)D_\alpha \exp(2g_sV_s)\bigr]\, .
\end{eqnarray}
Here $V' = Yv', V = T^a V^a, V_s = T^k V_s^k$, 
and $Y$, $T^a$ and $T^k$ are the generators of $U(1)$, $SU(2)$ and $SU(3)$,  
respectively. According to Section \ref{construction} the SUSY-invariant Lagrangian of the MSSM is written as 
\begin{eqnarray}
\mathcal{L}_{SUSY} & = & \int d^2\theta d^2\bar \theta \left (\bar{\hat L} e^{g'V' + 2gV}\hat L
+ \bar{\hat E}e^{g'V' + 2gV}\hat E + \right. \nonumber \\
&& {}\left. \bar{\hat Q}e^{g'V' + 2gV + 2g_sV_s}\hat Q + 
\bar{\hat U}e^{g'V' + 2gV + 2g_sV_s}\hat U+
 \bar{\hat D}e^{g'V' + 2gV + 2g_sV_s}\hat D + \right. \nonumber \\
&& {} \left. \bar{\hat H}_1 e^{g'V' + 2gV + 2g_sV_s}\hat H_1 + \bar{\hat H}_2 e^{g'V' + 2gV + 2g_sV_s}\hat H_2\right ) + \nonumber \\
&& {} \int d^2\theta\Bigl [\frac{1}{16{g'}^2}W_{\alpha}'{W^{\alpha}}' + \frac{1}{16{g}^2}W_{\alpha}^a{W^{\alpha}}^a
+  \frac{1}{16{g_s}^2}W_{s\alpha}^k{W_s^{\alpha}}^k + h.c.\Bigr] + \nonumber \\
&& {} \Biggl[\int d^2\theta \varepsilon_{ij}\left(\lambda_u \hat H_2^j \hat Q^i \hat U + \lambda_d \hat H_1^i \hat Q^j \hat D + 
\lambda_e \hat H_1^i \hat L^j \hat E
- \mu \hat H_1^i \hat H_2^j\right) + h.c. \Biggr], \label{eqn:MSSMSUSYLagrangian}
\end{eqnarray}
where we have suppressed the color and generation indices. 
The last line in (\ref{eqn:MSSMSUSYLagrangian}) is the F-component of the superpotential,
where $\varepsilon_{ij}$ is antisymmetric, i.e. $\varepsilon_{12}= -\varepsilon_{21} = 1$ and 
$\mu$ is the mass parameter mixing the two Higgs fields. The Yukawa coupling
constants $\lambda_u$, $\lambda_d$ and $\lambda_e$ are $3\times 3$ matrices in family space.

The superpotential in (\ref{eqn:MSSMSUSYLagrangian}) does not contain
some terms which are gauge invariant and analytic in the chiral superfields. These terms are
\begin{eqnarray}  
W' & = & \varepsilon_{ij}\left(\lambda \hat L^i \hat L^j \hat E + \lambda' \hat L^i \hat Q^j \hat D - \mu' \hat H_1^i \hat L^j\right)
+ \lambda''\hat D\hat D \hat U\, ,\label{eqn:Rviolation}
\end{eqnarray}
where $\lambda$, $\lambda'$, $\lambda''$, and $\mu'$ are coupling constants. 
The first and second term in \ref{eqn:Rviolation} violate the lepton ($L$) and baryon ($B$) numbers, respectively.
$B$- and $L$-violating processes have never been seen experimentally, hence these terms are not included in the MSSM potential.
In Section \ref{Rparity} we will analyze $R$-parity which is related to $B$ and $L$ number conservation.

In order to describe the MSSM completely, we need to specify the soft SUSY-breaking terms.
Girardello and Grisaru \cite{Girardello} found out that the allowed terms are the scalar mass terms, gaugino mass terms
and trilinear scalar interaction terms.  
The soft SUSY-breaking Lagrangian can be written as
\begin{eqnarray}
\mathcal{L}_{soft}& = & - M_{\tilde Q}^2\left(\tilde u_L^\dagger \tilde u_L + \tilde d_L^\dagger \tilde d_L\right ) 
- M_{\tilde u_R}^2 \tilde u_R^\dagger \tilde u_R -  M_{\tilde d_R}^2 \tilde d_R^\dagger \tilde d_R \nonumber \\
&& {}- M_{\tilde L}^2\left (\tilde{\nu}_L^\dagger \tilde{\nu}_L + \tilde e_L^\dagger  \tilde e_L\right )
- M_{\tilde e_R}^2\tilde e_R^\dagger \tilde e_R \nonumber \\
&& {} - m_1^2 H_1^\dagger H_1 -  m_2^2 H_2^\dagger H_2 + \left (m_3^2\varepsilon_{ij} H_1^i H_2^j + h.c.\right ) \nonumber \\
&& {} + \frac{1}{2}\left (M_1\lambda'\lambda' + M_2\lambda^a\lambda^a + M_3\lambda_s^k\lambda_s^k + h.c.\right )\nonumber \\
&& {} - \varepsilon_{ij}\left (\lambda_uA_uH_2^j\tilde Q^i\tilde u_R^\dagger + \lambda_dA_dH_1^i\tilde Q^j\tilde d_R^\dagger
+ \lambda_eA_eH_1^i\tilde L^j\tilde e_R^\dagger + h.c.\right )\, .\label{eqn:softLagrangian}
\end{eqnarray}
The first and second lines in (\ref{eqn:softLagrangian}) consist of slepton and squark mass terms. The parameters
$M_{\tilde Q}^2$, $M_{\tilde u_R}^2$, $M_{\tilde d_R}^2$, $M_{\tilde L}^2$ and $ M_{\tilde e_R}^2$ are $3\times 3$ matrices 
in family space. 
In the third line we have the soft SUSY-breaking contributions to the Higgs potential with the Higgs mass parameters
$m_1^2$, $m_2^2$ and $m_3^2$. The gaugino mass terms are in the fourth line with 
the bino, wino, and gluino mass parameters $M_1$, $M_2$, and $M_3$.
The last line in (\ref{eqn:softLagrangian}) are trilinear scalar interaction terms with parameters $A_u$, $A_d$, and $A_e$, which are   
$3\times 3$ matrices in family space. 
Similarly to the superpotential in (\ref{eqn:MSSMSUSYLagrangian}), 
there are no terms which violate the $B$ or $L$ numbers in the soft-breaking Lagrangian.

As discussed in Section \ref{superfields}, the superfields can be expanded in component form. 
The auxiliary fields (F- and D- fields) can be eliminated with the help of their equations of motion
which are similar to (\ref{eqn:auxiliarya}) and (\ref{eqn:auxiliaryb}). Explicit expressions can e.g. be found in Ref.\cite{MSSMLagrangian}. 

For the quantization of the classical Lagrangian and for higher-order calculations we need to specify the gauge.
A renormalizable 't Hooft gauge is chosen for this purpose. The principles can be found in Ref.\cite{Becchi, Faddeev}.
For the MSSM Lagrangian which contains the SUSY-invariant and soft SUSY-breaking terms,
the gauge fixing Lagrangian can be written as\cite{Tomas}
\begin{eqnarray}
\mathcal{L}_{fix} & = & -\frac{1}{2\xi_A}\left (\partial^\mu A_\mu\right )^2 - 
\frac{1}{2\xi_Z}\left (\partial^\mu Z_\mu + \xi_Zm_ZG^0\right )^2\nonumber \\
&& {}-  \frac{1}{\xi_W}\big |\partial^\mu W_\mu^+ + i \xi_Wm_WG^+\big |^2
- \frac{1}{2\xi_G}\left (\partial^\mu G_\mu^k\right )^2
\end{eqnarray}
with arbitrary parameters $\xi_A$, $\xi_Z$, $\xi_W$, and $\xi_G$. 
Here $A_\mu$, $Z_{\mu}$, and $W_\mu^\pm$ are the physical electroweak gauge fields,
$G^0$, $G^\pm$ are unphysical Higgs fields (Goldstone bosons). 
Since $\mathcal{L}_{fix}$ involves the unphysical components of the gauge fields,
one has to add the Fadeev-Popov ghost term $\mathcal{L}_{gh}$ for
compensating these effects \cite{Faddeev}.
The complete Lagrangian of the MSSM is
\begin{eqnarray}
\mathcal{L} & = & \mathcal{L}_{SUSY} + \mathcal{L}_{soft} + \mathcal{L}_{fix} +  \mathcal{L}_{gh}\, .
\end{eqnarray}
In the 't Hooft-Feynman gauge $\xi_A = \xi_Z= \xi_W= \xi_G = 1$. The propagators for
the unphysical Higgs fields then have poles at the masses of the corresponding physical
particles $W^\pm$ and $Z$.
\subsection{$R$-parity and Its Consequences}\label{Rparity}
In the SM the $B$ and $L$ numbers are conserved accidentally. 
In the MSSM, we can separate the most general gauge-invariant superpotential into
two parts, one which conserves $B$ and $L$ numbers and the another one which violates them. Since 
$B$ and $L$ violation is not observed in nature, these terms must be suppressed or excluded.
$B$ and $L$ number conservation together with spin conservation can be related to a new discrete symmetry, 
$R$-parity\cite{Rsymmetry} or equivalently matter-parity\cite{matterparity}. The $R$-parity is defined as
\begin{eqnarray}
P_R & = & (-1)^{3(B-L)+2s}\, ,
\end{eqnarray}
where $s$ is the spin of the particle.  Thus, all SM particles have $R$-parity $P_R = 1$, while
all superpartners have $R$-parity $P_R = -1$. Usually $R$-parity is assumed to be conserved in the MSSM,
hence the superpotential $W'$ which violates $R$-parity is forbidden.

If $R$-parity is assumed to be conserved, we have two important phenomenological consequences:
\begin{itemize}
\item The lightest supersymmetric particle (LSP) must be stable. This means that supersymmetric particles
other than 
the LSP must eventually decay into a state which contains an odd number of LSPs.
\item The supersymmetric particles can only be produced in even numbers in collider experiments.
\end{itemize}
Moreover, a stable LSP has to be electrically neutral and interacts only weakly with ordinary matter,
it can therefore make a good cold dark matter candidate \cite{neutralinodarkmatter}. 
\section{The Physical Fields of the MSSM}
In this section the electroweak-symmetry breaking is discussed briefly.
The Higgs fields in the MSSM have eight real scalar degrees of freedom. 
When the electroweak symmetry $SU(2)_L\times U(1)_Y$ is broken down to $U(1)_{EM}$,
three of them turn into Goldstone bosons which are subsequently absorbed by $W^\pm$ and $Z$. The remaining five degrees of freedom 
form the physical Higgs bosons $h^0, H^0, A^0$, and $H^\pm$~\cite{MSSMLagrangian, MSSMHiggsPotential}.    
After the electroweak symmetry is broken, fields with
the same $SU(3)_C\times U(1)_{EM}$ quantum numbers can mix.
Therefore we have sfermions mixing and higgsinos and gauginos mixing into charginos and neutralinos.
This will be studied in this section.    
\subsection{Higgs Bosons}
The scalar Higgs potential in the MSSM arises from the auxiliary F- and D-fields and the soft-SUSY breaking terms.
It can be written as
 \begin{eqnarray}
V & = & V_F + V_D + V_{soft}
\nonumber \\
& = & \left (m_1^2+|\mu|^2\right ) H_1^\dagger H_1 +  \left (m_2^2+|\mu|^2\right ) H_2^\dagger H_2 - m_3^2 \varepsilon_{ij}\left(H_1^i H_2^j + h.c.\right )+
\nonumber \\ 
&&{} \frac{1}{8}\left (g^2 + {g'}^2\right )\left ( H_1^\dagger H_1 -H_2^\dagger H_2\right )^2 + 
\frac{g^2}{2}| H_1^\dagger H_2 |^2 \, .
\label{eqn:MSSMHiggs1}
\end{eqnarray}
We can take advantage of the arbitrary nature of the soft SUSY-breaking parameters
$m_1^2, m_2^2$, and absorb $|\mu|^2$ into them, i.e. $m_1^2 + |\mu|^2\rightarrow m_1^2$,
 $m_2^2 + |\mu|^2\rightarrow m_2^2$.
Without losing generality, we can choose the vacuum expectation values of the Higgs fields as
\begin{eqnarray}
\langle H_1\rangle = \left(\begin{array}{c}
v_1\\
0 
\end{array}\right )\, ,
\ \ \ \
\langle H_2\rangle  = \left(\begin{array}{c}
0 \\
v_2
\end{array}\right)\, ,
\label{eqn:MSSMHiggsvev}
\end{eqnarray}
where $v_1, v_2$ are non-negative.
The electroweak symmetry $SU(2)_L\times U(1)_Y$ is broken down to $U(1)_{EM}$
if these conditions are satisfied,
\begin{eqnarray}
m_1^2 + m_2^2 - 2m_3^2&\ge& 0\, , \nonumber \\
m_1^2 m_2^2 - m_3^4 &<& 0\, .
\end{eqnarray}

The Higgs doublets $H_1$ and $H_2$ are decomposed in the following way:
\begin{eqnarray}
H_1 & = & \left(\begin{array}{c}
v_1 + \frac{1}{\sqrt{2}}(\phi_1^0 + i \chi_1^0 )\\
-\phi_1^{-} 
\end{array}\right )\, ,
\nonumber \\
H_2 & = & \left(\begin{array}{c}
\phi_2^+ \\
v_2 + \frac{1}{\sqrt{2}}(\phi_2^0 + i \chi_2^0 )
\end{array}\right)\, ,
\label{eqn:MSSMHiggs2}
\end{eqnarray}
where $\phi_1^0$ and $\phi_2^0$ denote the neutral CP-even Higgs fields,
$\chi_1^0$ and $\chi_2^0$ denote the neutral CP-odd Higgs fields,
and $\phi_1^\pm$ and $\phi_2^\pm$ denote the charged Higgs fields.
Inserting (\ref{eqn:MSSMHiggs2}) into the Higgs potential (\ref{eqn:MSSMHiggs1}),
one obtains
 \begin{eqnarray}
V & = & -T_{\phi_1^0}\phi_1^0 - T_{\phi_2^0}\phi_2^0 + \frac{1}{2}\left(\begin{array}{cc}
\phi_1^{0} & \phi_2^{0} 
\end{array}\right)M_{\phi^0}
\left(\begin{array}{c}
\phi_1^0 \\
\phi_2^0 
\end{array}\right) \nonumber \\
&& {} + \frac{1}{2}\left(\begin{array}{cc}
\chi_1^{0} & \chi_2^{0} 
\end{array}\right)M_{\chi^0}
\left(\begin{array}{c}
\chi_1^0 \\
\chi_2^0 
\end{array}\right)\, +  \frac{1}{2}\left(\begin{array}{cc}
\phi_1^{+} & \phi_2^{+} 
\end{array}\right)M_{\phi^\pm}
\left(\begin{array}{c}
\phi_1^- \\
\phi_2^- 
\end{array}\right)\, .
\label{eqn:MSSMHiggs3}
\end{eqnarray}
Here we only consider the linear and quadratic terms in the Higgs potential (\ref{eqn:MSSMHiggs1}).
$T_{\phi_1^0}$ and $T_{\phi_2^0}$ denote the tadpoles of the Higgs fields $\phi_1^0$ and $\phi_2^0$.
They must vanish since the Higgs potential should have a minimum,
\begin{eqnarray}
T_{\phi_1^0} & = & \frac{-4m_1^2 v_1 + 4m_3^2 v_2 - \left(g^2 + {g'}^2 \right )v_1 \left (v_1^2 - v_2^2\right )}{2\sqrt 2} = 0\, , \nonumber \\
T_{\phi_2^0} & = & \frac{4m_3^2 v_1 - 4m_2^2 v_2 + \left(g^2 + {g'}^2\right)v_2\left (v_1^2 - v_2^2 \right )}{2\sqrt 2}= 0\, .
\label{eqn:MSSMHiggsTad}
\end{eqnarray}
In (\ref{eqn:MSSMHiggs3}), $M_{\phi^0}$ is the mass matrix of the neutral CP-even Higgs fields,   
\begin{eqnarray}
M_{\phi^0}& = & 
\left(\begin{array}{cc}
m_1^2 + \frac{1}{4}\tilde g^2(3v_1^2 - v_2^2)& -m_3^2 - \frac{1}{2}\tilde g^2v_1 v_2 \\
- m_3^2 - \frac{1}{2}\tilde g^2v_1 v_2 &  m_2^2 + \frac{1}{4}\tilde g^2(3v_2^2 - v_1^2)
\end{array}\right)\, ,
\end{eqnarray}
$M_{\chi^0}$ is the mass matrix of the neutral CP-odd Higgs fields, 
\begin{eqnarray}
 M_{\chi^0}& = &\left(\begin{array}{cc}
m_1^2 + \frac{1}{4}\tilde g^2(v_1^2 - v_2^2)& - m_3^2 \\
- m_3^2 &  m_2^2 + \frac{1}{4}\tilde g^2(v_2^2 - v_1^2)
\end{array}\right)\, ,
\end{eqnarray}
and $M_{\phi^\pm}$ is the mass matrix of the charged Higgs fields,   
\begin{eqnarray}
M_{\phi^\pm}& = & 
\left(\begin{array}{cc}
m_1^2 + \frac{1}{4}\left (\tilde g^2v_1^2 + \bar g^2 v_2^2\right )& -m_3^2 - \frac{1}{2}g^2v_1 v_2 \\
- m_3^2 - \frac{1}{2}g^2v_1 v_2 & m_2^2 + \frac{1}{4}\left (\bar g^2v_1^2 + \tilde g^2 v_2^2\right )
\end{array}\right)\, ,
\end{eqnarray}
with $\tilde g ^2= g^2 + {g'}^2$, $\bar g ^2= g^2 - {g'}^2$.
These mass matrices can be diagonalized with the help of the unitary $2\times 2$ matrices $U_{\alpha}$, 
$U_{\beta_n}$, and $U_{\beta_c}$, respectively,
\begin{subequations}
\label{eqn:ReHiggs1}
\begin{eqnarray}
U_{\alpha}M_{\phi^0}U_{\alpha}^\dagger & = & M_{\phi^0}^D=\left(\begin{array}{cc}
m_{h^0}^2 & 0\\
 0 & m_{H^0}^2
\end{array} \right)\, , 
\hspace*{5mm}
U_{\alpha} = \left(\begin{array}{cc}
-\sin\alpha & \cos\alpha \\
\cos\alpha & \sin\alpha 
\end{array}\right )\, ,\label{eqn:ReHiggs1a}\\
U_{\beta_n}M_{\chi^0}U_{\beta_n}^\dagger & = &M_{\chi^0}^D= \left(\begin{array}{cc}
m_{A^0}^2 & 0\\
0 & 0
\end{array}\right)\, , 
\hspace{5mm} U_{\beta_n} = \left(\begin{array}{cc}
-\sin\beta_n & \cos\beta_n \\
\cos\beta_n & \sin\beta_n
\end{array}\right )\, , 
\label{eqn:ReHiggs1b} \\
U_{\beta_c}M_{\phi^\pm}U_{\beta_c}^\dagger & = &M_{\phi^\pm}^D= \left(\begin{array}{cc}
m_{H^\pm}^2 & 0\\
0 & 0
\end{array}\right)\, , 
\hspace{5mm} U_{\beta_c} = \left(\begin{array}{cc}
-\sin\beta_c & \cos\beta_c \\
\cos\beta_c & \sin\beta_c
\end{array}\right ).
\label{eqn:ReHiggs1c}
\end{eqnarray}
\end{subequations}
These transformations result in the Higgs boson mass eigenstates: two neutral CP-even Higgs bosons $h^0$ and $H^0$,
\begin{eqnarray}
\left(\begin{array}{c}
h^0 \\
H^0 
\end{array}\right)&=& U_{\alpha} \left(\begin{array}{c}
\phi_1^0 \\
\phi_2^0 
\end{array}\right)\, ,
\label{eqn:ReHiggs2}
\end{eqnarray}
the neutral CP-odd Higgs boson $A^0$ and Goldstone boson $G^0$, 
\begin{eqnarray}
\left(\begin{array}{c}
A^0 \\
G^0 
\end{array}\right)&=& U_{\beta_n} \left(\begin{array}{c}
\chi_1^0 \\
\chi_2^0 
\end{array}\right)\, ,
\label{eqn:ReHiggs3}
\end{eqnarray}
and the charged Higgs bosons
\begin{eqnarray}
\left(\begin{array}{c}
H^\pm \\
G^\pm 
\end{array}\right)&=& U_{\beta_c} \left(\begin{array}{c}
\phi_1^\pm \\
\phi_2^\pm 
\end{array}\right)\, .
\label{eqn:ReHiggscharged}
\end{eqnarray}
In terms of the Higgs boson mass eigenstates, the scalar Higgs potential(\ref{eqn:MSSMHiggs3}) can be written as
\begin{eqnarray}
V & = & -T_{h^0}h^0 - T_{H^0}H^0  + 
\frac{1}{2}\left(\begin{array}{cc}
h^{0} & H^{0} 
\end{array}\right)M_{\phi^0}^D
\left(\begin{array}{c}
h^0 \\
H^0 
\end{array}\right) + 
\\ \nonumber
&&{}\frac{1}{2}\left(\begin{array}{cc}
A^{0} & G^{0} 
\end{array}\right)M_{\chi^0}^D
\left(\begin{array}{c}
A^0 \\
G^0 
\end{array}\right) + \left(\begin{array}{cc}
H^{+} & G^{+} 
\end{array}\right)M_{\phi^\pm}^D
\left(\begin{array}{c}
H^- \\
G^- 
\end{array}\right)\, .
\label{eqn:ReHiggs4}
\end{eqnarray}
Here $T_{h^0}, T_{H^0}$ denote the tadpoles of the physical fields $h^0$, $H^0$, 
\begin{eqnarray}
\left(\begin{array}{c}
T_{h^0} \\
T_{H^0}
\end{array}\right) & = & U_\alpha \left(\begin{array}{c}
T_{\phi_1^0} \\
T_{\phi_2^0}
\end{array}\right) \, .
\end{eqnarray}

From (\ref{eqn:ReHiggs1b}) one gets,
 \begin{eqnarray}
m_{A^0}^2 & = & m_3^2\left (\cot\beta \cos^2\beta_n + \tan\beta \sin^2\beta_n + \sin 2\beta_n\right )\, ,
\label{eqn:MAmass}
\end{eqnarray}
where we have applied (\ref{eqn:MSSMHiggsTad}). 
Using the physical parameters $e$, $\theta_W (s_W = \sin\theta_W, c_W = \cos\theta_W)$,
$\tan\beta$, $T_{h^0}$, $T_{H^0}$, $m_Z$, $m_{A^0}$ instead of the parameters
$v_1, v_2, g^2, {g'}^2, m_1^2, m_2^2, m_3^3$ via (\ref{eqn:MAmass}) and the relations, 
\begin{eqnarray}
\tan\beta  = \frac{v_2}{v_1}, \hspace{2mm}s_W  = 1- m_W^2/m_Z^2, \hspace{2mm}e = g' c_W = g s_W,\hspace{2mm}
m_Z^2 = \frac{1}{2}(g_1^2 + g_2^2 )(v_1^2 + v_2^2)\, ,   
\end{eqnarray}
we obtain
\begin{subequations}
\label{eqn:ReHiggs5}
\begin{eqnarray}
\left(U_{\alpha}M_{\phi^0}U_{\alpha}^\dagger\right)_{11} & = & m_Z^2\sin^2(\alpha + \beta)+ m_{A^0}^2 \cos^2(\alpha - \beta)/\cos^2(\beta - \beta_n)
\nonumber \\
&&{}+ \frac{e}{2m_Z s_W c_W}T_{H^0} \cos(\alpha - \beta)\sin^2(\alpha - \beta_n)/\cos^2(\beta - \beta_n)
\nonumber \\
&&{}+ \frac{e}{2m_Z s_W c_W}T_{h^0} \frac{1}{2}\sin(\alpha - \beta_n)\left (\cos(2\alpha - \beta - \beta_n)\right.
\nonumber \\
&&{}\left.+ 3\cos(\beta - \beta_n)\right)/\cos^2(\beta - \beta_n)\, , \label{eqn:ReHiggs5a}
\\
\left(U_{\alpha}M_{\phi^0}U_{\alpha}^\dagger\right)_{12} & = & - m_Z^2\sin(\alpha + \beta)\cos(\alpha + \beta)+ m_{A^0}^2 \sin(\alpha - \beta)\cos(\alpha - \beta)/\cos^2(\beta - \beta_n)
\nonumber \\
&&{}+ \frac{e}{2m_Z s_W c_W}T_{H^0} \sin(\alpha - \beta)\sin^2(\alpha - \beta_n)/\cos^2(\beta - \beta_n)
\nonumber \\
&&{}- \frac{e}{2m_Z s_W c_W}T_{h^0}\cos(\alpha - \beta)\cos^2(\alpha -\beta_n)
/\cos^2(\beta - \beta_n)\, ,\label{eqn:ReHiggs5b}
\\
\left(U_{\alpha}M_{\phi^0}U_{\alpha}^\dagger\right)_{22} & = & m_Z^2\cos^2(\alpha + \beta)+ m_{A^0}^2 \sin^2(\alpha - \beta)/\cos^2(\beta - \beta_n)+
\nonumber \\
&&{}\frac{e}{2m_Z s_W c_W}\frac{T_{H^0}}{2}\cos(\alpha - \beta_n)\left (\cos(2\alpha - \beta - \beta_n)\, \right. \nonumber\\
&&{}\left. - 3\cos(\beta - \beta_n)\right)/\cos^2(\beta - \beta_n)
\nonumber \\
&&{}- \frac{e}{2m_Z s_W c_W}T_{h^0}\sin(\alpha - \beta)\cos^2(\alpha - \beta_n)/\cos^2(\beta - \beta_n)\, ,
\label{eqn:ReHiggs5c}
\end{eqnarray}
\begin{eqnarray}
\left(U_{\beta_n}M_{\chi^0}U_{\beta_n}^\dagger\right)_{12} & = & -m_{A^0}^2 \tan(\beta - \beta_n)
-\frac{e}{2m_Z s_W c_W}\left (T_{H^0} \sin(\alpha - \beta_n) + \, \right. \nonumber
\\
&&{}\left. T_{h^0} \cos(\alpha - \beta_n)\right )/\cos(\beta - \beta_n)\, ,\label{eqn:ReHiggs6a}
\\
\left(U_{\beta_n}M_{\chi^0}U_{\beta_n}^\dagger\right)_{22}& = & m_{A^0}^2 \tan^2(\beta - \beta_n) + \,  \nonumber
\\
&&{}\frac{e}{2m_Z s_W c_W}\left (- T_{H^0} \cos(\alpha + \beta - 2\beta_n) \, \right.  \nonumber \\
&&{}\left. + T_{h^0} \sin(\alpha + \beta - 2\beta_n)\right )/\cos^2(\beta - \beta_n))\, ,
\label{eqn:ReHiggs6b}
\end{eqnarray}
\begin{eqnarray}
\left(U_{\beta_c}M_{\phi^\pm}U_{\beta_c}^\dagger\right)_{11} & = & m_{A^0}^2 + m_W^2\, ,\nonumber\\
\left(U_{\beta_c}M_{\phi^\pm}U_{\beta_c}^\dagger\right)_{12} & = & -\left ( m_{A^0}^2 + m_W^2\right )\tan(\beta - \beta_c) \,  \nonumber \\
&&{}- \frac{e}{2m_Z s_W c_W}\left (T_{H^0}\sin(\alpha - \beta_c) +T_{h^0}\cos(\alpha - \beta_c)\right ) /\cos(\beta - \beta_c)\, , \nonumber\\
\left(U_{\beta_c}M_{\phi^\pm}U_{\beta_c}^\dagger\right)_{22} & = & \left ( m_{A^0}^2 + m_W^2\right )\tan^2(\beta - \beta_c)\,  \nonumber \\
&&{}- \frac{e}{2m_Z s_W c_W} T_{h^0}\cos(\alpha + \beta - 2\beta_c) /\cos^2(\beta - \beta_c)\,  \nonumber \\
&& {} + \frac{e}{2m_Z s_W c_W} T_{H^0}\sin(\alpha + \beta - 2\beta_c) /\cos^2(\beta - \beta_c)\, . 
\label{eqn:ChargeHiggsmass}
\end{eqnarray}
\end{subequations}
The matrices $U_{\alpha}M_{\phi^0}U_{\alpha}^\dagger$, $U_{\beta_n}M_{\chi^0}U_{\beta_n}^\dagger$ and 
$U_{\beta_c}M_{\phi^\pm}U_{\beta_c}^\dagger$ should be diagonal at tree level. This leads
to the following conclusions, 
\begin{eqnarray}
\beta_n = \beta_c  = \beta, \hspace{5mm}
\tan2\alpha = \tan2\beta \frac{m_A^2 + m_Z^2}{m_A^2 - m_Z^2}, \hspace{5mm}-\frac{\pi}{2}<\alpha <\frac{\pi}{2}\, .
\end{eqnarray}
Therefore, the tree-level masses of the physical Higgs bosons are
\begin{subequations}
\begin{eqnarray}
m_{h^0}^2 & = & \frac{1}{2}\left (m_{A^0}^2 + m_Z^2 - \sqrt{(m_{A^0}^2 + m_Z^2)^2 - 4m_{A^0}^2 m_Z^2 \cos^2 2\beta }\right )\, ,
\\
m_{H^0}^2 & = & \frac{1}{2}\left (m_{A^0}^2 + m_Z^2 + \sqrt{(m_{A^0}^2 + m_Z^2)^2 - 4m_{A^0}^2 m_Z^2 \cos^2 2\beta }\right )\, ,
\\
m_{H^{\pm}}^2& =& m_{A^0}^2 + m_W^2\, .
\end{eqnarray}
\end{subequations}
In the 't Hooft-Feynman gauge, the masses of the Goldstone bosons are 
\begin{equation}
m_{G^0} = m_Z\, ,\ \  m_{G^\pm} = m_W.  
\end{equation}
\newpage
\subsection{$W$ and $Z$ Gauge Bosons}
In analogy to the SM, the physical gauge bosons $W_\mu^\pm$, $Z_\mu$ and $A_\mu$
are obtained from the electroweak interaction eigenstates $W_\mu^a (a = 1, 2, 3)$ and $B_\mu$ via
the following definition,
\begin{eqnarray}
W_\mu^\pm & = & \frac{1}{\sqrt 2}\left (W_\mu^1 \mp W_\mu^2\right )\, ,\nonumber \\
Z_\mu & = & -s_W B_\mu + c_W W_\mu^3\, , \nonumber \\
A_\mu & = & c_W B_\mu + s_W W_\mu^3\, ,
\end{eqnarray}
and correspondingly their superpartners transform as
\begin{eqnarray}
\lambda^\pm & = & \frac{1}{\sqrt 2} \left(\lambda^1 \mp \lambda^2\right )\, , \nonumber \\
\lambda_Z & = & -s_W \lambda' + c_W \lambda^3 \, , \nonumber \\
\lambda_A & = & c_W \lambda' + s_W \lambda^3 \, .\label{eqn:gauginomixing}
\end{eqnarray}
The $SU(2)\times U(1)$ covariant derivative is
\begin{eqnarray}
D_\mu & = & \partial_\mu + i gT^aW_\mu^a + i g'\frac{Y}{2}B_\mu \nonumber \\
& = & \partial_\mu + \frac{i g}{\sqrt 2}T^+W_\mu^+ + \frac{i g}{\sqrt 2}T^-W_\mu^-
+ \, \nonumber \\
&& {}\frac{i g}{c_W}\bigl[T^3 - Qs_W^2\bigr]Z_\mu + i eQA_\mu\, . 
\end{eqnarray}
Here we introduced the new operators
 \begin{eqnarray}
T^\pm &= &T^1 \pm i T^2\, ,\nonumber \\
Q &= &T^3 + \frac{Y}{2}\, , 
\end{eqnarray}
where $Q$ is the charge operator.
Hence the kinetic terms for the Higgs fields are
\begin{eqnarray}
\mathcal{L}_{Higgs kinetic} & = & \left(D^\mu H_1\right )^\dagger \left (D_\mu H_1\right )
+ \left(D^\mu H_2\right )^\dagger \left (D_\mu H_2\right )\, .
\label{eqn:Higgskinetic}
 \end{eqnarray}
After gauge symmetry breaking the Higgs fields acquire their vacuum expectation values as in (\ref{eqn:MSSMHiggsvev}).
From (\ref{eqn:Higgskinetic}) we can obtain the masses of the gauge bosons,
\begin{eqnarray}
m_W^2 &= &\frac{g^2}{2}\left (v_1^2 + v_2^2\right )\, , \nonumber \\
m_Z^2 &= &\frac{g^2}{2c_W^2}\left (v_1^2 + v_2^2\right ) =  
\frac{1}{2}\left (g^2 + {g'}^2\right )\left (v_1^2 + v_2^2\right )\, ,
\end{eqnarray}
and the masses of the gluons and photons are zero.
This is consistent with the predictions of the SM.
\subsection{Fermions}
Considering the Yukawa coupling terms in the superpotential in the SUSY Lagrangian (\ref{eqn:MSSMSUSYLagrangian}),
\begin{eqnarray}
\mathcal{L}_{Yukawa} & = & - \varepsilon_{ij}\left(\lambda_u H_2^j Q^i u_R^c + \lambda_d H_1^i Q^j d_R^c + 
\lambda_e H_1^i L^j e_R^c\right ) + h.c.\, , 
\end{eqnarray}
the fermion mass terms can be obtained when the Higgs fields get their vacuum expectation values,
\begin{eqnarray}
\mathcal{L}_{f-mass} & = & - \lambda_u v_2 u_L u_R^c - \lambda_d v_1 d_L d_R^c - 
\lambda_e v_1 e_L e_R^c + h.c.\, .
\end{eqnarray}
Introducing four-component Dirac spinors,
\begin{eqnarray}
u = \left(\begin{array}{c}
u_L\\
\bar{u}_R^c
\end{array}\right )\, , \ \ \ d = \left(\begin{array}{c}
d_L\\
\bar{d}_R^c
\end{array}\right )\, , \ \ \ e = \left(\begin{array}{c}
e_L\\
\bar{e}_R^c
\end{array}\right )\, ,
\label{eqn:fermion-Dirac-spinors}
\end{eqnarray}
the fermion mass terms can be written as
\begin{eqnarray}
\mathcal{L}_{f-mass} & = & - \lambda_u v_2 u\bar u - \lambda_d v_1 d \bar d - 
\lambda_e v_1 e \bar e\, . 
\end{eqnarray}
Hence the fermion masses are 
\begin{eqnarray}
m_u = \lambda_u v_2\, , \ \ \ m_d = \lambda_d v_1\, , \ \ \  m_e = \lambda_e v_1\, , 
\end{eqnarray}
and the Yukawa coupling constants can be written as
\begin{eqnarray}
\lambda_u = \frac{m_u}{v_2} = \frac{m_u g}{\sqrt 2 \sin\beta m_W}\, , \ \ 
\lambda_d = \frac{m_d}{v_1} = \frac{m_d g}{\sqrt 2 \cos\beta m_W}\, , \ \ 
\lambda_e = \frac{m_e}{v_1} = \frac{m_e g}{\sqrt 2 \cos\beta m_W}\, . 
\label{eqn:Yukawaconstants}
\end{eqnarray}
Here we have used $\tan\beta = \frac{v_2}{v_1}$ and $m_W^2 = \frac{g^2}{2}\left (v_1^2 + v_2^2\right )$. 
\subsection{Sfermions}
The mass terms of the sfermions arise from the soft SUSY-breaking Lagrangian, the auxiliary F- and D-fields.
When the electroweak symmetry $SU(2)_L\times U(1)_Y$ is broken down to $U(1)_{EM}$, the Higgs fields get their vacuum expectation values,
the sfermion mass terms can be written as
\begin{eqnarray}
\mathcal{L}_{soft}^{\tilde f}& = & - M_{\tilde Q}^2\left(\tilde u_L^\dagger \tilde u_L + \tilde d_L^\dagger \tilde d_L\right ) 
- M_{\tilde u_R}^2 \tilde u_R^\dagger \tilde u_R -  M_{\tilde d_R}^2 \tilde d_R^\dagger \tilde d_R \nonumber \\
&& {}- M_{\tilde L}^2\left (\tilde{\nu}_L^\dagger \tilde{\nu}_L + \tilde e_L^\dagger  \tilde e_L\right )
- M_{\tilde e_R}^2\tilde e_R^\dagger \tilde e_R \nonumber \\
&& {} - \left (\lambda_uA_uv_2\tilde u_L\tilde u_R^\dagger + \lambda_dA_dv_1\tilde d_L\tilde d_R^\dagger
+ \lambda_eA_ev_1\tilde e_L\tilde e_R^\dagger + h.c.\right )\, ,
\end{eqnarray}
\begin{eqnarray}
\mathcal{L}_{Aux-F}^{\tilde f} & = & - \lambda_e^2 v_1^2 \left (\tilde e_L^\dagger \tilde e_L + \tilde e_R^\dagger \tilde e_R\right )
-  \lambda_d^2 v_1^2 \left (\tilde d_L^\dagger \tilde d_L + \tilde d_R^\dagger \tilde d_R\right ) \nonumber \\
&& {}- \lambda_u^2 v_2^2 \left (\tilde u_L^\dagger \tilde u_L + \tilde u_R^\dagger \tilde u_R\right )\, \nonumber \\
&& {}+ \left ( \lambda_e v_2 \mu \tilde e_R^\dagger \tilde e_L +  \lambda_d v_2 \mu \tilde d_R^\dagger \tilde d_L
+ \lambda_u v_1 \mu \tilde u_R^\dagger \tilde u_L + h.c.\right )\, , 
\end{eqnarray}
\begin{eqnarray}
\mathcal{L}_{Aux-D}^{\tilde f} & = & \frac{v_1^2 - v_2^2}{2}\Biggl [ g^2\left (\tilde \nu_L^\dagger T_{\nu_L}^3 \tilde \nu_L +
\tilde e_L^\dagger T_{e_L}^3 \tilde e_L + 
\tilde u_L T_{u_L}^3 \tilde u_L + \tilde d_L T_{d_L}^3 \tilde d_L
\right ) - \Biggr. \, \nonumber \\
&& {} \Biggl. {g'}^2\left (\tilde \nu_L^\dagger \frac{Y_{\nu_L}}{2} \tilde \nu_L +
\tilde e_L^\dagger \frac{Y_{e_L}}{2} \tilde e_L +   \tilde e_R^\dagger \frac{Y_{e_R}}{2}\tilde e_R +
\right. \Biggr. \, \nonumber \\
&& {} \left. \Biggl. 
\tilde u_L^\dagger \frac{Y_{u_L}}{2}\tilde u_L 
+ \tilde d_L^\dagger \frac{Y_{d_L}}{2}\tilde d_l + \tilde u_R^\dagger \frac{Y_{u_R}}{2}\tilde u_R + 
\tilde d_R^\dagger \frac{Y_{d_R}}{2}\tilde d_R \right )\Biggr]\nonumber \\
 & = & - m_Z^2\cos 2\beta \sum_f \Biggl[ \tilde f_L^\dagger (T_{f_L}^3 - Q_{f_L} s_W^2)\tilde f_L 
+ \tilde f_R^\dagger Q_{f_R} s_W^2 \tilde f_R \Biggr ] 
\end{eqnarray}
Here we have applied $\frac{g^2(v_1^2 - v_2^2)}{2c_W^2} = m_Z^2 \cos 2\beta $. $T^3$, $Y$ and $Q$ are 
the weak isospin, weak hypercharge and electric charge of the fermions, respectively. 
They satisfy the relation $Q = T^3 + \frac{Y}{2}$. 
For the right-handed fermions $T_{f_R}^3 = 0$, so $Q_{f_R} = \frac{Y_{f_R}}{2}$.

Adding together the Lagrangian above, the sfermion mass terms of the Lagrangian can be written as 
\begin{align}
{\mathcal{L}}_{\tilde f-mass} &=-\tilde \nu_L^\dagger M_{\tilde \nu}\tilde \nu_L - \begin{pmatrix}
{{\tilde{f}}_{L}}^\dagger, {{\tilde{f}}_{R}}^{\dagger} \end{pmatrix}
\mathcal{M}_{\tilde{f}}\begin{pmatrix}{\tilde{f}}_{L}\\{\tilde{f}}_{R}
\end{pmatrix} \, .
\end{align} 
After replacing the Yukawa coupling constants $\lambda_e, \lambda_u$, and $\lambda_d$ with
the expressions in (\ref{eqn:Yukawaconstants}), we get the sfermion mass matrix $\mathcal{M}_{\tilde{f}}$,   
\begin{align}\label{Sfermionmassenmatrix}
\mathcal{M}_{\tilde{f}} = \begin{pmatrix}  m_f^2 + M_{\tilde f_L}^2 + m_Z^2 \cos2\beta (T_f^3 - Q_{{f}} s_W^2) & 
m_f (A_f - \mu \kappa) \\
  m_f (A_f - {\mu}\kappa) &   m_f^2 + M_{\tilde{f}_R}^2 + m_Z^2 \cos2 \beta Q_{{f}} s_W^2 \end{pmatrix},
\end{align}
where $M_{\tilde e_L}^2 = M_{\tilde L}^2$,  $M_{\tilde q_L}^2 = M_{\tilde Q}^2$,
and the parameter $\kappa$ is defined as $\kappa = \cot \beta$ for up-type
squarks and $\kappa = \tan \beta$ for down-type squarks and sleptons.
The sneutrino mass matrix $M_{\tilde{\nu}}$ for a given flavor is 1-dimensional, and only has the left-handed entry of 
(\ref{Sfermionmassenmatrix}), 
\begin{eqnarray}
M_{\tilde{\nu}} = M_{\tilde L}^2 + \frac{1}{2}m_Z^2 \cos 2\beta\, .
\label{eqn:ReSle13}
\end{eqnarray}

The sfermion mass matrix(\ref{Sfermionmassenmatrix}) can be diagonalized by
a unitary $2\times 2$ matrix $U_{\tilde{f}}$,
\begin{equation}\label{Sfermiondigonal}
U_{\tilde{f}}M_{\tilde{f}}U_{\tilde{f}}^\dagger = M_{\tilde{f}}^D=
\left ( \begin{array}{cc}
m_{\tilde{f_1}}^2 & 0\\
0 & m_{\tilde{f_2}}^2
\end{array}
\right ), 
\end{equation}
with the mass eigenvalues $m_{\tilde{f_1}}^2, m_{\tilde{f_2}}^2$,
\begin{align}
\begin{split}\label{Sfermioneigenmass}
&m_{{\tilde{f}}_{1,2}}^2 = \frac{1}{2}(M_L^2
+M_{{\tilde{f}}_R}^2) +  m_f^2 + \frac{1}{2}T_f^3 m_Z^2 \cos2 \beta \\&
\qquad\ \quad \mp
\frac{1}{2} \sqrt{\bigl[M_{\tilde{f}_L}^2 - M_{{\tilde{f}}_R}^2 + m_Z^2 \cos2\beta
 (T_f^3 - 2 Q_{{f}} s_W^2)\bigr]^2 + 4  m_f^2 (A_f - {\mu} \kappa)^2}.
\end{split}
\end{align}
The sfermion mass eigenstates are given by
 \begin{equation}
 \left(\begin{array}{c}
\tilde{f}_1\\
\tilde{f}_2
\end{array}\right)= U_{\tilde{f}}\left(\begin{array}{c}
\tilde{f}_L\\
\tilde{f}_R
\end{array}\right),
\end{equation}
where the matrix $U_{\tilde{f}}$ is determined by (\ref{Sfermiondigonal}).
\subsection{Charginos and Neutralinos}
The mass terms of the higgsinos and gauginos arise from the SUSY Lagrangian 
\begin{eqnarray}
- \sqrt 2\Bigl[ i \bar{\tilde H}_1 \left (g\bar\lambda + \frac{1}{2}g'\bar\lambda' \right )H_1
+ i \bar{\tilde H}_2 \left (g\bar\lambda + \frac{1}{2}g'\bar\lambda' \right )H_2 + h.c.\Bigr ]\, , 
\label{eqn: higgsinoandgauginoa}
\end{eqnarray}
the superpotential 
\begin{eqnarray}
\varepsilon_{ij}\left (\mu \tilde H_1^i \tilde H_2^j + h.c.\right )\, ,\label{eqn: higgsinoandgauginob}
\end{eqnarray}
and the soft-breaking terms
\begin{eqnarray}
\frac{1}{2}\left (M_1 \lambda' \lambda' + M_2 \lambda^a \lambda^a+ h.c.\right )\, .\label{eqn: higgsinoandgauginoc}
\end{eqnarray}
The charginos are a mixture of the charged higgsinos and gauginos.
We introduce 2$\times 1$ matrices 
\begin{equation}
\psi^L = \left(\begin{array}{c}
-i \lambda^+\\
 \tilde H_2^1
\end{array}\right)\, , \hspace*{5mm}
\psi^R = \left(\begin{array}{c}
-i \lambda^-\\
 \tilde H_1^2
\end{array}\right)
\label{eqn:charginonotation}
\end{equation}
for the charged higgsinos and gauginos,
where $\lambda_\pm $ is defined via (\ref{eqn:gauginomixing}).
Apply (\ref{eqn:charginonotation}) to the mass terms of the higgsinos and gauginos Lagrangian 
(\ref{eqn: higgsinoandgauginoa}, \ref{eqn: higgsinoandgauginob}, \ref{eqn: higgsinoandgauginoc}),
one obtains the mass term of the charginos after the Higgs fields acquire their vacuum expectation values,
\begin{eqnarray}
\mathcal{L}_{\chi^c-mass}& = & -
\bigl[{\psi^R}^\top X \,\psi^L + 
{\overline{\psi}^L}^\top X^\dagger\,{\overline{\psi}^R}
\bigr] ,
\label{eqn:ReCha1}
\end{eqnarray}
with the mass matrix
\begin{eqnarray}
X & = & \left(\begin{array}{cc}
M_2 & \sqrt{2}\,m_W\,\sin\beta\\
\sqrt{2}\,m_W\,\cos\beta & \mu
\end{array}\right)\, . 
\label{eqn:ReCha2}
\end{eqnarray}
The mass matrix (\ref{eqn:ReCha2}) can be diagonalized by two unitary $2\times 2$ matrices $U$ and $V$,
\begin{eqnarray}
U\,X\,V^\top &=& M_{\tilde{\chi}^+}^D=
\left(\begin{array}{cc}
m_{\tilde{\chi}_1^+} & 0 \\
0 & m_{\tilde{\chi}_2^+}
\end{array}\right),
\label{eqn:ReCha3}
\end{eqnarray}
with the eigenvalues $m_{\tilde{\chi}_1^+}^2, m_{\tilde{\chi}_2^+}^2$,
\begin{eqnarray}
m_{\tilde{\chi}_1^+}^2, m_{\tilde{\chi}_2^+}^2 & = & \frac{1}{2}\, \bigg\{ 
       M_2^2 + \mu^2 + 2m_W^2 \mp \Big[ (M_2^2-\mu^2)^2 \nonumber \\ 
&&{}+ 4m_W^4\cos^2 2\beta
       + 4m_W^2(M_2^2+\mu^2+2\,\mu\, M_2\, \sin 2\beta) 
       \Big]^{\frac{1}{2}} \bigg\}~.
\label{eqn:Charmass}
\end{eqnarray}
Hence the chargino mass eigenstates can be written as
\begin{eqnarray}
\chi^R = U\,\psi^R , \quad
\chi^L = V\,\psi^L \, ,
\label{eqn:ReCha19}
\end{eqnarray}
with the unitary matrices $U$ and $V$ which are determined by (\ref{eqn:ReCha3}). 
The four-component chargino spinors are defined as 
\begin{eqnarray}
\tilde{\chi}_i^+ = \left(\begin{array}{c}
\chi^L_i \\ \overline{\chi}^R_i
\end{array}\right), i = 1,\hspace*{2mm}2\, .
\label{eqn:ReCha4}
\end{eqnarray}

Similarly to the chargino case, the neutralinos arise from mixing of the neutral higgsinos and 
gauginos.
In the following we introduce a $1\times 4$ matrix 
\begin{equation}
{\psi^0}^\top = \left(\begin{array}{cccc}
-i \lambda_A\, , & -i \lambda_Z\, , & \tilde H_1^1\, , & \tilde H_2^2
\end{array}\right),
\end{equation}
to the Lagrangian of the higgsinos and gauginos (\ref{eqn: higgsinoandgauginoa}, \ref{eqn: higgsinoandgauginob}, 
\ref{eqn: higgsinoandgauginoc}), where $\lambda_A, \lambda_Z$ are defined in (\ref{eqn:gauginomixing}).
After the electroweak symmetry is broken, one obtains the mass term of the neutralino Lagrangian 
\begin{eqnarray}
\mathcal{L}_{\chi^0-mass} & = 
-\frac{1}{2}
\bigl[{\psi^0}^\top Y \,\psi^0 + 
{\overline{\psi}^0}^\top Y^\dagger\,{\overline{\psi}^0}
\bigr] 
\label{eqn:ReNeu1}
\end{eqnarray}
with the mass matrix
\begin{eqnarray}
Y = \left(\begin{array}{cccc}
M_1 & 0 & -m_Z\,s_W\,\cos\beta & m_Z\,s_W\,\sin\beta \\
0 & M_2 &  m_Z\,c_W\,\cos\beta & -m_Z\,c_W\,\sin\beta \\
-m_Z\,s_W\,\cos\beta & m_Z\,c_W\,\cos\beta & 0 & -\mu \\
m_Z\,s_W\,\sin\beta & -m_Z\,c_W\,\sin\beta & -\mu & 0
\end{array}\right) \, .
\label{eqn:ReNeu2}
\end{eqnarray}
In order to diagonalize the mass matrix and get the neutralino mass eigenstates, the
following transformations must be performed:
\begin{eqnarray}
N^\ast Y N^\dagger & = & M_{\tilde{\chi}^0}^D = \left(\begin{array}{cccc}
m_{\tilde{\chi}^0_{1}} & 0 & 0 & 0 \\ 
0 & m_{\tilde{\chi}^0_{2}} & 0 & 0 \\ 
0 & 0 & m_{\tilde{\chi}^0_{3}} & 0 \\ 
0 & 0 & 0 & m_{\tilde{\chi}^0_{4}} \\ 
\end{array}\right) \, ,
\label{eqn:ReNeu3}
\end{eqnarray}
\begin{eqnarray}
\chi^0 = N\,\psi^0\, ,
\label{eqn:ReNeu4}
\end{eqnarray}
where $N$ is a unitary $4 \times 4$ matrix and $\chi^0$ are the neutralino mass eigenstates.
One of the  eigenvalues in (\ref{eqn:ReNeu3}) might be negative if the matrix $N$ is real. Therefore,
this matrix should be complex for positive neutralino masses even if all the elements in $Y$ are real.
The neutralinos are Majorana fermions, their Majorana spinors are defined by
\begin{eqnarray}
\tilde{\chi}^0_{i}& = &\left(\begin{array}{c}
\chi^0_i \\ \overline{\chi}^0_i
\end{array}\right), \hspace*{3mm}i=1,\ldots,4 \, .
\label{eqn:ReNeu5}
 \end{eqnarray}
We have four neutralinos $\tilde{\chi}^0_{1}, \tilde{\chi}^0_{2}, \tilde{\chi}^0_{3}, \tilde{\chi}^0_{4}$.
They are labeled in ascending order, $m_{\tilde{\chi}^0_{1}}< m_{\tilde{\chi}^0_{2}}< m_{\tilde{\chi}^0_{3}}<
m_{\tilde{\chi}^0_{4}}$. The lightest neutralino $\tilde{\chi}^0_{1}$ is the only MSSM particle which can make a good 
cold dark matter candidate.
\subsection{Gluinos}
The gluinos $\lambda_s^k$ are the superpartners to the gluons $G_\mu^k$.  
Their mass terms arise from the soft-breaking terms of the Lagrangian,
\begin{eqnarray}
\mathcal{L}& = \frac{1}{2}\left ( M_3 \lambda_s^k \lambda_s^k  + h.c.\right )\, . 
\end{eqnarray}
Since the gluinos are Majorana fermions, their Majorana spinors are 
defined by
\begin{eqnarray}
\tilde g_s^k & = & \left (\begin{array}{c}
\lambda_s^k\\
\bar \lambda_s^k 
\end{array}\right )
\end{eqnarray}
with the Majorano mass $M_3$.
\chapter{Renormalization of the MSSM}\label{ReMSSM}
In order to calculate higher-order corrections, one must renormalize the
parameters and fields of the MSSM. Several approaches for the
renormalization of the MSSM have been developed \cite{ReNeu, ReSle,
MSSMRE, ReHiggs, RetanBOS, RetanBMS}. Here we follow the strategy of Refs.\cite{ReNeu, ReSle},
i.e. we employ on-shell renormalization. 
We assume here that all relevant parameters are real quantities. This
amounts to the assumption that the soft supersymmetry breaking terms conserve
CP. In this chapter the basic ideas of regularization and renormalization are presented briefly. 
The $\overline{\rm DR}$ and on-shell renormalization scheme are introduced and explicit 
on-shell renormalization conditions for the different fields are formulated.
The renormalization of the SM sector, the chargino and neutralino sector, the sfermion sector, and 
the Higgs sector are discussed in detail.  
\section{Principles of Regularization and Renormalization}\label{REprinci}
At tree level, the parameters of the Lagrangian are directly related to 
the physical quantities. In higher-order perturbation theory, these direct relations are destroyed. 
Furthermore, the divergent loop integrals make the calculation ambiguous.
The theory has to be \mbox{regularized} for mathematical consistence.
One usually employs dimensional regularization~\cite{DR} for higher-order calculations of the SM, where 
the integrals are calculated in D dimensions.
Dimensional regularization preserves Lorentz and gauge invariance of the theory, but 
breaks chiral symmetry and SUSY. In supersymmetric theories, 
if we treat the vector fields in D dimensions, this will lead to a mismatch 
between the fermionic and bosonic degrees of freedom, hence SUSY is broken.
In order to avoid the disadvantage of \mbox {dimensional} \mbox {regularization}, 
dimensional reduction was developed in Ref.\cite{DR-Siegel}, where only the momenta are calculated in 
D dimensions, while the fields and the Dirac algebra are kept 4-dimensional.
It is commonly used in supersymmetric theories.

After regularization the parameters in the original Lagrangian (the so-called bare \mbox{parameters}), 
which are physically meaningless, are different from the corresponding physical quantities by UV-divergent 
contributions. These divergences cancel in relations between physical quantities.
Therefore, one may replace the bare parameters by the renormalized ones and the 
renormalization constants (counterterms) in the Lagrangian,
\begin{eqnarray}
g_0  &\longrightarrow g + \delta g\, ,
\end{eqnarray}
where the renormalized parameters $g$ are UV finite and measurable, while the counterterms $\delta g$
are UV divergent and absorb the divergent parts of the loop integrals. 
They are fixed by renormalization conditions, 
which determine the relation between the renormalized parameters and 
the physical quantities. One must choose a set of independent parameters in order to make predictions from the theory.

Parameter renormalization is sufficient to get finite S-matrix elements when vave function renormalization for external
on-shell particles is included, but the off-shell Green functions are not finite.
In order to obtain finite propagators and vertices, 
we must renormalize the field by the multiplicative renormalization
\begin{eqnarray}
\psi_0  &\longrightarrow \sqrt{\delta Z_{\psi}}\psi\overset{\textrm{one-loop}}{=} \psi + \frac{1}{2}\delta Z_{\psi}\psi\, .
\label{eqn:OSre3}
\end{eqnarray}

After the parameters and fields are renormalized,  the Lagrangian can be written as 
\begin{eqnarray}
\mathcal{L}_0(g_0, \psi_0)& = &\mathcal{L}(g, \psi) + \delta\mathcal{L}(g, \delta g, \psi, \delta \psi)\, ,
\end{eqnarray}
where the renormalized Lagrangian $\mathcal{L}$ has the same form as the bare Lagrangian $\mathcal{L}_0$ but 
depends on the renormalized parameters and fields and
$\delta\mathcal{L}$ contains the counterterms.
\section{Renormalization Scheme}
The renormalization constants can be chosen arbitrarily. Their divergent parts 
are determined by the structure of the relevant loop integrals, while their finite parts depend on the choice of 
the renormalization conditions. Here we focus on two different renormalization schemes:
\begin{itemize}
\item On-shell renormalization scheme\\
The counterterms are fixed in a way such that the finite renormalized parameters are equal to 
some physical quantities.
\item $\overline {DR}$ renormalization scheme\\
The counterterms are defined such that they only contain the UV-divergent parts of the bare parameters or fields.
The UV-divergent parts are proportional to $\Delta$, which can be written as 
\begin{eqnarray}
\Delta & =& \frac{2}{4-D} - \gamma_E + \log 4\pi\, ,
\end{eqnarray}
where $\gamma_E$ is Euler's constant. 
\end{itemize}
\section{On-shell Renormalization Scheme}\label{REos}
The basic idea of the on-shell renormalization scheme is:
\begin{itemize}
\item The counterterms for parameters of the physical particles are chosen in a way such that 
the renormalized masses are equal to the physical masses, which are the real parts of the 
poles of the corresponding renormalized propagators. 
\item  The diagonal entries of the field renormalization matrix are fixed by the requirement 
that the real parts of the renormalized propagators have unity residues.
\item The renormalized 1PI two-point function (the inverse of the renormalized propagator) 
is diagonal for on-shell external particles. 
This determines the non-diagonal entries of the field renormalization matrix.
\end{itemize}
In the on-shell renormalization scheme all renormalization conditions are formulated for on-shell
external particles. Now we come to the on-shell renormalization conditions for different types of fields.
\subsection{On-shell Renormalization Conditions for Scalars}\label{REosScalar}
Consider the scalar one-particle irreducible(1PI) diagram $i \Sigma(p^2)$. Via the Dyson summation the full propagator can be
written as 

\begin{fmffile}{Reselfa}
\begin{eqnarray}
 i \Delta(p^2) & = & \parbox{18mm}{\begin{fmfgraph*}(25,15)
\fmfkeep{born}
  \fmfleft{i}\fmfright{o}
  \fmf{dashes}{i,o}
\end{fmfgraph*}} \quad + \quad
\parbox{25mm}{\begin{fmfgraph*}(35,15)
\fmfkeep{self}
  \fmfleft{i}\fmfright{o}
  \fmf{dashes}{i,v1}\fmf{dashes}{v2,o}
\fmf{plain,left,tension=0.45}{v1,v2,v1}
\end{fmfgraph*}} \quad +\quad 
\parbox{35mm}{\begin{fmfgraph*}(50,15)
\fmfkeep{selfa}
  \fmfleft{i}\fmfright{o}
  \fmf{dashes}{i,v1}
\fmf{dashes}{v2,v3}\fmf{dashes}{v4,o}
\fmf{plain,left,tension=0.45}{v1,v2,v1}
\fmf{plain,left,tension=0.45}{v3,v4,v3}
\end{fmfgraph*}} \quad + \quad \cdots \nonumber \\
& = & \frac{i}{p^2 - m^2}+ \frac{i}{p^2 - m^2}i \Sigma(p^2)
\frac{i}{p^2 - m^2} + \frac{i}{p^2 - m^2}(i \Sigma(p^2)
\frac{i}{p^2 - m^2})^2 + \cdots  \nonumber \\
& = & \frac{i}{p^2 - m^2+ \Sigma(p^2)}.
\end{eqnarray}
\end{fmffile}
If there are $n$ mass eigenstates $\phi_i (i = 1, \cdots n)$, the renormalized 1PI two-point function can be written as
\begin{eqnarray}
\hat{\Gamma}_{ij}(p)& = & i\left(p^2 - m_j^2\right)\delta_{ij} + i\hat{\Sigma}_{ij}(p^2),
\label{eqn:OSre1}
\end{eqnarray}
 $\hat{\Sigma}_{ij}(p^2)$ is  the renormalized 1PI self-energy.

The on-shell renormalization conditions require that 
the poles in the renormalized propagators occur at $p^2 = m_j^2$ and the renormalized 1PI two-point function is diagonal, 
which are equivalent to
\begin{eqnarray}
\tilde{\rm Re}\hat{\Gamma}_{ij}(p)\big|_{p^2 = m_j^2}=0 &\Longrightarrow &\tilde{\rm Re}\hat{\Sigma}_{ij}(m_j^2)= 0\, .
\label{eqn:OSre2a}
\end{eqnarray}
$\tilde{\rm Re}$ takes the real parts of the loop integrals in the self-energies. It can be replaced by ${\rm Re}$
if all the relevant couplings are real.
Furthermore the real parts of the renormalized propagators have unity residues,
\begin{eqnarray}
\lim_{p^2\to m_i^2}\frac{1}{p^2-m_i^2}\tilde{\rm Re}\hat{\Gamma}_{ii}(p)=i 
&\Longrightarrow & \tilde{\rm Re} \hat{\Sigma}_{ii}'(m_i^2)=0\, ,
\label{eqn:OSre2b}
\end{eqnarray}
where $\hat{\Sigma}_{ii}'(m_i^2) = \frac{d}{dp^2}\hat{\Sigma}_{ii}(p^2)\Big |_{p^2 = m_i^2}$.
\subsection{On-shell Renormalization Conditions for Fermions}\label{REosFermion}
Similarly to the scalar case, the renormalized 1PI two-point functions for the fermions can be written as
\begin{eqnarray}
 \hat{\Gamma}_{ij}^f(p) & = &i\delta_{ij}(\not\!{p} - m_i)+ i\hat{\Sigma}_{ij}(p^2)\, .
\end{eqnarray}
$\hat{\Sigma}_{ij}(p^2)$ is the renormalized 1PI self-energy for the fermions, which can be decomposed via
\begin{eqnarray}
\hat{\Sigma}_{ij}(p^2) & = &
\not\!{p}\omega_L\hat{\Sigma}_{ij}^L(p^2) 
+\not\!{p}\omega_R\hat{\Sigma}_{ij}^R(p^2)
+ \omega_L\hat{\Sigma}_{ij}^{SL}(p^2)+ \omega_R\hat{\Sigma}_{ij}^{SR}(p^2)\, .
\label{eqn:ReCha12}
\end{eqnarray}
where $\omega_{L,R}=(1 \mp \gamma_5)/2$.

The on-shell renormalization conditions for fermions are
\begin{eqnarray}
 \tilde{\rm Re}\hat{\Gamma}_{ij}^f(p)\mu_j(p)\big|_{p^2= m_j^2}  =  0, &&
\bar{\mu}_i(p^\prime)\tilde{\rm Re}\hat{\Gamma}_{ij}^f(p^{\prime})\big|_{p^{\prime^2} = m_i^2} = 0\, ,
\label{eqn:On-shellFermiona}
\end{eqnarray}
\begin{eqnarray}
\lim_{p^2\to m_i^2}\frac{\not\!{p} + m_i}{p^2-m_i^2}\tilde{\rm Re}\hat{\Gamma}_{ii}^f(p)\mu_i(p) = i\mu_i(p),
\hspace{2mm}
 \lim_{p^{\prime^2}\to m_i^2}\bar{\mu}_i(p^\prime)\tilde{\rm Re}\hat{\Gamma}_{ii}^f(p^{\prime})
\frac{\not\!{p^\prime} + m_i}{p^{\prime^2}-m_i^2}= i\bar{\mu}_i(p^\prime),
\label{eqn:On-shellFermionb}
\end{eqnarray}
here $\mu(p), \bar{\mu}(p)$ are Dirac spinors of the external fermion fields.
The diagonal equations of (\ref{eqn:On-shellFermiona}) ensure that the renormalized fermion masses are 
the poles of the corresponding propagators,
while its non-diagonal equations make the renormalized 1PI two-point function diagonal for on-shell external particles.
The conditions that the renormalized propagators have unity residues are satisfied by (\ref{eqn:On-shellFermionb}). 
These on-shell renormalization conditions are translated into the relations between 
the renormalized self-energies for the fermions,
\begin{subequations}
\begin{eqnarray}
m_j\tilde{\rm Re}\hat{\Sigma}_{ij}^L(m_j^2) + \tilde{\rm Re}\hat{\Sigma}_{ij}^{SR}(m_j^2)& = & 0\, ,
\label{eqn:OSre14a} \\
m_j\tilde{\rm Re}\hat{\Sigma}_{ij}^R(m_j^2) + \tilde{\rm Re}\hat{\Sigma}_{ij}^{SL}(m_j^2)& = & 0\, ,
\label{eqn:OSre14b}\\
\tilde{\rm Re} \hat{\Sigma}_{ii}^L(m_i^2)
+ m_i^2\left(\tilde{\rm Re}\hat{\Sigma}_{ii}^{L'}(m_i^2)+\tilde{\rm Re}\hat{\Sigma}_{ii}^{R^{\prime}}(m_i^2)\right)
 \nonumber \\
+ m_i\left(\tilde{\rm Re}\hat{\Sigma}_{ii}^{SL^{\prime}}(m_i^2) + 
\tilde{\rm Re}\hat{\Sigma}_{ii}^{SR^{\prime}}(m_i^2)\right ) & = & 0\, ,
\label{eqn:OSre14c}\\
\tilde{\rm Re} \hat{\Sigma}_{ii}^R(m_i^2)
+ m_i^2\left(\tilde{\rm Re}\hat{\Sigma}_{ii}^{L^{\prime}}(m_i^2)+\tilde{\rm Re}\hat{\Sigma}_{ii}^{R^{\prime}}(m_i^2)\right)
 \nonumber \\
+ m_i\left(\tilde{\rm Re}\hat{\Sigma}_{ii}^{SL^{\prime}}(m_i^2) + 
\tilde{\rm Re}\hat{\Sigma}_{ii}^{SR^{\prime}}(m_i^2)\right ) & = & 0\, .
\label{eqn:OSre14d}
\end{eqnarray}
\label{eqn:OSre14}
\end{subequations}
\subsection{On-shell Renormalization Conditions for Gauge Bosons}\label{REosGauge}
Using the 't Hooft-Feynman gauge, the renormalized 1PI two-point function for the gauge fields can be written as
\begin{eqnarray}
 \hat{\Gamma}^{ij}_{\mu\nu}(p) & = &-i g_{\mu\nu}\left(p^2 - m_i^2\right )\delta_{ij}+ i\left (g_{\mu\nu} - 
\frac{p_\mu p_\nu}{p^2}\right )\hat{\Sigma}_{T}^{ij}(p^2) - i\frac{p_\mu p_\nu}{p^2}\hat{\Sigma}_{L}^{ij}(p^2) \, .
\label{eqn:On-shellGauge1}
\end{eqnarray}
Here $\hat{\Sigma}^{ij}(p^2)$ is the renormalized self-energy for the gauge fields, and the indices $T$ and $L$ denote
the transversal and longitudinal parts, respectively.  The on-shell renormalization conditions require that
the renormalized masses are equal to the physical masses and the renormalized 1PI two-point function 
is diagonal when the external particles are on their mass shell, 
\begin{eqnarray}
 \tilde{\rm Re}\hat{\Gamma}^{ij}_{\mu\nu}(p)\varepsilon_j^{\nu}(p)\big|_{p^2 = m_j^2}  =  0\, , \hspace*{3mm}
\tilde{\rm Re}\hat{\Gamma}^{ij}_{\mu\nu}(p)\varepsilon_i^{\nu}(p)\big|_{p^2 = m_i^2}  =  0\, , 
\label{eqn:On-shellGaugea}
\end{eqnarray}
and the renormalized propagators have unity residues, 
\begin{eqnarray}
\lim_{p^2\to m_i^2}\frac{1}{p^2-m_i^2}\tilde{\rm Re}\hat{\Gamma}^{ii}_{\mu\nu}(p)\epsilon_i^{\nu}(p) = -i\epsilon_{i, \mu}(p)\, ,
\label{eqn:On-shellGaugeb}
\end{eqnarray}
where $\epsilon^{\nu}$ are the polarization vectors and satisfy $p_\mu \epsilon_i^{\mu}(p) = 0$ when $p^2 = m_i^2$.
Applying (\ref{eqn:On-shellGauge1}) to the on-shell conditions (\ref{eqn:On-shellGaugea}, \ref{eqn:On-shellGaugeb}),
one obtains the on-shell conditions for the renormalized self-energies,
\begin{eqnarray}
 \tilde{\rm Re}\hat{\Sigma}_T^{ij}(m_i^2) = 0\, , \ \ 
 \tilde{\rm Re}\hat{\Sigma}_T^{{ii}'}(m_i^2) = 0\, .\label{eqn:OSconditiongauge}
\end{eqnarray}
Note that the longitudinal part of the gauge boson self-energies are dropped since they are always finite. 
\section{Renormalization of the SM-like Sector}
In the SM the input parameters are chosen to be the electric charge $e$, the fermion masses $m_f$, and the masses of the $W$ and $Z$ gauge bosons.
The on-shell renormalization of the Standard Model has been performed in \cite{SMDenner, SMHollik}. Here we follow the 
conventions of Ref.~\cite{SMDenner}.
\subsection{Fermion Sector Renormalization}
The bilinear part of the Lagrangian for SM fermions is 
 \begin{eqnarray}
\mathcal{L} &= & \bar f_i \left(\not\!{p}-m_{f_i}\right) f_i\, ,
\label{eqn: RefermionLagrangian}
\end{eqnarray}
where $f = \nu, e, u, d$ denotes the four-component Dirac spinors
and  $i= 1, 2, 3$  denotes the generation index. The fermion masses and fields are renormalized via 
\begin{eqnarray}
m_f &\longrightarrow &m_f + \delta m_f\, ,\label{eqn: Reconstantfermionmass}\\
\omega_L f_i &\longrightarrow & \left (\delta_{ij} + \frac{1}{2}\delta Z_{ij}^{f, L}\right )\omega_L f_j\, , \nonumber \\
\omega_R f_i &\longrightarrow & \left (\delta_{ij} + \frac{1}{2}\delta Z_{ij}^{f, R}\right )\omega_R f_j\, ,
\label{eqn: Reconstantfermionfield}
\end{eqnarray}
where $\delta m_f$ is the counterterm for the fermion masses $m_f$, and $\delta Z_{ij}^{f, L}$ and $\delta Z_{ij}^{f, R}$
are the field renormalization constants for the left- and right-handed fermion fields, respectively. 
Applying the transformations (\ref{eqn: Reconstantfermionmass}, \ref{eqn: Reconstantfermionfield}) to 
(\ref{eqn: RefermionLagrangian}), one obtains the counterterm Lagrangian,
\begin{eqnarray}
\delta \mathcal{L} & = & \bar f_i \not\!{p}\Bigl [\delta Z_{ij}^{f, L}\omega_L + \delta Z_{ij}^{f, R}\omega_R\Bigr ] f_j
-\Bigl [ m_{f_i}\left (\delta Z_{ij}^{f, L} +  \delta Z_{ij}^{f, R}\right) + \delta m_{f_i} \bigr ]\bar f_i f_j \, .
\end{eqnarray}

In general the renormalized self-energy is equal to the unrenormalized self-energy
$\Sigma_{ij}(p^2)$ plus the corresponding counterterms, which are the derivatives of the 
counterterm  Lagrangian $\delta\mathcal{L}$ with respect to the fields $\bar f_i$ and $f_j$,
\begin{eqnarray}
\hat{\Sigma}_{ij}(p^2)& = & \Sigma_{ij}(p^2) + \frac{\partial}{\partial \bar f_i}
\frac{\partial}{\partial f_j}\delta\mathcal{L}\, .
\label{eqn:OSre6}
\end{eqnarray}
Therefore the renormalized self-energies for the SM fermions can be written as
\begin{subequations}
\begin{eqnarray}
\hat{\Sigma}_{ij}^{f, L}(p^2)& = &\Sigma_{ij}^{f, L}(p^2)+ \delta Z_{ij}^{f, L}\, , \\
\hat{\Sigma}_{ij}^{f, R}(p^2)& = &\Sigma_{ij}^{f, R}(p^2)+ \delta Z_{ij}^{f, R}\, , \\
\tilde{\rm Re}\hat{\Sigma}_{ij}^{f, SL}(p^2) &=& \tilde{\rm Re}\hat{\Sigma}_{ij}^{f, SR}(p^2)=
\hat{\Sigma}_{ij}^{f, S}(p^2) \, \nonumber \\
&=& \Sigma_{ij}^{f, S}(p^2)+  
m_{f_i}\left (\delta Z_{ij}^{f, L} +  \delta Z_{ij}^{f, R}\right) + \delta m_{f_i}\, .
\end{eqnarray}
\label{eqn:SMfermion-selfenergy} 
\end{subequations}

Assuming the CKM matrix as an identity matrix $V_{ij} = \delta_{ij}$, all the field renormalization constants and
the self-energies are diagonal,
\begin{eqnarray}
\delta Z_{ij}^{f, L} = \delta Z_{ij}^{f, R} = 0\, , \ \ {\rm for}\ \  i\neq j\, .
\end{eqnarray}
Applying the renormalized fermion self-energies (\ref{eqn:SMfermion-selfenergy}) to the on-shell renormalization conditions
(\ref{eqn:OSre14}), one can fix the renormalization constants 
\begin{subequations}
\begin{eqnarray}
\delta m_{f_i} & = & \frac{m_{f_i}}{2}\left (\tilde{\rm Re} \Sigma_{ii}^{f, L}(m_{f_i}^2) + 
\tilde{\rm Re} \Sigma_{ii}^{f, R}(m_{f_i}^2)\right )+ 
\tilde{\rm Re}\Sigma_{ii}^{f, S}(m_{f_i}^2)\, ,\label{eqn:SMfermion-Reconstanta}\\
\delta Z_{ii}^{f, L} & = & - \tilde{\rm Re} \Sigma_{ii}^{f, L}(m_{f_i}^2) - m_{f_i}^2\left (
\tilde{\rm Re} \Sigma_{ii}^{{f, L}'}(m_{f_i}^2) + \tilde{\rm Re} \Sigma_{ii}^{{f, R}'}(m_{f_i}^2)\right )\, \nonumber \\
&& {}- 2m_{f_i}\tilde{\rm Re}\Sigma_{ii}^{{f, S}'}(m_{f_i}^2)\, ,\label{eqn:SMfermion-Reconstantb} \\
\delta Z_{ii}^{f, R} & = & - \tilde{\rm Re} \Sigma_{ii}^{f, R}(m_{f_i}^2) - m_{f_i}^2\left (
\tilde{\rm Re} \Sigma_{ii}^{{f, L}'}(m_{f_i}^2) + \tilde{\rm Re} \Sigma_{ii}^{{f, R}'}(m_{f_i}^2)\right )\, \nonumber \\
&& {}- 2m_{f_i}\tilde{\rm Re}\Sigma_{ii}^{{f, S}'}(m_{f_i}^2)\, .
\label{eqn:SMfermion-Reconstantc} 
\end{eqnarray}
\label{eqn:SMfermion-Reconstant} 
\end{subequations}
\subsection{Gauge Sector Renormalization}
The bilinear part of the Lagrangian describing the gauge fields is
\begin{eqnarray}
\mathcal{L}& = & W_{\mu}^-\left (p^2 - m_W^2 \right )W^{+ \ \mu} + Z_{\mu}\left (p^2 - m_Z^2\right )Z^{\mu}
+ A_{\mu}p^2A^{\mu}\, .  
\end{eqnarray}
The gauge sector is renormalized via the transformations
\begin{subequations}
\begin{eqnarray}
m_W^2 &\longrightarrow & m_W^2 + \delta m_W^2\, ,\label{eqn:Gaugeboson-Reconstanta} \\
m_Z^2 &\longrightarrow & m_Z^2 + \delta m_Z^2\, ,\label{eqn:Gaugeboson-Reconstantb} \\
W^\pm &\longrightarrow & W^\pm  + \frac{1}{2}\delta Z_W  W^\pm\, \label{eqn:Gaugeboson-Reconstantc} \\
\left(\begin{array}{c} Z \\ A \end{array}\right)
&\longrightarrow & \left(\begin{array}{cc}
1+ \frac{1}{2}\delta Z_{ZZ} &  \frac{1}{2}\delta Z_{ZA}\\
 \frac{1}{2} \delta Z_{AZ} & 1+ \frac{1}{2}\delta Z_{AA}
\end{array}\right)\left(\begin{array}{c} Z \\ A \end{array}\right)\, .\label{eqn:Gaugeboson-Reconstantd} 
\end{eqnarray}
\label{eqn:Gaugeboson-Reconstant} 
\end{subequations}
After the renormalization one obtains the Lagrangian which gives us counterterms
\begin{eqnarray}
\delta \mathcal{L} & = & W_{\mu}^-\bigl [\delta Z_W (p^2 - m_W^2) - \delta m_W^2 \bigr ]W^{+ \ \mu}\, \nonumber \\
&&{} + Z_{\mu}\bigl [\delta Z_{ZZ} (p^2 - m_Z^2) - \delta m_Z^2 \bigr ]Z^{\mu} + 
A_{\mu}p^2A^{\mu}\delta Z_{AA}\, \nonumber \\
&&{} + \frac{1}{2}A_{\mu}\bigl [\delta Z_{ZA} (p^2 - m_Z^2) + \delta Z_{AZ} p^2 \bigr ]Z^{\mu}\, .
\end{eqnarray}
The renormalized self-energies for gauge fields can be written as follows,
\begin{subequations}
\begin{eqnarray}
\hat\Sigma_T^{W}(p^2) & = & \Sigma_T^{W}(p^2) + \delta Z_W (p^2 - m_W^2) - \delta m_W^2\, ,\\
\hat\Sigma_T^{ZZ}(p^2) & = & \Sigma_T^{ZZ}(p^2) + \delta Z_{ZZ} (p^2 - m_Z^2) - \delta m_Z^2\, ,\\
\hat\Sigma_T^{AA}(p^2) & = & \Sigma_T^{AA}(p^2) + \delta Z_{AA} p^2 \, ,\\
\hat\Sigma_T^{AZ}(p^2) & = & \Sigma_T^{AZ}(p^2) + \frac{1}{2}\left (\delta Z_{ZA} (p^2 - m_Z^2) + \delta Z_{AZ}p^2\right )\, .
\end{eqnarray}
\end{subequations}
Using the on-shell renormalization conditions (\ref{eqn:OSconditiongauge}), 
we can determine the renormalization constants for the gauge sector of the SM,
 \begin{subequations}
\begin{eqnarray}
\tilde{\rm Re}\hat{\Sigma}_T^{W}(m_W^2) = 0 & \Longrightarrow & \delta m_W^2 = \tilde{\rm Re}\Sigma_T^{W}(m_W^2)\, , 
\label{eqn:ReconstantW}\\
\tilde{\rm Re}\hat{\Sigma}_T^{ZZ}(m_Z^2) = 0 & \Longrightarrow & \delta m_Z^2 = \tilde{\rm Re}\Sigma_T^{ZZ}(m_Z^2)\, , 
\label{eqn:ReconstantZ}\\
\tilde{\rm Re}\hat{\Sigma}_T^{AZ}(m_Z^2) = 0 & \Longrightarrow & \delta Z_{AZ} =- \frac{2 \tilde{\rm Re}\Sigma_T^{AZ}(m_Z^2)}
{m_Z^2}\, ,\label{eqn:fieldconstantAZ}\\
\tilde{\rm Re}\hat{\Sigma}_T^{AZ}(0) = 0 & \Longrightarrow & \delta Z_{ZA} = \frac{2 \tilde{\rm Re}\Sigma_T^{AZ}(0)}
{m_Z^2}\, ,\label{eqn:fieldconstantZA}\\
\tilde{\rm Re}\hat{\Sigma}_T^{W'}(m_W^2) = 0 & \Longrightarrow &\delta Z_W = - \tilde{\rm Re}\Sigma_T^{W'}(m_W^2)\, ,
\label{eqn:fieldconstantW}\\
\tilde{\rm Re}\hat{\Sigma}_T^{{ZZ}'}(m_Z^2) = 0 & \Longrightarrow &\delta Z_{ZZ} = - \tilde{\rm Re}\Sigma_T^{{ZZ}'}(m_Z^2)\, ,
\label{eqn:fieldconstantZZ}\\
\tilde{\rm Re}\hat{\Sigma}_T^{{AA}'}(0) = 0 & \Longrightarrow &\delta Z_{AA} = - \tilde{\rm Re}\Sigma_T^{{AA}'}(0)\, .
\label{eqn:fieldconstantAA}
\end{eqnarray}
\end{subequations}
 The on-shell definition of the weak mixing angle
$\theta_W(s_W =\sin\theta_W, c_W = \cos\theta_W)$ is \cite{sirlin}
\begin{eqnarray}
s_W^2 &= &1- \frac{m_W^2}{m_Z^2}\, .
\end{eqnarray}
Hence its counterterm is directly related to the counterterms of the gauge
boson masses,
\begin{eqnarray}
\frac{\delta s_W}{s_W} &= & - \frac{1}{2} \frac {c_W^2} {s_W^2}
\left( \frac {\delta m_W^2} {m_W^2} - \frac {\delta m_Z^2} {m_Z^2} \right)\, ,\nonumber \\
\frac{\delta c_W}{c_W} &= & \frac{1}{2}\left( \frac {\delta m_W^2} {m_W^2} - \frac {\delta m_Z^2} {m_Z^2} \right)\, .
\end{eqnarray}
%
\subsection{Electric Charge Renormalization}
The three-point function $\gamma \bar ff$ vertex at one-loop level can be depicted as\\
 \vspace*{1cm}
\begin{fmffile}{Charge}
\begin{eqnarray}
\hat \Gamma^{\gamma \bar ff}_{\mu}(p, p')
& = & \ \ \ \
\parbox{20mm}{\begin{fmfgraph*}(20,15)
\fmfpen{thin}
  \fmfleft{i}\fmfright{o1,o2}
 \fmf{fermion,tension=0.3}{o2,v1}
\fmf{fermion,tension=0.3}{o1,v1}
\fmflabel{$\bar {f},\ \ p'$}{o2}
\fmflabel{$f,\ \ p$}{o1}
\fmflabel{$A_\mu$}{i}
\fmf{photon}{v1,i}
\end{fmfgraph*}}\hspace*{1cm} \quad + \quad \hspace*{3mm}
\parbox{30mm}{\begin{fmfgraph*}(25,15)
\fmfpen{thin}
  \fmfleft{i}\fmfright{o1,o2}
 \fmf{fermion,tension=0.3}{o2,v1}
\fmf{fermion,tension=0.3}{o1,v1}
\fmf{photon}{v1,i}
\fmfv{decor.shape=circle,decor.filled=empty,decor.size=0.2w}{v1}
\end{fmfgraph*}} \quad + \quad
\parbox{30mm}{\begin{fmfgraph*}(25,15)
\fmfpen{thin}
  \fmfleft{i}\fmfright{o1,o2}
 \fmf{fermion,tension=0.3}{o2,v1}
\fmf{fermion,tension=0.3}{o1,v1}
\fmf{photon}{v1,i}
\fmfv{decor.shape=cross}{v1}
\end{fmfgraph*}}\nonumber \\
&&{}\nonumber \\
&&{}\nonumber \\
& = & \hspace*{5mm}-i eQ_f\gamma_{\mu} \hspace*{1.8cm}+ \hspace*{1.0cm} i e \Lambda_{\mu}^{\gamma \bar ff}(p, p')
\hspace*{0.8cm}
+ \hspace*{0.2cm} i e \delta\Lambda_{\mu}^{\gamma \bar ff}\, .
\end{eqnarray}
\end{fmffile}
The on-shell renormalization condition for the electric charge requires that all corrections to the $\gamma \bar ff$ vertex should vanish 
for the on-shell external particles in the Thomson limit ($p = p'$),
\begin{eqnarray}
\bar{u}(p)\hat \Gamma^{\gamma \bar ff}_{\mu}(p, p)u(p)\Big |_{p^2 = m_f^2} = -ieQ_f\bar{u}(p)\gamma_{\mu}u(p)\, \nonumber \\
\Longrightarrow \bar{u}(p)\left (\Lambda_{\mu}^{\gamma \bar ff}(p, p) +\delta\Lambda_{\mu}^{\gamma \bar ff}\right ) u(p) = 0\, .
\label{eqn:ChargeOScondition}
\end{eqnarray}
The electric charge is renormalized via
\begin{eqnarray}
e & \longrightarrow &(1 + \delta Z_e) e\, .
\label{eqn:Chargerenormal}
\end{eqnarray}
Together with the fermion field (\ref{eqn: Reconstantfermionfield}) and photon field transformation (\ref{eqn:Gaugeboson-Reconstantd}) 
one obtains the counterterm for the $\gamma \bar ff$ vertex, 
\begin{eqnarray}
\delta\Lambda_{\mu}^{\gamma \bar ff}& = & -Q_f\gamma_{\mu}\left (\delta Z_e + \frac{1}{2}\delta Z_{AA} + \delta Z^{f, L}
\omega_L +  \delta Z^{f, R} \omega_R \right)\, \nonumber \\
&&{}  + \gamma_\mu (v_f - a_f\gamma_5) \frac{1}{2}\delta Z_{ZA}\, ,
\end{eqnarray}
where $v_f$ and $a_f$ are the vector and axial vector coupling of the $Z$ boson to the fermion $f$.\\
Inserting the Ward-identity
\begin{eqnarray}
\bar{u}(p)\Lambda_{\mu}^{\gamma \bar ff}(p, p)u(p)& = & -Q_f\bar{u}(p)\Bigl [ \frac{\partial}{\partial p^\mu} 
\Sigma_{ff}(p)\Bigr ]u(p) - 2a_f\bar{u}(p)\gamma_\mu \omega_L u(p) \frac{\Sigma_T^{AZ}(0)}{m_Z^2}\, ,
\end{eqnarray}
where 
\begin{eqnarray}
\bar{u}(p)\Bigl [ \frac{\partial}{\partial p^\mu} 
\Sigma_{ff}(p)\Bigr ]u(p) = - \bar{u}(p)\gamma_\mu \left (\delta Z^{f, L}\omega_L +  \delta Z^{f, R} \omega_R \right)u(p)\, ,
\end{eqnarray}
into the charge renormalization condition (\ref{eqn:ChargeOScondition}), we can fix the charge renormalization constant
$\delta Z_e$. Its explicit expression is
\begin{eqnarray}
\delta Z_e & = & -\frac{1}{2}\delta Z_{AA} - \frac{1}{2}\frac{s_W}{c_W}\delta Z_{ZA}\, ,
\label{eqn:chargeconstant}
\end{eqnarray}
where we have used the relation $v_f -a_f = -\frac{s_W}{c_W}Q_f$.

The fermion-loop contributions to the photon field renormalization constant $\delta Z_{AA}$ in (\ref{eqn:chargeconstant}) give rise to
large logarithm $\ln m_f$ ($f$ denotes the light fermion).
The fine structure constant $\alpha = \frac{e^2}{4\pi}$ and the 
Fermi constant $G_\mu$ have the relation as
\begin{eqnarray}
\alpha & = & \frac{m_W^2 s_W^2}{\pi}\frac{\sqrt{2}G_\mu}{1 + \Delta r} \, ,
\end{eqnarray}
where $\Delta r$ summarizes all the radiative corrections to the muon decay~\cite{SMDenner, Deltar, DeltarSUSY, Deltar2}. 
One therefore can parameterize the Born matrix element by $G_\mu$, i.e.  
\begin{eqnarray}
e & = & 2 m_W s_W\left (\frac{\sqrt{2}G_\mu}{1 + \Delta r}\right )^{\frac{1}{2}} \, .
\label{eqn:chargeDeltar}
\end{eqnarray}
Combine the charge renormalization constant (\ref{eqn:chargeconstant}) and the $\Delta r$ contributions in (\ref{eqn:chargeDeltar}),
one obtains
\begin{eqnarray}
\delta \tilde Z_e & = & \delta Z_e - \frac{1}{2}\Delta r\, \nonumber \\
& = &\frac{1}{2}\frac{c_W^2}{s_W^
2}(\frac{\delta m_Z^2}{m_Z^2}-
\frac{\delta m_W^2}{m_W^2})-\frac{1}{2}\frac{\Sigma_T^W (0)- \delta m_W^2}{m_W^2}-
\frac{1}{s_Wc_W}\frac{\Sigma_T^{AZ}(0)}{m_Z^2} -\frac{1}{2}\Delta r_{\rm box}\, ,
\label{eqn:chargemuon}
\end{eqnarray}
where $\Delta r_{\rm box}$ denotes the box corrections to the muon decay.
Thus the large logarithm $\ln m_f$ disappears in (\ref{eqn:chargemuon}).
\newpage
\section{Renormalization of the Chargino and Neutralino Sector}
\subsection*{Renormalization of the Chargino Sector}
The kinetic and mass terms of the chargino Lagrangian in terms of the four-component chargino spinors $\tilde{\chi}_i^+$ can be written as
 \begin{eqnarray}
\mathcal{L}=\overline{\tilde{\chi}_i^+}\bigl[\not\!{p}\delta_{ij} 
- \omega_L\,(M_{\tilde{\chi}^+}^D)_{ij} - \omega_R\,(M_{\tilde{\chi}^+}^D)^\top_{ij}
\bigr]\tilde{\chi}_j^+\, ,
\label{eqn:ReCha5}
\end{eqnarray}
where $M_{\tilde{\chi}^+}^D = UXV^\top$ is the diagonalized mass matrix of the charginos.
In order to renormalize the chargino sector, we introduce the counterterm for the chargino mass matrix $X$,
\begin{eqnarray}
X &\longrightarrow & X + \delta X \, ,
\label{eqn:ReCha22}
\end{eqnarray}
\begin{eqnarray}
\delta X & = & \left(\begin{array}{cc}
\delta M_2 & \sqrt{2}\,\delta\big(m_W\,\sin\beta\big)\\
\sqrt{2}\,\delta\big(m_W\,\cos\beta\big) & \delta\mu
\end{array}\right)\;\; ,
\label{eqn:ReCha7}
\end{eqnarray}
in which the counterterms $\delta M_2$ and $\delta\mu$ are determined in the chargino sector renormalization.
The counterterm for the $W$ boson mass has been determined in (\ref{eqn:ReconstantW})
and the renormalization of $\tan\beta$ will be discussed in Section \ref{ReHiggs}.

The chargino fields are renormalized via the transformations
\begin{eqnarray}
\omega_{L}\tilde{\chi}^+_i &\longrightarrow &\left(\delta_{ij} + \frac{1}{2}\left (\delta Z^L\right )_{ij}\right)
\omega_{L}\tilde{\chi}^+_j\, ,
\nonumber \\
\omega_{R}\tilde{\chi}^+_i& \longrightarrow &\left(\delta_{ij} + \frac{1}{2}\left (\delta Z^R\right )^{\ast}_{ij}\right)
\omega_{R}\tilde{\chi}^+_j\, ,
\label{fieldchargino}
\end{eqnarray}
where the field renormalization constants $\delta Z^L$, $\delta Z^R$ are general $2\times 2$-matrices.
Applying the transformations (\ref {fieldchargino}) and (\ref{eqn:ReCha22}) to the Lagrangian (\ref{eqn:ReCha5}),
one gets the counterterm  Lagrangian
\begin{equation}
\begin{split}
\Delta \mathcal{L}&=\overline{\tilde{\chi}_i^+}\not\!{p}\bigl[ \frac{1}{2}\left(\delta Z^L + 
{\delta Z^L}^{\dagger}\right )_{ij}\omega_{L} + \frac{1}{2}\left( {\delta Z^{R}}^\ast + 
{\delta Z^R}^{\top}\right )_{ij}\omega_{R}\bigr]\tilde{\chi}_j^+\\
& \quad - \overline{\tilde{\chi}_i^+}\biggl[\bigl[U\delta X V^{\top} + 
\frac{1}{2}{\delta Z^{R}}^{\top}M_{\tilde{\chi}^+}^D
+  \frac{1}{2}M_{\tilde{\chi}^+}^D\delta Z^L \bigr]_{ij}\omega_L \bigr.\\
& \quad  \bigl. \bigl[ (U\delta XV^{\top})^\top + \frac{1}{2}{M_{\tilde{\chi}^+}^D}^\top{\delta Z^R}^{\ast}
+ \frac{1}{2}{\delta Z^L}^{\dagger}{M_{\tilde{\chi}^+}^D}^{\top} \bigr]_{ij}\omega_R\biggr]\tilde{\chi}_j^+.
\end{split}
\label{eqn:ReCha11}
\end{equation}
According to (\ref{eqn:ReCha12}) and (\ref{eqn:OSre6}), the renormalized self-energies $\hat{\Sigma}_{ij}(p)$ for the charginos
can be written as
\begin{eqnarray}
\hat{\Sigma}_{ij}^L(p^2) & = & \Sigma_{ij}^L(p^2)+
\frac{1}{2}\left(\delta Z^L+ {\delta Z^L}^\dagger
\right)_{ij}\, ,
\nonumber \\
\hat{\Sigma}_{ij}^R(p^2) & = & \Sigma_{ij}^R(p^2)+
\frac{1}{2}\left({\delta Z^R}^{\ast}+ {\delta Z^R}^\top
\right)_{ij}\, ,
\nonumber \\
\hat{\Sigma}_{ij}^{SL}(p^2)&=&\Sigma_{ij}^{SL}(p^2) -
\left[
U\delta X V^\top+\frac{1}{2}{\delta Z^{R}}^{\top}M_{\tilde{\chi}^+}^D
+  \frac{1}{2}M_{\tilde{\chi}^+}^D\delta Z^L\right ]_{ij}\, ,
\nonumber \\
\hat{\Sigma}_{ij}^{SR}(p^2) & = &\Sigma_{ij}^{SR}(p^2) -
\left[
(U\delta X V^\top)^\top+\frac{1}{2}{M_{\tilde{\chi}^+}^D}^\top{\delta Z^R}^{\ast}
+ \frac{1}{2}{\delta Z^L}^{\dagger}{M_{\tilde{\chi}^+}^D}^{\top}\right]_{ij} \, .
\label{eqn:ReCha13}
\end{eqnarray}

The counterterms $\delta M_2$ and $\delta\mu$ are determined by renormalizing the two charginos via
the on-shell renormalization scheme. According to the on-shell renormalization conditions for the fermions in Section \ref{REosFermion}, we obtain
the on-shell renormalization conditions for the charginos,
\begin{subequations}
\begin{eqnarray}
 m_{\tilde{\chi}_{j}^{+}}\tilde{Re}\hat{\Sigma}_{ij}^{L}(m_{\tilde{\chi}_{j}^{+}}^2)+ 
\tilde{Re}\hat{\Sigma}_{ij}^{SR}(m_{\tilde{\chi}_{j}^{+}}^2)& = &0 \, ,
\label{eqn:ReCha14b} 
\\
 m_{\tilde{\chi}_{j}^{+}}\tilde{Re}\hat{\Sigma}_{ij}^{R}(m_{\tilde{\chi}_{j}^{+}}^2)+ 
\tilde{Re}\hat{\Sigma}_{ij}^{SL}(m_{\tilde{\chi}_{j}^{+}}^2)& = &0 \, ,
\label{eqn:ReCha14c} 
\\ 
\tilde{Re}\hat{\Sigma}_{ii}^{L}(m_{\tilde{\chi}_{i}^{+}}^2) +
m_{\tilde{\chi}_{i}^{+}}^2 \left (\tilde{Re}\hat{\Sigma}_{ii}^{L^{\prime}}(m_{\tilde{\chi}_{i}^{+}}^2) + 
\tilde{Re}\hat{\Sigma}_{ii}^{R^{\prime}}(m_{\tilde{\chi}_{i}^{+}}^2)\right )\nonumber \\
+ 2 m_{\tilde{\chi}_{i}^{+}}\tilde{Re}\hat{\Sigma}_{ii}
^{SL^{\prime}}(m_{\tilde{\chi}_{i}^{+}}^2)& = & 0 \, ,
\label{eqn:ReCha14d} 
\\ 
\tilde{Re}\hat{\Sigma}_{ii}^{R}(m_{\tilde{\chi}_{i}^{+}}^2) +
m_{\tilde{\chi}_{i}^{+}}^2 \left (\tilde{Re}\hat{\Sigma}_{ii}^{L^{\prime}}(m_{\tilde{\chi}_{i}^{+}}^2) + 
\tilde{Re}\hat{\Sigma}_{ii}^{R^{\prime}}(m_{\tilde{\chi}_{i}^{+}}^2)\right )\nonumber \\
+ 2 m_{\tilde{\chi}_{i}^{+}}\tilde{Re}\hat{\Sigma}_{ii}
^{SL^{\prime}}(m_{\tilde{\chi}_{i}^{+}}^2)& = & 0 \, .
\label{eqn:ReCha14e}
\end{eqnarray}
\end{subequations}
(\ref{eqn:ReCha14d}) and (\ref{eqn:ReCha14e}), which make the renormalized chargino propagators 
have the residues $1$, fix the diagonal entries of the chargino field renormalization matrices,
\begin{eqnarray}
\delta Z^L_{ii} & = & - \tilde{Re}\Sigma_{ii}^{L}(m_{\tilde{\chi}_{i}^{+}}^2) -
2 m_{\tilde{\chi}_{i}^{+}}\tilde{Re}\Sigma_{ii}^{SL^{\prime}}(m_{\tilde{\chi}_{i}^{+}}^2)\, \nonumber \\
&&{}- m_{\tilde{\chi}_{i}^{+}}^2 \left (\tilde{Re}\Sigma_{ii}^{L^{\prime}}(m_{\tilde{\chi}_{i}^{+}}^2) + 
\tilde{Re}\Sigma_{ii}^{R^{\prime}}(m_{\tilde{\chi}_{i}^{+}}^2)\right ) \, ,
\label{eqn:ReCha15}\\
\delta Z^R_{ii} & = & - \tilde{Re}\Sigma_{ii}^{R}(m_{\tilde{\chi}_{i}^{+}}^2) -
2 m_{\tilde{\chi}_{i}^{+}}\tilde{Re}\Sigma_{ii}^{SL^{\prime}}(m_{\tilde{\chi}_{i}^{+}}^2)\, \nonumber \\
&&{}- m_{\tilde{\chi}_{i}^{+}}^2 \left (\tilde{Re}\Sigma_{ii}^{L^{\prime}}(m_{\tilde{\chi}_{i}^{+}}^2) + 
\tilde{Re}\Sigma_{ii}^{R^{\prime}}(m_{\tilde{\chi}_{i}^{+}}^2)\right ) \, .
\label{eqn:ReCha21}
\end{eqnarray}

The diagonal equations of (\ref{eqn:ReCha14b}) and (\ref{eqn:ReCha14c}),
which ensure that the renormalized chargino masses are the poles of the corresponding propagators, 
determine the counterterms $\delta M_2 $ and $\delta \mu $,
\begin{eqnarray}
\delta M_2 & = &
\Bigl[U_{22} V_{22} \Big( m_{\tilde{\chi}_{1}^{+}}\big [\tilde{Re}\Sigma_{11}^L
(m_{\tilde{\chi}_{1}^{+}}^2)+ 
\tilde{Re}\Sigma_{11}^R(m_{\tilde{\chi}_{1}^{+}}^2)\big] 
+2 \tilde{Re}\Sigma_{11}^{SL}(m_{\tilde{\chi}_{1}^{+}}^2)\Big) \nonumber \\
&& \hspace{0 cm}
{}- U_{12} V_{12}\Big(
m_{\tilde{\chi}_{2}^{+}}\big[\tilde{Re}\Sigma_{22}^L(m_{\tilde{\chi}_{2}^{+}}^2)+ 
\tilde{Re}\Sigma_{22}^R(m_{\tilde{\chi}_{2}^{+}}^2)\big] 
+2 \tilde{Re}\Sigma_{22}^{SL}(m_{\tilde{\chi}_{2}^{+}}^2)
\Big) \nonumber \\
&& \hspace{0 cm}
{} + 2\,\big(U_{12}U_{21}-U_{11}U_{22}\big)V_{12}V_{22}\,\delta(\sqrt{2}m_W\sin\beta)
\nonumber \\ &&
{} + 2\,U_{12}U_{22}\big(V_{12}V_{21}-V_{11}V_{22}\big)\,\delta(\sqrt{2}m_W\cos\beta)\Bigr] /\Delta \, ,
\label{eqn:ReCha16}\\
\delta \mu & = & 
\Bigl[U_{11}V_{11}\Big(
m_{\tilde{\chi}_{2}^{+}}\big[\tilde{Re}\Sigma_{22}^L(m_{\tilde{\chi}_{2}^{+}}^2)+
\tilde{Re}\Sigma_{22}^R(m_{\tilde{\chi}_{2}^{+}}^2)\big]
+2 \tilde{Re}\Sigma_{22}^{SL}(m_{\tilde{\chi}_{2}^{+}}^2)
\Big)\nonumber \\
&& \hspace{0cm}
{}- U_{21} V_{21} \Big(
m_{\tilde{\chi}_{1}^{+}}\big[\tilde{Re}\Sigma_{11}^L(m_{\tilde{\chi}_{1}^{+}}^2)+
\tilde{Re}\Sigma_{11}^R(m_{\tilde{\chi}_{1}^{+}}^2)\big]
+2 \tilde{Re}\Sigma_{11}^{SL}(m_{\tilde{\chi}_{1}^{+}}^2)
\Big) \nonumber \\
&& \hspace{0cm}
{} + 2\,U_{11}U_{21}\big(V_{12} V_{21} - V_{11} V_{22}
\big)\,\delta(\sqrt{2}m_W\sin\beta)
\nonumber \\ &&
{} + 2\,\big( U_{12} U_{21} - U_{11}U_{22}\big) V_{11}
V_{21}\,\delta(\sqrt{2}m_W\cos\beta)\Bigr]  /\Delta \, ,
\label{eqn:ReCha17}
\end{eqnarray}
with $\Delta =  2(U_{11}U_{22}V_{11}V_{22}- U_{12}U_{21}V_{12}V_{21})$. 
In contrast their non-diagonal equations, which make the renormalized chargino 1PI two-point functions 
diagonal for on-shell external particles, 
determine the non-diagonal entries of the chargino field renormalization matrices,
\begin{eqnarray}
\label{eqn:ReCha18}
\delta Z^L_{ij} & = & 
\frac{2}{m_{\tilde{\chi}_{i}^{+}}^2-m_{\tilde{\chi}_{j}^{+}}^2} \, \Big[
m_{\tilde{\chi}_{j}^{+}}^2\,\tilde{Re}\Sigma_{ij}^{L}(m_{\tilde{\chi}_{j}^{+}}^2) +
m_{\tilde{\chi}_{i}^{+}}\,m_{\tilde{\chi}_{j}^{+}}\,\tilde{Re}\Sigma_{ij}^{R}(m_{\tilde{\chi}_{j}^{+}}^2) +
m_{\tilde{\chi}_{i}^{+}}\,\tilde{Re}\Sigma_{ij}^{SL}(m_{\tilde{\chi}_{j}^{+}}^2)
\nonumber \\
& & 
{}+m_{\tilde{\chi}_{j}^{+}}\,\tilde{Re}\Sigma_{ij}^{SR}(m_{\tilde{\chi}_{j}^{+}}^2)
- m_{\tilde{\chi}_{i}^{+}}\,(U \delta X V^\top)_{ij} 
- m_{\tilde{\chi}_{j}^{+}}\,(U \delta X V^\top)_{ji}\Big] \, ,
\nonumber \\
\delta Z^R_{ij} & = & 
\frac{2}{m_{\tilde{\chi}_{i}^{+}}^2-m_{\tilde{\chi}_{j}^{+}}^2}\, \Big[
m_{\tilde{\chi}_{j}^{+}}^2\,\tilde{Re}\Sigma_{ij}^{R}(m_{\tilde{\chi}_{j}^{+}}^2) +
m_{\tilde{\chi}_{i}^{+}}\,m_{\tilde{\chi}_{j}^{+}}\,\tilde{Re}\Sigma_{ij}^{L}(m_{\tilde{\chi}_{j}^{+}}^2) +
m_{\tilde{\chi}_{j}^{+}}\,\tilde{Re}\Sigma_{ij}^{SL}(m_{\tilde{\chi}_{j}^{+}}^2)
\nonumber \\
& & 
{}+m_{\tilde{\chi}_{i}^{+}}\,\tilde{Re}\Sigma_{ij}^{SR}(m_{\tilde{\chi}_{j}^{+}}^2)
- m_{\tilde{\chi}_{j}^{+}}\,(U \delta X V^\top)_{ij} 
- m_{\tilde{\chi}_{i}^{+}}\,(U \delta X V^\top)_{ji}\Big] \, .
\end{eqnarray}
\newpage
\subsection*{Renormalization of the Neutralino Sector}\label{Reneu}
In terms of the neutralino Majorana spinors the Lagrangian which describes the kinematic and mass terms of the neutralinos
can be written as 
\begin{eqnarray}
\mathcal{L} & = &\frac{1}{2}\;\overline{\tilde{\chi}^0_i}\,\bigl[
\not\!{p}\,\delta_{ij} - \big(N^\ast\,Y\,N^\dagger\big)_{ij}\,\omega_L  - 
\big(N\,Y^\dagger\,N^\top\big)_{ij}\,\omega_R \bigr]\, \tilde{\chi}^0_j \, .
\label{eqn:ReNeu6}
\end{eqnarray}
In analogy to the chargino case, we introduce the counterterm for the neutralino mass matrix $Y$ and
the field renormalization constants for the neutralino fields by the transformations 
\begin{eqnarray}
Y & \to & Y + \delta Y\, , 
\label{eqn:ReNeu7a}\\
\omega_{L}\tilde{\chi}^0_i &=&\left(\delta_{ij} + \frac{1}{2}\left (\delta Z^0\right )_{ij}\right)
\omega_{L}\tilde{\chi}^0_j\, ,
\nonumber \\
\omega_{R}\tilde{\chi}^0_i& =&\left(\delta_{ij} + \frac{1}{2}\left (\delta Z^0\right )^{\ast}_{ij}\right)
\omega_{R}\tilde{\chi}^0_j\, ,
\label{eqn:ReNeu7b}
\end{eqnarray}
where the field renormalization constant $\delta Z^0$ is a general complex $4\times 4$-matrix.
One does not need to renormalize the left and right components of neutralinos independently
due to the definition of their Majorana spinors in (\ref{eqn:ReNeu5}).
The elements of the matrix $\delta Y$ are the counterterms for the parameters in
the mass matrix (\ref{eqn:ReNeu2}),
{\small 
\begin{eqnarray}
\delta Y  = \left(\begin{array}{cccc}
\delta M_1 & 0 & -\delta \left (m_Z\,s_W\,\cos\beta\right ) & \delta \left (m_Z\,s_W\,\sin\beta \right ) \\
0 & \delta M_2 &  \delta \left (m_Z\,c_W\,\cos\beta \right ) & -\delta \left (m_Z\,c_W\,\sin\beta\right) \\
-\delta \left (m_Z\,s_W\,\cos\beta\right ) & \delta \left (m_Z\,c_W\,\cos\beta \right ) & 0 & -\delta \mu \\
\delta \left (m_Z\,s_W\,\sin\beta\right ) & -\delta \left (m_Z\,c_W\,\sin\beta\right ) & -\delta \mu & 0
\end{array}\right) \, .
\label{eqn:ReNeudeltamass}
\end{eqnarray}}
Applying the transformations (\ref{eqn:ReNeu7a}, \ref{eqn:ReNeu7b}) to the Lagrangian(\ref{eqn:ReNeu6}),
one arrives at the counterterm Lagrangian
\begin{eqnarray}
\mathcal{L}_{\rm CT} & = & \frac{1}{2}\, \overline{\tilde{\chi}^0_i}\, \not\!{p}\,
\Bigl[\frac{1}{2}\left({\delta Z^0}^\ast + {\delta Z^0}^\top \right)_{ij} \omega_R  
+ \frac{1}{2}\left(\delta Z^0 + {\delta Z^0}^\dagger \right)_{ij} \omega_L
\Bigr]\,\tilde{\chi}^0_j  \nonumber\\
&&{}-
\frac{1}{2}\,\overline{\tilde{\chi}^0_i}\,\Bigl[\left( N^\ast\delta Y N^\dagger + 
\frac{ {\delta Z^0}^\top M_{\tilde{\chi}_0}^D + M_{\tilde{\chi}_0}^D {\delta Z^0}
}{2} \right)_{ij}\omega_L
\nonumber \\ && \hspace{1.4cm} +\,  
\left( N {\delta Y}^\dagger N^\top+ \frac{M_{\tilde{\chi}_0}^D{\delta Z^0}^\ast +
{\delta Z^0}^\dagger M_{\tilde{\chi}_0}^D}{2} \right)_{ij}\omega_R \Bigr]\,\tilde{\chi}^0_j \, ,
\label{eqn:ReNeu12}
\end{eqnarray}
here $ M_{\tilde{\chi}_0}^D$ is the diagonalized neutralino mass matrix which has been defined in (\ref{eqn:ReNeu3}).
Similarly to the chargino case, the renormalized neutalino self-energies can be written as 
\begin{eqnarray}
\hat{\Sigma}_{ij}^R(p^2) & = & \Sigma_{ij}^R(p^2)+\frac{1}{2}\left({\delta Z^0}^\ast + 
{\delta Z^0}^\top \right)_{ij}
\nonumber \\
\hat{\Sigma}_{ij}^L(p^2) & = & \Sigma_{ij}^L(p^2)+\frac{1}{2}\left(\delta Z^0 + 
{\delta Z^0}^\dagger \right)_{ij}
\nonumber \\
\hat{\Sigma}_{ij}^{SR}(p^2) & = &\Sigma_{ij}^{SR}(p^2) -
\left( N {\delta Y}^\dagger N^\top+\frac{M_{\tilde{\chi}_0}^D {\delta Z^0}^\ast +
{\delta Z^0}^\dagger M_{\tilde{\chi}_0}^D}{2} \right)_{ij}
\nonumber \\
\hat{\Sigma}_{ij}^{SL}(p^2)&=&\Sigma_{ij}^{SL}(p^2) -
\left( N^\ast\delta Y N^\dagger + \frac{ {\delta Z^0}^\top M_{\tilde{\chi}_0}^D +  
M_{\tilde{\chi}_0}^D {\delta Z^0}
}{2} \right)_{ij}\;\;\; .
\label{eqn:ReNeu13}
\end{eqnarray}
Obviously they obey the relations
\begin{equation}
\hat{\Sigma}^L_{ij}(p^2) = \hat{\Sigma}^R_{ji}(p^2) \hspace*{2mm},
\hat{\Sigma}^{SR}_{ij}(p^2) = \hat{\Sigma}^{SR}_{ji}(p^2) \hspace*{2mm},
\hat{\Sigma}^{SL}_{ij}(p^2)  =  \hat{\Sigma}^{SL}_{ji}(p^2) \hspace*{2mm},
\hat{\Sigma}^{SL}_{ij}(p^2)  =\hat{\Sigma}^{SR}_{ij}(p^2)^\dagger \, .
\end{equation}

Only the counterterm for the parameter $M_1$ in the chargino/neutralino sector is not determined so far.
We can fix it by renormalizing one of the four neutralinos via the on-shell renormalization scheme.
Conventionally, the lightest neutralino $\tilde{\chi}_1^0$ is chosen for this task.
From the previous discussions, 
the on-shell renormalization conditions for the neutralino sector can be expressed as follows,
\begin{subequations}
\begin{eqnarray}
m_{\tilde{\chi}_j^0} \tilde{Re}\hat{\Sigma}_{ij}^L(m_{\tilde{\chi}_j^0}^2) +\tilde{Re}\hat{\Sigma}_{ij}^{SR}(m_{\tilde{\chi}_j^0}^2)
& = & 0\, , 
\label{eqn:ReNeu14a}\\
m_{\tilde{\chi}_j^0} \tilde{Re}\hat{\Sigma}_{ij}^R(m_{\tilde{\chi}_j^0}^2) +\tilde{Re}\hat{\Sigma}_{ij}^{SL}(m_{\tilde{\chi}_j^0}^2)
& = & 0\, ,
\label{eqn:ReNeu14b}\\
\mbox{for  }(i\neq j) \vee (i=j=1)\, \nonumber \\ 
\tilde{Re}\hat{\Sigma}^L_{ii}(m_{\tilde{\chi}_i^0}^2) +
2 m_{\tilde{\chi}_i^0}^2\, \tilde{Re}\hat{\Sigma}^L_{ii}\,\!'(m_{\tilde{\chi}_i^0}^2) 
+2 m_{\tilde{\chi}_i^0} \tilde{Re}\hat{\Sigma}^{SL}_{ii}\,\!'(m_{\tilde{\chi}_i^0}^2)
& = & 0 \, .
\label{eqn:ReNeu14c}
\end{eqnarray}
\end{subequations}
The diagonal equations ($i = j =1$) of (\ref{eqn:ReNeu14a}) and (\ref{eqn:ReNeu14b}), which ensure that the renormalized lightest 
neutralino mass is the pole of the corresponding propagator, determine the counterterm $\delta M_1$. 
In contrast their non-diagonal equations, which fix the non-diagonal entries of
the neutralino field renormalization matrix, make the renormalized neutralino 1PI two-point function diagonal 
when the external particles are on their mass shell.
The diagonal entries of the neutralino field renormalization matrix are determined by (\ref{eqn:ReNeu14c}), 
which make the renormalized neutralino propagators have the 
residues $1$.  
 
Inserting the renormalized neutralino self-energies (\ref{eqn:ReNeu13}) into the on-shell renormalization conditions,
one obtains the expressions for the renormalization constants $\delta M_1$, $\delta Z^0_{ii}$ and 
$\delta Z^0_{ij}$ which are as follows,
\begin{subequations}
\begin{eqnarray}
\delta M_1 & = & \frac{1}{N_{11}^2}\Bigl[
2 N_{11}\big[N_{13}\,\delta(m_Z\sin\theta_W\cos\beta)-
N_{14}\,\delta(m_Z\sin\theta_W\sin\beta) \big] - \nonumber \\
&&\hspace{1cm}
2 N_{12}\big[N_{13}\, \delta(m_Z\cos\theta_W\cos\beta)
-N_{14}\, \delta(m_Z\cos\theta_W\sin\beta)\big]
- N_{12}^2\,\delta M_2 \nonumber \\
&& \hspace{1cm} + 2 N_{13} N_{14}\, \delta\mu + m_{\tilde{\chi}_1^0}
\tilde{Re}\Sigma^L_{11}(m_{\tilde{\chi}_1^0}^2) + 
\tilde{Re}\Sigma^{SR}_{11}(m_{\tilde{\chi}_1^0}^2)
\Bigr] \, ,
\end{eqnarray}
\begin{eqnarray}
\delta Z^0_{ii} & = & -\tilde{Re}\Sigma_{ii}^L(m_{\tilde{\chi}_i^0}^2)-
2 m_{\tilde{\chi}_i^0}\big[m_{\tilde{\chi}_i^0}\, \tilde{Re}\Sigma_{ii}^{L'}(m_{\tilde{\chi}_i^0}^2)+
\tilde{Re}\Sigma_{ii}^{SL'}(m_{\tilde{\chi}_i^0}^2)\big] \, ,  \\[0.4cm]
\delta Z^0_{ij} & = & 
\frac{2}{m_{\tilde{\chi}_i^0}^2 - m_{\tilde{\chi}_j^0}^2} \big[
m_{\tilde{\chi}_j^0}^2\, \tilde{Re}\Sigma_{ij}^{L}(m_{\tilde{\chi}_j^0}^2)
+ m_{\tilde{\chi}_i^0}m_{\tilde{\chi}_j^0}\, \tilde{Re}\Sigma_{ij}^{R}(m_{\tilde{\chi}_j^0}^2)
+  m_{\tilde{\chi}_i^0}\tilde{Re}\Sigma_{ij}^{SL}(m_{\tilde{\chi}_j^0}^2) \nonumber \\
& & 
{}
+  m_{\tilde{\chi}_j^0}\tilde{Re}\Sigma_{ij}^{SR}(m_{\tilde{\chi}_j^0}^2)
 - m_{\tilde{\chi}_j^0}(N \delta Y N^\top)_{ij} - m_{\tilde{\chi}_i^0}(N^\ast \delta Y N^\dagger)_{ij}
\big]  \, .
\end{eqnarray}
\end{subequations}
\section{Renormalization of the Sfermion Sector}
The kinetic and mass terms of the sfermion Lagrangian are 
\begin{eqnarray}
\mathcal{L}& = & \left( \begin{array}{cc}
\tilde{f}_1^\dagger & \tilde{f}_2^\dagger 
\end{array}\right)
\left (p^2 - M_{\tilde{f}}^D\right ) \left(\begin{array}{c}
\tilde{f}_1\\
\tilde{f}_2
\end{array}\right) + \tilde{\nu}_l^\dagger (p^2 - M_{\tilde{\nu}_l})\tilde{\nu}_l\, ,
\label{eqn:ReSle1}
\end{eqnarray}
where $M_{\tilde{f}}^D= U_{\tilde{f}}M_{\tilde{f}}U_{\tilde{f}}^\dagger$ is the diagonalized sfermion mass matrix.
At one-loop level, the counterterms for the sfermion mass matrices $M_{\tilde{f}}$ and $M_{\tilde{\nu}_l}$ are introduced via
\begin{eqnarray}
M_{\tilde{f}}  \to  M_{\tilde{f}} + \delta M_{\tilde{f}}\, , \ \ 
M_{\tilde{\nu}_l} \to M_{\tilde{\nu}_l} + \delta M_{\tilde{\nu}_l}\, .
\label{eqn:ReSle5b}
\end{eqnarray}
The elements of the matrix $\delta M_{\tilde{f}}$ and $\delta M_{\tilde{\nu}_l}$ are the counterterms for the parameters in
the mass matrices(\ref{Sfermionmassenmatrix}), (\ref{eqn:ReSle13}), respectively,
\begin{subequations}
\begin{eqnarray}
\delta \mathcal{M}_{\tilde{f}} &=&\begin{pmatrix} \delta M_{\tilde f_L}^2 + \delta C_{\tilde f_{11}}
 & m_f \delta A_f + \delta C_{\tilde f_{12}} \\
  m_f\delta A_f +\delta C_{\tilde f_{12}}   & \delta M_{\tilde{f}_R}^2 + \delta C_{\tilde f_{22}} 
 \end{pmatrix},\label{sneutrinodeltamassa}\\
\delta M_{\tilde{\nu}_l}&=&\delta M_{\tilde l_L}^2 + \delta \left (\frac{1}{2}m_Z^2 \cos 2\beta \right )\, ,
\label{sneutrinodeltamassb}
\end{eqnarray}
\end{subequations}
where 
\begin{eqnarray}
\delta C_{\tilde f_{11}}& =& \delta \left ( m_f^2 + m_Z^2 \cos2\beta (T_f^3 - Q_{{f}} s_W^2)\right )\, , \nonumber \\
\delta C_{\tilde f_{12}}& = & A_f\delta m_f-\delta \left (m_f \mu \kappa \right )\, , \nonumber \\
\delta C_{\tilde f_{22}}& = &\delta \left (m_f^2 + m_Z^2 \cos2 \beta Q_{{f}} s_W^2\right )\, .
\label{eqn:ReSleCfactor}
\end{eqnarray} 
In order to get finite Green functions, the field renormalization constants are introduced via
the transformations 
\begin{subequations}
\label{eqn:ReSle6}
\begin{eqnarray}
\left(\begin{array}{c}
\tilde{f}_1\\
\tilde{f}_2
\end{array}\right) & \to & \left(1 +\frac{1}{2}\delta Z_{\tilde{f}}\right)\left(\begin{array}{c}
\tilde{f}_1\\
\tilde{f}_2
\end{array}\right)\, ,\label{eqn:ReSle6a}\\
\tilde{\nu_l} & \to & \left(1 +\frac{1}{2}\delta Z_{\tilde{\nu_l}}\right)\tilde{\nu_l}\, .
\label{eqn:ReSle6b}
\end{eqnarray}
\end{subequations}
The field renormalization constants $\delta \tilde Z_{\tilde{f}}$ are general $2\times 2$ matrices. 
Inserting (\ref{eqn:ReSle5b}) and (\ref{eqn:ReSle6}) into the Lagrangian (\ref{eqn:ReSle1}), 
one obtains the counterterm Lagrangian
\begin{eqnarray}
\mathcal{L}_{\rm CT} & = & \left( \begin{array}{cc}
\tilde{f}_1^\dagger & \tilde{f}_2^\dagger 
\end{array}\right)
\frac{1}{2}p^2\left (\delta Z_{\tilde{f}}^\dagger + \delta Z_{\tilde{f}}\right )
\left(\begin{array}{c}
\tilde{f}_1\\
\tilde{f}_2
\end{array}\right)\nonumber\\
&&{} - \left (\begin{array}{cc}
\tilde{f}_1^\dagger & \tilde{f}_2^\dagger 
\end{array}\right)
\Biggl [ \frac{1}{2}\left (\delta Z_{\tilde{f}}^\dagger M_{\tilde{f}}^D +
M_{\tilde{f}}^D\delta Z_{\tilde{f}}\right ) +
 U_{\tilde{f}}\delta M_{\tilde{f}}U_{\tilde{f}}^\dagger\Biggr ]\left(\begin{array}{c}
\tilde{f}_1\\
\tilde{f}_2
\end{array}\right) \nonumber\\
&&{}+ \tilde{\nu}_l^\dagger \Biggl [ \frac{1}{2}\left (p^2 - M_{\tilde{\nu}_l}\right )
\left (\delta Z_{\tilde{\nu_l}}^\dagger + \delta Z_{\tilde{\nu_l}} \right )- \delta M_{\tilde{\nu_l}} \Biggr ]\tilde{\nu}_l\, .
\end{eqnarray}
Hence the renormalized self-energies for the sfermions are
\begin{eqnarray}
\hat{\Sigma}_{\tilde{f}_{ij}}(p^2) & = & \Sigma_{\tilde{f}_{ij}}(p^2)+\frac{1}{2}p^2
\left (\delta Z_{\tilde{f}}^\dagger + \delta Z_{\tilde{f}}\right )_{ij}
- \frac{1}{2}\left (\delta Z_{\tilde{f}}^\dagger M_{\tilde{f}}^D +
M_{\tilde{f}}^D\delta Z_{\tilde{f}}\right )_{ij} \nonumber \\
&&{}- \left( U_{\tilde{f}}\delta M_{\tilde{f}}U_{\tilde{f}}^\dagger \right )_{ij} \, , 
\label{eqn:ReSle11a}\\
\hat{\Sigma}_{\tilde{\nu_l}}(p^2) & = & \Sigma_{\tilde{\nu_l}}(p^2)+\frac{1}{2}\left (p^2 -  M_{\tilde{\nu}_l}\right )
\left (\delta Z_{\tilde{\nu_l}}^\dagger + \delta Z_{\tilde{\nu_l}}\right ) - \delta M_{\tilde{\nu_l}}\, .
\label{eqn:ReSle11b}
\end{eqnarray}
For convenience we define
\begin{eqnarray}
\delta m_{\tilde f_i}^2 = \left (U_{\tilde{f}}\delta M_{\tilde{f}}U_{\tilde{f}}^\dagger \right )_{ii}\, , \ \ 
\delta m_{\tilde f_{12}} = \left (U_{\tilde{f}}\delta M_{\tilde{f}}U_{\tilde{f}}^\dagger \right )_{12}\, .
\end{eqnarray}
The independent parameters in the sfermion sector are the soft-SUSY breaking parameters 
$M_{\tilde{f}_L}^2$, $M_{\tilde{f}_R}^2$, and $A_{f}$. Their counterterms are determined in the sfermion sector.
Here we treat squarks and sleptons separately.   
\subsection{Renormalization Constants for the Squarks}
There are five independent parameters:  $M_{\tilde{Q}_L}^2$, $M_{\tilde{u}_R}^2$,  $M_{\tilde{d}_R}^2$,
$A_u$ and $ A_{d}$ in every generation of squarks. In order to fix their counterterms, 
one can renormalize two up- and one of the down-type squarks via the on-shell renormalization scheme.
Here we choose the lighter down-type squark $\tilde d_1$. The on-shell renormalization conditions
can be written as follows,
\begin{eqnarray}
\tilde{Re}\hat{\Sigma}_{\tilde{u}_{ii}}(m_{\tilde{u}_i}^2) =  0 \, , \ \
\tilde{Re}\hat{\Sigma}_{\tilde{d}_{11}}(m_{\tilde{d}_1}^2) =  0 \, , \nonumber \\
\tilde{Re}\hat{\Sigma}_{\tilde{f}_{12}}(m_{\tilde{f}_1}^2)= 0 \, , \ \ 
\tilde{Re}\hat{\Sigma}_{\tilde{f}_{12}}(m_{\tilde{f}_2}^2) =  0 \, , \ \
\tilde{Re}\hat{\Sigma}_{\tilde{f}_{ii}}^{\prime}(m_{\tilde{f}_i}^2) = 0 \, ,
\label{eqn:ReSle10}
\end{eqnarray}
where $i = 1, 2$ is the index of the squarks.  $f$ can be an up- or a down-quark.  
Inserting the renormalized self-energy (\ref{eqn:ReSle11a}) into the on-shell conditions above, and choosing
 \begin{eqnarray}
\delta Z_{\tilde{f}_{12}}=\delta Z_{\tilde{f}_{21}}\, ,
\end{eqnarray}
 one gets the counterterms of the mass matrices and fields for the squarks. They can be written as 
\begin{subequations}
\begin{eqnarray}
\delta m_{\tilde u_i}^2 &= &
 \tilde{Re}\Sigma_{\tilde{u}_{ii}}(m_{\tilde{u}_i}^2)\, ,
\label{eqn:ReSle3a}\\
\delta m_{\tilde d_1}^2&= &
 \tilde{Re}\Sigma_{\tilde{d}_{11}}(m_{\tilde{d}_1}^2)\, ,
\label{eqn:ReSle3b}\\
\delta m_{\tilde f_{12}} &= &
 \frac{\tilde{Re}\Sigma_{\tilde{f}_{12}}(m_{\tilde{f}_1}^2) +\tilde{Re}\Sigma_{\tilde{f}_{12}}(m_{\tilde{f}_2}^2)}{2}\, ,
\label{eqn:ReSle3e}\\
\delta Z_{\tilde{f}_{12}}& = & - \frac{\tilde{Re}\Sigma_{\tilde{f}_{12}}(m_{\tilde{f}_1}^2) -
\tilde{Re}\Sigma_{\tilde{f}_{12}}(m_{\tilde{f}_2}^2)}{m_{\tilde{f}_1}^2-m_{\tilde{f}_2}^2}\, ,
\label{eqn:ReSle3d} \\
\delta Z_{\tilde{f}_{ii}}& = & - \tilde{Re}\Sigma_{\tilde{f}_{ii}}^{\prime}(m_{\tilde{f}_i}^2)\, .
\label{eqn:ReSle3c}
\end{eqnarray}
\end{subequations}

The counterterms $\delta \! M_{\tilde{Q}_L}^2$, $\delta \!M_{\tilde{u}_R}^2$, $\delta \! A_{u}$ are formulated from
(\ref{eqn:ReSle3a}) and (\ref{eqn:ReSle3e}) $\left(f = u\right )$, which can be expressed as 
\begin{subequations}
\begin{align}\label{deltaML}
\delta \! M_{\tilde Q_L}^2 &= U_{\tilde{u}_{11}}^2
\delta  m_{\tilde{u}_1}^2 + U_{\tilde{u}_{12}}^2
\delta  m_{\tilde{u}_2}^2 - 2
U_{\tilde{u}_{12}}U_{\tilde{u}_{22}} \delta m_{\tilde{u}_{12}}
 - \delta C_{\tilde{u}_{11}}\,,
 \\[4mm] 
 \delta \! M_{{\tilde{u}}_R}^2 &= U_{\tilde{u}_{12}}^2
 \delta  m_{\tilde{u}_1}^2 + U_{\tilde{u}_{11}}^2
 \delta  m_{\tilde{u}_2}^2 + 2
 U_{\tilde{u}_{12}}U_{\tilde{u}_{22}} \delta  m_{\tilde{u}_{12}} - \delta C_{\tilde{u}_{22}}\, ,
 \label{deltaMuR}\\[4mm] 
 \delta \! A_u &= \frac{1}{m_u}\Bigl[U_{\tilde{u}_{11}}
 U_{\tilde{u}_{12}}\bigl(\delta  m_{\tilde{u}_1}^2 
 - \delta  m_{\tilde{u}_2}^2\bigr)
 + (U_{\tilde{u}_{11}} U_{\tilde{u}_{22}} 
 +U_{\tilde{u}_{12}}
 U_{\tilde{u}_{21}}) \delta m_{\tilde{u}_{12}}
  - \delta C_{\tilde{u}_{12}} \Bigr]\, .
 \label{deltaAu}
\end{align}
\end{subequations}
The other countertems are derived from (\ref{eqn:ReSle3b}) and (\ref{eqn:ReSle3e})~$\left(f = d\right )$. 
Their expressions are 
\begin{subequations}
\begin{align}
\delta \! M_{{\tilde{d}}_R}^2 &= \frac{U_{\tilde{d}_{12}}^2 - U_{\tilde{d}_{11}}^2}{U_{\tilde{d}_{12}}^2}
\delta  m_{\tilde{d}_1}^2 + 2
\frac{U_{\tilde{d}_{11}} U_{\tilde{d}_{12}}}{U_{\tilde{d}_{12}}^2}
\delta m_{\tilde{d}_{12}} +
\frac{U_{\tilde{d}_{11}}^2 U_{\tilde{u}_{11}}^2}{U_{\tilde{d}_{12}}^2}
\delta  m_{\tilde{u}_1}^2  \\[1.5mm]&
\nonumber \quad\
+ \frac{U_{\tilde{d}_{11}}^2  U_{\tilde{u}_{12}}^2}{U_{\tilde{d}_{12}}^2}
\delta  m_{\tilde{u}_2}^2 - 2 \frac{U_{\tilde{d}_{11}}^2 U_{\tilde{u}_{12}} U_{\tilde{u}_{22}}}{U_{\tilde{d}_{12}}^2}
\delta m_{\tilde{u}_{12}}
 - \delta C_{\tilde{d}_{22}}  +
\frac{U_{\tilde{d}_{11}}^2}{U_{\tilde{d}_{12}}^2} (\delta 
C_{\tilde{d}_{11}} - \delta C_{\tilde{u}_{11}})\,,
 \\[4mm]\label{deltaAd} 
\delta \! A_d &= \frac{1}{m_d}\Bigl[
\frac{U_{\tilde{d}_{22}}}{U_{\tilde{d}_{12}}} 
\delta  m_{\tilde{d}_1}^2 +
\frac{U_{\tilde{d}_{21}}}{U_{\tilde{d}_{12}}} \delta m_{\tilde{d}_{12}} 
- \frac{U_{\tilde{d}_{11}} U_{\tilde{u}_{11}}^2}{U_{\tilde{d}_{12}}}
\delta  m_{\tilde{u}_1}^2
\\[1.5mm]& \nonumber \quad\ \quad\ - \frac{U_{\tilde{d}_{11}}
  U_{\tilde{u}_{12}}^2}{U_{\tilde{d}_{12}}} 
\delta  m_{\tilde{u}_2}^2 + 2
\frac{U_{\tilde{d}_{11}} U_{\tilde{u}_{12}} U_{\tilde{u}_{22}}}{ 
  U_{\tilde{d}_{12}}} \delta m_{\tilde{u}_{12}}
- \delta C_{\tilde{d}_{12}} -
\frac{U_{\tilde{d}_{11}}}{U_{\tilde{d}_{12}}} (\delta 
C_{\tilde{d}_{11}} - \delta C_{\tilde{u}_{11}})\Bigr]\, .
\end{align} 
\end{subequations}

The counter\-term  $\delta  m_{\tilde{d}_2}^2$ can be 
expressed by the counter\-terms of the soft-breaking para\-meters,
\begin{align}\label{Countermassed1}
\begin{split}
\delta  m_{\tilde{d}_2}^2 &= U_{\tilde{d}_{21}}^2 \delta \!
M_{\tilde Q_L}^2  + 2
U_{\tilde{d}_{21}}U_{\tilde{d}_{22}} m_d \delta \! A_d +
U_{\tilde{d}_{22}}^2 \delta \! M_{{\tilde{d}}_R}^2 \\& \quad\ + U_{\tilde{d}_{21}}^2
\delta C_{\tilde{d}_{11}} +2 U_{\tilde{d}_{21}}U_{\tilde{d}_{22}} \delta C_{\tilde{d}_{12}} +
U_{\tilde{d}_{22}}^2 \delta C_{\tilde{d}_{22}} \, .
\end{split}
\end{align}
\subsection{Renormalization Constants for the Sleptons}\label{ReSle}
The independent parameters in each generation of the sleptons are 
$M_{\tilde l_L}^2$,  $M_{\tilde l_R}^2$, and $A_l$. Their counterterms are determined by renormalizing
the two charged sleptons in the on-shell renormalization scheme. From the previous discussion, one gets 
the on-shell renormalization conditions for the slepton sector,
\begin{eqnarray}
\tilde{Re}\hat{\Sigma}_{\tilde{l}_{ii}}(m_{\tilde{l}_i}^2) =  0 \, , \ \ \
\tilde{Re}\hat{\Sigma}_{\tilde{l}_{12}}(m_{\tilde{l}_1}^2) =  0 \, , \ \ \ 
\tilde{Re}\hat{\Sigma}_{\tilde{l}_{12}}(m_{\tilde{l}_2}^2) = 0 \, ,\nonumber \\
\tilde{Re}\hat{\Sigma}_{\tilde{l}_{ii}}^{\prime}(m_{\tilde{l}_i}^2) =  0 \, , \ \ \ 
\tilde{Re}\hat{\Sigma}_{\tilde{\nu_l}}^{\prime}(m_{\tilde{\nu_l}}^2) =  0\, .
\label{eqn:ReSlecondition}
\end{eqnarray}
Similar to the squark renormalization, the counterterms and the field renormalization constants 
are obtained by solving the equations (\ref{eqn:ReSlecondition}),
\begin{subequations}
\begin{eqnarray}
\delta  m_{\tilde{l}_i}^2 & = & 
 \tilde{Re}\Sigma_{\tilde{l}_{ii}}(m_{\tilde{l}_i}^2)\, ,
\label{eqn:ReSleconstantc}\\
\delta m_{\tilde{l}_{12}}^2 &=&
 \frac{\tilde{Re}\Sigma_{\tilde{l}_{12}}(m_{\tilde{l}_1}^2) +\tilde{Re}\Sigma_{\tilde{l}_{12}}(m_{\tilde{l}_2}^2)}
{2}\, ,
\label{eqn:ReSleconstantd}\\
\delta Z_{\tilde{l}_{ii}}& = & - \tilde{Re}\Sigma_{\tilde{l}_{ii}}^{\prime}(m_{\tilde{l}_i}^2)\, ,
\label{eqn:ReSleconstanta} \\
\delta Z_{\tilde{l}_{12}}& = & - \frac{\tilde{Re}\Sigma_{\tilde{l}_{12}}(m_{\tilde{l}_1}^2) -\tilde{Re}\Sigma_{\tilde{l}_{12}}(m_{\tilde{l}_2}^2)}{m_{\tilde{l}_1}^2-m_{\tilde{l}_2}^2}\, ,
\label{eqn:ReSleconstante} \\
\delta Z_{\tilde{\nu_l}}& = & - \tilde{Re}\Sigma_{\tilde{\nu_l}}^{\prime}(m_{\tilde{\nu_l}}^2)\, .
\label{eqn:ReSleconstantb}
\end{eqnarray}
\end{subequations}
From(\ref{eqn:ReSleconstantc}) and (\ref{eqn:ReSleconstantd}), one finds the expressions for the counterterms  
$\delta M_{\tilde{l}_L}^2, \delta M_{{\tilde{l}}_R}^2$ and $\delta A_l$,
\begin{subequations}
\begin{align}\label{deltaMLl}
\delta \! M_{\tilde l_L}^2 &= U_{\tilde{l}_{11}}^2
\delta  m_{\tilde{l}_1}^2 + U_{\tilde{l}_{12}}^2
\delta  m_{\tilde{l}_2}^2 - 2
U_{\tilde{l}_{12}}U_{\tilde{l}_{22}} \delta m_{\tilde{l}_{12}}
 - \delta C_{\tilde{l}_{11}}\,,
 \\[4mm] 
 \delta \! M_{{\tilde{l}}_R}^2 &= U_{\tilde{l}_{12}}^2
 \delta  m_{\tilde{l}_1}^2 + U_{\tilde{l}_{11}}^2
 \delta  m_{\tilde{l}_2}^2 + 2
 U_{\tilde{l}_{12}}U_{\tilde{l}_{22}} \delta m_{\tilde{l}_{12}} - \delta C_{\tilde{l}_{22}}\, ,
 \label{deltaMlR}\\[4mm] 
 \delta \! A_l &= \frac{1}{m_l}\Bigl[U_{\tilde{u}_{11}}
 U_{\tilde{l}_{12}}\bigl(\delta  m_{\tilde{l}_1}^2 
 - \delta  m_{\tilde{l}_2}^2\bigr)
 + (U_{\tilde{l}_{11}} U_{\tilde{l}_{22}} 
 +U_{\tilde{l}_{12}}
 U_{\tilde{l}_{21}}) \delta m_{\tilde{l}_{12}}
  - \delta C_{\tilde{l}_{12}} \Bigr]\, .
 \label{deltaAl}
\end{align}
\end{subequations}
The counter\-term for the sneutrino mass $M_{\nu_l}$ is not independent. It can be 
expressed by $\delta M_{\tilde{l}_L}^2$ as in (\ref{sneutrinodeltamassb}).
\newpage
\section{Renormalization of the Higgs sector}\label{ReHiggs}
In the CP-conserving MSSM the bilinear terms of the Lagrangian which describe the MSSM Higgs sector 
can be written as
 \begin{eqnarray}
\mathcal{L}& =&\left(\begin{array}{cc}
h_1^\dagger & h_2^\dagger 
\end{array}\right) \left (p^2 - M_{h}^D\right )\left(\begin{array}{c}
h_1 \\
 h_2 
\end{array}\right)\, ,
\label{HiggsLagrangian}
\end{eqnarray}
where the Higgs multiplet $\left(\begin{array}{cc}
h_1 & h_2 
\end{array}\right)$ can be
$\left(\begin{array}{cc}
A^0 & G^0 
\end{array}\right)\, , \, \left(\begin{array}{cc}
h^0 & H^0 
\end{array}\right)$\, or\, $\left(\begin{array}{cc}
H^\pm & G^\pm 
\end{array}\right)$.  
 $M_h^D$ is the mass matrix of the Higgs bosons, 
$M_h^D = U_{\beta}M_{\chi^0}U_{\beta}^\dagger$, \ \ $U_{\alpha}M_{\phi^0}U_{\alpha}^\dagger$\, or
\ \ $U_{\beta}M_{\phi^\pm}U_{\beta}^\dagger$, respectively.
Their definitions and expressions can be found in (\ref{eqn:ReHiggs1}) and (\ref{eqn:ReHiggs5}).

In order to renormalize the Higgs sector we introduce the renormalization constants for their mass matrices 
and fields by the transformations
\begin{eqnarray}
M_{\chi^0} \to M_{\chi^0} + \delta M_{\chi^0}&,&\ \ \ 
M_{\phi^0}\to  M_{\phi^0} + \delta M_{\phi^0}\, ,\nonumber \\
M_{\phi^\pm}\to  M_{\phi^\pm} + \delta M_{\phi^\pm}&,&\ \ \ 
\left(\begin{array}{c}
h_1 \\
 h_2 
\end{array}\right) \longrightarrow \left (1 + \frac{1}{2}\delta Z_h\right )\left(\begin{array}{c}
h_1 \\
 h_2 
\end{array}\right )\, ,
\label{ReconstantHiggs}
\end{eqnarray}
where $\delta Z_h$ is a general $2\times 2$ matrix. 
Inserting these transformations into the Lagrangian (\ref{HiggsLagrangian}), one can get the counterterm Lagrangian
 \begin{eqnarray}
\delta \mathcal{L}=\left(\begin{array}{cc}
h_1^\dagger & h_2^\dagger 
\end{array}\right)\Bigl [\frac{1}{2}p^2 
\left (\delta Z_h + \delta Z_h^\dagger \right )-
\frac{1}{2}\left (M_h^D \delta Z_h + \delta Z_h^\dagger M_h^D \right )
 -  \delta M_h^D \Bigr ]\left(\begin{array}{c}
h_1 \\
 h_2 
\end{array}\right)\, .
\label{HiggsLagrangianCT}
\end{eqnarray}
The renormalized self-energies for the Higgs bosons can be derived from the counterterm Lagrangian:
 \begin{eqnarray}
\hat{\Sigma}_{ij}(p^2)& = & \Sigma_{ij}(p^2) + \Bigl [\frac{1}{2}p^2\left (\delta Z + \delta Z^\dagger\right )
-\frac{1}{2}\left (M_h^D \delta Z_h + \delta Z_h^\dagger M_h^D \right )
 -  \delta M_h^D \Bigr ]_{ij}\, .
\end{eqnarray}
The independent parameters in the Higgs sector are chosen to be the tadpoles
$T_{h^0}, \, T_{H^0}$, 
the mass $m_{A^0}^2$, and the ratio of the vacuum expectation values $\tan\beta$.
Their counterterms are fixed in the Higgs sector.
\newpage
\subsection{Counterterms for the Tadpoles}
The tadpoles $T_{h^0}$ and $T_{H^0}$ are equal to zero at tree-level.
Their counterterms can be fixed by requiring that the renormalized tadpoles,
which are defined as the unrenormalized ones plus the counterterms, are equal to zero at one-loop order as well,
\begin{eqnarray}
\tilde{Re}\hat{T}_{h^0} & = & \tilde{Re}T_{h^0} + \delta T_{h^0} = 0\, , \label{eqn:ReHiggs13a}\\
\tilde{Re}\hat{T}_{H^0} & = & \tilde{Re}T_{H^0} + \delta T_{H^0} = 0\, .
\label{eqn:ReHiggs13b}
\end{eqnarray} 
Hence the counterterms for the Tadpoles can be expressed as
\begin{eqnarray}
 \delta T_h = - \tilde{Re}T_{h^0}\, ,\ \ 
\delta T_H = - \tilde{Re}T_{H^0}\, .
\end{eqnarray}
\subsection{Counterterm for $\tan\beta$}
Since $\tan\beta$ is the ratio of the vacuum expectation values, $\tan\beta = \frac{v_2}{v_1}$,
we introduce the renormalization constants for the vacuum expectation values $v_1$ and $v_2$,
 \begin{eqnarray}
v_1 &\to& \left (1 + \frac{1}{2}\delta Z_{H_1}\right )\left (v_1 + \delta v_1\right )\, ,
\nonumber \\
v_2 &\to& \left (1 + \frac{1}{2}\delta Z_{H_2}\right )\left (v_2 + \delta v_2\right )\, .
\end{eqnarray}
$\delta Z_{H_1}$ and $\delta Z_{H_2}$ are the field renormalization constants for the Higgs doublets $H_1$ and $H_2$, respectively.
\begin{eqnarray}
\frac{v_2}{v_1}& \to & \frac{\left (1 + \frac{1}{2}\delta Z_{H_2}\right )\left (v_2 + \delta v_2\right )}
{\left (1 + \frac{1}{2}\delta Z_{H_1}\right )\left (v_1 + \delta v_1\right )}
\overset{\textrm{one-loop}}{=} \frac{v_2}{v_1}\frac{1 + \frac{1}{2}\delta Z_{H_2}+ \frac{\delta v_2}{v_2}}
{1 + \frac{1}{2}\delta Z_{H_1}+ \frac{\delta v_1}{v_1}}\nonumber \\
&\approx &\frac{v_2}{v_1}\left (1 + \frac{1}{2}\delta Z_{H_2}+ \frac{\delta v_2}{v_2}\right )
\left (1 - \frac{1}{2}\delta Z_{H_1}- \frac{\delta v_1}{v_1}\right )\nonumber \\
& = & \frac{v_2}{v_1}\left (1 + \frac{1}{2}\delta Z_{H_2} - \frac{1}{2}\delta Z_{H_1}\right )\, ,
\label{eqn:tranformationVEV}
\end{eqnarray}
where we take $\frac{\delta v_2}{v_2} - \frac{\delta v_1}{v_1}=0$. The transformation (\ref{eqn:tranformationVEV})
can also be expressed as, 
\begin{eqnarray}
\delta \tan\beta = \tan\beta \frac{\delta Z_{H_2} -\delta Z_{H_1}}{2}.
\end{eqnarray}

The Lagrangian of the coupling of the neutral Higgs boson to $Z$ boson is  
\begin{eqnarray}
\mathcal{L}_{\chi Z}& = & i m_Z \left (p^{\mu}\chi_1 Z_{\mu}\cos\beta + p^{\mu}\chi_2 Z_{\mu}\sin\beta \right )\, ,
\end{eqnarray}
After renormalization, its counterterm Lagrangian can be written as
 \begin{eqnarray}
 \mathcal{L}_{\chi Z_{\rm CT}}& = & \mathcal{L}_{A^0Z {\rm CT}} +  \mathcal{L}_{G^0Z{\rm CT}}\, ,
\nonumber \\
 \mathcal{L}_{A^0Z_{\rm CT}} &= &- im_Z p^{\mu}A^0 Z_{\mu}\sin\beta 
\cos\beta\left(\delta Z_{H_2} -\delta Z_{H_1}\right )\, ,
\nonumber \\
  \mathcal{L}_{G^0Z_{\rm CT}}& = &  im_Z p^{\mu}G^0 Z_{\mu}\frac{\delta Z_{G^0G^0} + \delta Z_{ZZ}}{2}
+  i\delta m_Z p^{\mu}G^0 Z_{\mu}\, .
\end{eqnarray}
Hence the renormalized self-energies are expressed as
\begin{eqnarray}
\hat{\Sigma}_{A^0Z}(p^2) & = & \Sigma_{A^0Z}(p^2)- im_Z \sin\beta \cos\beta\left(\delta Z_{H_2} -\delta Z_{H_1}\right )\, ,
\nonumber \\
\hat{\Sigma}_{G^0Z}(p^2) & = & \Sigma_{G^0Z}(p^2) +i\left (m_Z\frac{\delta Z_{G^0G^0} + \delta Z_{ZZ}}{2}+  \delta m_Z\right )\, .
\end{eqnarray}
According to \cite{RetanBOS} one can determine $\delta \tan\beta$ by the requirement that the $A^0-Z$ mixing vanishes 
for an on-shell $A^0$ boson, 
\begin{eqnarray}
\tilde{Re}\hat{\Sigma}_{A^0Z}(m_{A^0}^2) = 0 &\Rightarrow & 
\frac{\delta \tan\beta^{\rm OS}}{\tan\beta} = \frac{1}{2m_Z \sin\beta \cos\beta}{\rm Im}\left[\tilde{Re}\Sigma_{AZ}(m_A^2)\right ].
\end{eqnarray}
A convenient choice is the $\overline{\rm DR}$ renormalization of $\tan\beta$ \cite{RetanBMS},
which means that the counterterm only contains the UV-divergent parts,
\begin{eqnarray}
\frac{\delta\tan\beta^{\overline{\rm DR}}}{\tan\beta}& = &
\frac{1}{2m_Z\sin\beta\cos\beta} \bigl[ \mathrm{Im} \Sigma_{A^0Z}(m_A^2)
\bigr]_{\rm div}\, .
 \end{eqnarray}
Here the subscript "div' means that only the UV-divergent parts are considered.
Since this choice has the advantage of providing the gauge invariant and process independent counterterms,
it has been assessed to be the best choice of defining $\tan\beta~$\cite{Dominik}.

\subsection{Renormalization Constants for the Neutral CP-odd Higgs Bosons}
The renormalized self-energies for the CP-odd Higgs bosons can be expressed as
\begin{subequations}
\label{eqn:ReHiggs12}
\begin{eqnarray}
\hat{\Sigma}_{A^0A^0}(p^2) & = & \Sigma_{A^0A^0}(p^2)+ \left (p^2 - m_{A^0}^2 \right )\delta Z_{A^0A^0}-\delta m_{A^0}^2\, ,
\label{eqn:ReHiggs12a} \\
\hat{\Sigma}_{G^0G^0}(p^2) & = & \Sigma_{G^0G^0}(p^2)+ \left (p^2 - m_{G^0}^2\right )\delta Z_{G^0G^0}- 
\delta m_{G^0}^2 \, ,
\label{eqn:ReHiggs12b} \\
\hat{\Sigma}_{A^0G^0}(p^2) & = & \Sigma_{A^0G^0}(p^2)+ \frac{1}{2}p^2\left (\delta Z_{A^0G^0} + \delta Z_{G^0A^0}
\right )\, \nonumber \\ 
&&{}-\frac{1}{2} m_{A^0}^2\delta Z_{A^0G^0}-\frac{1}{2} m_{G^0}^2\delta Z_{G^0A^0} -\delta m_{A^0G^0}^2\, .
\label{eqn:ReHiggs12c}
\end{eqnarray}
\end{subequations}
From (\ref{eqn:ReHiggs6a}) and (\ref{eqn:ReHiggs6b}), one gets the explicit expressions for the counterterms $\delta m_{A^0G^0}^2$ 
and $\delta m_{G^0}^2$, 
\begin{subequations}
\begin{eqnarray}
\delta m_{A^0G^0}^2 &=&  - 
\frac{e}{2m_Z s_W c_W}\left (\delta T_{H^0} \sin(\alpha - \beta) + \delta T_{h^0} \cos(\alpha - \beta)\right )\, ,\\
\delta m_{G^0}^2 &=& \frac{e}{2m_Z s_W c_W}\left 
(-\delta T_{H^0} \cos(\alpha - \beta) + \delta T_{h^0} \sin(\alpha - \beta)\right )\, ,
\end{eqnarray}
\end{subequations}
which are dependent on the counterterms $\delta T_{h^0}$ and  $\delta T_{H^0}$.

The counterterm for $m_{A^0}^2$ is determined by renormalizing the neutral CP-odd Higgs boson $A^0$ via the 
on-shell renormalization scheme,
\begin{eqnarray}
\tilde{Re}\hat{\Sigma}_{A^0A^0}(m_{A^0}^2)&=&0\, ,
\label{eqn:ReHiggs14}
\end{eqnarray}
which makes the renormalized mass equals to the pole of the propagator. The diagonal entries of the field renormalization matrix
are fixed such that the residues of the renormalized propagators are equal to $1$, 
\begin{eqnarray}
\tilde{Re}\hat{\Sigma}_{A^0A^0}^{\prime}(m_{A^0}^2)=0\, ,\ \ \ 
\tilde{Re}\hat{\Sigma}_{G^0G^0}^{\prime}(m_{G^0}^2)=0 \, .
\label{eqn:ReHiggs15b}
\end{eqnarray}
The on-shell renormalization scheme also requires that the renormalized 1PI two-point function for the CP-odd Higgs boson
is diagonal for the external on-shell particles, which determine the non-diagonal entries of the field renormalization matrix,
\begin{eqnarray}
\tilde{Re}\hat{\Sigma}_{A^0G^0}(m_{A^0}^2)=0\, , \ \ \ 
\tilde{Re}\hat{\Sigma}_{A^0G^0}(m_{G^0}^2)=0\, .  
\label{eqn:ReHiggs16b}
\end{eqnarray}
Inserting the expressions for the renormalized self-energies (\ref{eqn:ReHiggs12})
into the on-shell renormalization conditions above, we obtain the expressions for the renormalization constants,
\begin{subequations}
\begin{eqnarray}
\delta m_{A^0}^2 &=&\tilde{Re}\Sigma_{A^0A^0}(m_{A^0}^2)\, ,
\\
\delta Z_{A^0A^0}& =& - \tilde{Re}\Sigma_{A^0A^0}^{\prime}(m_{A^0}^2)\, ,
\\
\delta Z_{G^0G^0} &= &- \tilde{Re}\Sigma_{G^0G^0}^{\prime}(m_{G^0}^2)\, ,
\\
 \delta Z_{G^0A^0} &=& \frac{2\left(-\tilde{Re}\Sigma_{A^0G^0}(m_{A^0}^2)+\delta m_{A^0G^0}^2 
\right)}{m_{A^0}^2- m_{G^0}^2}\, ,
\\
 \delta Z_{A^0G^0} & = &\frac{2\left(-\tilde{Re}\Sigma_{A^0G^0}(m_{G^0}^2)+\delta m_{A^0G^0}^2 
\right)}{m_{G^0}^2- m_{A^0}^2}\, .
\end{eqnarray}
\end{subequations}
\subsection{Renormalization Constants for the Neutral CP-even Higgs Bosons}
The renormalized self-energies for the CP-even Higgs bosons can be expressed as
\begin{eqnarray}
\hat{\Sigma}_{h^0h^0}(p^2) & = & \Sigma_{h^0h^0}(p^2)+ \left (p^2 - m_{h^0}^2 \right )\delta Z_{h^0h^0}- \delta m_{h^0}^2\, ,
\nonumber\\
\hat{\Sigma}_{H^0H^0}(p^2) & = & \Sigma_{H^0H^0}(p^2)+ \left (p^2 - m_{H^0}^2 \right )\delta Z_{H^0H^0}-\delta m_{H^0}^2\, ,
\nonumber\\
\hat{\Sigma}_{h^0H^0}(p^2) & = & \Sigma_{h^0H^0}(p^2)+ \frac{1}{2}p^2\left (\delta Z_{h^0H^0}+ 
\delta Z_{H^0h^0} \right ) - \nonumber \\
&&{}\frac{1}{2}\left( m_{h^0}^2 \delta Z_{h^0H^0} + m_{H^0}^2 \delta Z_{H^0h^0}\right )
 - \delta m_{h^0H^0}^2\, .
\label{eqn:Reself-EvenHiggs}
\end{eqnarray}
From (\ref{eqn:ReHiggs5a}), (\ref{eqn:ReHiggs5b}) and (\ref{eqn:ReHiggs5c}) one derives the expressions
for the counterterms $\delta m_{h^0}^2$, $\delta m_{H^0}^2$ and $\delta m_{h^0H^0}^2$,
\begin{subequations}
\label{eqn:deltamassevenH}
\begin{eqnarray}
\delta m_{h^0}^2 &=& \delta m_{A^0}^2\cos^2(\alpha - \beta) + 
\delta m_Z^2\sin^2(\alpha + \beta) +
\nonumber \\ 
&&{} \frac{e}{2m_Z s_W c_W}\delta T_{H^0} \cos(\alpha - \beta)\sin^2(\alpha - \beta) + \nonumber \\
&&{} \frac{e}{2m_Z s_W c_W}\delta T_{h^0} \sin(\alpha - \beta)(1 +\cos^2(\alpha - \beta)) + \nonumber \\
&&{} \delta \tan\beta \cos^2\beta \left(m_{A^0}^2\sin 2(\alpha - \beta) + m_Z^2\sin 2(\alpha + \beta)\right ) \, ,\\
\delta m_{H^0}^2 &=& \delta m_{A^0}^2\sin^2(\alpha - \beta) + \delta m_Z^2\cos^2(\alpha + \beta) -
\nonumber \\ 
&&{} \frac{e}{2m_Z s_W c_W}\delta T_{H^0} \cos(\alpha - \beta)\left (1 + \sin^2(\alpha - \beta)\right ) - \nonumber \\
&&{} \frac{e}{2m_Z s_W c_W}\delta T_{h^0} \sin(\alpha - \beta)\cos^2(\alpha - \beta) \nonumber - \\
&&{}  \delta \tan\beta \cos^2\beta \left(m_{A^0}^2\sin 2(\alpha - \beta) + m_Z^2\sin 2(\alpha + \beta)\right ) \, ,\\
\delta m_{h^0H^0}^2 &=& \frac{1}{2}\left (\delta m_{A^0}^2\sin2(\alpha - \beta) -
\delta m_Z^2\sin2(\alpha + \beta)\right ) +
\nonumber \\ 
&&{} \frac{e}{2m_Z s_W c_W}\left (\delta T_{H^0} \sin^3(\alpha - \beta) - \delta T_{h^0} \cos^3(\alpha - \beta)\right ) \nonumber - \\
&&{}  \delta \tan\beta \cos^2\beta \left(m_{A^0}^2\cos 2(\alpha - \beta) + m_Z^2\cos 2(\alpha + \beta)\right )\, ,
\end{eqnarray}
\end{subequations}
where all the counterterms appearing on right-hand side are already determined.
Hence in the neutral CP-even Higgs sector, only the field renormalization constants are to be fixed.
Similarly to case of the neutral CP-odd Higgs boson, the on-shell renormalization conditions for the
 CP-even Higgs sector can be formulated as
\begin{eqnarray}
\tilde{Re}\hat{\Sigma}_{h^0h^0}^{\prime}(m_{h^0}^2)=0 \, , \ \ \
\tilde{Re}\hat{\Sigma}_{H^0H^0}^{\prime}(m_{H^0}^2)=0 \, ,\nonumber \\
\tilde{Re}\hat{\Sigma}_{h^0H^0}(m_{h^0}^2)=0 \, ,\ \ \
\tilde{Re}\hat{\Sigma}_{h^0H^0}(m_{H^0}^2)=0 \, .
\end{eqnarray}
Applying the expressions for the renormalized self-energies (\ref{eqn:Reself-EvenHiggs}) to the on-shell conditions above, one 
obtains the field renormalization constants, 
\begin{eqnarray}
\delta Z_{h^0h^0} &=& - \tilde{Re}\Sigma_{h^0h^0}^{\prime}(m_{h^0}^2)\, ,\ \
\delta Z_{H^0H^0} = - \tilde{Re}\Sigma_{H^0H^0}^{\prime}(m_{H^0}^2)\, ,
\nonumber\\
\delta Z_{H^0h^0} &=& 2\frac{-\tilde{Re}\Sigma_{h^0H^0}(m_{h^0}^2)+\delta m_{h^0H^0}^2}
{m_{h^0}^2 - m_{H^0}^2}\, , \nonumber\\ 
\delta \tilde{Z}_{h^0H^0} &=& 2\frac{- \tilde{Re}\Sigma_{h^0H^0}(m_{H^0}^2)+\delta m_{h^0H^0}^2}
{m_{H^0}^2 - m_{h^0}^2}\, .
\end{eqnarray}
\subsection{Renormalization Constants for the Charged Higgs Bosons}
The renormalized self-energies for the charged Higgs bosons can be expressed as
\begin{subequations}
\label{eqn:ReCHiggs12c}
\begin{eqnarray}
\hat{\Sigma}_{H^-H+}(p^2) & = & \Sigma_{H^-H^+}(p^2)+ \left (p^2 - m_{H^\pm}^2 \right )\delta Z_{H^-H^+}-\delta m_{H^\pm}^2\, ,
\label{eqn:ReCHiggs12a} \\
\hat{\Sigma}_{G^-G^+}(p^2) & = & \Sigma_{G^-G^+}(p^2)+ \left (p^2 - m_{G^\pm}^2\right )\delta Z_{G^-G^+}- 
\delta m_{G^\pm}^2 \, ,
\label{eqn:ReCHiggs12b} \\
\hat{\Sigma}_{H^-G^+}(p^2) & = & \Sigma_{H^-G^+}(p^2)+ \frac{1}{2}p^2\left (\delta Z_{H^-G^+} + \delta Z_{H^-G^+}
\right )\, \nonumber \\ 
&&{}-\frac{1}{2} m_{H^\pm}^2\delta Z_{H^-G^+}-\frac{1}{2} m_{G^+}^2\delta Z_{G^+H^-} -\delta m_{H^-G^+}^2\, .
\end{eqnarray}
\end{subequations}
From (\ref{eqn:ChargeHiggsmass}) one gets the explicit expressions for the counterterms 
$\delta m_{H^\pm}^2$, $\delta m_{H^-G^+}^2$ and $\delta m_{G^\pm}^2$, 
\begin{subequations}
\begin{eqnarray}
\delta m_{H^\pm}^2 & = & \delta m_{A^0}^2 + \delta m_W^2\, , \\
\delta m_{H^-G^+}^2 &=&  - 
\frac{e}{2m_Z s_W c_W}\left (\delta T_{H^0} \sin(\alpha - \beta) + \delta T_{h^0} \cos(\alpha - \beta)\right )\, \nonumber \\
&&{}- \delta\tan\beta m_{H^\pm}^2 \cos^2\beta \, ,\\
\delta m_{G^\pm}^2 &=& \frac{e}{2m_Z s_W c_W}\left 
(-\delta T_{h^0} \cos(\alpha - \beta) + \delta T_{h^0} \sin(\alpha - \beta)\right )\, .
\end{eqnarray}
\end{subequations}
The field renormalization constants are determined by the on-shell conditions, 
\begin{subequations}
\begin{eqnarray}
\tilde{Re}\hat{\Sigma}_{H^-H^+}^{\prime}(m_{H^\pm}^2)=0 &\Longrightarrow & Z_{H^-H^+} =- \tilde{Re}\Sigma_{H^-H^+}^{\prime}(m_{H^\pm}^2)\, , \\
\tilde{Re}\hat{\Sigma}_{G^-G^+}^{\prime}(m_{G^\pm}^2)=0 &\Longrightarrow & Z_{G^-G^+} =- \tilde{Re}\Sigma_{G^-G^+}^{\prime}(m_{G^\pm}^2)\, ,\\
\tilde{Re}\hat{\Sigma}_{H^-G^+}(m_{H^\pm}^2)=0 &\Longrightarrow & 
\delta Z_{G^+H^-} = 2\frac{-\tilde{Re}\Sigma_{H^-G^+}(m_{H^-}^2)+\delta m_{H^-G^+}^2}{m_{H^\pm}^2 - m_{G^\pm}^2}\, ,\\
\tilde{Re}\hat{\Sigma}_{H^-G^+}(m_{G^\pm}^2)=0 &\Longrightarrow & \delta \tilde{Z}_{H^-G^+} =2\frac{- \tilde{Re}\Sigma_{H^-G^+}(m_{G^\pm}^2)+
\delta m_{H^-G^+}^2}
{m_{G^\pm}^2 - m_{H^\pm}^2}\, .
\end{eqnarray}
\end{subequations}
One can also fix the field renormalization constants for the Higgs fields via the $\overline{\rm DR}$ scheme, where the counterterms 
contain only the UV-divergent parts, 
\begin{eqnarray}
\delta Z^{\overline{\rm DR}} & = & \left (\delta Z^{\rm OS}\right )_{\rm div}\, .
\end{eqnarray}
\chapter[Calculations for $\tilde \chi_2^0$ Decay]{Calculations for the Next-to-lightest Neutralino $\tilde \chi_2^0$ Decay}\label{calculations}
The Minimal Supersymmetric Standard Model (MSSM) is the supersymmetric extension of 
the Standard Model (SM) with the minimal particle content,
the details have been presented in Chapter~\ref{MSSM}. 
In the MSSM with conserved
$R$-parity, the lightest supersymmetric particle (LSP), which in many scenarios
is the lightest neutralino $\tilde{\chi}_1^0$, appears at the end of the decay
chain of each supersymmetric particle. The LSP escapes the detector, giving
the characteristic SUSY signature of missing energy. While this helps to
suppress backgrounds from SM processes, it also makes the measurement of
supersymmetric particle masses more difficult. 

At the LHC, the total SUSY production-cross section is expected to be
dominated by the production of gluinos and squarks, which decay into lighter
charginos or neutralinos.
Of particular interest are decay chains leading to
the next-to-lightest neutralino $\tilde{\chi}_2^0$.  $\tilde \chi_2^0$ in
turn can always decay into the LSP $\tilde\chi_1^0$ and two fermions $f \bar f$, 
at least for light SM fermions $f$. 
The leptonic final states
are of particular interest, since they can be identified relatively easily
even at the LHC. Depending on neutralino, slepton and Higgs boson masses, the possible leptonic decays
of $\tilde \chi_2^0$ are three-body decays $\tilde\chi_2^0\rightarrow \tilde \chi_1^0 l^- l^+$,
cascade two-body decays $\tilde\chi_2^0 \rightarrow \tilde l^\pm l^\mp \rightarrow \tilde \chi_1^0 l^- l^+$ 
\begin{figure}[htb]
\begin{center}
\begin{tabular}{c}
\includegraphics[width=0.4\linewidth]{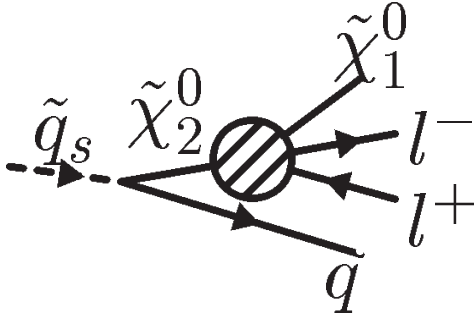}
\end{tabular}
\caption{The Feynman diagram for the squark decay 
$\tilde q_s \to q \tilde\chi_2^0 \to q\tilde\chi_1^0 l^- l^+$ at the LHC, $s = 1,2$ labels the squark
 mass eigenstates. \label{fig.squarkdecay}}
\end{center}
\end{figure}
and/or $\tilde \chi_2^0 \rightarrow \tilde \chi_1^0 Z (\phi^0) \rightarrow
\tilde \chi_1^0 l^- l^+$,
where $\phi^0$ stands for one of the three neutral Higgs bosons or the neutral Goldstone boson of the MSSM. 
The dilepton invariant-mass distribution of these decays has a specific shape with a sharp edge near the
endpoint which only depends on the kinematics. 
This distribution can be measured accurately at the LHC. 
In particular, 
its endpoint is used in several analyses that aim to reconstruct (differences of) 
supersymmetric particle masses \cite{LHC, LHC/LC}.
Under favorable circumstances it has been shown
that the endpoint can be measured to an accuracy of $0.1\%$ at the LHC
\cite{LHC}. In order to match this accuracy in the theoretical prediction, at least one-loop corrections to
$\tilde \chi_2^0$ decays have to be included.

Turning to the planned $e^+e^-$ linear collider ILC, $\tilde \chi_1^0 \tilde
\chi_2^0$ production is often the first process that is kinematically
accessible \cite{ddk1} (other than $\tilde \chi_1^0$ pair production, which
leads to an invisible final state). The detailed analysis of $\tilde \chi_2^0$
decays can then yield information about heavier supersymmetric particles. Under favorable
circumstances, ${\cal O}(10^4)$ $\chi_2^0 \rightarrow \chi_1^0 l^+ l^-$ decays
may be observed at the ILC, again making the inclusion of quantum corrections
mandatory to match the experimental precision. 

In this chapter, we calculate leptonic $\tilde \chi_2^0$ decays at one-loop
level. Cases where $\tilde \chi_2^0$ has two-body decays
$\tilde\chi_2^0 \rightarrow \tilde l_1^\pm l^\mp \rightarrow \tilde \chi_1^0 l^- l^+$  
\ ($\tilde l_1$ stands for the lighter one of the two charged sleptons) are treated 
both completely and in a single-pole approximation.
In the complete calculation one has to employ complex slepton masses in the relevant propagators and
one-loop integrals. The single-pole approximation in this case is performed in the way that
the $\tilde{\chi}_2^0$ decays are treated as the production and decay of the sleptons
$\tilde{l}_1$. 
We compare the results from the complete and approximate calculations 
and find a good agreement.
We also analyze a scenario where $\tilde \chi_2^0$ only has three-body decays.
The virtual photonic contributions are infrared (IR) divergent, hence
the contributions of the real photon bremsstrahlung 
must be added to the one-loop corrrections in order to cancel these IR divergences.
In addition to calculating the integrated partial widths, we study the
differential decay width of $\tilde{\chi}_2^0$ as a function of the dilepton
invariant mass. In order to obtain the total decay width of $\tilde\chi_2^0$ and hence the 
branching ratios of the leptonic decays, the invisible decays 
$\tilde\chi_2^0 \rightarrow \tilde \chi_1^0 \nu_l \bar \nu_l$ and the hadronic decays
$\tilde\chi_2^0 \rightarrow \tilde \chi_1^0 q \bar q$ are also calculated.

This chapter is organized as follows. Section \ref{tree} gives
the tree-level calculations for $\tilde\chi_2^0$ leptonic decays. 
In Section \ref{complete} we discuss how to calculate these decays completely at one-loop level,
where the virtual corrections and the real photon bremsstrahlung are considered in detail.
When the lighter sleptons $\tilde l_1$ can be on shell, these decays are calculated approximately
in Section \ref{approx}.
The total decay width of $\tilde\chi_2^0$ and the branching ratios of the leptonic decays 
are discussed in Section \ref{brlepton}, where the invisible decays and the hadronic decays are calculated.
The numerical results and discussions are given in Section \ref{numerical},
where the SPS1a parameter set \cite{SPS1aa, SPS1ab} is presented in detail.

\newpage
\section{Tree-level Calculations for $\tilde\chi_2^0\rightarrow \tilde \chi_1^0 l^- l^+$}\label{tree}
\begin{figure}[b]
\begin{center}
\begin{tabular}{cccc}
\includegraphics[width=0.22\linewidth]{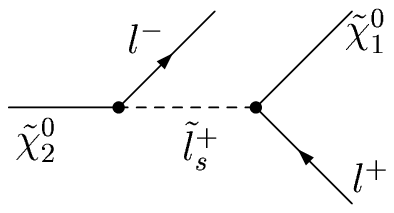} &
\includegraphics[width=0.22\linewidth]{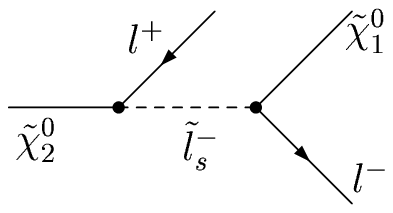}& 
\includegraphics[width=0.22\linewidth]{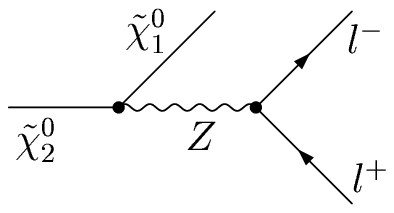} &
\includegraphics[width=0.22\linewidth]{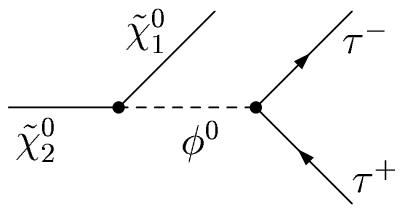} \\
(a) & (b) & (c) & (d)
\end{tabular}
\end{center}
\caption{The Born Feynman diagrams for $\tilde{\chi}_2^0 \longrightarrow
  \tilde{\chi}_1^0 l^- l^+ (l = e, \mu, \tau)$, $s = 1,2$ labels the slepton
  mass eigenstates, $\phi^0$ denotes the MSSM neutral Higgs boson $h^0, H^0, A^0$
and neutral Goldstone boson $G^0$. Since the Yukawa coupling $\phi^0 l^- l^+$ is proportional
to the lepton mass, the Higgs intermediate states are neglected when $l = e$ and $\mu$. 
\label{treediagram}}
\end{figure}

The Born Feynman diagrams for $\tilde{\chi}_2^0(k_1) \longrightarrow
\tilde{\chi}_1^0(k_2) l^-(k_3) l^+(k_4) (l = e, \mu, \tau)$ are displayed in
Figure~\ref{treediagram}. The Mandelstam variables are defined as
\begin{eqnarray}
T = \left (k_1 - k_3 \right )^2 \, , \ \ \ T_{12} = \left (k_1 - k_2 \right )^2\, , \ \ \ 
S_{23} = \left (k_2 + k_3 \right )^2 \, .
\end{eqnarray}
The decay width of this process can be written as (see Appendix~\ref{Kinematics})
\begin{eqnarray}
\Gamma^{(0)} & = & \frac{1}{(2\pi)^5}\frac{1}{2m_{\tilde \chi_2^0}}\int \sum \bigl |M^{(0)}\bigr |^2 d\Phi_{1\to3}\, ,
\label{eqn:bornGamma}
\end{eqnarray}
where $M^{(0)}$ is the matrix element of the Born diagrams, it is squared and averaged over the spin of
the external particles. The expressions for the phase-space element $d\Phi_{1\to3}$ can be found 
in Appendix~\ref{Kinematics}.

If the two-body decays $\tilde{\chi}_2^0\rightarrow \tilde{l}^\pm_1 l^\mp \rightarrow
\tilde{\chi}_1^0 l^- l^+$ are kinematically allowed, i.e. the sleptons
$\tilde{l}_1$ can be on shell at some points in the phase space, 
a finite width of $\tilde{l}_1$ is necessary. 
It arises from the imaginary part of the slepton self-energy.
A finite width is introduced via Dyson summation
\begin{eqnarray}
\frac{i}{k^2 -  m_{\tilde{l}_1}^2}+ \frac{i}{k^2 - m_{\tilde{l}_1}^2}
 i \hat{\Sigma}(k^2) \frac {i} {k^2 - m_{\tilde{l}_1}^2 }
 + \cdots & = & \frac{i}{k^2 -  m_{\tilde{l}_1}^2+ \hat{\Sigma}(k^2)}\, ,
\label{eqn:completetree1a}
 \end{eqnarray}
 where $\hat{\Sigma}(k^2)$ is the renormalized $\tilde{l}_1$
 self-energy.
Following Ref.\cite{douple-pole}, a gauge invariant matrix element is obtained 
 by a Laurent expansion around the complex pole:
\begin{eqnarray}
\frac{1}{k^2 -  m_{\tilde{l}_1}^2+ \hat{\Sigma}(k^2)}& = &  
\frac{1}{k^2 -  m_p^2} 
\left(1 - \frac{{\rm Re}\hat{\Sigma}(k^2)}{k^2 -
    m_{\tilde{l}_1}^2}\right)\, , 
\label{eqn:tree2a}
\end{eqnarray}
were $m_p^2$ denotes the position of the complex pole in (\ref{eqn:completetree1a}).
It is obtained as the solution of
\begin{eqnarray}
m_p^2 - m_{\tilde{l}_1}^2 +\hat{\Sigma}(m_p^2)=0\, .
\label{eqn:complexpole-eq}
\end{eqnarray}

For the tree-level amplitude the complex pole $m_p^2$ is calculated at one-loop level.
Its explicit expression is 
\begin{eqnarray}
m_p^2 = m_{\tilde{l}_1}^2 -  i {m_{\tilde{l}_1}} \Gamma_{\tilde{l}_1}^{(0)}\, ,
\end{eqnarray}
where we have employed on-shell renormalization scheme as in Section \ref{ReSle},
and $\Gamma_{\tilde{l}_1}^{(0)}$ is the tree-level decay width of
$\tilde{l}_1$, $m_{\tilde{l}_1} \Gamma_{\tilde{l}_1}^{(0)}$ is the
imaginary part of the slepton self-energy $\Sigma(m_{\tilde{l}_{1}}^2)$.
Since the second term in the parentheses in (\ref{eqn:tree2a}) is at one-loop level,
we do not need it in the tree-level calculations.
Therefore, the gauge-invariant tree-level amplitude for the decays $\tilde\chi_2^0 \rightarrow \tilde \chi_1^0 l^- l^+$
can be written as

\begin{eqnarray}\label{eqn:propM0}
M^{(0)}& = & 
\frac{V^{(0)}_{\tilde{\chi}_2^0\tilde{l}_1^{\pm}l^{\mp}}(k^2) 
V^{(0)}_{\tilde{l}_1^{\pm}\tilde{\chi}_1^0 l^{\pm}}(k^2)}{k^2 -m_{\tilde{l}_1}^2 + i
  m_{\tilde{l}_1} \Gamma_{\tilde{l}_1}^{(0)}} + B(k^2)\, , 
\end{eqnarray}
where $V^{(0)}_{\tilde{\chi}_2^0\tilde{l}_1^{\pm}l^{\mp}}$ and $V^{(0)}_{\tilde{l}_1^{\pm}\tilde{\chi}_1^0 l^{\pm}}$ 
represent the $\tilde{\chi}_2^0\tilde{l}_1^{\pm}l^{\mp}$ and $\tilde{l}_1^{\pm}\tilde{\chi}_1^0 l^{\pm}$ vertices,
respectively, $B(k^2)$ denotes the non-resonant part of the matrix element, i.e. 
the matrix element of the diagram (a) and (b) for $s=2$ and
diagram (c) and (d) in Figure \ref{treediagram}.


Using the residue theorem one can easily obtain the relation for an analytic function $f(k^2)$,
\begin{equation} \label{eqn:propid}
\int_{-\infty}^{\infty} dk^2 \frac {f(k^2)} {\left| k^2 - m_{\tilde l_1}^2 + i
    \Gamma_{\tilde l_1}^{(0)} m_{\tilde l_1} \right|^2} \simeq \frac {\pi} {m_{\tilde
    l_1} \Gamma_{\tilde l_1}^{(0)}}f( m_{\tilde l_1}^2)\, , \ \ \ {\rm if} \ \ \Gamma_{\tilde l_1}^{(0)} \ll  m_{\tilde l_1}\, .
\end{equation}
This means that the function $f(k^2)$ in (\ref{eqn:propid}) 
will be dominated by the regions of $k^2$ close to $m^2_{\tilde l_1}$
if $\Gamma^{(0)}_{\tilde l_1} \ll m_{\tilde l_1}$ . 
The non-resonant part of the matrix element is much smaller
than the resonant one (diagram (a) and (b) for $s=1$ in Figure \ref{treediagram}),
hence it can be neglected approximately.  
Applying (\ref{eqn:propM0}, \ref{eqn:propid}) to (\ref{eqn:bornGamma}),
neglecting the contributions from the non-resonant diagrams, the decay width of 
$\tilde\chi_2^0 \rightarrow \tilde \chi_1^0 l^- l^+$ can be written approximately as
\begin{eqnarray}
\Gamma^{(0)}(\tilde\chi_2^0 \rightarrow \tilde \chi_1^0 l^- l^+)& \simeq & 
\frac{1}{(2\pi)^2}\frac{1}{2m_{\tilde \chi_2^0}}\int 
\sum \bigl |M^{(0)}(\tilde{\chi}_2^0 \to \tilde{l}_1^{\pm}l^{\mp})\bigr |^2 
d\Phi_{\tilde{\chi}_2^0\to \tilde{l}_1^{\pm}l^{\mp}}\, \nonumber \\
&&{}\cdot \frac{1}{(2\pi)^2}\frac{1}{2m_{\tilde l_1}\Gamma_{\tilde l_1}^{(0)}} 
\int \sum \bigl |M^{(0)}(\tilde{l}_1^{\pm}\to \tilde{\chi}_1^0 l^{\pm})\bigr |^2 
d\Phi_{\tilde{l}_1^{\pm}\to \tilde{\chi}_1^0 l^{\pm}}\, \nonumber \\
& = & \Gamma^{(0)}( \tilde{\chi}_2^0 \rightarrow \tilde{l}_1^{\pm} l^{\mp})
 Br^{(0)}(\tilde{l}_1^{\pm}\rightarrow \tilde{\chi}_1^0  l^{\pm})\, , 
\end{eqnarray}
where the branching ratio of the decay $\tilde{l}_1^{\pm}\rightarrow \tilde{\chi}_1^0  l^{\pm}$ is defined by 
\begin{eqnarray}
Br^{(0)}(\tilde{l}_1^{\pm}\rightarrow \tilde{\chi}_1^0 l^{\pm}) & =
&\frac{ {\Gamma^{(0)}(\tilde{l}_1^{\pm} \rightarrow  
\tilde{\chi}_1^0 l^{\pm})}} {\Gamma_{\tilde{l}_1}^{(0)}}\, .
\label{eqn:tree5}
\end{eqnarray}
Hence, when the lighter sleptons $\tilde{l}_1$ can be on shell,
we can compute the relevant partial widths in the single-pole approximation,
where the decays $\tilde{\chi}_2^0 \rightarrow \tilde{\chi}_1^0 l^- l^+$ are
treated as the production and decay of the sleptons $\tilde{l}_1$.
\section{Complete One-loop Calculation for $\tilde\chi_2^0\rightarrow \tilde \chi_1^0 l^- l^+$}\label{complete}
\subsection{Virtual Corrections}
\begin{figure}[htb]
\begin{center}
\vspace*{1.5cm}
\begin{tabular}{llll}
\includegraphics[width=0.20\linewidth]{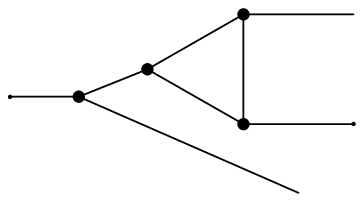} &
\includegraphics[width=0.20\linewidth]{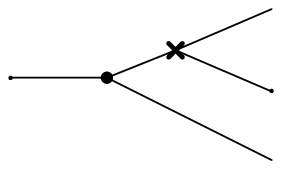} & \includegraphics[width=0.20\linewidth]{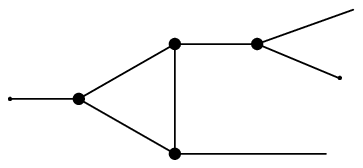} &
\includegraphics[width=0.20\linewidth]{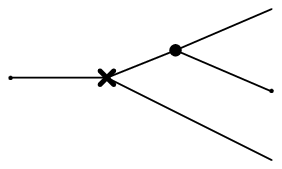}
\end{tabular}\\
\vspace*{2cm}
\begin{tabular}{lll}
 \hspace*{1.5cm}\includegraphics[width=0.15\linewidth]{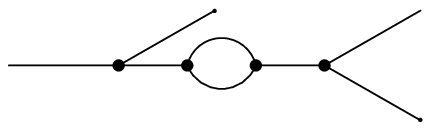} & \hspace*{2.5cm}
\includegraphics[width=0.10\linewidth]{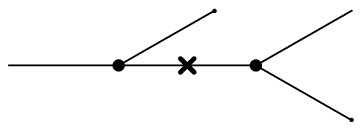} & \hspace*{1cm}\includegraphics[width=0.2\linewidth]{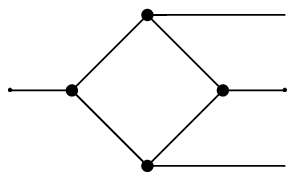} 
\end{tabular}
\end{center}
\caption{Classification of the one-loop virtual diagrams
\label{virtdiagram}}
\end{figure}

In general the virtual one-loop corrections to three-body decays can be
classified as {\em vertex contributions}, {\em self-energy contributions} and
{\em box contributions}. The first two classes are UV finite after adding
the contributions from the counterterms that originate from the
renormalization of the MSSM, as discussed in Chapter~\ref{ReMSSM}. The box
diagrams are by themselves UV finite. Different types of diagrams and their counterterm diagrams
are shown in Figure~(\ref{virtdiagram}). 
The MSSM Feynman rules, as well as the
resulting counterterms, are implemented in the {\em FeynArts} package of
computer program \cite{FeynArts}, which allows an automated generation of the Feynman diagrams.
The matrix element and the one-loop integrals 
are calculated with the help of the packages {\em FormCalc} and {\em LoopTools} \cite{FormCLT}, respectively.
The squared matrix element at one-loop level can be written as
\begin{eqnarray}
\bigl |M^{(1)}\bigr |^2 & = & \bigl |M^{(0)} + M_{\rm virt}\bigr |^2 \simeq \bigl |M^{(0)}\bigr |^2
+ 2 {\rm Re} \bigl [M^{(0)} M_{\rm virt}^\dagger \bigr ]\label{eqn:Amplitudevirtual}\, ,
\end{eqnarray}
where $M_{\rm virt}$ is matrix element of the virtual one-loop diagrams 
for the process $\tilde\chi_2^0\rightarrow \tilde \chi_1^0 l^- l^+$. 
The generic virtual one-loop diagrams are shown in
Appendix \ref{non-photonicdiagrams}.
Applying (\ref{eqn:Amplitudevirtual})
to the expressions for the width (\ref{eqn:bornGamma}) one obtains
\begin{eqnarray}
\Gamma^{(1)} & = & \Gamma^{(0)} + \Gamma_{\rm virt}\, .
\end{eqnarray}

Similarly to the tree-level case, diagrams with a slepton $\tilde l_1$
propagator have singularities when $\tilde l_1$ can be on shell.
The single-pole approximation can also be used at the one-loop level, which will be discussed in Section \ref{approx}.
Here we focus on the complete calculation.
Following the strategy in Section \ref{tree}, one can obtain a gauge invariant matrix element at one-loop level.  
In order to obtain $\mathcal{O}(\alpha)$
accuracy near the $\tilde{l}_1$ resonance, one needs to calculate the complex pole $m_p^2$ to two-loop level \cite{douple-pole},
\begin{eqnarray}
m_p^2 = m_{\tilde{l}_1}^2 -  i m_{\tilde{l}_1} \Gamma_{\tilde{l}_1}^{(1)}\, ,
\end{eqnarray}
where we have applied the on-shell renormalization scheme at two-loop level, 
and $\Gamma_{\tilde{l}_1}^{(1)}$ denotes the one-loop-level width of $\tilde l_1$.
Then the gauge invariant matrix element at one-loop level can be written as
\begin{eqnarray}
M^0 + M_{\rm virt} & = &\frac{A(k^2)}
{k^2 -  m_{\tilde{l}_1}^2 + i m_{\tilde{l}_1}\Gamma_{\tilde{l}_1}^{(1)}} + C(k^2)\, , 
\label{eqn:complete3a}
\end{eqnarray}
where $ C(k^2)$ denotes the non-resonant part of the matrix element,
the residue $A(k^2)$ can be expressed as
\begin{eqnarray}
A(k^2) &= & V^{(0)}_{\tilde{\chi}_2^0\tilde{l}_1^{\pm}l^{\mp}}(k^2)
 V^{(0)}_{\tilde{l}_1^{\pm}\tilde{\chi}_1^0 l^{\pm}}(k^2)\left(1 - \frac{{\rm Re}\hat{\Sigma}(k^2)}{k^2 -
    m_{\tilde{l}_1}^2}\right)+ \, \nonumber \\
&&{} V^{(0)}_{\tilde{\chi}_2^0\tilde{l}_1^{\pm}l^{\mp}}(k^2)
\hat V^{(1)}_{\tilde{l}_1^{\pm}\tilde{\chi}_1^0 l^{\pm}}(k^2)
+\hat V^{(1)}_{\tilde{\chi}_2^0\tilde{l}_1^{\pm}l^{\mp}}(k^2)
 V^{(0)}_{\tilde{l}_1^{\pm}\tilde{\chi}_1^0 l^{\pm}}(k^2)\, ,
\end{eqnarray}
where $\hat V^{(1)}_{\tilde{\chi}_2^0\tilde{l}_1^{\pm}l^{\mp}}$ and 
$\hat V^{(1)}_{\tilde{l}_1^{\pm}\tilde{\chi}_1^0 l^{\pm}}$ 
represent the renormalized $\tilde{\chi}_2^0\tilde{l}_1^{\pm}l^{\mp}$ and $\tilde{l}_1^{\pm}\tilde{\chi}_1^0 l^{\pm}$ 
vertices at one-loop level, respectively.

\begin{figure}[htb]
\begin{center}
\begin{tabular}{ccc}
\includegraphics[width=0.25\linewidth]{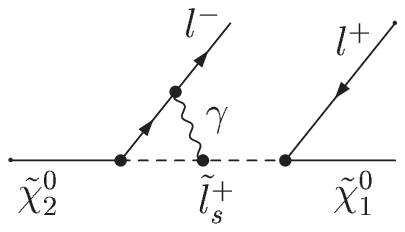} & 
\includegraphics[width=0.25\linewidth]{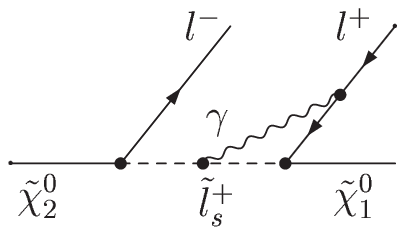} & 
\includegraphics[width=0.25\linewidth]{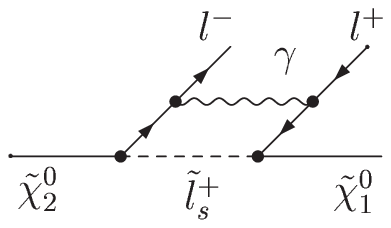} \\
(a1) & (a2) & (a3)\\
& & \\
\includegraphics[width=0.25\linewidth]{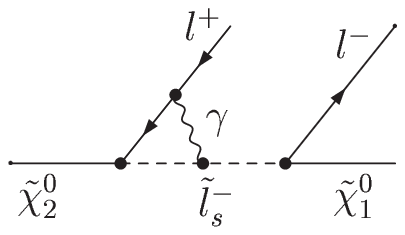} & 
\includegraphics[width=0.25\linewidth]{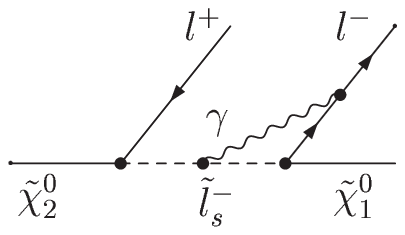} & 
\includegraphics[width=0.25\linewidth]{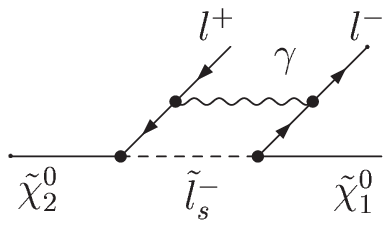} \\
(b1) & (b2) & (b3)\\
& & \\
\includegraphics[width=0.25\linewidth]{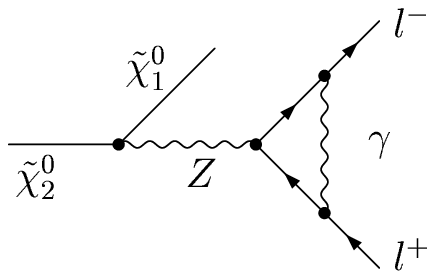} & 
\includegraphics[width=0.25\linewidth]{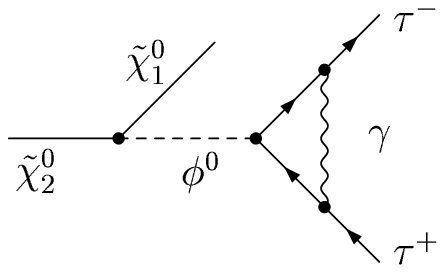} &  \\
(c) & (d) & 
\end{tabular}
\end{center}
\caption{Virtual photonic corrections in the complete calculation, $s = 1,2$ labels the slepton
  mass eigenstates, $\phi^0$ denotes the neutral Higgs boson $h^0, H^0, A^0$ and the neutral Goldstone boson
 $G^0$.\label{photonvirt}}
\end{figure}

Moreover, the one-loop integrals in the diagrams shown in the first two lines of Figure~\ref{photonvirt} also give
singularities when the sleptons $\tilde{l}_1$ are on shell. 
One should therefore use complex slepton masses, 
\begin{eqnarray}
\frac{1}{k^2 - m_{\tilde{l}_1}^2}& \longrightarrow & \frac{1}{k^2 -
  m_{\tilde{l}_1}^2 + i m_{\tilde{l}_1}\Gamma_{\tilde{l}_1}^{(1)}}\, ,
\label{eqn:complete3}
\end{eqnarray}
in the one-loop integrals from these diagrams.
The one-loop integrals with complex masses can be calculated automatically by {\em LoopTools}.
The analytical expressions for scalar three-point and four-point functions with real
arguments can be found in Refs.~\cite{SMDenner, soft, loopfunctiona, loopfunctionb}. 
We generalized the scalar four-point function to allow for complex arguments. 
The scalar three-point functions from the diagrams (a1), (a2), (b1) and (b2) in Figure~\ref{photonvirt}
are calculated analytically.
The analytical results are presented in Appendix~\ref{one-loopintegrals}. 
\subsection{Analytical Results for Virtual Photonic Corrections
 {\small $(l = e, \mu)$}}\label{virtual}
The virtual photonic diagrams are shown in  Figure~\ref{photonvirt}, where
the diagrams (a3), (b3), (c) and (d) have the property that 
the virtual photons are attached to external on-shell charged particles. 
This results in IR divergences, which we regularized by introducing a
fictitious photon mass $\lambda$. The IR divergences cancel after we add
contributions from real photon bremsstrahlung, which will be discussed in Section (\ref{real}).
The masses of the light leptons, i.e. $m_l~(l = e,\mu)$, are neglected except 
when they appear in the one-loop integrals. 
The virtual photonic corrections are calculated analytically,
where the terms which contain the soft singularity $\ln \lambda $,
the mass singularity $\ln m_l$ and the large logarithm
$\ln (m_{\tilde{l_1}}/\Gamma_{\tilde{l_1}})$ (only exits when $\tilde l_1$ can be on shell), are
treated separately as the singular part of the virtual photonic corrections.

The photonic corrections from the diagrams in the first two lines of Figure~\ref{photonvirt} 
(with their counterterms) can be written as
\begin{eqnarray}
d\Gamma_{\rm virt}^{\tilde l_s, \gamma}& = & \frac{\alpha}{\pi}Q_l^2\left (
\delta_{{\rm virt}, a} \sum |M_a^{(0)}|^2 + \delta_{{\rm virt}, b}\sum |M_b^{(0)}|^2\right )d\Phi_{1\to3}\, ,
\label{eqn:virtualslepton}
\end{eqnarray}
where $M_a^{(0)}$ and $M_b^{(0)}$ are the matrix element 
of the Born diagram (a) and (b) in Figure \ref{treediagram}, respectively, $Q_l$ denotes the lepton charge,  $Q_l^2 =1$.
The factor $\delta_{{\rm virt}, a}$ can be expressed as
 \begin{eqnarray}
\delta_{{\rm virt}, a} & = &
\frac{m_{\chi_2^0}^2}{T - m_{\chi_2^0}^2}\left (B_0(m_{\chi_2^0}^2, m_l^2, m_{\tilde{l_s}}^2)
-B_0(T, 0, m_{\tilde{l_s}}^2) \right )+\nonumber \\
&&{} \frac{m_{\chi_1^0}^2}{T - m_{\chi_1^0}^2}\left (B_0(m_{\chi_1^0}^2, m_l^2, m_{\tilde{l_s}}^2)
-B_0(T, 0, m_{\tilde{l_s}}^2) \right ) + \nonumber \\
&& {}
2 \left (B_0(m_l^2, m_l^2, 0)- \frac{1}{2}B_0(T, 0, m_{\tilde{l_s}}^2)\right )- 
(m_{\chi_2^0}^2-  m_{\tilde{l_s}}^2)C_0^a - \, \nonumber \\
&&{} (m_{\chi_1^0}^2-  m_{\tilde{l_s}}^2)C_0^b- T_{12}(T - m_{\tilde{l_s}}^2)D_0 
+ 2 ~\delta Z_{ll}^{\gamma}\, ,
\label{eqn:virt-slepton-factor}
\end{eqnarray}
where $C_0^a$, $C_0^b$ and $D_0$ are scalar three- and four-point functions, 
$\delta Z_{ll}^{\gamma}$ is the photonic part of the lepton field renormalization constant
$\delta Z_{ii}^{l, L}$ or $\delta Z_{ii}^{l, R}$ which have been expressed in (\ref{eqn:SMfermion-Reconstantb}) and
(\ref{eqn:SMfermion-Reconstantc}), respectively.  
The general definitions of the scalar one-loop integrals can be found in Appendix~\ref{one-loopintegrals}.
The arguments of $C_0^a$, $C_0^b$ and $D_0$ in (\ref{eqn:virt-slepton-factor}) are expressed as follows,
\begin{align}
C_0^a &=  C_0(m_l^2, m_{\chi_2^0}^2, T, 0, m_l^2, m_{\tilde{l_s}}^2)\, , \nonumber \\
C_0^b &=  C_0(m_l^2, m_{\chi_1^0}^2, T, 0, m_l^2, m_{\tilde{l_s}}^2)\, ,\nonumber \\
D_0 & = D_0(m_l^2, m_{\chi_2^0}^2, m_{\chi_1^0}^2, m_l^2, T, T_{12}, 0,  m_l^2, m_{\tilde{l_s}}^2, m_l^2)\, .
\label{eqn:sleptonfunction}
\end{align}
As discussed before, one should use complex masses  
$m_{\tilde{l}_s}^2 \to m_{\tilde{l}_s}^2 + i m_{\tilde{l}_s}\Gamma_{\tilde{l}_s}^{(1)}$ in the 
scalar three- and four-point functions shown in (\ref{eqn:sleptonfunction})
when the slepton $\tilde l_s$ can be on shell.
The scalar two-, three- and four-point functions as well as the lepton field renormalization constant
in (\ref {eqn:virt-slepton-factor}) are calculated in Appendix~\ref{one-loopintegrals}, where the   
three- and four-point functions with complex arguments are also presented. 
One finds that the expressions in (\ref {eqn:virt-slepton-factor})
are UV divergent. This indicates that the virtual photonic corrections are
UV divergent. Adding these UV-divergent part to the non-photonic virtual corrections, one obtains
UV-finite results. 

The terms in (\ref {eqn:virt-slepton-factor}) which contain the soft singularity $\ln \lambda $,
the mass singularity $\ln m_l$ and the large logarithm
$\ln (m_{\tilde{l_1}}/\Gamma_{\tilde{l_s}})$ are
treated separately as the singular part of $\delta_{{\rm virt}, a}$, which 
can be expressed as
\begin{eqnarray}
\delta_{{\rm virt}, a}^{\rm sing} & = &
 - \ln\left (\frac{m_l^2}{T_{12}}\right )\ln\left (\frac{\lambda^2}{m_l^2}\right ) - 
\frac{1}{2}\ln^2 \left (\frac{m_l^2}{T_{12}}\right ) - \frac{3}{2}\ln\left (\frac{m_l^2}{T_{12}}\right )
- \ln\left (\frac{\lambda^2}{m_l^2}\right ) + \nonumber \\
&&{}\ln\left (\frac{m_{\tilde{l_s}}^2-T-i\Gamma_{\tilde{l_s}}^{(1)} m_{\tilde{l_s}}}{m_{\tilde{l_s}}^2}\right )
 \Biggl [ \ln\left (\frac{m_{\chi_2^0}^2-T}{T_{12}}\right )+ \Biggr. \nonumber\\
 && {} \Biggl. \ln\left (\frac{T-  m_{\chi_1^0}^2}{T_{12}}\right ) +
\ln\left (\frac{m_{\chi_2^0}^2 - T}{T}\right )+ \ln\left (\frac{T-  m_{\chi_1^0}^2}{T}\right )\Biggr ] \, ,
\label{eqn:virt-slepton-factora}
\end{eqnarray}
where the terms proportional to $\ln\bigl [(m_{\tilde{l_s}}^2-T-i\Gamma_{\tilde{l_s}}^{(1)} m_{\tilde{l_s}})/m_{\tilde{l_s}}^2\bigr ]$
only exist when the slepton $\tilde l_s$ can be on shell.
The factor $\delta_{{\rm virt}, b}$ in (\ref{eqn:virtualslepton}) is calculated in the same way as $\delta_{{\rm virt}, a}$.
Its singular part can be obtained by replacing $T$ with $S_{23}$ in (\ref{eqn:virt-slepton-factora}),
\begin{eqnarray}
\delta_{{\rm virt}, b}^{\rm sing} & = &\delta_{{\rm virt}, a}^{\rm sing}{\Big |}_{T\to ~S_{23}}\, .
\label{eqn:virt-slepton-factorb}
\end{eqnarray}

The photonic virtual diagram (c) in Figure~\ref{photonvirt} is calculated and its contribution can be written as
\begin{eqnarray}
d\Gamma_{\rm virt}^{Z,\gamma} & = & \frac{\alpha}{\pi}Q_l^2\delta_{{\rm virt}, c}~
\sum |M_c^{(0)}|^2 d\Phi_{1\to3}\, ,
\label{eqn:virtualZ}
\end{eqnarray}
where $M_c^{(0)}$ is the matrix element of the Born diagram (c) in Figure \ref{treediagram}.
The factor $\delta_{{\rm virt}, c}$ reads                  
\begin{eqnarray}
\delta_{{\rm virt}, c} & = & -C_0^c~T_{12} + 2 B_0 (m_l^2, 0, m_l^2) - \frac{3}{2}B_0(T_{12}, m_l^2, m_l^2)
 + 2 ~\delta Z_{ll}^{\gamma}\, ,
\label{eqn:virtualZa}
\end{eqnarray}
where  $C_0^c$ is the scalar three-point function $C_0^c = C_0\left (m_l^2, m_l^2, T_{12}, 0, m_l^2, m_l^2\right )$.
The scalar two-point and three-point functions as well as the lepton field renormalization constant
in (\ref{eqn:virtualZa}) are calculated in Appendix~\ref{one-loopintegrals}.
One finds that the factor $\delta_{{\rm virt}, c}$ is UV finite. 
The singular part of $\delta_{{\rm virt}, c}$, which 
contains the singularities $\ln \lambda $ and $\ln m_l$, are 
\begin{eqnarray}
\delta_{{\rm virt}, c}^{\rm sing} & = & - \ln \left (\frac{m_l^2}{T_{12}}\right )\ln\left (\frac{\lambda^2}{m_l^2}\right ) -
 \frac{1}{2}\ln^2\left(\frac{m_l^2}{T_{12}}\right ) - \frac{3}{2}\ln\left (\frac{m_l^2}{T_{12}}\right ) 
 - \ln\left (\frac{\lambda^2}{m_l^2}\right )\, .
\label{eqn:virtualZb}
\end{eqnarray}

Combining the expressions for the photonic virtual corrections (\ref{eqn:virtualslepton}), 
(\ref{eqn:virt-slepton-factora}), (\ref{eqn:virt-slepton-factorb}), (\ref{eqn:virtualZ}) and (\ref{eqn:virtualZb}),
the singular part of virtual photonic corrections ($l = e, \mu$) can be written as
\begin{eqnarray}
d\Gamma_{\rm virt}^{\gamma}{\big |}_{\rm sing} &=& \frac{\alpha}{\pi}Q_l^2
\Biggl [\left ((\delta_{{\rm virt}, a}^{\rm sing}-\delta_{{\rm virt}, c}^{\rm sing}) \sum |M_a^{(0)}|^2  + 
(\delta_{{\rm virt}, b}^{\rm sing}- \delta_{{\rm virt}, c}^{\rm sing})\sum |M_b^{(0)}|^2 \right )_{s=1}
\, \Biggr.\nonumber \\
&&{} \Biggl. + \delta_{{\rm virt}, c}^{\rm sing}~\sum |M^{(0)}|^2\Biggr ]d\Phi_{1\to3}\, .
\label{eqn:virtualphoton}
\end{eqnarray}
Here we have concentrated on the case where the lighter slepton $\tilde l_1$ can be on shell. 
The factor $\delta_{{\rm virt}, a}^{\rm sing} -\delta_{{\rm virt}, c}^{\rm sing}$ is proportional to
$\ln\bigl [(m_{\tilde{l_1}}^2-T-i\Gamma_{\tilde{l_1}}^{(1)} m_{\tilde{l_1}})/m_{\tilde{l_1}}^2\bigr ]$.
After integration over the Mandelstam variable $T$, one obtains
\begin{eqnarray}
\int \ln\left (\frac{m_{\tilde{l_1}}^2-T-i\Gamma_{\tilde{l_1}}^{(1)} m_{\tilde{l_1}}}{m_{\tilde{l_1}}^2}\right ) d~T 
& = & i \Gamma_{\tilde{l_1}}^{(1)} m_{\tilde{l_1}}\ln \left (\frac{- i \Gamma_{\tilde{l_1}}^{(1)}}{m_{\tilde{l_1}}}\right )
+ \cdots\, ,
\label{integrateT}
\end{eqnarray}
where we have set $T = m_{\tilde{l_1}}^2$ because in this region the integrated result is dominant.
The ellipses in (\ref {integrateT}) represent terms which have nothing with $\Gamma_{\tilde{l_1}}^{(1)}$.
Since
\begin{eqnarray}
\lim_{ \Gamma_{\tilde{l_1}}^{(1)}\to 0}  i \Gamma_{\tilde{l_1}}^{(1)} m_{\tilde{l_1}}
\ln \left (\frac{- i \Gamma_{\tilde{l_1}}^{(1)}}{m_{\tilde{l_1}}}\right ) & = & 0\, , 
\end{eqnarray}
the large logarithm $\ln\left ( \Gamma_{\tilde{l_1}}^{(1)}/m_{\tilde{l_1}}\right )$ disappears in the integrated result 
when $\Gamma_{\tilde{l_1}}^{(1)}\to 0$.
Similarly there is a large logarithm $\ln\left ( \Gamma_{\tilde{l_1}}^{(1)}/m_{\tilde{l_1}}\right )$
in the factor $\delta_{{\rm virt}, b}^{\rm sing} -\delta_{{\rm virt}, c}^{\rm sing}$ in (\ref{eqn:virtualphoton}),
which also disappears in the integrated result when  $\Gamma_{\tilde{l_1}}^{(1)}$ goes to $0$.

Since we keep the $\tau$ mass everywhere, 
the virtual photonic corrections with $\tau^-\tau^+$ final states are calculated numerically. 
\subsection{Real Photon Bremsstrahlung}\label{real}
\begin{figure}[htb]
\begin{center}
\begin{tabular}{cccc}
\includegraphics[width=0.22\linewidth]{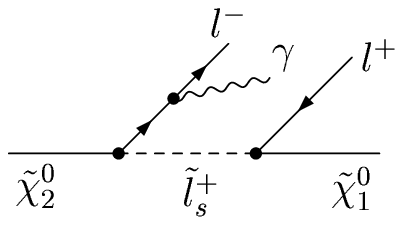} &
\includegraphics[width=0.22\linewidth]{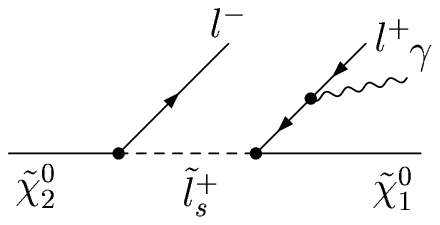}& 
\includegraphics[width=0.22\linewidth]{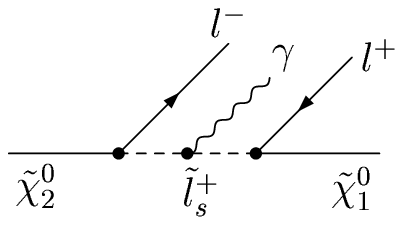} & 
 \includegraphics[width=0.22\linewidth]{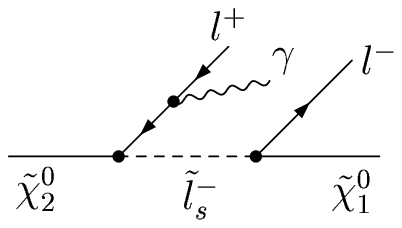} \\
\includegraphics[width=0.22\linewidth]{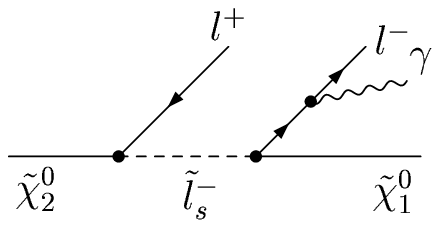} &
\includegraphics[width=0.22\linewidth]{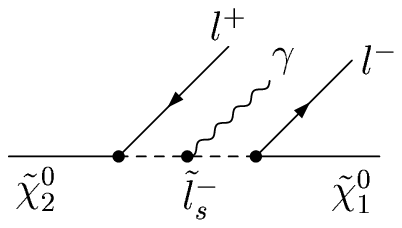} & 
\includegraphics[width=0.22\linewidth]{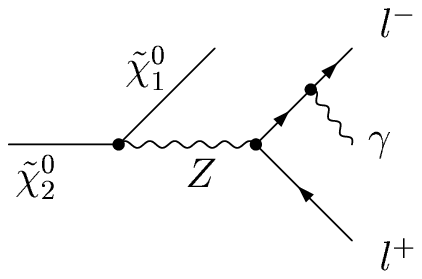} &
\includegraphics[width=0.22\linewidth]{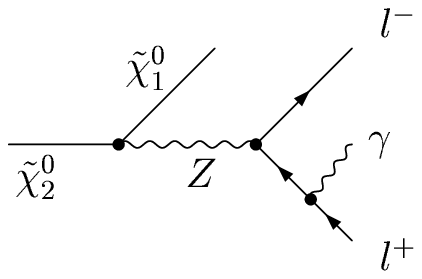}\\
\includegraphics[width=0.22\linewidth]{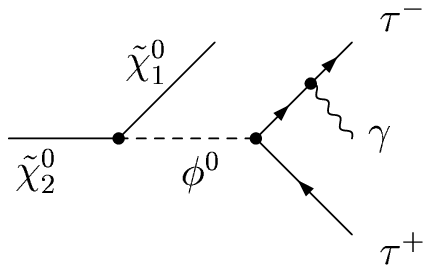} &
\includegraphics[width=0.22\linewidth]{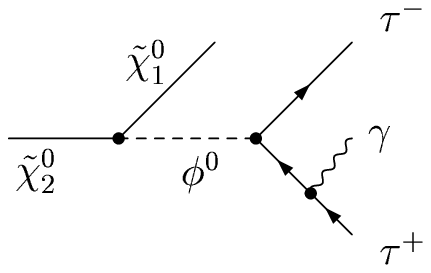} 
\end{tabular}
\end{center}
\caption{The Feynman diagrams for the real photon bremsstrahlung $\tilde{\chi}_2^0 (k_1) \longrightarrow
  \tilde{\chi}_1^0 (k_2) l^- (k_3) l^+ (k_4) \gamma (q)$, $s = 1,2$ labels the slepton
  mass eigenstates, $\phi^0$ denotes the neutral Higgs boson $h^0, H^0, A^0$ and the neutral Goldston boson
  $G^0$.
\label{realdiagram}}
\end{figure}

In order to cancel the IR divergences in the virtual corrections, we have to 
add contributions from real photon bremsstrahlung to the one-loop corrections.
The diagrams for the process $\tilde{\chi}_2^0(k_1) \longrightarrow
\tilde{\chi}_1^0(k_2) l^-(k_3) l^+(k_4) \gamma (q)~ (l = e, \mu, \tau)$ are displayed in
Figure~\ref{realdiagram}. 
Generally the decay width of the real photon bremsstrahlung can be written as
\begin{eqnarray}
\Gamma_{\rm brems} & = & \frac{1}{(2\pi)^8}\frac{1}{2m_{\tilde \chi_2^0}}\int \sum \bigl |M_{\rm brems}\bigr |^2 d\Phi_{1\to4}\, ,
\label{eqn:realGamma}
\end{eqnarray}
where $M_{\rm brems}$ denotes the matrix element of the diagrams in Figure~\ref{realdiagram}.
The definition for the phase-space element $d\Phi_{1\to4}$ can be found in Appendix ~\ref{Kinematics}.
One must use the complex slepton masses in the propagators as (\ref{eqn:complete3}) when 
the lighter sleptons $\tilde l_1$ can be on shell.

From Figure~\ref{realdiagram} we know that the real photon can be emitted from the charged leptons and sleptons.
In the case of photon emission from a charged lepton, the amplitude for the real photon emission has the property\\
\begin{minipage}{0.55\linewidth}
\begin{eqnarray}
M_{\rm brems}\propto \frac{1}{(k_i+q)^2 -m_i^2} = \frac{1}{2 k_i\cdot q }\, .
\label{properity}
\end{eqnarray}
\end{minipage}
\hspace*{4.5cm}
\begin{minipage}{0.12\linewidth}
\vspace*{3mm}
\includegraphics[width=\linewidth]{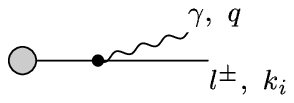} \\
\vspace*{3mm}
\end{minipage}\\
This will give rise to IR divergences which are regularized by a photon mass. 
The ``soft photon bremsstrahlung'' contributions are defined via the condition that 
the photon energy $E_{\gamma}\le \Delta E$, where the
cutoff parameter $\Delta E$ should be small compared to the relevant physical energy scale.
The complementary contributions are called ``hard photon bremsstrahlung'', 
which are defined as real emission contributions with $E_\gamma > \Delta E$. 
The contributions of soft photon bremsstrahlung are IR divergent,
which are sufficient to cancel the IR divergences in the virtual corrections,
while the contributions of hard photon bremsstrahlung are IR finite.
The contributions of the real emission can be expressed as
\begin{eqnarray} \label{brems1}
\Gamma_{\rm brems}& = & \Gamma_{\rm soft}(\Delta E) + \Gamma_{\rm hard}(\Delta E)\, .  
\end{eqnarray}
The dependence on the largely arbitrary parameter $\Delta E$ cancels after
summing soft and hard contributions, provided it is sufficiently small.
The contributions of the photon radiation from charged sleptons
are always finite because the sleptons are internal particles.

If we neglect the light lepton masses, the factor $\frac{1}{2 k_i\cdot q }$ in (\ref{properity}) can be written as
\begin{eqnarray}
\frac{1}{2 k_i\cdot q } = \frac{1}{2k_{i0} E_\gamma (1 - \cos \theta_{l\gamma})}\, ,
\end{eqnarray}
where $\theta_{l\gamma}$ is the angle between momentum of the photon and momentum of the emitting particle.
If $ \cos \theta_{l\gamma} \approx 1$, i.e.  the photon and its emitter are collinear, 
the contributions of the hard photon bremsstrahlung also contain a divergence. It is
regularized by the masses of the leptons in the final state.
However, since the lepton masses, i.e. $m_e$ and $m_\mu$, are very small, it is very
difficult to get stable numerical results from a direct numerical evaluation of hard photon bremsstrahlung
even we keep the light lepton masses everywhere.
This can be overcome by dividing hard photon bremsstrahlung into a collinear
part, where the angle between the photon and the radiating particle is smaller
than a very small angle $\Delta \theta$, and the complementary non-collinear
part,
\begin{eqnarray} \label{brems2}
\Gamma_{\rm hard}(\Delta E)& = & \Gamma_{\rm coll}(\Delta E, \Delta\theta) +
\Gamma_{\rm non-coll}(\Delta E, \Delta\theta)\, . 
\end{eqnarray}
The angular cutoff $\Delta \theta$ should be so small that 
we can assume that the collinear-photon emission does not change the {\em direction} of the three-momentum of
the emitting lepton.

So far we have divided the four-particle phase space into a soft, a collinear and a non-collinear region. 
This is called phase-space-slicing method. The soft and collinear contributions can be calculated analytically, while the
non-collinear contributions are calculated numerically using the multi-channel-approach in the Monte Carlo program
\cite{mutichannela, mutichannelb}. The details about this technique are presented in Appendix~\ref{mcamcp}.
   
\newpage
\subsubsection{Soft Photon Bremsstrahlung}\label{softbrems}
Since the energy of the emitted soft photon is by definition very small, this emission essentially does not change the momenta
of the other final state particles. 
In the soft region, the squared amplitude $\bigl |M_{\rm soft}\bigr |^2$ can be written
as the Born squared amplitude $\bigl |M^{(0)}\bigr |^2$ multiplied by a soft factor, 
\begin{eqnarray}
\bigl |M_{\rm soft}\bigr |^2 & = &-Q_l^2 e^2 \left ( \frac{k_3^{\mu}} {k_3\cdot q} - \frac{k_4^{\mu}} {k_4\cdot q} \right )^2 
\bigl |M^{(0)}\bigr |^2\, ,
\end{eqnarray}
where we have neglected all the terms proportional to the photon momentum $q$ in the numerator.
The four-particle phase space can also be factorized into a three-particle phase space and a soft part.
Therefore the contributions of the soft photon bremsstrahlung can be written as
\begin{equation}
\begin{split}
d\Gamma_{soft}&= -\frac{\alpha}{2\pi^2}Q_l^2\int_{|\vec{q}|\leq \Delta E}\frac{d^{3}|\vec{q}|}{2E_\gamma}
\left ( \frac{k_3^{\mu}} {k_3\cdot q} - \frac{k_4^{\mu}} {k_4\cdot q} \right )^2\sum |M^{(0)}|^2 d\Phi_{1\to 3}\\
&= - \frac{\alpha}{2\pi^2}Q_l^2 (I_{33} + I_{44}-  2 I_{34})\bigl |M^{(0)}\bigr |^2 d\Phi_{1 \to 3}\, , 
\label{softapprox}
\end{split}
\end{equation}
where 
\begin{equation}
I_{ij} =  \int_{|\vec{q}|\leq \Delta E}\frac{d^{3}|\vec{q}|}{2E_{\gamma}}
 \frac{k_i \cdot k_j}{(k_i\cdot q)(k_j\cdot q)}\, ,
\end{equation}
$E_{\gamma} = \sqrt{|\vec{q}|^2 + \lambda^2}$.

The general analytical expressions for the integrals $I_{ij}$ can be found in Refs. \cite{SMDenner, soft}. 
This has been implemented in the package of {\em FormCalc}. 
The light lepton mass, i.e. $m_l~(l = e, \mu)$ is kept only when it acts as a regulator for the mass singularity,
hence the analytical expressions for the soft contributions can be written as
\begin{equation}
\begin{split}
d\Gamma_{soft}&= \frac{\alpha}{\pi}Q_l^2 \Biggl [\ln \left (\frac{m_l^2}{T_{12}}\right )\ln\left (\frac{\lambda^2}{4\Delta E^2}\right ) -
\frac{1}{4}\ln^2\left (\frac{m_l^2}{4k_{30}^2}\right )  - \frac{1}{4}\ln^2\left (\frac{m_l^2}{4k_{40}^2}\right)  +
\ln \left (\frac{\lambda^2}{4\Delta E^2}\right ) \Biggr. \\ 
& \quad \Biggl. - \frac{\pi^2}{3}- \frac{1}{2}\ln\left (\frac{m_l^2}{4k_{30}^2}\right )- 
\frac{1}{2}\ln\left (\frac{m_l^2}{4k_{40}^2}\right )- 
Li_2 \left (1- \frac{4k_{30}k_{40}}{T_{12}}\right )\Biggr ]d \Gamma^{(0)}\, ,
\end{split}
\end{equation}
where $k_{i0}$ denotes the energy of the charged lepton whose four-momentum is defined as $k_i$,
the dilogarithm $Li_2 (x)$ is defined in Appendix \ref{one-loopintegrals}.
\newpage
\subsubsection{Collinear Photon Bremsstrahlung}\label{hardbrems}
From the discussions above we define the collinear region: $E_{\gamma} > \Delta E$ and 
$1-\delta_{\theta} < |\cos\theta_{\gamma l}| < 1 $, where $\delta_{\theta} = 1 - \cos\Delta \theta$.
This means that the collinear part describes real photon radiation outside the soft photon region
and collinear to the emitter. 
We consider a final-state radiation,  
\begin{align}
l (\tilde k_i) &\to \gamma (q) + l (k_i), \ \ \ \ i = 3, 4\ \ {\rm (see \ Figure \ \ref{realdiagram})}\, . 
\end{align}
The squared matrix element of the collinear photon bremsstrahlung can be written as
\begin{equation}
\begin{split}
\sum |M|^2_{\rm coll}&= \sum_{i = 3}^4\frac{Q_l^2e^2}
{k_i \cdot q}\left (P(z_i)-\frac{m_i^2}{k_i \cdot q}\right )\sum|M^{(0)}(\tilde{k_i})|^2\, , 
\end{split}
\end{equation}
where $z_i= \frac{k_i}{k_i + q}$, $P(z_i)= \frac{1+ z_i^2}{1-z_i}$, $\tilde{k_i} = \frac{k_i}{z_i}$.
The phase space in the collinear region can be separated into a three-particle phase space multiplied by a
collinear factor,
\begin{eqnarray}
d\Phi_{k_1 \to k_2 + k_3 + k_4 + q}& =& \frac{d^3 k_2}{2k_{20}} \frac{d^3 k_3}{2k_{30}} 
\frac{d^4 k_4}{2k_{40}} \frac{d^3 q}{2E_\gamma} \delta^{(4)}(k_1 - k_2 - k_3 - k_4 - q)\, \nonumber\\
&=& \frac{d^3 k_2}{2k_{20}} \frac{d^2 \tilde{k_3}}{2\tilde{k_{30}}} 
\frac{d^3 k_4}{2k_{40}} z^2 \frac{d^3 q}{2E_\gamma}\delta^{(4)}(k_1 - k_2 - \tilde{k_3} - k_4)\, \nonumber \\
&=& d\Phi_{k_1 \to k_2 + \tilde{k_3} + k_4}z^2 \frac{d^3 q}{2E_\gamma}\, .
\end{eqnarray}
This is corresponding to the phase space of the radiation $l (\tilde k_3) \to \gamma (q) + l (k_3)$.
Therefore the differential collinear contributions can be expressed as 
\begin{equation}
d\Gamma_{\rm coll}= \frac{\alpha}{2\pi}Q_l^2 \sum_{i = 3}^4 G_i~ d\Gamma^{(0)}(\tilde{k_i}),
\label{collinearapprox}
\end{equation}
where 
\begin{eqnarray}
G_i&= & \frac{1}{\pi}
\int \frac{1}{k_i \cdot q}\left (P(z_i)-\frac{m_i^2}{k_i \cdot q}\right )z_i^2 \frac{d^3 q}{2E_\gamma}\, .
\label{constantcollinear}
\end{eqnarray}
It can be written as
\begin{eqnarray}
G_i&=&\int_{0}^{1- \frac{\Delta E}{\tilde{k}_{i0}}}\Bigg [P(z_i)
\ln\left (\frac{4\tilde{k_{i0}}^2}{m_i^2}\frac{\delta_\theta}{2}z_i^2 \right )-\frac{2z_i}{1-z_i}\Bigg ]dz_i\, 
\label{constantcollineara}
\end{eqnarray}
after integrating out the solid angle of the momentum $q$ analytically.

If we treat a charged lepton and a collinear photon inclusively, i.e. the momentum
of collinear photon is added to that of emitting lepton, 
the variable $z_i$ in (\ref{constantcollineara}) can be integrated out analytically. 
Hence the differential contributions of the collinear emissions are written as
the differential tree-level decay width multiplied by a universal function \cite{collinear, axel}, 
\begin{equation}
d\Gamma_{\rm coll}= \frac{\alpha}{2\pi}Q_l^2 d\Gamma^{(0)}\sum_{i = 3}^4 
\Biggl [\Bigl [\frac{3}{2} +  2\ln \left (\frac{\Delta E}{k_{i0}}\right )\Bigr ]
\Bigl [1- \ln \left (\frac{4k_{i0}^2}{m_i^2}\frac{\delta_\theta}{2}\right )\Bigr ]+ 3 - \frac{2}{3}\pi^2\Biggr ] \, .
\label{eqn:collinearsafe}
\end{equation}
This approach is for collinear-safe observables \cite{axel}. 
If one adds the soft and collinear contributions to the virtual corrections,
all singularities ($\ln m_l$ and $\ln \lambda$) cancel. This is in accordance with 
the Kinoshita-Lee-Nauenberg theorem \cite{kln}.
At the LHC the electron energy is determined calorimetrically.
In this case a collinear photon would hit the same cell of the calorimeter as
the electron, so the two energies cannot be disentangled. Hence the electron observables are defined as collinear-safe observables
in our calculation.

We also consider the non-collinear-safe observables \cite{axel}, 
i.e. the lepton and its collinear photon are not treated inclusively. Since the phase space depends on the variables $z_i$,
the integration over $z_i$ cannot be performed analytically. The differential contributions of the collinear photon
bremsstrahlung are written as
\begin{equation}
d\Gamma_{\rm coll}= \frac{\alpha}{2\pi}Q_l^2 \sum_{i = 3}^4 ~\int_{0}^{1- \frac{\Delta E}{\tilde{k}_{i0}}}dz_i\Bigg [P(z_i)
\ln\left (\frac{4\tilde{k_{i0}}^2}{m_i^2}\frac{\delta_\theta}{2}z_i^2 \right )-\frac{2z_i}{1-z_i}\Bigg ]
 d\Gamma^{(0)}(\tilde{k_i})\, .
\label{eqn:collinearsafeNO}
\end{equation}
In this case the mass singularity $\ln m_l$ cannot be canceled in the differential width and hence become visible.
At the LHC, muons pass through the calorimeter, where the photons are detected, and measured forther outside in the muon 
detector. Hence the muon observables are treated as non-collinear-safe observables in our calculation.
\subsection{QED Corrections}\label{QED}
The virtual photonic corrections by themselves are UV divergent,
 hence one cannot meaningfully separate the QED corrections from
the one-loop contributions by simply selecting diagrams which contain a
photon. Since the light lepton ($e$ and $\mu$) final states and 
the $\tau$ final states are treated differently,
the corresponding QED corrections are defined differently as following.
 
In the case of the light lepton final states, the photonic virtual corrections and the soft
photon bremsstrahlung are calculated analytically.
We can pick out potentially large QED terms from the the sum
of virtual and soft photon bremsstrahlung corrections, $d\Gamma_{\rm virt}+ d\Gamma_{\rm soft}$,
\begin{eqnarray} \label{rem}
d\Gamma_{\rm virt}+ d\Gamma_{\rm soft} & = & d\tilde{\Gamma} + d\Gamma_{\rm remainder}\, .
\end{eqnarray}
Here $d\tilde{\Gamma}$ contains all the potentially large QED terms, 
\begin{eqnarray}
d\tilde \Gamma & = & \frac{\alpha}{\pi}Q_l^2\left \{
\Biggl [\left (\ln\left (\frac{m_l^2}{T_{12}}\right ) + 1\right )\ln\left (\frac{k_{30}k_{40}}{\Delta E^2}\right ) 
-\frac{3}{2}\ln\left (\frac{m_l^2}{T_{12}}\right ) \Biggr ]~d\Gamma^{(0)} + \right. \, \nonumber \\
&& {}\left . \ln\left (\frac{m_{\tilde{l_1}}^2-T-i\Gamma_{\tilde{l_1}}^{(1)} m_{\tilde{l_1}}}{m_{\tilde{l_1}}^2}\right )
 \Biggl [ \ln\left (\frac{m_{\chi_2^0}^2-T}{T_{12}}\right )+ 
\ln\left (\frac{T-  m_{\chi_1^0}^2}{T_{12}}\right ) +\Biggr.\right. \nonumber\\
 && {} \left. \Biggl. 
\ln\left (\frac{m_{\chi_2^0}^2 - T}{T}\right )+ \ln\left (\frac{T-  m_{\chi_1^0}^2}{T}\right )\Biggr ]
\sum |M_a^{(0)}|^2d\Phi_{1\to3} 
+ \right. \, \nonumber \\
&& {}\left . \ln\left (\frac{m_{\tilde{l_1}}^2-S_{23}-i\Gamma_{\tilde{l_1}}^{(1)} m_{\tilde{l_1}}}{m_{\tilde{l_1}}^2}\right )
 \Biggl [ \ln\left (\frac{m_{\chi_2^0}^2-S_{23}}{T_{12}}\right )+ 
\ln\left (\frac{S_{23}-  m_{\chi_1^0}^2}{T_{12}}\right ) +\Biggr.\right. \nonumber\\
 && {} \left. \Biggl. 
\ln\left (\frac{m_{\chi_2^0}^2 - S_{23}}{S_{23}}\right )+ \ln\left (\frac{S_{23}-  m_{\chi_1^0}^2}{S_{23}}\right 
)\Biggr ]~\sum |M_b^{(0)}|^2d\Phi_{1\to3} \right \}\, ,
\label{eqn:largeterms}
\end{eqnarray}
where we have assumed that the lighter slepton $\tilde l_1$ can be on shell. The matrix element $M_a^{(0)}$ and
$M_b^{(0)}$ are obtained from the Born diagrams (a) and (b) in Figure \ref{treediagram}, respectively. 
As discussed in Section \ref{real} 
the terms proportional to $\ln\left ( \Gamma_{\tilde{l}_1}^{(1)}/m_{\tilde{l}_1}\right )$ in (\ref{eqn:largeterms})
disappear in the integrated width.
$d \Gamma_{\rm remainder}$ defined in (\ref{rem}) 
is IR and UV finite and free of such large QED logarithms in (\ref{eqn:largeterms}). 
The ``QED contributions" can then be defined as
\begin{eqnarray} \label{qed}
d\Gamma_{\rm QED}& = &d\tilde{\Gamma} + d\Gamma_{\rm hard} \, .
\end{eqnarray}

Note that the QED corrections defined in this way do not depend on the cutoff
parameters $\Delta E$ and $\Delta \theta$. Moreover, terms proportional to $\log m_l$
in (\ref{qed}) cancel in the integrated width and in the differential width for the collinear-safe observables.
Using the
definitions (\ref{rem}) and (\ref{qed}), the complete one-loop contributions
can be written as
\begin{eqnarray} \label{llrem}
d\Gamma^{(1)} & = & d\Gamma^{(0)} + d\Gamma_{\rm virt} + d\Gamma_{\rm
  brems}\, 
\nonumber \\ & = & d\Gamma^{(0)} + d\Gamma_{\rm remainder} + d\Gamma_{\rm
  QED}\, . 
\end{eqnarray}

When $\tau^- \tau^+$ are the final states of $\tilde \chi_2^0$ decay,  
the $\tau$ mass $m_{\tau}$ is kept everywhere. 
This mass is so large that
a stable numerical result can be obtained from the hard photon bremsstrahlung, 
hence we do not need to divide the hard photon bremsstrahlung contribution into collinear and non-collinear parts.
We follow a slightly different procedure to define the ``QED part'' of the
correction. The virtual corrections contain photonic and non-photonic
contributions,
\begin{eqnarray}
d\Gamma_{\rm virt} & = & d\Gamma_{\rm virt}^{\gamma} + d\Gamma_{\rm
  virt}^{\rm non-\gamma}\, , 
\end{eqnarray}
both of which are UV divergent, while the sum is finite.
The photonic virtual corrections can be split into an UV-finite
part $d\tilde{\Gamma}$ and an UV-divergent part $d\Gamma_{\rm UV-div}^{\gamma}$,
\begin{eqnarray}
d\Gamma_{\rm virt}^{\gamma} & = & d\tilde{\Gamma} + d\Gamma_{\rm
  UV-div}^{\gamma}\, .  
\end{eqnarray}
Here $d\Gamma_{\rm UV-div}^\gamma$ contains the terms that would be subtracted
in the dimensional reduction regularization of $d\Gamma_{\rm virt}^\gamma$. After
this rearrangement, the virtual corrections can be written as
\begin{eqnarray}
d\Gamma_{\rm virt} & = & d\tilde{\Gamma} + d\Gamma_{\rm UV-div}^{\gamma} +
d\Gamma_{\rm virt}^{\rm non-\gamma}\, \nonumber \\ & = &  d\tilde{\Gamma} +
d\Gamma_{\rm remainder}\, , 
\end{eqnarray}
where $d\Gamma_{\rm remainder} = d\Gamma_{\rm UV-div}^{\gamma}+ d\Gamma_{\rm
  virt}^{\rm non-\gamma}$ as well as $d\tilde\Gamma$ are UV finite. The ``QED
corrections" are finally defined as
\begin{eqnarray}
d\Gamma_{\rm QED} & =& d\tilde{\Gamma}+ d\Gamma_{\rm brems}\, ,
\end{eqnarray}
where $d\Gamma_{\rm brems}$ stands for the contribution from all diagrams with
real photon emission. By construction, $d\Gamma_{\rm QED}$ is UV and
IR finite. 
\section{Approximate One-loop Calculation for \mbox{$\tilde \chi_2^0 \to \tilde \chi_1^0 l^- l^+$}}\label{approx}
\begin{figure}[htb]
\begin{center}
\begin{tabular}{cc}
\includegraphics[width=0.3\linewidth]{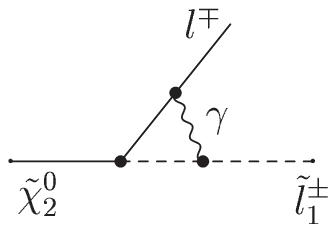} &
\includegraphics[width=0.3\linewidth]{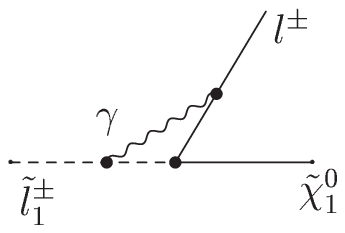} \\
 (a) & (b)
\end{tabular}
\end{center}
\caption{Virtual photonic corrections in the approximate calculation.\label{photonvirtApp}}
\end{figure}

\begin{figure}[htb]
\begin{center}
\begin{tabular}{cccc}
\includegraphics[width=0.22\linewidth]{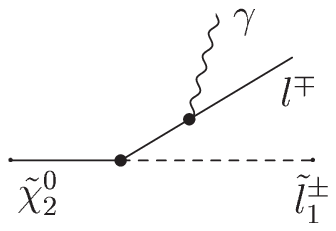} & 
\includegraphics[width=0.22\linewidth]{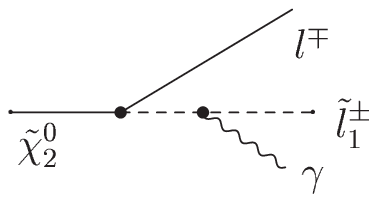} & \includegraphics[width=0.22\linewidth]{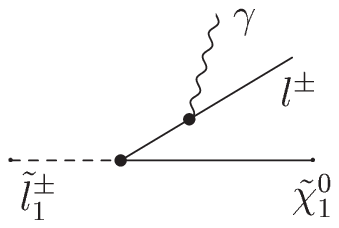} & 
\includegraphics[width=0.22\linewidth]{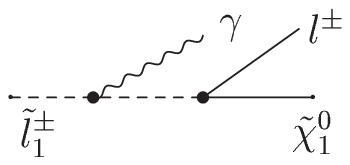} \\
 (a1) & (a2) &  (b1) & (b2)
\end{tabular}
\end{center}
\caption{Diagrams for the real photon bremsstrahlung in the approximate calculation.\label{realdiagramAPP}}
\end{figure}

If $\tilde \chi_2^0 \rightarrow \tilde l_1^\pm l^\mp$ two-body decays are allowed and
$\tilde \chi_2^0$ does not have other two-body decay modes, at one-loop
level, just like at tree-level, the decays $\tilde{\chi}_2^0 \rightarrow
\tilde{\chi}_1^0 l^- l^+ \ (l = e, \mu, \tau)$ can be approximately treated as
production and decays of the slepton $\tilde{l}_1^\pm$,
\begin{eqnarray}
\Gamma^{(1)}(\tilde{\chi}_2^0 \rightarrow  \tilde{\chi}_1^0 l^+ l^-)
& \simeq &\Gamma^{(1)}(\tilde{\chi}_2^0 \rightarrow \tilde{l}_1^{\pm} l^{\mp})
 Br^{(1)}(\tilde{l}_1^{\pm}\rightarrow \tilde{\chi}_1^0 l^{\pm})\, ,
\end{eqnarray}
with
\begin{eqnarray}
Br^{(1)}(\tilde{l}_1^{\pm}\rightarrow \tilde{\chi}_1^0 l^{\pm}) & =
&\frac{ {\Gamma^{(1)}(\tilde{l}_1^{\pm} \rightarrow \tilde{\chi}_1^0 l^{\pm})}} {\Gamma_{\tilde{l}_1^{\pm}}^{(1)}}\, .
\end{eqnarray}
The virtual contributions of the production and decay of the slepton
$\tilde{l}_1^\pm$, which now only contain vertex type corrections, 
are again calculated with the help of the programs {\em  FeynArts, FormCalc} and {\em LoopTools}.
The virtual photonic diagrams are shown in Figure \ref{photonvirtApp}, which are IR and UV divergent.
The scalar three-point functions corresponding to these diagrams are expressed 
analytically in Appendix \ref{one-loopintegrals} in the case of $l= e, \mu$.
An IR finite result is obtained by adding the contributions from the real photon bremsstrahlung (see Figure
\ref{realdiagramAPP}),
where both the diagrams (a2) and (b2) must be included in order to \mbox {preserve} the gauge invariance. 
The contributions of the real photon bremsstrahlung 
are again \mbox {separated} into an IR-divergent soft part and an IR-finite hard part. The division of
the hard photon bremsstrahlung contribution into a collinear part, which can
be calculated analytically, and a non-collinear part, which is calculated
numerically, proceeds along the lines described in Section \ref{real}.
As discussed in Section \ref{QED}, the UV-divergent photonic
contributions cannot be treated separately as ``QED corrections". We define
the ``QED corrections" in the same way as the complete calculation in Section \ref{QED}. One
finally arrives at a total one-loop contribution which is independent of the
cutoff parameters.
\section
[Total Width of $\tilde\chi_2^0$ and the
Branching \mbox {Ratios} of $\tilde\chi_2^0 \to \tilde\chi_1^0 l^+ l^-$]
{Total Decay Width of $\tilde\chi_2^0$ and the
 Branching \mbox{Ratios} of the Decays $\tilde\chi_2^0 \to \tilde\chi_1^0 l^+ l^-$}\label{brlepton}
As discussed in the beginning of this chapter, the next-to-lightest neutralino $\tilde \chi_2^0$
can decay into the LSP $\tilde \chi_1^0$ and two fermions $f\bar f$. The leptonic final states 
are important because they can be identified at the LHC. Moreover, 
the endpoint of the dilepton invariant-mass distribution is used to determine the mass relations
of supersymmetric particles. The invisible $\tilde \chi_2^0$ decay modes, i.e. $\tilde \chi_2^0 \to
\tilde \chi_1^0 \nu_l \bar \nu_l$, do not effect the dilepton invariant-mass distribution. 
But they contribute to the total width of $\tilde \chi_2^0$. 
Since it is very difficult to identify quarks at the LHC, 
the hadronic decays $\tilde \chi_2^0 \to \tilde \chi_1^0 q \bar q$ are less interesting than leptonic decays.
In order to obtain the total decay width of $\tilde \chi_2^0$,
these hadronic decays must be calculated . The  total decay width of $\tilde \chi_2^0$
can be written as
\begin{equation}
\Gamma_{\tilde{\chi}_2^0} = \sum_{l=e,\mu,\tau} \Bigl [ \Gamma(\tilde{\chi}_2^0
\rightarrow l^- l^+ \tilde{\chi}_1^0) + \Gamma(\tilde{\chi}_2^0 \rightarrow
\nu_l \bar \nu_l \tilde{\chi}_1^0) \Bigr ] 
+ \sum_{q = u, d, c, s, b} \Gamma(\tilde{\chi}_2^0
\rightarrow q \bar q \tilde{\chi}_1^0)\, .
\end{equation}
Here we assume that the decay  $\tilde \chi_2^0 \to \tilde \chi_1^0 \bar t t$ is not kinematically allowed. The branching ratios 
of the leptonic decays  $\tilde \chi_2^0 \to \tilde \chi_1^0 l^+ l^-$ are defined as
\begin{equation}
Br(\tilde \chi_2^0 \to \tilde \chi_1^0 l^+ l^-) = \frac{\Gamma(\tilde \chi_2^0 \to \tilde \chi_1^0 l^+ l^-)}{\Gamma_{\tilde{\chi}_2^0}}\, .
\end{equation}
\subsection{The Invisible Decays $\tilde \chi_2^0 \to\tilde \chi_1^0 \nu_l \bar \nu_l$}
\begin{figure}[htb]
\begin{center}
\begin{tabular}{ccc}
\includegraphics[width=0.22\linewidth]{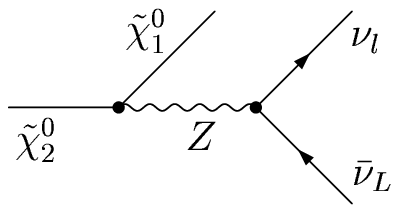} &
\includegraphics[width=0.22\linewidth]{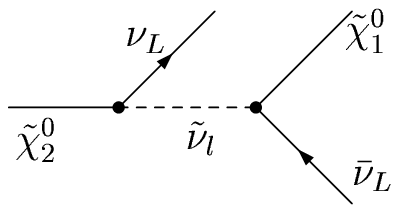}& 
\includegraphics[width=0.22\linewidth]{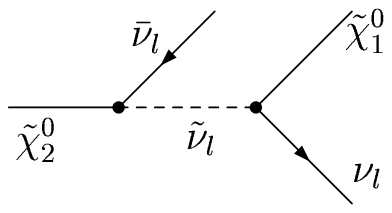} 
\end{tabular}
\end{center}
\caption{The Born Feynman diagrams for $\tilde{\chi}_2^0 \longrightarrow
  \tilde{\chi}_1^0 \nu_l \bar\nu_l $.\label{treediagramnu}}
\end{figure}

The invisible decays $\tilde \chi_2^0 \to\tilde \chi_1^0 \nu_l \bar\nu_l$ are calculated at tree and one-loop level. 
The Born Feynman diagrams are shown in Figure \ref{treediagramnu}. 
Here we focus on the case where
the decays $\tilde \chi_2^0 \to\tilde \chi_1^0 \nu_l \nu_l$ are pure three-body decays. 
Similarly to the calculations for the leptonic three-body decays
$\tilde \chi_2^0 \to\tilde \chi_1^0 l^- l^+$, these decays are also calculated with the help of
{\em FeynArts, FormCalc} and {\em LoopTools}. The one-loop corrections are again classified as vertex, self, and box 
contributions.
Since none of the external particles carries electric charge, there are no corrections involving
real or virtual photons, and hence no IR divergences.
Therefore, there are also no QED corrections in these decays. This makes the calculations 
for the invisible decays much simpler than for the leptonic decays.
\subsection{The Hadronic Decays $\tilde \chi_2^0 \to\tilde \chi_1^0 q \bar q~(q \neq t)$}
\begin{figure}[htb]
\begin{center}
\begin{tabular}{cccc}
\includegraphics[width=0.22\linewidth]{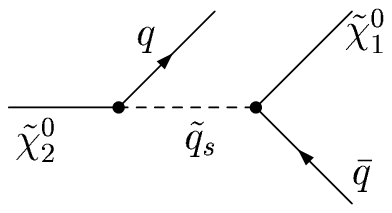} &
\includegraphics[width=0.22\linewidth]{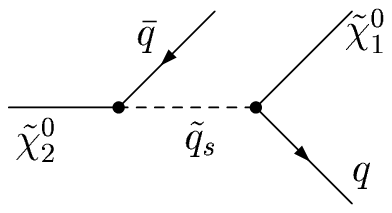}& 
\includegraphics[width=0.22\linewidth]{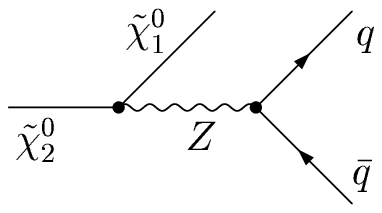} &
\includegraphics[width=0.22\linewidth]{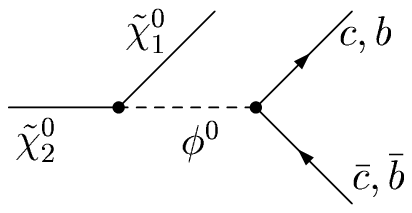}
\end{tabular}
\end{center}
\caption{The Born Feynman diagrams for $\tilde{\chi}_2^0 \longrightarrow
  \tilde{\chi}_1^0 q \bar q ~(q\neq t)$, $s = 1,2$ labels the squark
  mass eigenstates, $\phi^0$ denotes the neutral Higgs boson $h^0, H^0, A^0$ and the neutral Goldstone boson $G^0$.
\label{treediagramqq}}
\end{figure}

The hadronic decays of $\tilde \chi_2^0$ are calculated in order to obtain the total width of $\tilde \chi_2^0$.
The Born Feynman diagrams for the decays $\tilde \chi_2^0 \to\tilde \chi_1^0 q \bar q~(q \neq t)$ are shown in
Figure \ref{treediagramqq} where the masses of the light quarks, i.e. $u, d$ and $s$, are neglected.
Here we only consider the case where
the hadronic decays $\tilde \chi_2^0 \to\tilde \chi_1^0 q \bar q$ are pure three-body decays. Since
the SUSY-QCD corrections are not considered in our calculations, the decays $\tilde \chi_2^0 \to\tilde \chi_1^0 q \bar q$ 
can be treated in the same way as $\tilde \chi_2^0 \to\tilde \chi_1^0 l^- l^+$.

The one-loop virtual corrections for the decays $\tilde \chi_2^0 \to\tilde \chi_1^0 q \bar q$
are also classified into three types: 
vertex contributions, self-energy contributions and box contributions, where the vertex and self-energy contributions have been 
combined with their counterterms. The calculations for these contributions are performed with the help of 
{\em FeynArts, FormCalc} and {\em LoopTools}. 
The virtual photonic corrections (the diagrams are similar to the ones of the leptonic decays, i.e. Figure \ref{photonvirt})
are IR divergent. The contributions of the real photon bremsstrahlung are necessary for the cancellation of the IR divergences.
We neglect the light quark masses, i.e. $m_q (q = u, d, s)$, except when they appear in the one-loop integrals. 
This gives a mass singularity $\ln m_q$. The virtual photonic corrections can be calculated analytically in the same way as
in Section \ref{virtual}              . 

The diagrams for the real photon bremsstrahlung  $\tilde \chi_2^0 \to\tilde \chi_1^0 q \bar q \gamma$ are similar to the diagrams
in Figure \ref{realdiagram}. In analogy to Section \ref{real}, the contributions of the real photon bremsstrahlung
are also splitted into an IR-divergent soft part and an IR-finite hard part. For the light quark final states, 
we separate the hard photon bremsstrahlung into a collinear part and a non-collinear part in order to 
obtain stable numerical results. As presented in Section \ref{real}, the soft and collinear contributions are calculated analytically.
The analytical expressions for the singular part of the virtual photonic corrections, the soft and collinear contributions
can be obtained by performing the replacements
\begin{align}
Q_l^2 & \to Q_q^2\, ,
\ \ \ \  \ln m_l \to \ln m_q \, 
\end{align}
in the corresponding expressions in Section \ref{virtual} and \ref{real}. 
We treat the decays with heavy quark final states in the same way as $\tau^-\tau^+$ final states.
The QED corrections are defined in the same way 
as in Section \ref{QED} since the photonic contributions are UV divergent and cannot be treated separately.
\newpage
\section{Numerical Results and Discussions}\label{numerical}
In this section we present the numerical results both for 
a scenario where $\tilde \chi_2^0$ can undergo two-body
decays $\tilde \chi_2^0 \rightarrow \tilde l^\pm l^\mp \to \tilde \chi_1^0 l^- l^+$,
and for a scenario where $\tilde \chi_2^0$ only has three-body decays $\tilde \chi_2^0 \rightarrow \tilde \chi_1^0 l^- l^+$.
The two-body decays are calculated in a complete and an approximate way, and the corresponding numerical results are compared. 
We discuss the total decay width of $\tilde \chi_2^0$ and
the branching ratios of its leptonic decays. The dilepton invariant-mass $M_{l^+ l^-}$ distribution 
is also presented and discussed, where the dilepton invariant mass $M_{l^+ l^-}$
is defined as
\begin{equation}
M_{l^+l^-} = \sqrt {(k_{l^+} + k_{l^-})^2}\, .
\end{equation}
As discussed in Section \ref{real}, the dilepton invariant mass $M_{e^+e^-}$ is defined as collinear-safe observable,
i.e. we add the momentum of a collinear photon to that of the emitting electron, since
it is difficult to separate their energies at the LHC.
The energies of a muon and its collinear photon can be disentangled easily at the LHC,
hence the dilepton invariant mass $M_{\mu^+\mu^-}$ is defined as non-collinear-safe observable,
i.e. the momentum of a collinear photon is not added to that of its emitter muon.  
In this case the large logarithm $\ln m_{\mu}$ can not cancel in the distribution,
so the mass effect can be seen in the dilepton invariant-mass distribution. 
In our calculations the selectrons and smuons have equal masses and the light lepton mass $m_l~(l = e, \mu)$ is neglected 
except when it appears in the one-loop integrals,
so one will obtain identical distributions for $M_{e^+e^-}$ and $M_{\mu^+ \mu^-}$ if
both of them are defined as collinear-safe observables.
In order to see the differences of the two treatments (adding and not adding the momentum of a collinear photon
to the emitting lepton), we also show the comparison of dilepton invariant mass $M_{\mu^+\mu^-}$ and  $M_{e^+e^-}$ distributions.
\subsection{SPS1a Parameter Set}\label{sec:SPS1a}
In the MSSM, soft-SUSY breaking is implemented by adding all possible soft terms 
to the Lagrangian instead of assuming a particular SUSY-breaking mechanism. 
All the parameters in the soft-SUSY breaking Lagrangian (\ref{eqn:softLagrangian}) are
general matrices in flavor space and may be complex. This leads to
more than a hundred unknown free parameters in the MSSM. 
Therefore, it is not practicable to scan over the entire parameter space. 
The "Snowmass Points and Slopes" (SPS), where several ``benchmark scenarios'' 
have been suggested \cite{SPS1aa, SPS1ab}, are meant to illustrate characteristic 
features of various scenarios of SUSY breaking. Among those, 
the so-called SPS1a parameter set has been studied particularly widely.

The SPS1a parameter set is defined in the framework of the mSUGRA scenario \cite{MSSM},
where the SUSY-breaking mechanism is supposed to be minimal supergravity.
This scenario is characterized by four parameters and a sign, the scalar mass
parameter $M_0$, the gaugino mass parameter $M_{1/2}$, the scalar trilinear coupling 
$A_0$, the ratio of the Higgs vacuum expectation values $\tan\beta$, and the sign of the supersymmetric 
Higgs mass parameter $\mu$.  These parameters are defined in the $\overline{\rm DR}$ 
scheme at the GUT scale $M_{\rm GUT}$. The SPS1a benchmark point is defined by setting~\cite{SPS1aa}
\begin{equation}
M_0 = 100 \ {\rm GeV}\, , \ \ \ M_{1/2} = 250 \ {\rm GeV}\, , \ \ \ A_0 =  -100 \ {\rm GeV}\, , \ \ \ \tan\beta = 10\,
\ \ \ \mu > 0\, .  
\end{equation}
The low-energy parameters in the MSSM are obtained via renormalization group running from the high-energy scale 
to the weak scale. This can be performed with various programs \cite{isajet, softsusy, suspect}.

In Ref.~\cite{SPS1ab} the corresponding low-energy parameters for the SPS1a benchmark point are obtained with 
{\sl ISAJET 7.58}~\cite{isajet}, which read
\begin{eqnarray}
&&M_{\tilde g} = 595.2 \ {\rm GeV}, \quad m_{A^0} = 393.6 \ {\rm GeV}, \quad \mu = 352.4 \ {\rm GeV}, \nonumber \\
&&\tan\beta =  10, \quad \hspace*{11mm}M_1 = 99.1 \ {\rm GeV}, \quad \hspace*{3mm}M_2 = 192.7 \ {\rm GeV}\, ,
\label{eqn:parametersSPS1a}
\end{eqnarray}
where $M_1$ and $M_2$ are the gaugino mass parameters, $M_{\tilde g}$ denotes the mass of the gluino,
$ m_{A^0}$ is the mass of the neutral CP-odd Higgs boson $A^0$. 
The soft SUSY-breaking parameters in the diagonal entries of the squark
and slepton mass matrices have been chosen to be the same for the first 
and second generation. The off-diagonal entries have been neglected for the
first two generations, i.e. there are no sfermion mixing.  
The soft SUSY-breaking parameters in the diagonal entries of the the squark
and slepton mass matrices are  
\begin{eqnarray}
&&M_{\tilde Q1_L} = M_{\tilde Q2_L} = 539.9 \ {\rm GeV}, \quad
M_{\tilde{d}_R} = 519.5 \ {\rm GeV},  \quad
M_{\tilde{u}_R} = 521.7 \ {\rm GeV},\nonumber \\
&& M_{\tilde{l}_L} = 196.6 \ {\rm GeV}, \quad \hspace*{17mm}
M_{\tilde{l}_R} = 136.2  \ {\rm GeV}\, , 
\label{eqn:parametersfermion2}
\end{eqnarray}
where the index $i$ in $M_{\tilde Qi_L}$ denotes the generation, $\tilde{u}$ and $\tilde{d}$
denote the up- and the down-squarks for the first two generations and $\tilde{l}$
stands for the first and second generation sleptons.
The soft SUSY-breaking parameters in the diagonal entries of the squark and
slepton mass matrices of the third generation have the following values,
\begin{eqnarray}
&&M_{\tilde Q3_L} = 495.9 \ {\rm GeV} , \quad 
M_{\tilde{b}_R} = 516.9 \ {\rm GeV}, \quad 
M_{\tilde{t}_R} = 424.8 \ {\rm GeV}, \nonumber \\
&&M_{\tilde{\tau}_L} = 195.8 \ {\rm GeV}, \quad \hspace*{2mm}
M_{\tilde{\tau}_R} = 133.6 \ {\rm GeV}, 
\label{eqn:parametersfermion3}
\end{eqnarray}
while the trilinear couplings of the third generation read
\begin{equation}
A_t = -510.0 \ {\rm GeV}, \quad
A_b = -772.7 \ {\rm GeV}, \quad
A_{\tau} = -254.2 \ {\rm GeV}.
\label{eqn:parameterstrilinear}
\end{equation}

\begin{figure}[htb]
\begin{center}
\begin{tabular}{c}
\includegraphics[width=0.7\linewidth]{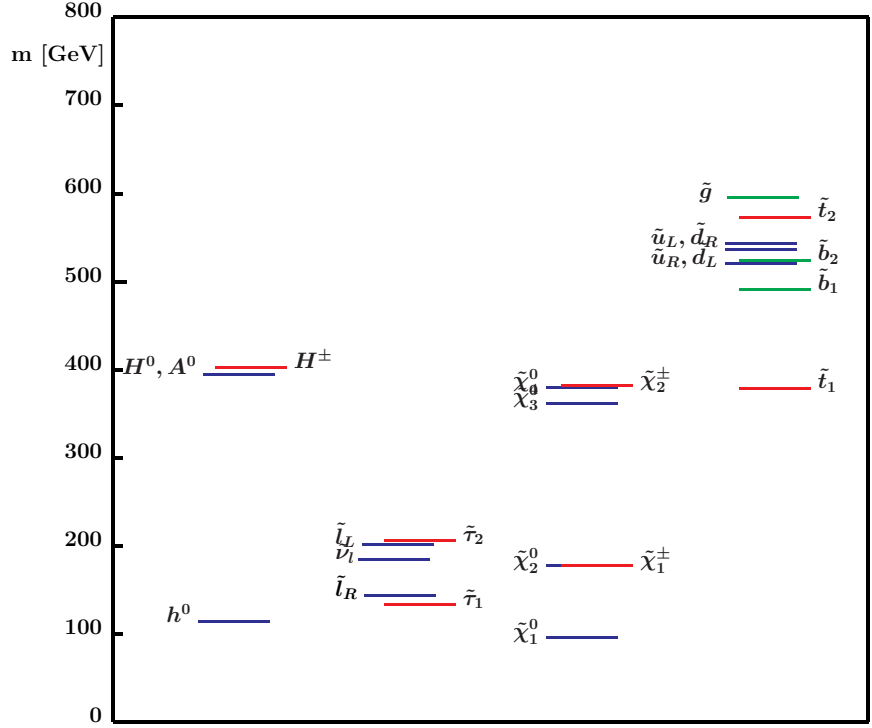}
\end{tabular}
\caption{The spectrum corresponding to SPS1a benchmark point\label{Figure:SPS1a}}
\end{center}
\end{figure}

The SPS1a benchmark point gives rise to a particle spectrum where many
states are accessible both at the LHC and at the ILC
\cite{LHC/LC}. The spectrum of supersymmetric particles at this benchmark point
is shown in Figure \ref{Figure:SPS1a}~\cite{SPS1aa, SPS1ab, SPS1ac}.
Note in particular that the
two-body decays $\tilde{\chi}_2^0\rightarrow \tilde{l}^\pm_1 l^\mp
\rightarrow \tilde{\chi}_1^0 l^- l^+$ are kinematically allowed. No
other two-body decay mode for $\tilde{\chi}_2^0$ is open. 
Moreover, the sleptons have the same masses for the first two generations,
i.e. $m_{\tilde \nu_e} = m_{\tilde \nu_\mu}$, $m_{\tilde e_L} = m_{\tilde \mu_L}$,  $m_{\tilde e_R} = m_{\tilde \mu_R}$.

Another set of parameter is called SPS1a$'$ which is proposed in the SUSY Parameter Analysis (SPA) project \cite{SPA}.
Its parameters are close to the snowmass point SPS1a except for a small change of the scalar mass parameter and
the trilinear coupling to comply with the measured dark matter density.  
The low-energy parameters for the SPS1a benchmark point  
which have been shown in (\ref{eqn:parametersSPS1a}, \ref{eqn:parametersfermion2}, \ref{eqn:parametersfermion2},
\ref{eqn:parameterstrilinear}) are used as the input parameters in calculating the decays of
$\tilde \chi_2^0$.
\subsection{Numerical Results for the SPS1a Parameter Set}\label{numericalSPS1a}
The SPS1a parameter set has been discussed in Section~\ref{sec:SPS1a}, where the mass spectrum of the MSSM particles
was given in Figure \ref{Figure:SPS1a}. 
Since the two-body decays $\tilde \chi_2^0 \rightarrow \tilde l_1 l$ are kinematically allowed,
the decays $\tilde\chi_2^0 \to \tilde\chi_1^0 l^- l^+$ are calculated in both complete and approximate way 
as discussed in Section \ref{complete} and \ref{approx}. The numerical results of both methods are compared.

\begin{figure}[htb]
\begin{center}
\psfrag{coll}{{\hspace{-8mm}\tiny $\tilde \Gamma + \Gamma_{\rm coll}$}}
\psfrag{noncoll}{{\hspace{-5mm}\tiny $\Gamma_{\rm non-coll}$}}
\psfrag{QED}{{\hspace{-3mm}\tiny $\Gamma_{\rm QED}$}}
\begin{tabular}{cc}
\includegraphics[width=0.45\linewidth]{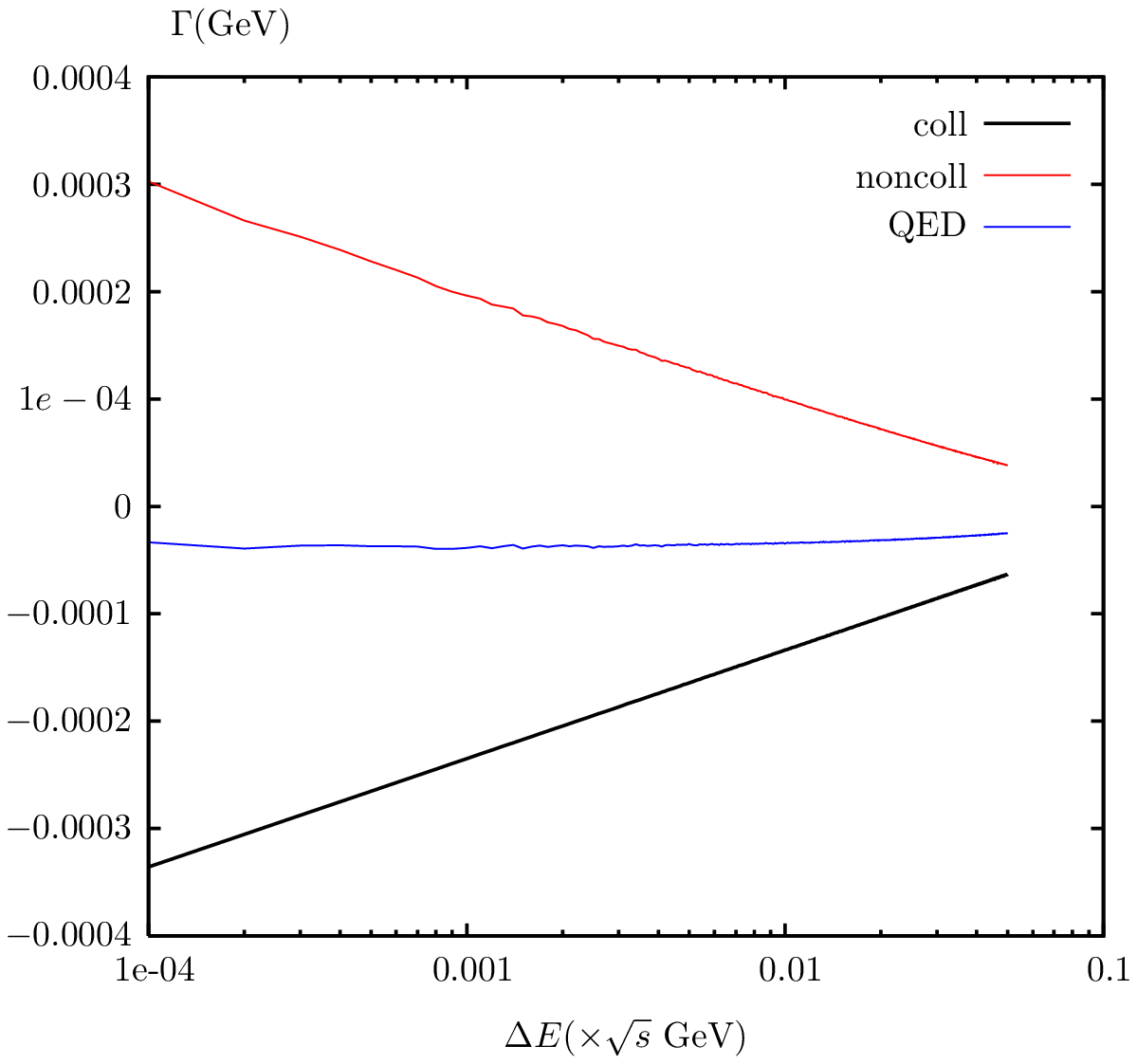} & \includegraphics[width=0.45\linewidth]{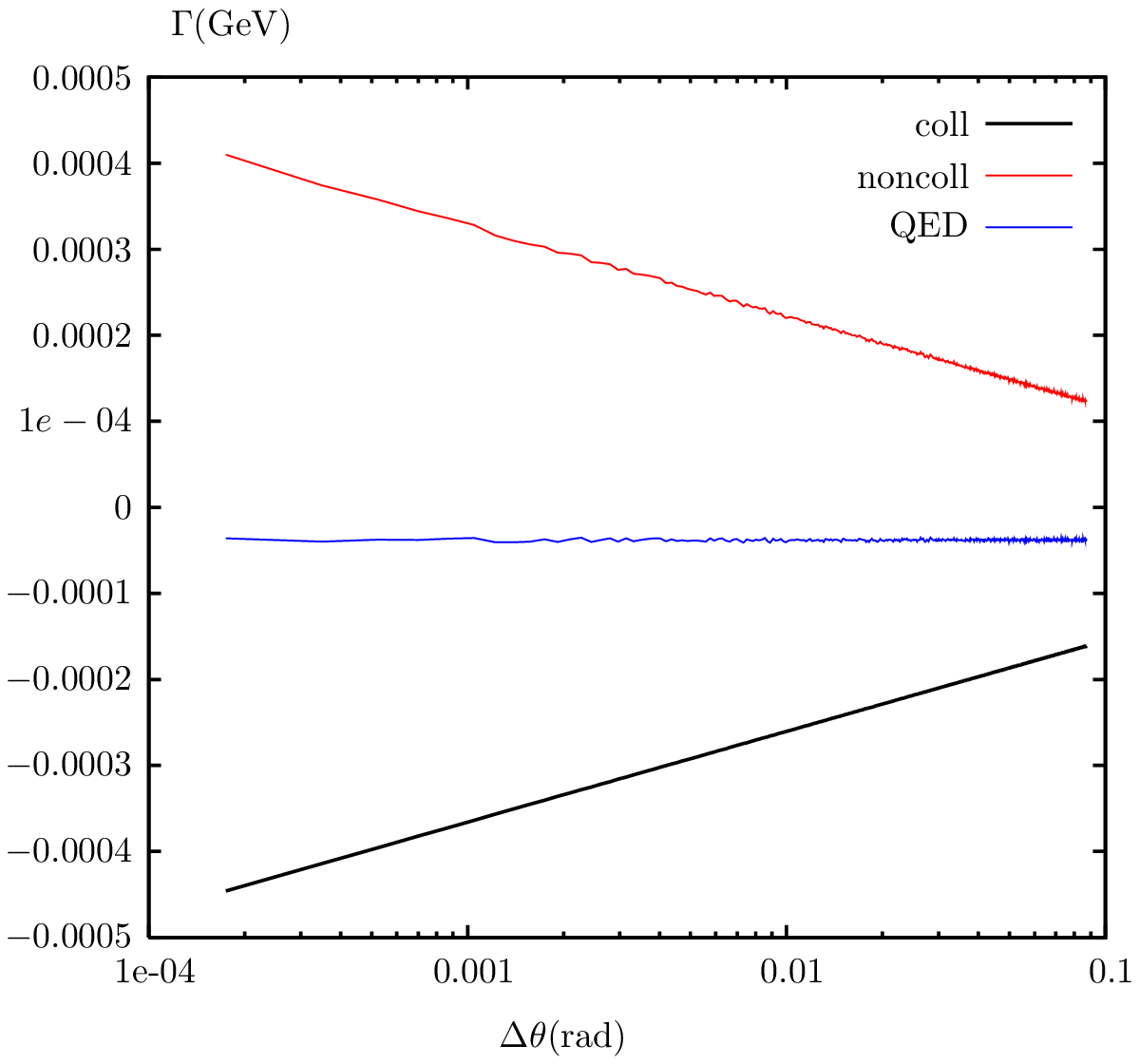}
\end{tabular}
\caption{The dependence of $\tilde{\Gamma} + \Gamma_{\rm coll}$, $\Gamma_{\rm non-coll}$ and $\Gamma_{\rm QED}$ on the cutoff parameters 
$\Delta E$ and $\Delta \theta$, where we have used $\Delta \theta = 1^\circ$ 
and $\Delta E = 0.001\sqrt s$ in the left and right plot, 
respectively. Here $\sqrt s = m_{\tilde \chi_2^0}$ denotes the center-of-mass energy.
The results are from the decay $\tilde \chi_2^0 \to \tilde \chi_1^0 e^- e^+$ for the SPS1a parameter set.\label{check}}
\end{center}
\end{figure}

We calculate the QED corrections of the decay $\tilde\chi_2^0 \to \tilde\chi_1^0 e^- e^+$ for the SPS1a parameter set 
and show the contributions $\tilde{\Gamma} + \Gamma_{\rm coll}$, $\Gamma_{\rm non-coll}$ and $\Gamma_{\rm QED}$ 
as function of $\Delta E$ and $\Delta \theta$ in Figure \ref{check}.
From these figures one obtains that the QED corrections do not depend on these cut-off parameters 
when they are very small. This is in accordance with the discussions in Section~\ref{real} 
that the cut-off parameters should be small enough so that the soft and collinear contributions
can be treated approximately as in (\ref{softapprox}) and (\ref{collinearapprox}).
The instabilities of QED corrections can be seen when the values of the cut-off parameters become too small. 
As a result, we use $\Delta \theta = 1^\circ$ and $\Delta E = 0.001\sqrt s$ in our calculations.
\newpage
\subsubsection{Dilepton Invariant Mass $M_{l^+ l^-}$ Distribution}
\begin{figure}[htb]
\begin{tabular}{ll}
\includegraphics[width=0.48\linewidth]{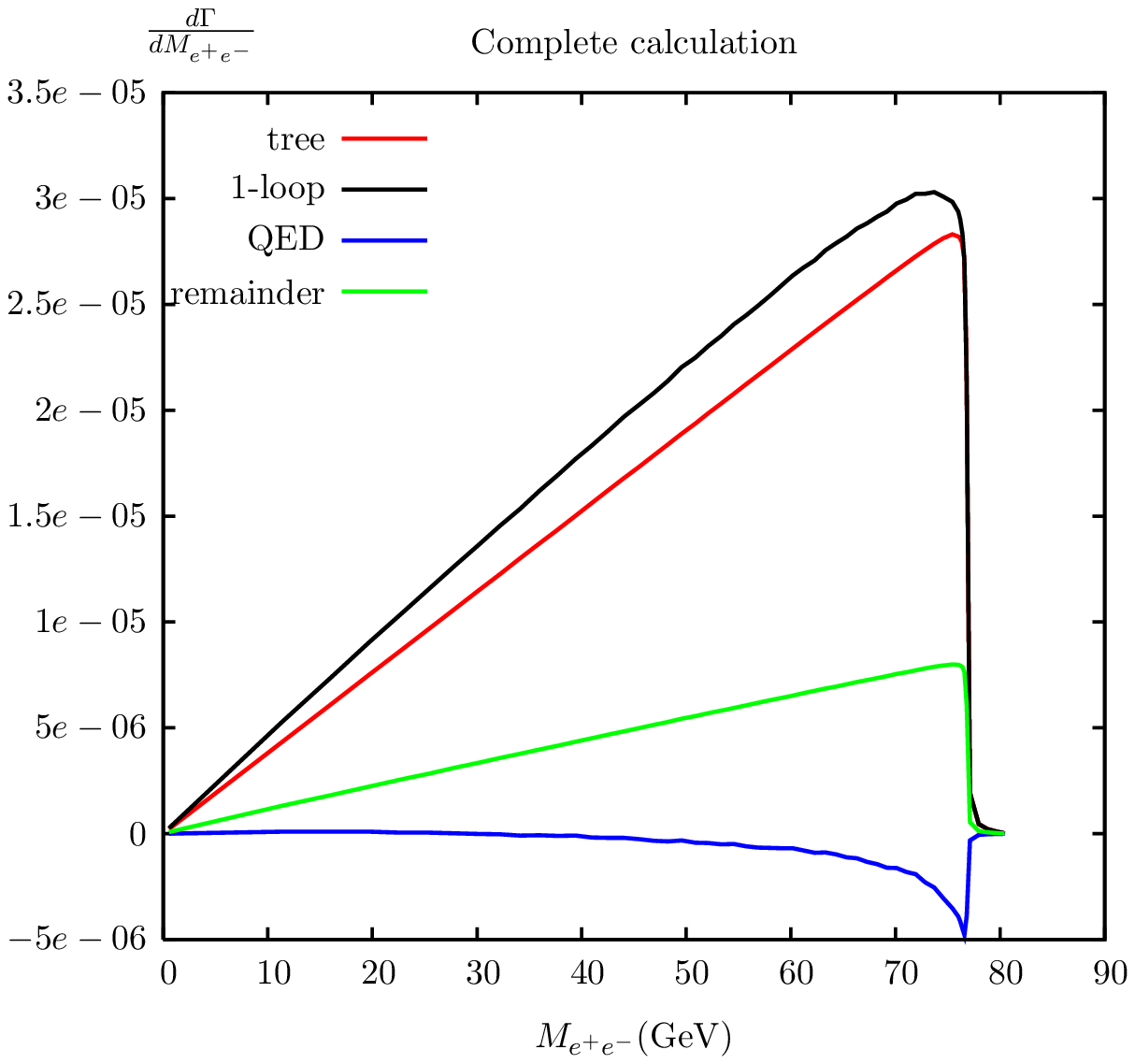}&
\includegraphics[width=0.465\linewidth]{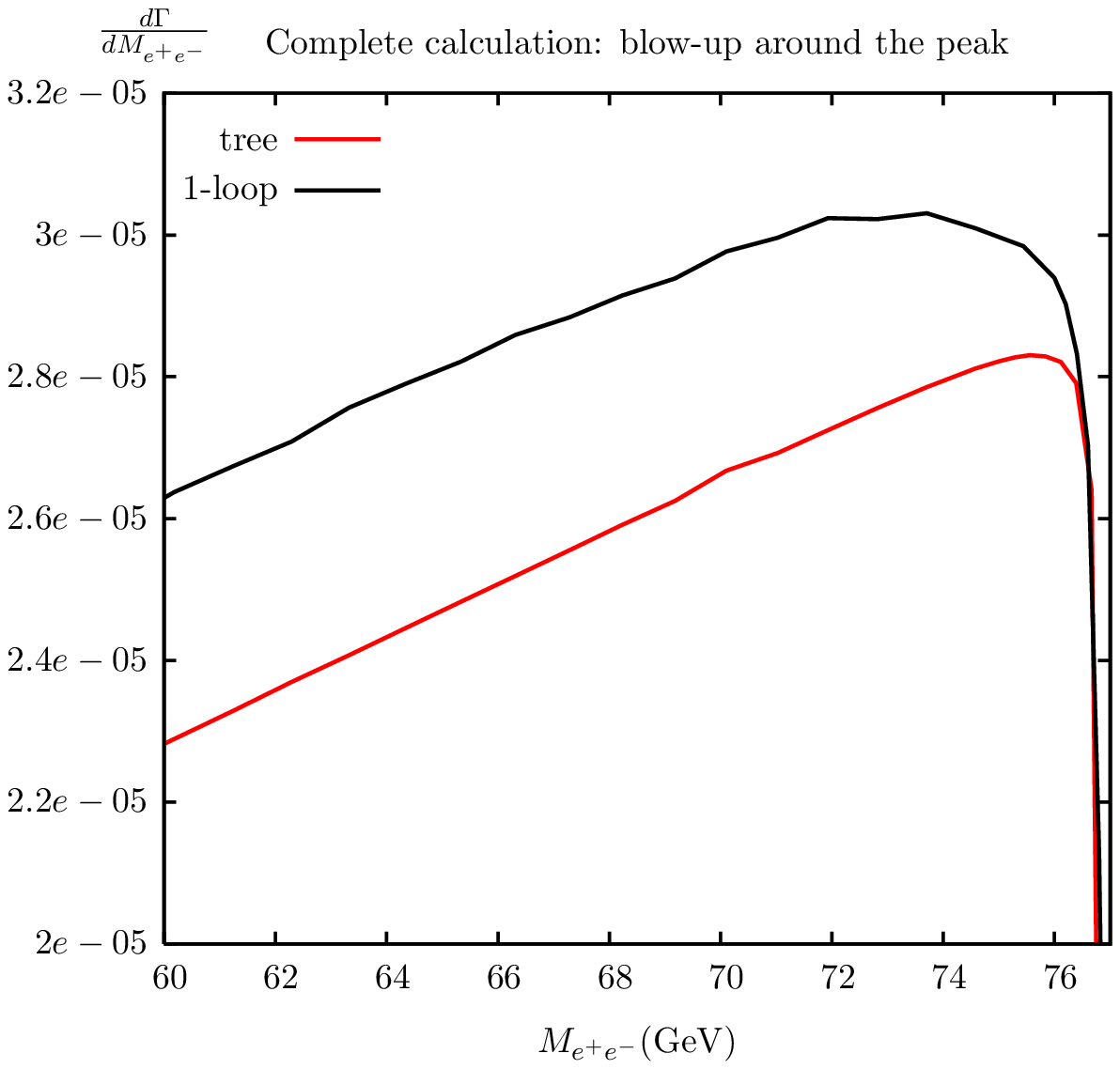}
\end{tabular}
\caption{The dilepton invariant mass $M_{e^+ e^-}$ distribution for the SPS1a
  parameter set (complete calculation). \label{SPS1aMee}}
\end{figure}

The dilepton invariant mass $M_{e^+ e^-}$ distribution from the
complete calculation is shown in Figure~\ref{SPS1aMee}. In the left frame we
show not only the tree- and one-loop-level predictions, but also
the separate QED and ``remainder'' corrections, see (\ref{llrem}). We
see that the non-QED contributions are positive and quite large everywhere,
whereas the QED contribution is large and negative near the endpoint of the
distribution, but small elsewhere. In full three-body kinematics this
endpoint is simply given by 
\begin{equation}
\left. M_{e^+e^-}^{\rm max}\right|_{\rm 3-body} = m_{\tilde \chi_2^0} -
m_{\tilde \chi_1^0} = 80.4 {\rm GeV}\, , 
\end{equation}
\begin{figure}[htb]
\begin{tabular}{ll}
\includegraphics[width=0.48\linewidth]{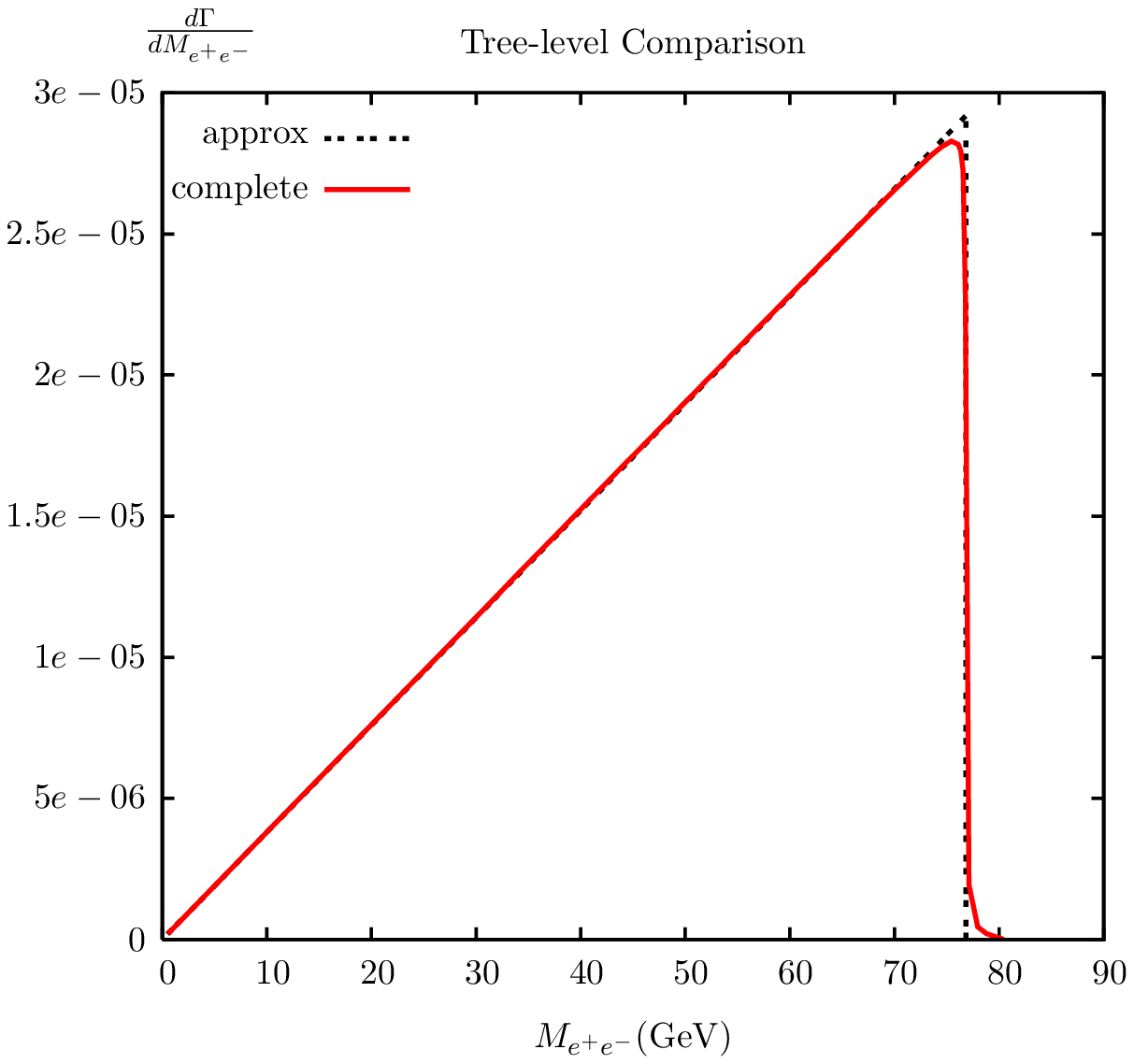}
\includegraphics[width=0.465\linewidth]{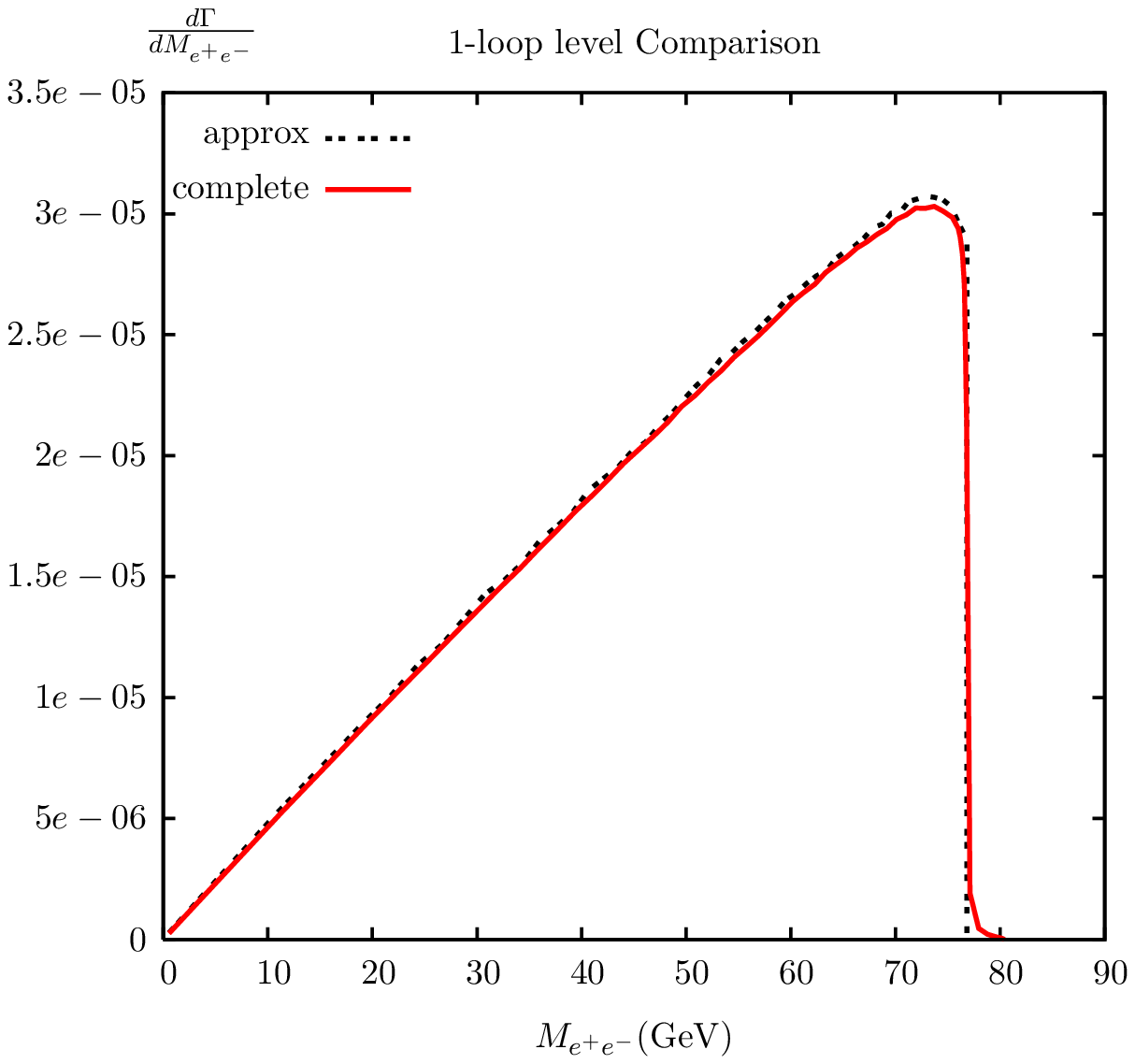}
\end{tabular}
\caption{The dilepton invariant mass $M_{e^+ e^-}$ distribution for the SPS1a
  parameter set (comparison of the complete calculation with the approximate
  calculation). \label{comparisonEE}}
\end{figure}
where the numerical value holds for the SPS1a scenario. 
Moreover, in this scenario $\tilde \chi_2^0$
decays are dominated by contributions with on-shell $\tilde l_1$ in the
intermediate state. The endpoint for this two-body configuration is given by
\begin{equation}
\left. M_{e^+ e^-}^{\rm max}\right|_{\rm 2-body} =  m_{\tilde{\chi}_2^0}
  \sqrt{1- \frac{ m_{\tilde{e}^{\pm}_1}^2} {m_{\tilde{\chi}_2^0}^2}} 
\sqrt{1- \frac{m_{\tilde{\chi}_1^0}^2} {m_{\tilde{e}^{\pm}_1}^2}}  \simeq 76.8
  \ {\rm GeV} \, .
\label{Mee_end}
\end{equation}
Note that this is only
3.6 GeV below the endpoint of the three-body decays. At tree level, 
the $M_{e^+e^-}$ distribution peaks
at the region which is a little below the endpoint of the two-body contribution. 
The right panel in
Figure~\ref{SPS1aMee}, which shows a blow-up of the peak region, shows that
the shape of $M_{e^+e^-}$ distribution altered once higher order corrections are included, 
i.e. the peak of this distribution is moved downwards by about $4$ GeV.
This is almost entirely due to contributions where a hard photon is emitted,
which takes away energy from the $e^+e^-$ system. This change of the shape of
the invariant-mass distribution near the endpoint is important, since in
(simulated) experiments one needs a fitting function describing this
distribution in order to determine the location of the endpoint \cite{meefit}.

In Figure~\ref{comparisonEE} we compare the numerical results of the complete
calculation and the single-pole approximation at tree (left) and one-loop
level (right). At tree level the $M_{e^+e^-}$ distribution computed in the
single-pole approximation has an exactly triangular shape, with a sharp edge
at the endpoint (\ref{Mee_end}). This edge is smeared out a bit in the
complete tree-level calculation, 
which includes complex slepton masses in the propagators as (\ref{eqn:propM0}).
As noted above, this edge is also softened considerably once
hard photon emission is included. The single-pole approximation therefore
works even better in the one-loop calculation. \mbox {However}, this excellent
agreement even for the differential decay width is partially accidental. The
agreement would become somewhat worse if the endpoints in two- and three-body
kinematics were further apart. This would happen if the mass of $\tilde l_1$
was close to the mass of either $\tilde \chi_2^0$ or $\tilde
\chi_1^0$, since then one of the two square roots in (\ref{Mee_end}) would
become small.

\begin{figure}[htb]
\psfrag{u}{{\tiny $\mu$}}
\begin{tabular}{cc}
\includegraphics[width=0.48\linewidth]{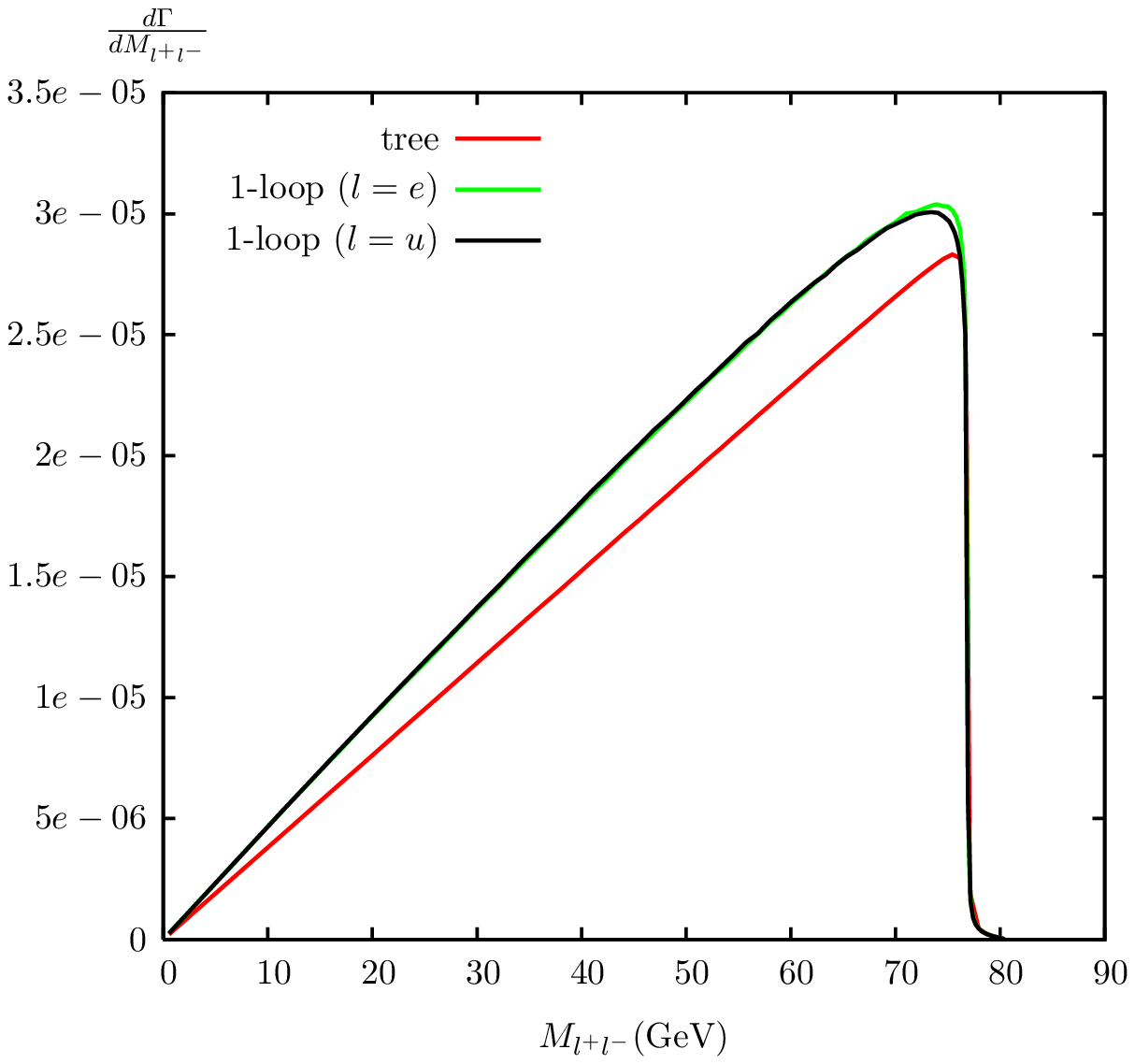}&
\includegraphics[width=0.465\linewidth]{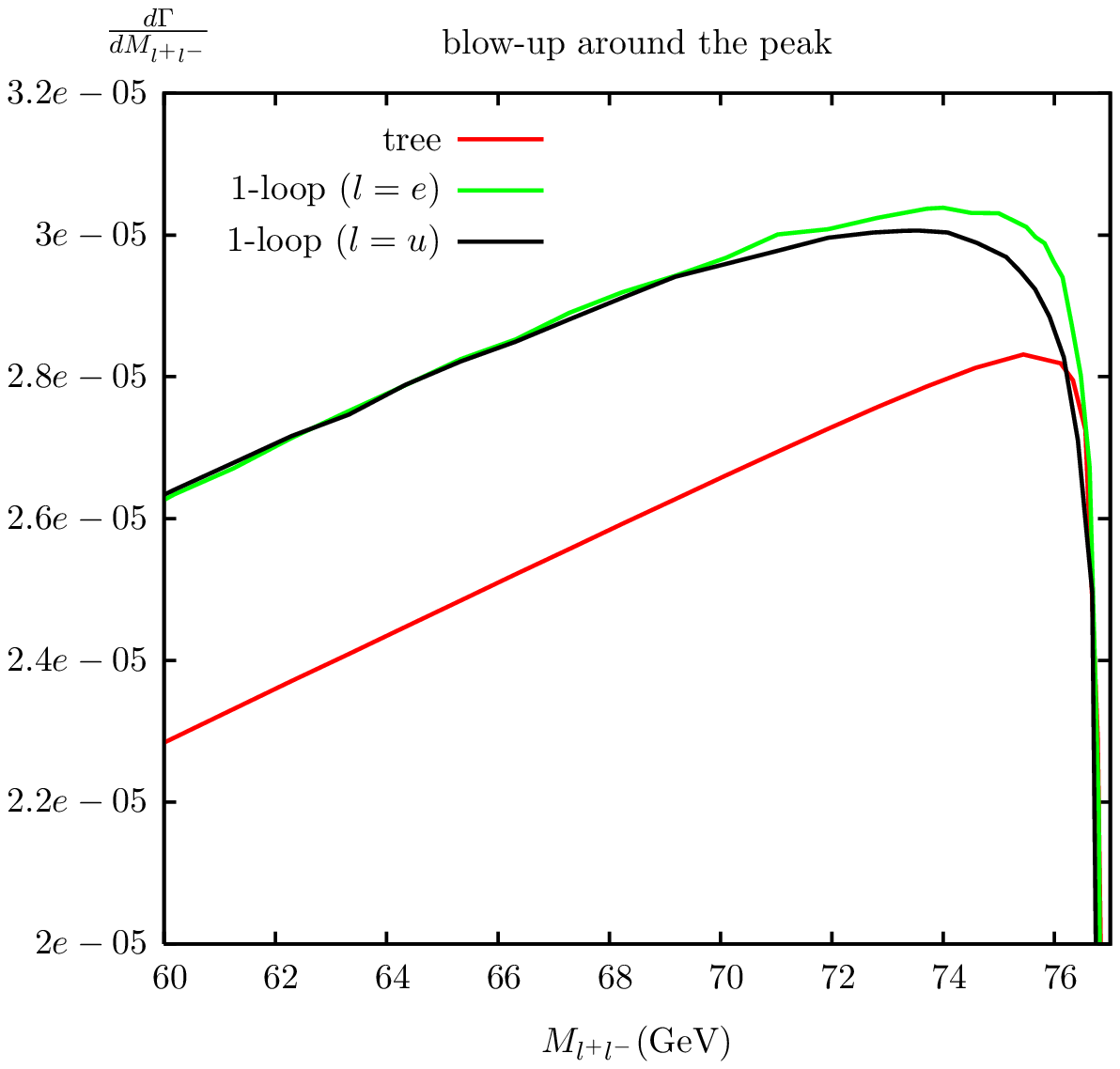}\\
\includegraphics[width=0.48\linewidth]{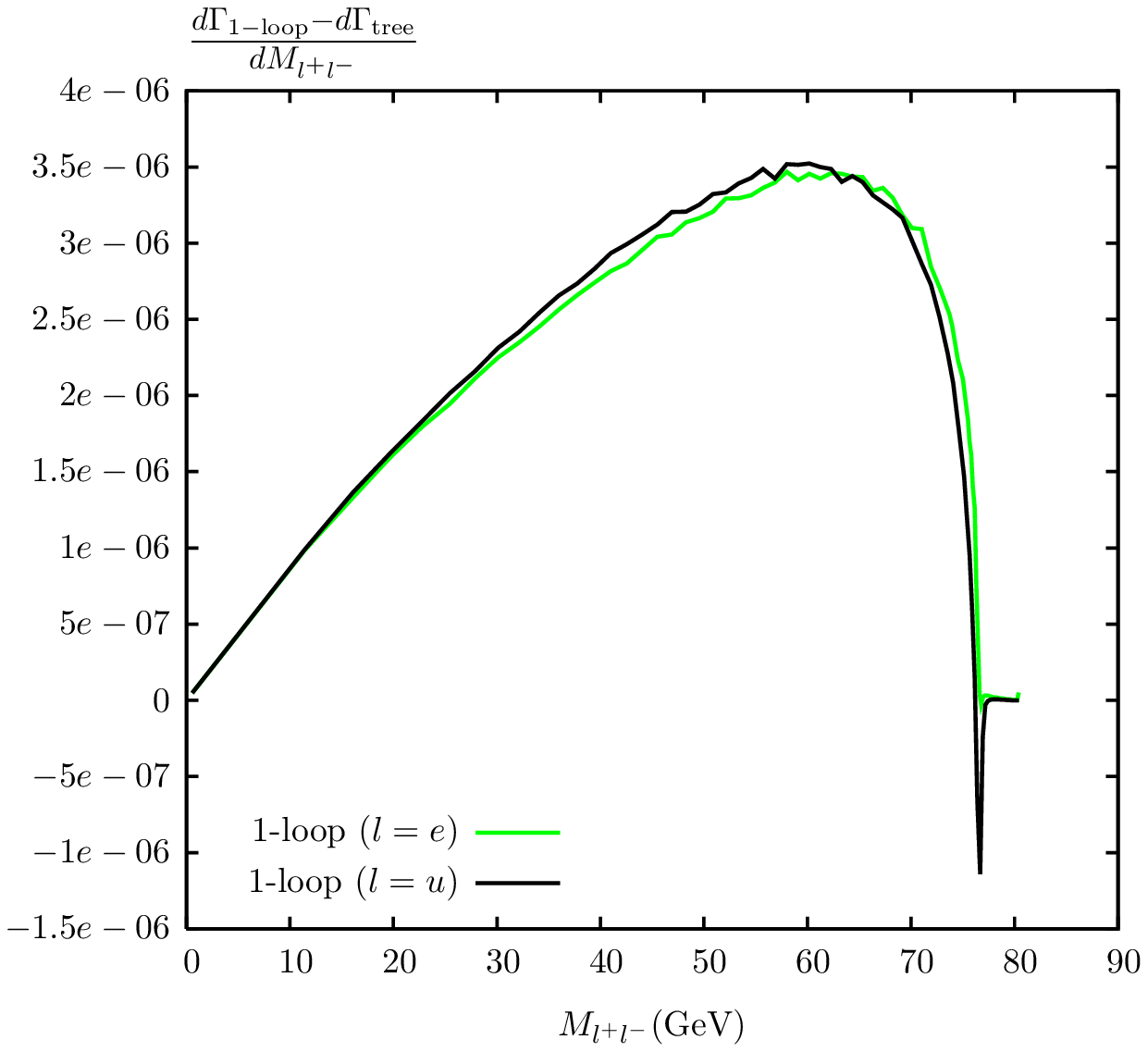}&
\includegraphics[width=0.465\linewidth]{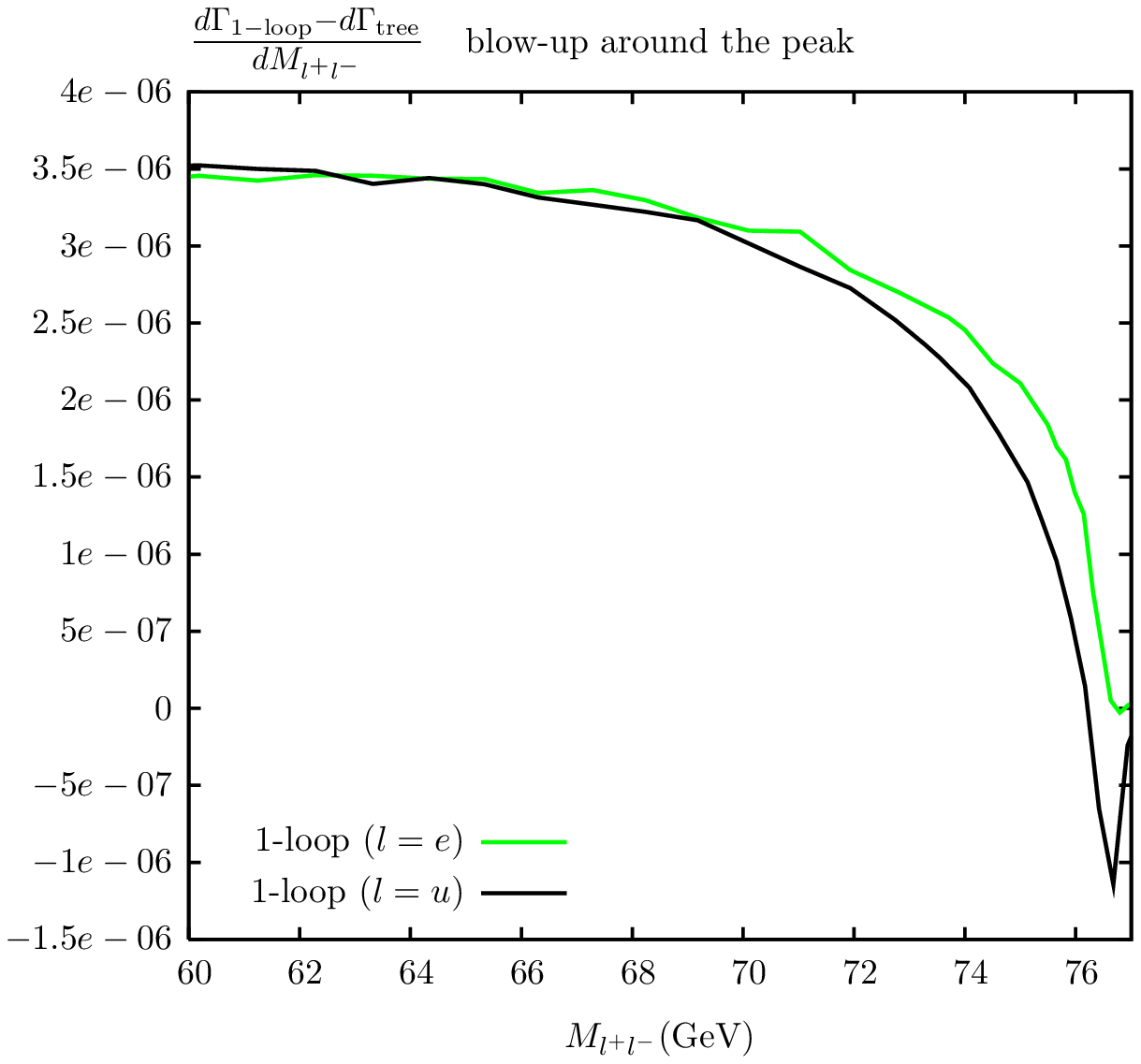}
\end{tabular}
\caption{The comparison of the dilepton invariant mass $M_{\mu^+ \mu^-}$ and $M_{e^+e^-}$ distribution for the SPS1a
  parameter set. \label{SPS1aMuu}}
\end{figure}

The comparison of the dilepton invariant mass $M_{\mu^+\mu^-}$ and  $M_{e^+e^-}$ distributions
are shown in Figure \ref{SPS1aMuu}. 
In the upper frames we show the dilepton invariant mass $M_{l^+l^-} (l = e, \mu)$ 
distribution both at tree and one-loop level.
Since the selectrons and smuons have equal masses and the light lepton mass $m_l~(l = e, \mu)$ is neglected 
except when it appears in the one-loop integrals,
their distributions are identical at tree level and different at one-loop level
due to the different treatment of the collinear-photon radiation.
From these figures one obtains that at one-loop level the mass effect is larger near the endpoint than 
in other regions and the peak of the $M_{\mu^+\mu^-}$ distribution
is shifted to lower invariant-mass values in comparison with the $M_{e^+e^-}$ distribution. 
We also show the relative one-loop corrections in the lower frames in Figure \ref{SPS1aMuu}. 
The relative one-loop corrections from the $\mu^+\mu^-$ final state is smaller than
that of the $e^+e^-$ final state in the upper invariant-mass region, while it is larger in 
the lower invariant-mass region.
The main reason is that we add the momenta of collinear photons to that of emitting electrons, 
but we do not do this for the collinear-photon radiation from muons. 
Hence the invariant mass $M_{\mu^+\mu^-}$ is reduced in comparison with $M_{e^+e^-}$. 
This leads to the shifting of events 
from the upper invariant-mass region to the lower invariant-mass region.  
%

\begin{figure}[htb]
\begin{tabular}{ll}
\includegraphics[width=0.48\linewidth]{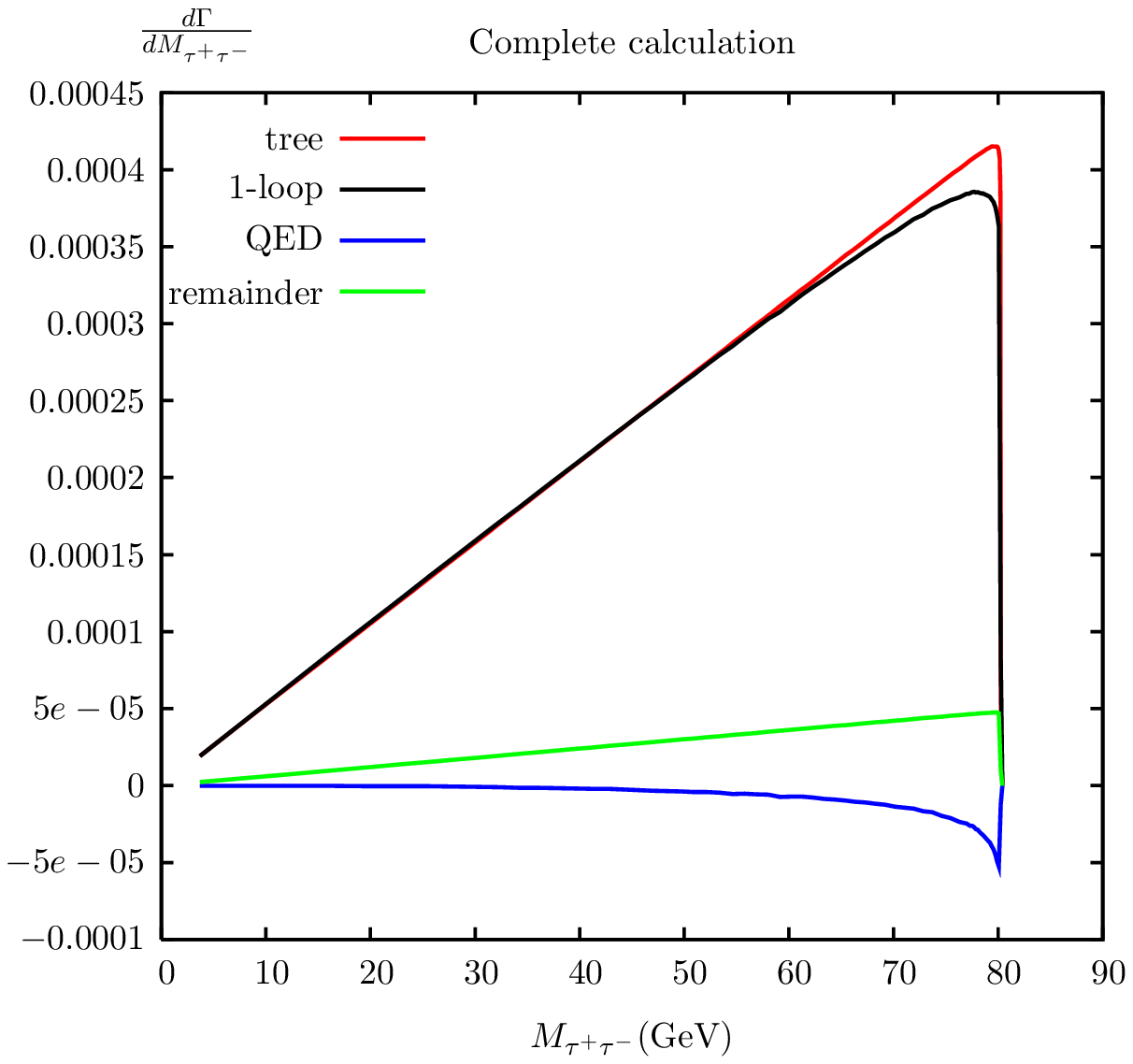}&
\includegraphics[width=0.45\linewidth]{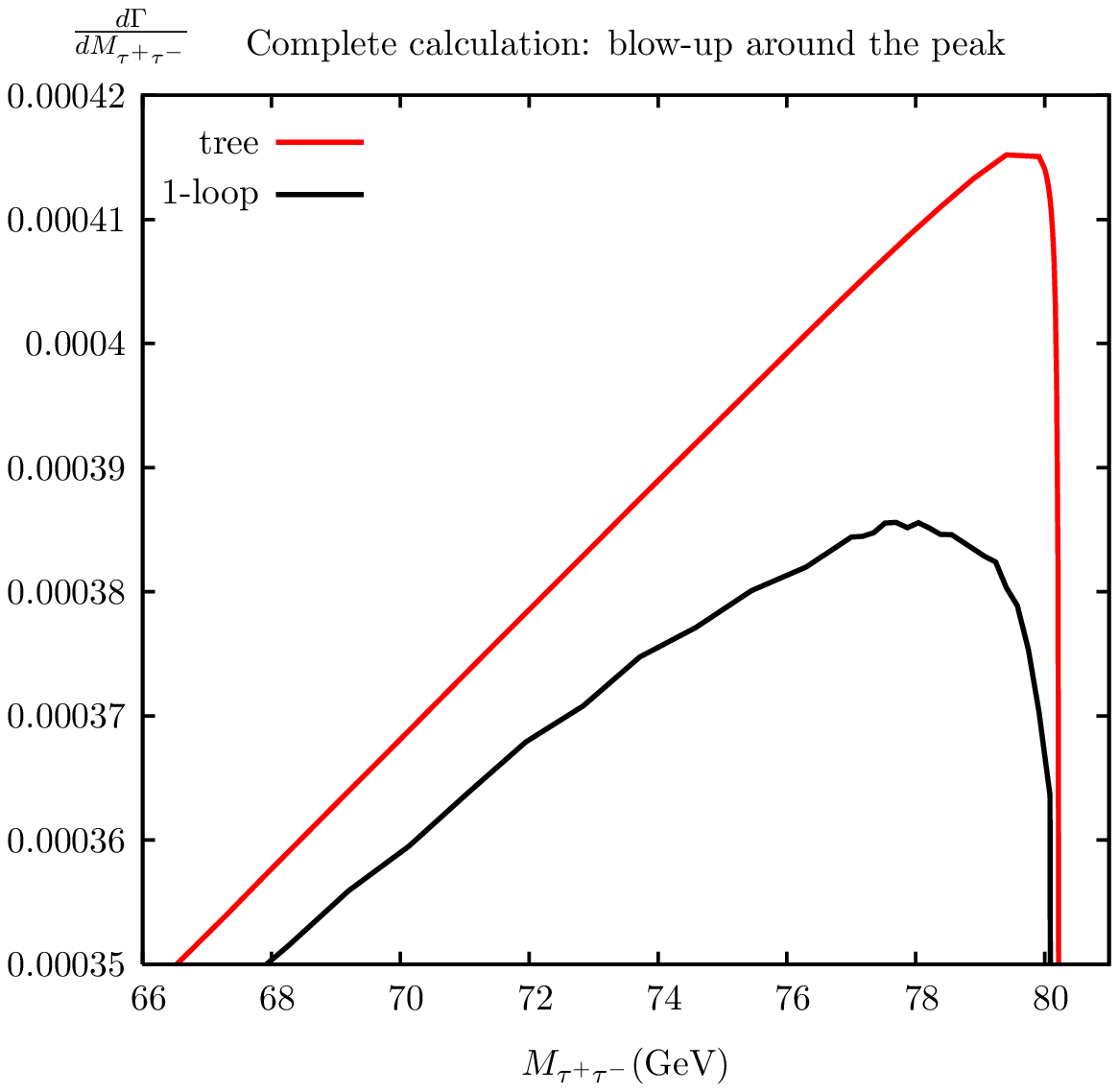}
\end{tabular}
\caption{The dilepton invariant mass $M_{\tau^+ \tau^-}$ distribution for the
  SPS1a parameter set (complete calculation). \label{SPS1aMll}}
\end{figure}
\begin{figure}[!b!]
\begin{tabular}{ll}
\includegraphics[width=0.475\linewidth]{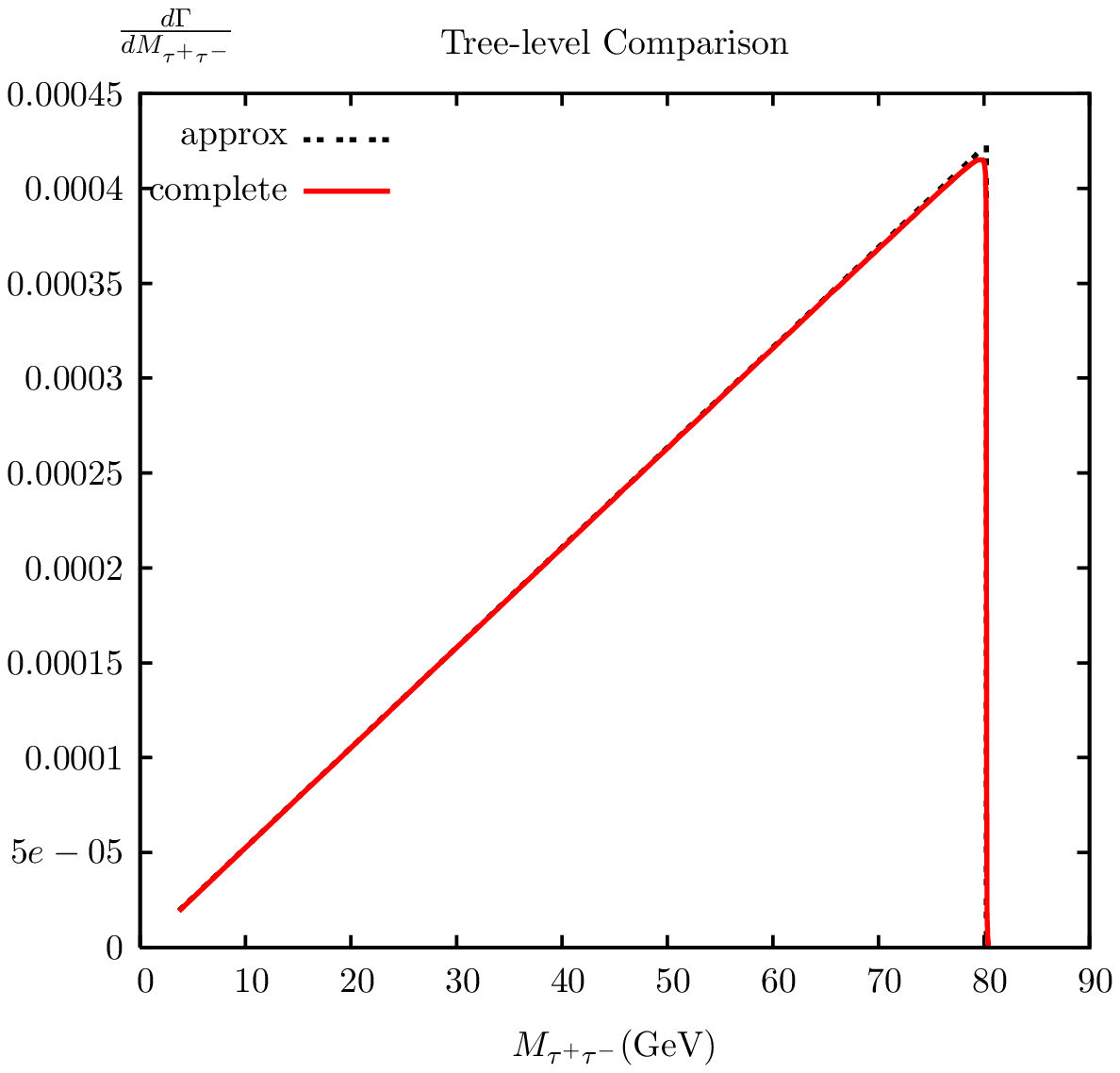}&
\includegraphics[width=0.475\linewidth]{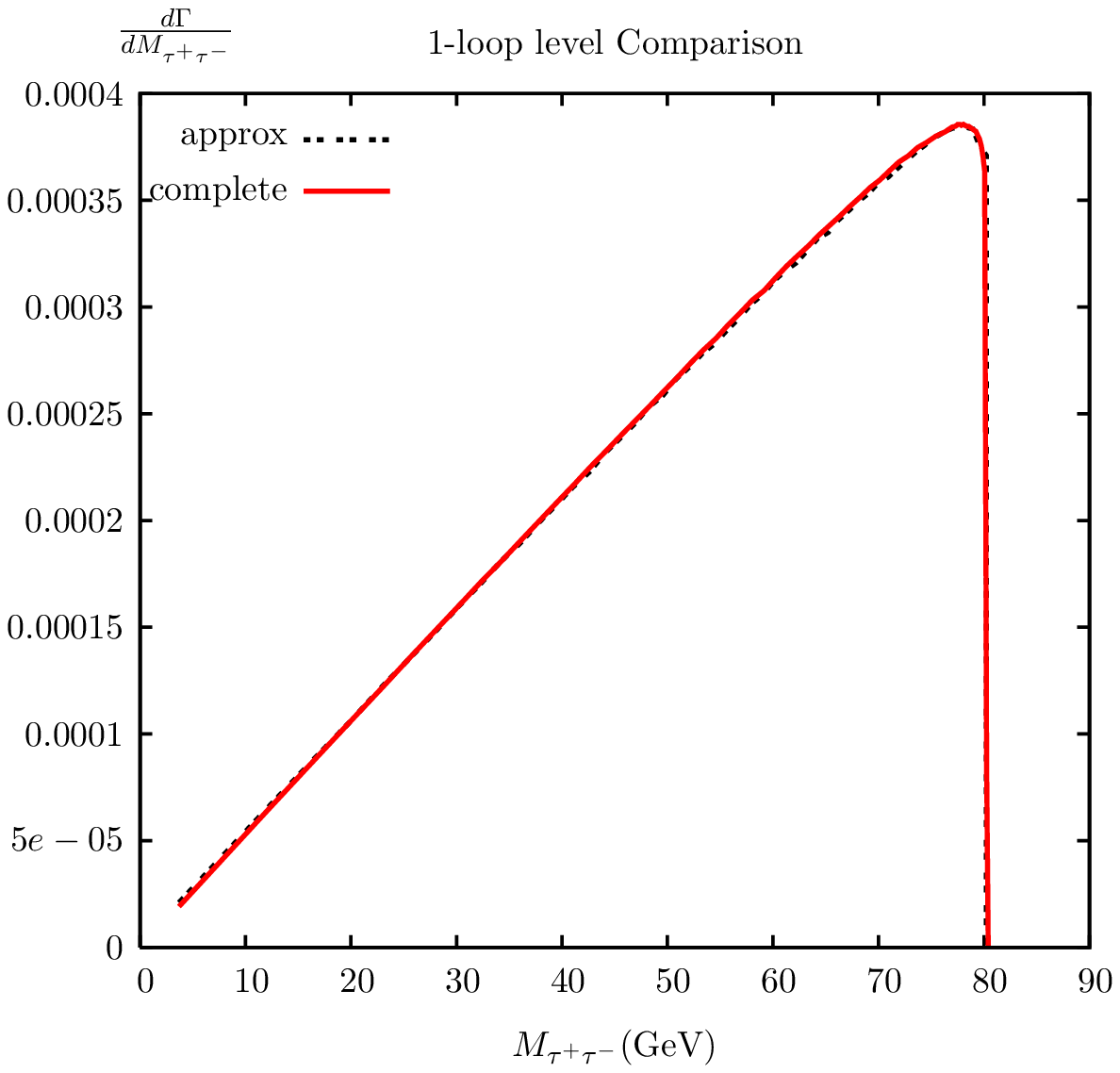}
\end{tabular}
\caption{The dilepton invariant mass $M_{\tau^+ \tau^-}$ distribution for the
  SPS1a parameter set (comparison of the complete calculation with the
  approximate calculation). \label{comparisonLL}} 
\end{figure}

The corresponding results for the dilepon invariant mass $M_{\tau^+ \tau^-}$ distribution   
are shown in Figures.~\ref{SPS1aMll} and \ref{comparisonLL}. The left plot in Figure~\ref{SPS1aMll}
shows not only the tree- and one-loop-level results, but also the QED and "remainder" corrections.
The definitions were presented in Section \ref{QED}. 
The endpoint region of the $M_{\tau^+ \tau^-}$ distribution from the complete
calculation is shown in the right panel of Figure~\ref{SPS1aMll}. We see that
the peak of the distribution is shifted downwards by about $2$ GeV once
higher-order corrections are included. A shift of this magnitude may be
significant, even though the $\tau^+ \tau^-$ invariant mass is in general
difficult to measure accurately, due to the presence of $\nu_\tau$
(anti-)neutrinos in the $\tau$ decay products, which escape detection.
In Figure~\ref{comparisonLL} predictions from the complete calculation are
compared to those from the single-pole approximation. In this case we find
almost perfect agreement even in the endpoint region, both at tree level and
after including one-loop corrections. The reason is that for the SPS1a parameter set,
$m_{\tilde \tau_1}$ happens to be very close to $\sqrt{ m_{\tilde \chi_1^0}
  m_{\tilde \chi_2^0} }$. Performing the replacement $m_{\tilde e_1} \rightarrow
m_{\tilde \tau_1}$ in (\ref{Mee_end}), one finds that the endpoint of the $\tau^+ \tau^-$
distribution in two- and three-body kinematics practically coincide.
\newpage
\newpage
\newpage
\subsubsection{Total Decay Width of $\tilde\chi_2^0$ and the
 Branching Ratios of the Decays $\tilde\chi_2^0 \to \tilde\chi_1^0 l^+ l^-$}
The partial widths of the different $\tilde{\chi}_2^0$ decay modes and the branching
ratios of its visible leptonic decays are listed in Table \ref{twobodytotal},
where the numbers in the parentheses are obtained from the approximate calculations.
We find $\sum_l \Gamma(\tilde \chi_2^0 \rightarrow \tilde \chi_1^0 \nu_l \bar
\nu_l) \ll \sum_l \Gamma(\tilde \chi_2^0 \rightarrow
\tilde \chi_1^0 l^+ l^-)$. This is not surprising, since the charged lepton
final state is accessible via on-shell $\tilde l_1$ intermediate state,
whereas for the neutrino final state all exchanged particles are
off shell. Since squark masses are near 500 GeV in SPS1a scenario, hadronic
final states contribute even less than neutrinos do.
\begin{table}[htb]
\begin{center}
\begin{tabular}{|l|l|l|}\hline
decay mode & tree-level width(MeV), \ \ Br&  one loop-level width(MeV), Br\\
\hline 
$e^{-} e^{+}\tilde{\chi}_1^0$ & $1.123$ (1.122), \hspace*{2.cm}$5.9\%$ &
$1.297$ (1.294), \hspace*{2.cm}$6.7\%$  \\ \hline 
$\mu^{-} \mu^{+}\tilde{\chi}_1^0$ & $1.123$ (1.122), \hspace*{2.cm}$5.9\%$ &
$1.297$ (1.294), \hspace*{2.cm}$6.7\%$\\ \hline 
$\tau^{-} \tau^{+}\tilde{\chi}_1^0$ & 16.870 (16.933), \hspace*{1.6cm}$88.0\%$
& 16.595 (16.646), \hspace*{1.6cm}$86.2\%$\\ \hline 
$\nu_e \bar \nu_e\tilde{\chi}_1^0$ & $0.012$ \hspace*{2.8cm}$$ & $0.012$
\hspace*{1.8cm}$$\\ \hline 
$\nu_\mu \bar \nu_\mu \tilde{\chi}_1^0$  & $0.012$ \hspace*{2.8cm}$$ & 0.012
\hspace*{1.8cm}$$  \\ \hline 
$\nu_\tau \bar \nu_\tau\tilde{\chi}_1^0$ & 0.013 \hspace*{2.8cm}$$ & $0.013$
\hspace*{1.8cm}$$\\ \hline 
$q \bar q\tilde{\chi}_1^0~(q\neq t)$& 0.015 \hspace*{2.8cm}$$ & $0.015$
\hspace*{1.8cm}$$\\ \hline
total width & 19.168 & 19.241\\ \hline
\end{tabular}
\end{center}
\caption{Partial widths of different $\tilde{\chi}_2^0$ decay modes and
  the branching ratios of its visible decays for the SPS1a parameter set. The
  numbers in parentheses give the corresponding partial widths calculated in
  the single pole approximation.
\label{twobodytotal}}
\end{table}

From the results in Table~\ref{twobodytotal} one concludes:
\begin{itemize}
  
\item The main decay mode of $\tilde{\chi}_2^0$ is $\tilde{\chi}_2^0
  \rightarrow \tau^{-} \tau^{+}\tilde{\chi}_1^0$. Its branching ratio is about
  $88.0\%$ at tree-level, $86.2\%$ at one-loop level. This mode dominates
  partly because of the lower mass of $\tilde \tau_1$ as compared to $\tilde
  e_1$ (133.0 GeV vs 142.7 GeV). Even more important is that 
 $\tilde l_1$ is a pure $SU(2)$ singlet for $l=e,\, \mu$, since we neglect terms $\propto m_l$ in the
mass matrices of these sleptons. In contrast, $\tilde \tau_L - \tilde \tau_R$
mixing is quite significant, leading to a sizable $SU(2)$ doublet component of
$\tilde \tau_1$. Therefore $\tilde \chi_2^0$ decays into (real or virtual)
$\tilde l_1$ can only proceed through its small $U(1)_Y$ gaugino (bino)
component for $l=e, \, \mu$, while the large $SU(2)$ gaugino (neutral wino)
component also contributes for $l=\tau$.
  
\item The total $\tilde \chi_2^0$ decay width is enhanced by 0.4\% when one-loop
  corrections are included. Such modest corrections are typical in the absence
  of large enhancement factors (e.g., large logarithms). 

\item One-loop corrections enhance the partial width and the branching ratio
  of $\tilde{\chi}_2^0 \rightarrow l^- l^+ \tilde{\chi}_1^0 \ (l = e, \mu)$
  decays by $15.5\%$ and $13.6\%$, respectively.
  
\item The single pole approximation reproduces the integrated partial widths
  to about 0.3\% accuracy. 
  This agreement is even better than in the $M_{l^+l^-}$ distribution
  shown in Figures \ref{comparisonEE} and \ref{comparisonLL}. 
  From (\ref{eqn:propid}) and the discussions of the large logarithm 
  $\ln\left (\Gamma_{\tilde l_1}^{(1)}/m_{\tilde l_1}\right )$ in 
  Section \ref{virtual} one might expect better agreement for the integrated
  partial width than for the kinematical distributions.

\end{itemize}
\subsection{Numerical Results for the Pure Three-body Decays}
We also investigated the effect of higher-order corrections on leptonic
$\tilde \chi_2^0$ decays for a scenario where $\tilde \chi_2^0$ does not have
any two-body decay modes. To that end we again use the SPS1a parameter set,
except that the soft SUSY-breaking parameters in the slepton mass matrix are set to
\begin{eqnarray} \label{newslep}
m_{\tilde{l}_L} = 230 GeV\, , \ \ m_{\tilde{l}_R} = 183 \ {\rm GeV}, \hspace*{3mm}l = e, \,
\mu,\, \tau\, . 
\end{eqnarray}
The masses of the relevant neutralinos and sleptons in this modified SPS1a parameter set 
are listed in Table~\ref{spstab} where one finds that $\tilde{\chi}_2^0$ has to undergo a pure three-body decay. 
Therefore we do not have to introduce complex slepton masses in the one-loop functions. 
\begin{table}[htb] 
\begin{center}
\begin{tabular}{|l|l|l|l|l|l|l|l|}\hline
particle & \ $\tilde{\chi}_2^0$ & $\ \tilde{\chi}_1^0$ &
$\tilde{e}_1 \, (\tilde{\mu}_1)$ & $\tilde{e}_2 \, (\tilde{\mu}_2)$ &
\ \ $\tilde{\tau}_1$ & \ \ $\tilde{\tau}_2$ & $\tilde{\nu}_l~(l = e\, ,\mu\, ,\tau)$  \\ \hline
mass (GeV) & 176.6 & 96.2 & 187.9 &  234.9 & 182.3 & 239.2 & \ \ \ 221.0  \\
\hline 
\end{tabular}
\end{center}
\caption{Masses of the relevant neutralinos and sleptons for the modified SPS1a \mbox {parameter~set}\label{spstab}}
\end{table}

\begin{figure}
\begin{tabular}{ll}
\includegraphics[width=.475\linewidth]{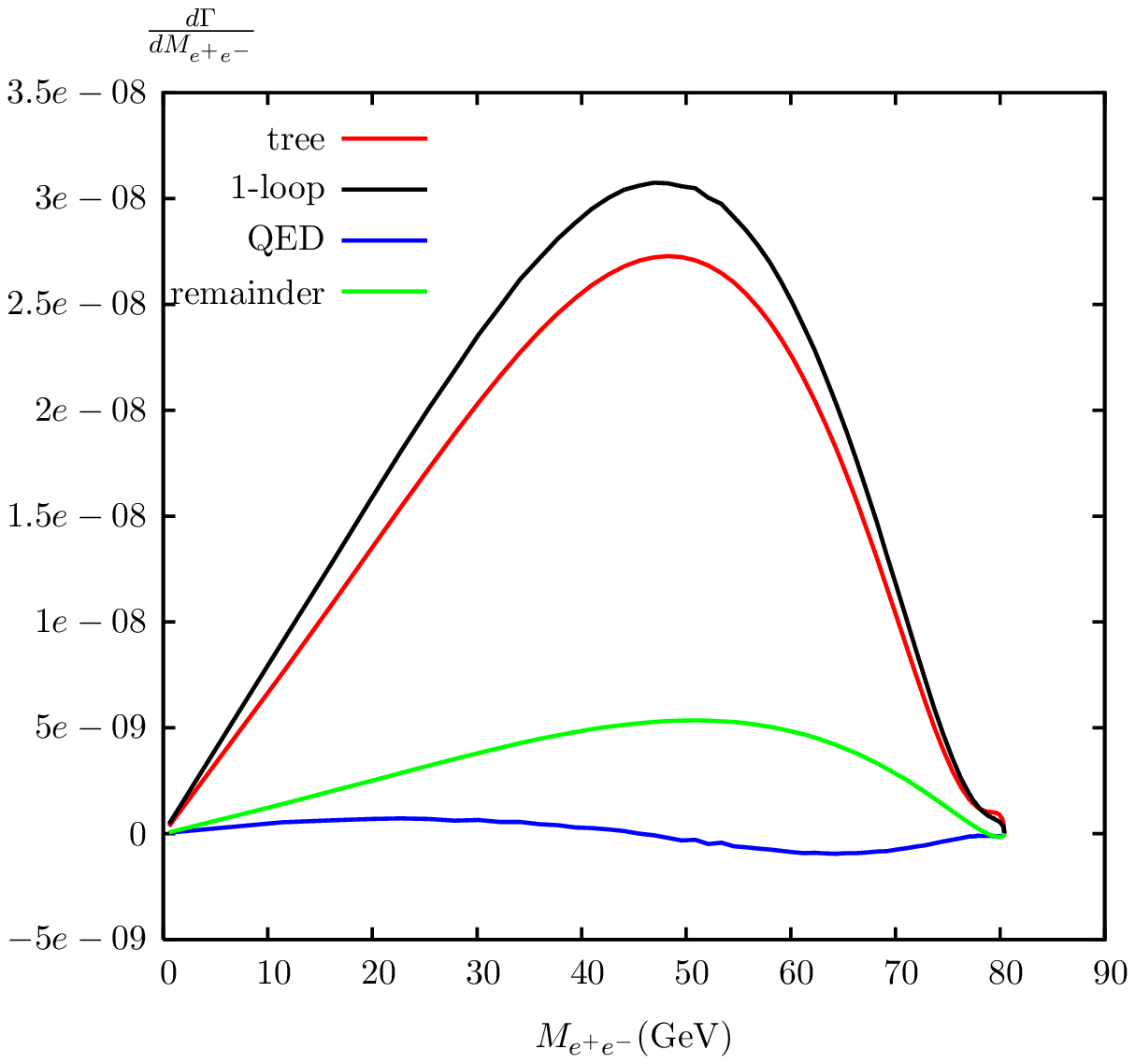} &
\includegraphics[width=.47\linewidth]{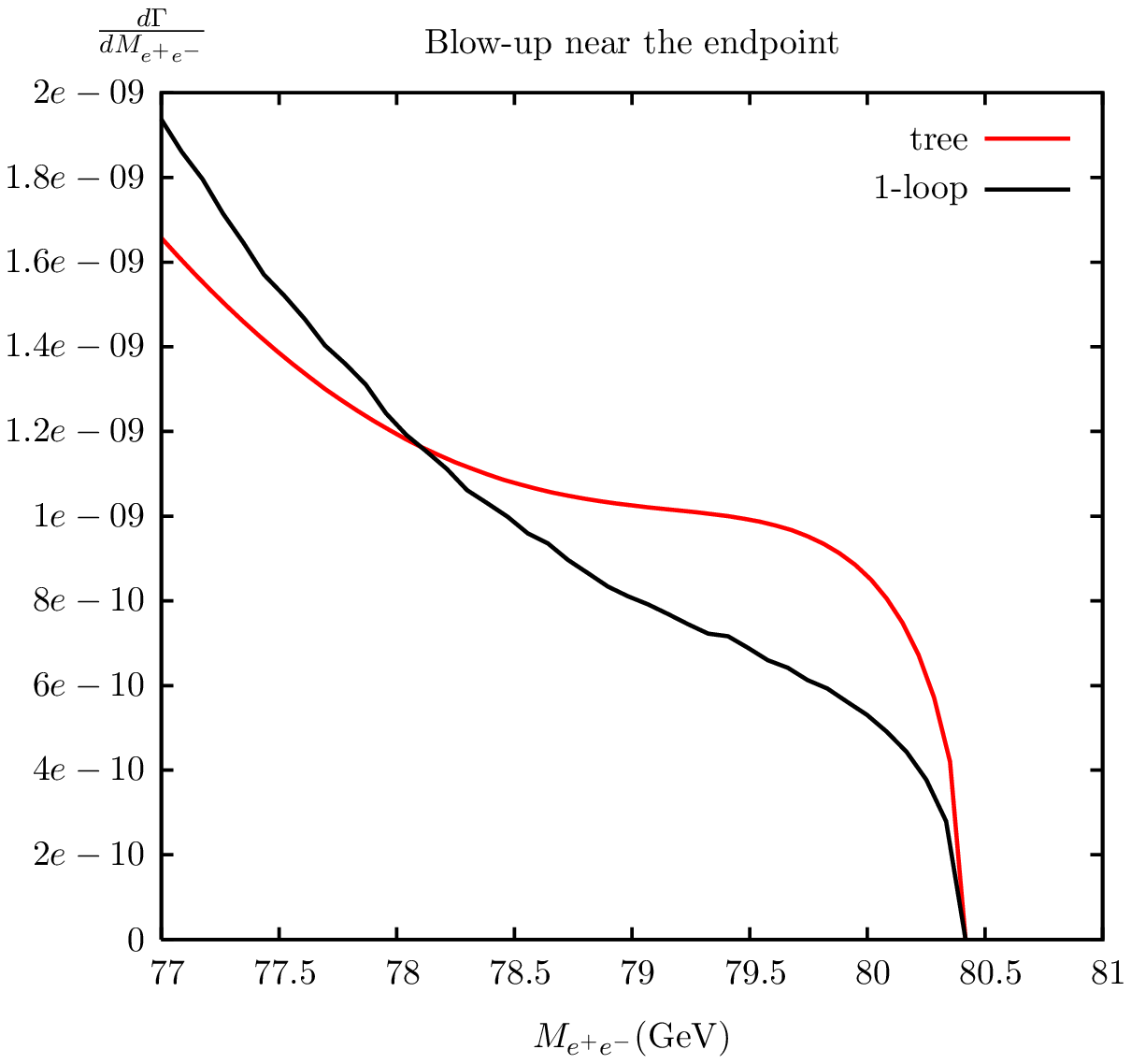}
\end{tabular}
\caption{The dilepton invariant mass $M_{e^+ e^-}$ distribution 
in the case of a genuine three-body decay. \label{Mee}} 
\end{figure}

\begin{figure}
\begin{tabular}{ll}
\includegraphics[width=.47\linewidth]{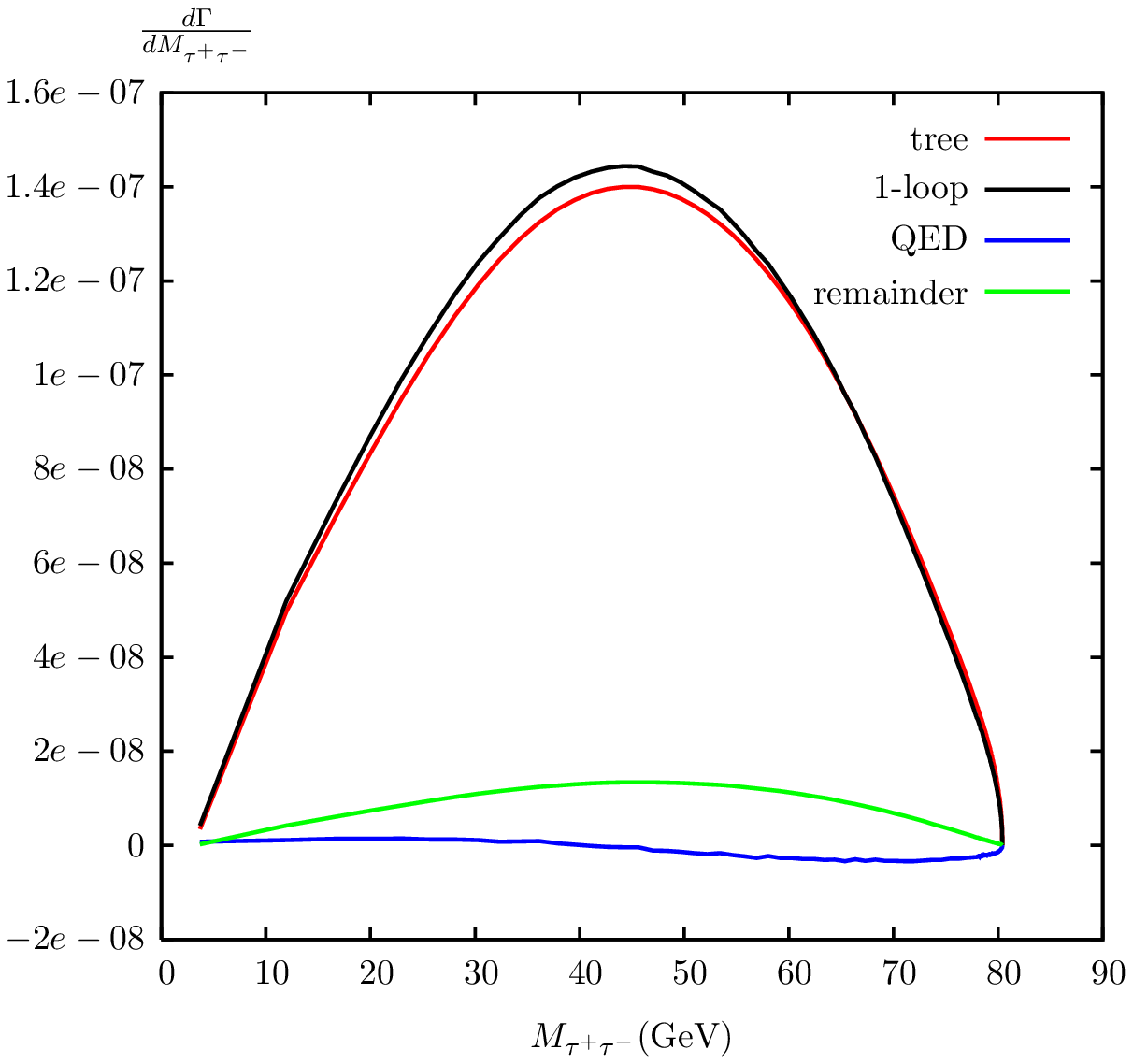} &
\includegraphics[width=.47\linewidth]{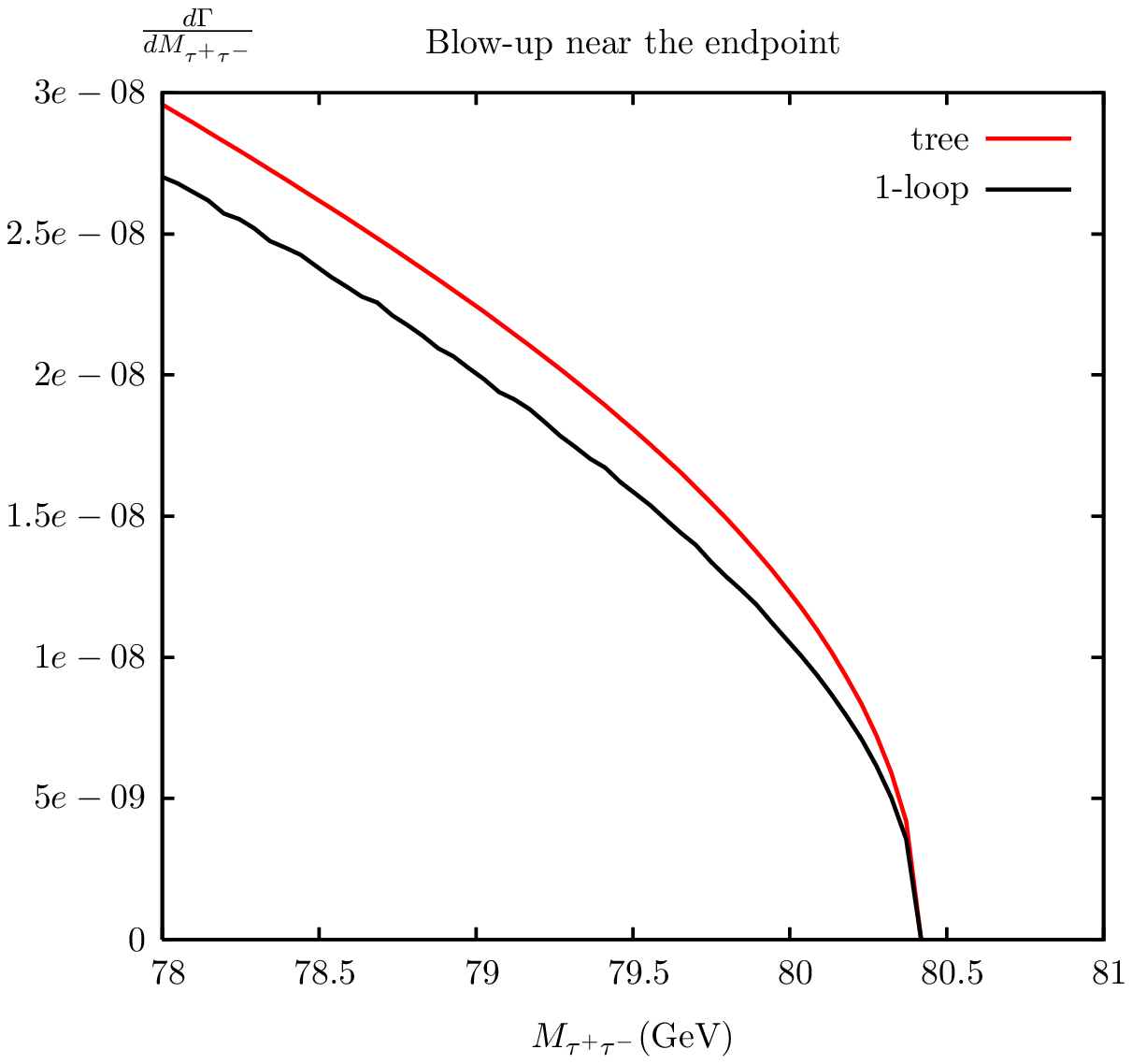}
\end{tabular}
\caption{The dilepton invariant mass $M_{\tau^+ \tau^-}$ distribution 
in the case of a genuine three-body decay.\label{Mll}}
\end{figure}

The dilepton invariant mass $M_{e^+ e^-}$ and $M_{\tau^+ \tau^-}$ distributions
are shown in Figures~\ref{Mee} and \ref{Mll}, respectively. \
At tree level the $M_{e^+e^-}$ distribution shows
a small peak near its upper endpoint from the exchange of nearly on-shell $Z$
bosons. Since the QED and non-QED corrections are very small and negative in this region, this peak is
less pronounced once one-loop corrections are included. This is of some
significance, since the shape of this distribution can now be used to infer
the strengths of various contributing diagrams, which in turn provides
information on slepton masses and neutralino mixing \cite{noya,matchev}. Since
$\tilde \tau$ exchange is much enhanced relative to $\tilde e$ exchange, 
one cannot see any contributions of $Z-$exchange even at tree-level
from the $M_{\tau^+ \tau^-}$ distribution.
Moreover we can observe that the invariant mass $M_{e^+ e^-}$ and $M_{\tau^+ \tau^-}$ distributions have a rather
sharp edge at their endpoints. These edges are again softened by real photon
emission, but remain quite distinct. This should facilitate the experimental
determination of the endpoint, and hence the measurement of $m_{\tilde
  \chi_2^0} - m_{\tilde \chi_1^0}$.

\begin{figure}[htb]
\psfrag{u}{{\tiny $\mu$}}
\begin{tabular}{cc}
\includegraphics[width=0.485\linewidth]{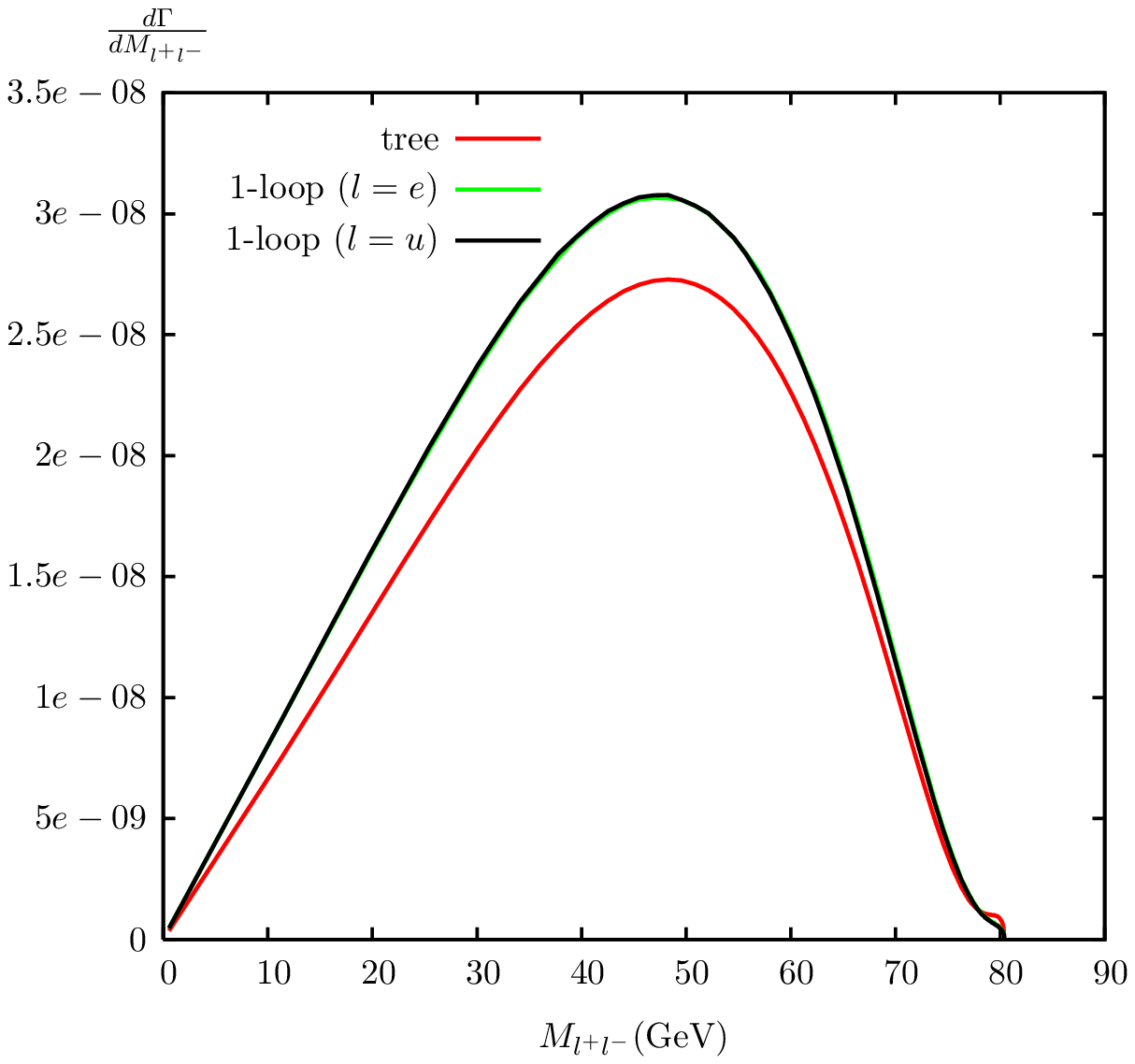}&
\includegraphics[width=0.46\linewidth]{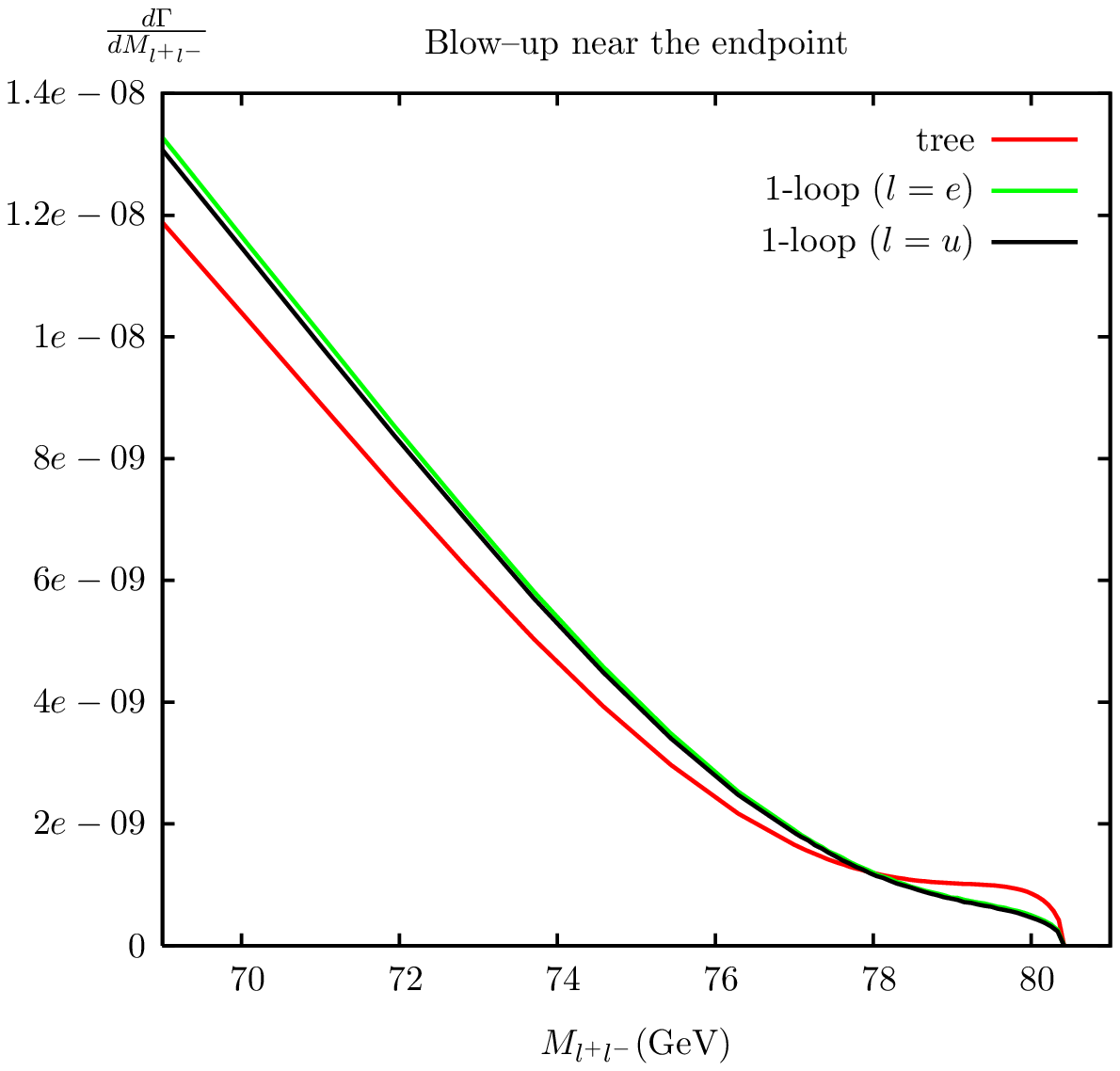}\\
\includegraphics[width=0.485\linewidth]{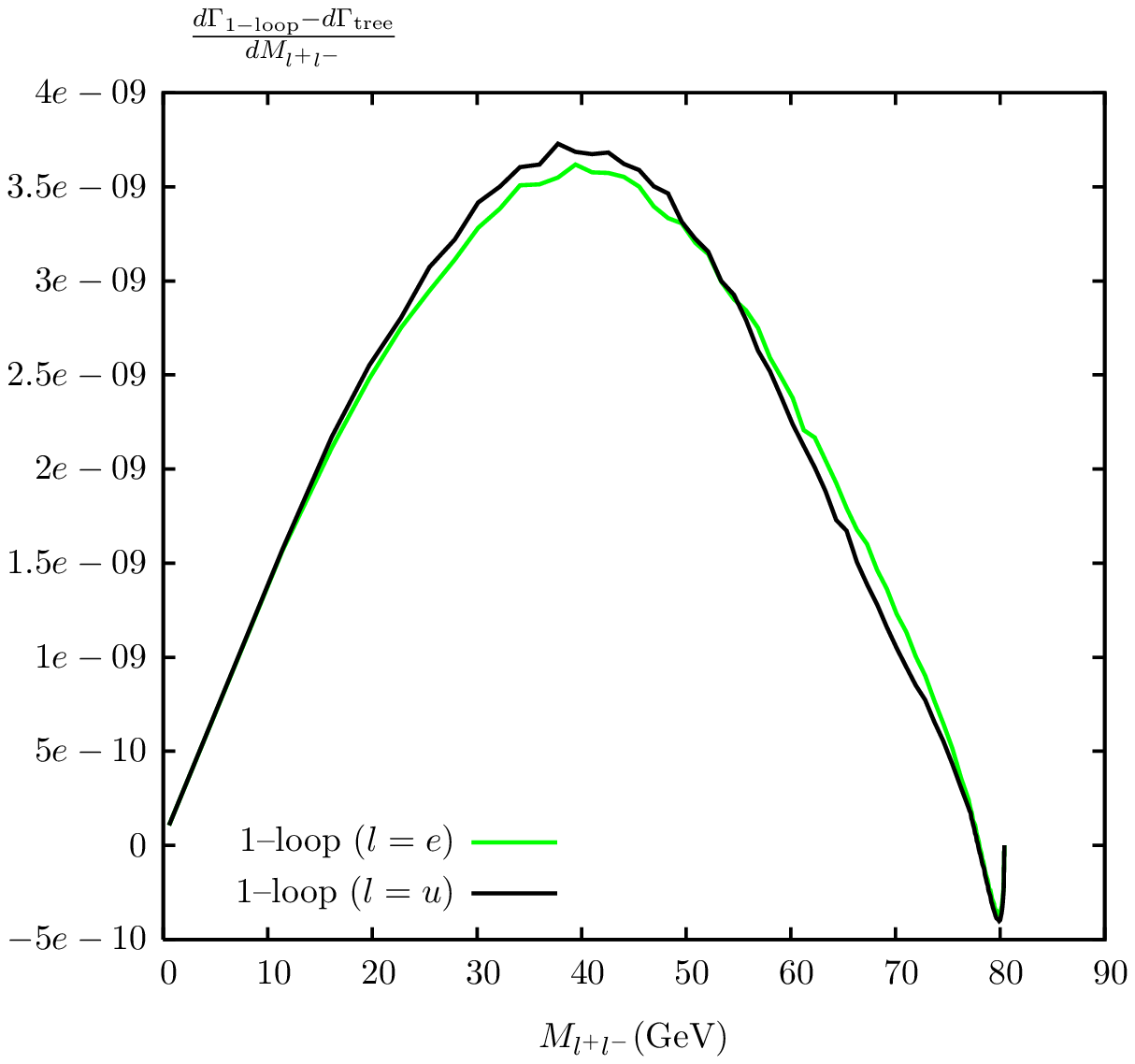}&
\includegraphics[width=0.46\linewidth]{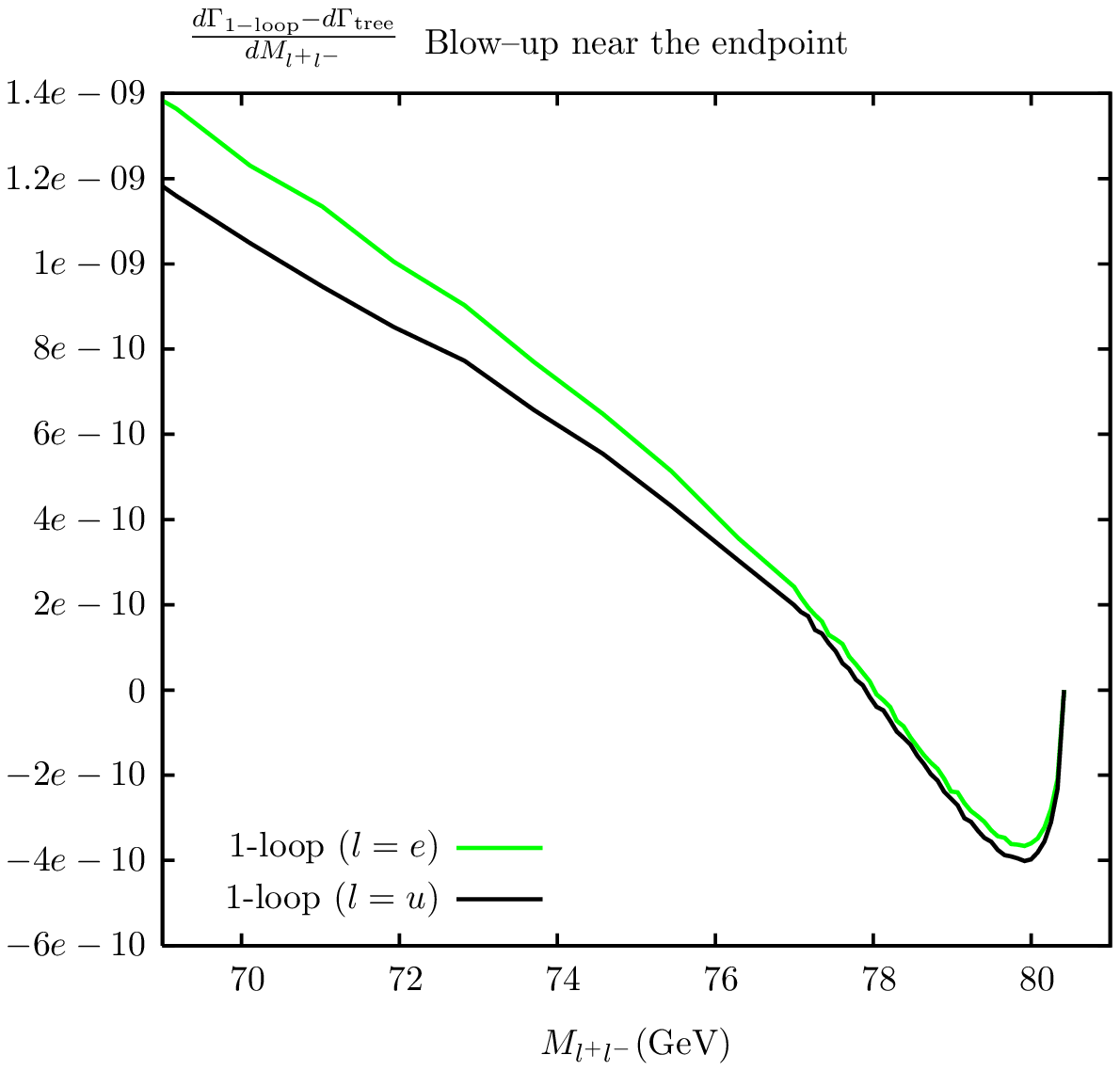}
\end{tabular}
\caption{The comparison of the dilepton invariant mass $M_{\mu^+ \mu^-}$ and $M_{e^+e^-}$ 
distribution in the case of a genuine three-body decay. \label{Muuoff}}
\end{figure}

We compare the dilepton invariant mass $M_{\mu^+\mu^-}$ and  $M_{e^+e^-}$ distributions
in Figure \ref{Muuoff}, where the tree- and one-loop-level results, the blow-up of the endpoint region
and the relative one-loop corrections are shown.
From these figures one obtains that
the shapes of the $M_{\mu^+\mu^-}$ and $M_{e^+e^-}$ distributions are identical at tree level and different at one-loop level
due to the different treatment of collinear-photon radiations. 
In contrast to the numerical results from the SPS1a parameter set (see Figure \ref{SPS1aMuu}),
the mass effect is small in Figure \ref{Muuoff}, but it is still distinct, especially in the relative one-loop corrections.
In the calculations for the invariant-mass distribution, the momentum of a collinear photon is added to that of the emitting electron,
but it is not added to that of the emitting muon. Hence 
the invariant mass $M_{\mu^+\mu^-}$ is reduced in comparison with $M_{e^+e^-}$. It leads to the shifting of events 
from the upper invariant-mass region to the lower invariant-mass region. 
This effect can be seen in the lower frames in Figure \ref{Muuoff}, i.e. 
in the lower invariant-mass region the relative one-loop corrections of the $\mu^+\mu^-$ final state
is larger than that of $e^+e^-$ final state, while the inverse relation holds  
in the upper invariant-mass region.

\newpage
The partial widths of different $\tilde{\chi}_2^0$ decay modes and the branching
ratios of its visible decays are shown in Table~\ref{threebodytotal}. 
\begin{table}[hpt]
\begin{center}
\begin{tabular}{|l|l|l|}\hline
decay mode & tree-level width(keV), Br& 1loop-level width(keV), Br\\ \hline
$e^{-} e^{+}\tilde{\chi}_1^0$ &$1.270$, \hspace*{1.8cm}\ \ \ \ \ $4.4\%$ &
$1.451$, \hspace*{1.8cm}$5.0\%$  \\ \hline 
$\mu^{-} \mu^{+}\tilde{\chi}_1^0$ & $1.270$, \hspace*{1.8cm}\ \ \ \ \ $4.4\%$ &
$1.451$, \hspace*{1.8cm}$5.0\%$\\ \hline 
$\tau^{-} \tau^{+}\tilde{\chi}_1^0$ & $7.209$, \hspace*{1.8cm}\ \ \ \ $25.1\%$ &
$7.383$, \hspace*{1.8cm}$25.4\%$\\ \hline 
$\nu_e \bar \nu_e \tilde{\chi}_1^0$ & $1.273$ \hspace*{1.8cm}$$ &
$1.355$ \hspace*{1.8cm}$$\\ \hline 
$\nu_\mu \bar \nu_\mu \tilde{\chi}_1^0$  & $1.273$ \hspace*{1.8cm}$$ &
1.355 \hspace*{1.8cm}$$  \\ \hline 
$\nu_\tau \bar \nu_\tau \tilde{\chi}_1^0$ & $1.273$  \hspace*{1.8cm}$$ &
$1.354$  \hspace*{1.8cm}$$ \\ \hline 
$u \bar u \tilde{\chi}_1^0$ & $2.480$  \hspace*{1.8cm}$$ &
$2.386$  \hspace*{1.8cm}$$ \\ \hline 
$d \bar d \tilde{\chi}_1^0$ & $3.330$  \hspace*{1.8cm}$$ &
$3.298$  \hspace*{1.8cm}$$ \\ \hline 
$c \bar c \tilde{\chi}_1^0$ & $2.475$  \hspace*{1.8cm}$$ &
$2.378$  \hspace*{1.8cm}$$ \\ \hline 
$s \bar s \tilde{\chi}_1^0$ & $3.330$  \hspace*{1.8cm}$$ &
$3.298$  \hspace*{1.8cm}$$ \\ \hline 
$b \bar b \tilde{\chi}_1^0$ &3.595  \hspace*{1.8cm}$$ &
$3.405$  \hspace*{1.8cm}$$ \\ \hline 
total width & $28.778$& $29.114$\\ \hline
\end{tabular}
\end{center}
\caption{The decay width of different $\tilde{\chi}_2^0$ decay modes and the
  branching ratios of its visible leptonic decays in the modified SPS1a
  scenario. \label{threebodytotal}}
\end{table}
The $\tau^+  \tau^- \tilde \chi_1^0$ final state is still the largest decay mode of $\tilde \chi_2^0$ 
($25.1\%$ at tree level, $25.4\%$ at one-loop level),
but it no longer dominates.
The hadronic final states have very large partial decay
widths and branching ratios: $\Gamma_{\rm hadronic}^{\rm tree} = 15.210$keV ($52.9\%$), 
$\Gamma_{\rm hadronic}^{\rm 1-loop} = 14.765$keV 
($50.7\%$), though the squark masses are much heavier than the slepton masses.
The part reason is 
that the $Z$-exchange diagrams give larger contributions to hadronic final states than to leptonic ones. 
Moreover the interference between $Z$ and squark exchanges is 
large and positive for the hadronic 
final states, while the interference between $Z$ and slepton exchanges is also large but negative for the leptonic
final states. This is the main reason why the hadronic decays of $\tilde \chi_2^0$ obtain so large branching ratios.

Note that exchange of the
$SU(2)$ doublet sleptons now dominates for $l=e, \, \mu$. This
dominance of $\tilde e_L$ exchange also explains why the $e^+ e^- \tilde
\chi_1^0$ and $\nu_e \bar \nu_e \tilde \chi_1^0$ final states now have quite
similar partial widths. We have assumed three exactly degenerate sneutrinos here, unlike in the
original SPS1a scenario, where $\tilde \nu_\tau$ is slightly lighter than
$\tilde \nu_e$. In the modified scenario a tiny difference between the one-loop partial
widths for $\nu_\tau \bar \nu_\tau \tilde \chi_1^0$ and $\nu_e \bar \nu_e
\tilde \chi_1^0$ final states nevertheless results from one-loop corrections
involving the $\tau$ mass or Yukawa coupling (e.g. from the $\tilde \nu$ and
$\nu$ two-point functions).
The total $\tilde \chi_2^0$ decay width is increased by $1.2\%$ when one-loop corrections are considered.
The partial width of the $\tilde \chi_2^0$ decay into electron and muon pairs is now 
enhanced by about $14.3\%$, leading to an increase of the corresponding branching
ratios by $13.6\%$ at one-loop level.

\chapter{Conclusions}\label{conclusions}
In the MSSM with conserved R-parity, the decays of the next-to-lightest neutralino $\tilde \chi_2^0$
into the LSP $\tilde \chi_1^0$ and two fermions are always involved in the decay chains of supersymmetric particles.
Moreover, $\tilde \chi_2^0$ is one of the
lightest visible supersymmetric particles that can be produced directly at future $e^+e^-$
colliders and plays a prominent role in the analysis of cascade decays of
gluinos and squarks at the LHC. An accurate understanding of its decays is therefore of considerable importance.
The leptonic decays of $\tilde \chi_2^0$ are particularly interesting 
since the endpoint of the dilepton invariant mass distribution
can be used to reconstruct the mass differences of the supersymmetric particles.
In this thesis, we have investigated $\tilde \chi_2^0$ leptonic decays 
$\tilde{\chi}_2^0\rightarrow \tilde{\chi}_1^0 l^- l^+$ at one-loop level.

For the cases where the intermediate charged sleptons $\tilde l_1$ can be on shell,
these decays were calculated both completely 
and in a single-pole approximation at one-loop level.
In the complete calculation one has to employ complex slepton masses in the relevant propagators and
one-loop integrals. The single-pole approximation in this case is performed in the way that
the $\tilde{\chi}_2^0$ decays are treated as a sequence of two two-body decays.
For the numerical evaluation we use the SPS1a parameter set.
We compare the results from the complete and approximate calculations 
and find that  
this approximation reproduces the integrated partial widths to
better than 0.5\% accuracy even after one-loop corrections are included.
From these calculations one obtains a rather small one-loop correction
to the total $\tilde \chi_2^0$ decay width, but the branching ratios for the
electron and muon final states are increased by about
13.6\% at one-loop level.

The dilepton invariant mass $M_{l^+l^-}$ distributions were also studied. 
The shape of these distributions is found to be altered by real photon
emission contributions, i.e. its peak is shifted by several GeV below the endpoint. 
This is very important since the shape of the distribution near the endpoint should be known if
the endpoint is to be determined accurately from real data.
In our calculation 
we define collinear photons as being
emitted at an angle $\Delta \theta < 1^\circ$ relative to the emitting lepton.
Since the selectrons and smuons have equal masses and the light lepton mass $m_l~(l = e, \mu)$ is neglected 
except when it appears in the one-loop integrals,
one will obtain identical distributions for $M_{e^+e^-}$ and $M_{\mu^+ \mu^-}$ if
the momentum of a collinear photon is added to that of the emitting lepton.
The actual effect of the collinear-photon radiation depends
on details of the measurement apparatus, and therefore has to be calculated
anew for each experiment.
We have focused on the LHC experiment in our calculation.
At the LHC the electron energy is determined calorimetrically.
In this case a collinear photon would hit the same cell of the calorimeter as
the electron, so the two energies cannot be disentangled. Hence
we add the momentum of a collinear photon to the one of the emitting electron in our calculation.
Since muons pass through the calorimeter, where the photons are detected, and measured forther outside
in the muon detector at the LHC,
the momentum of a collinear photon is not added to the one of its emitter muon in our calculation.  
In this case 
the mass effect can be seen in the dilepton invariant-mass distribution. 
We find that the peak of $M_{e^+e^-}$ distribution is moved downwards by about 
$4$ GeV once the one-loop corrections are added.
In contrast to the $M_{e^+e^-}$ distribution, the peak of the $M_{\mu^+\mu^-}$ distribution 
is a little shifted to lower invariant-mass values at one-loop level. 
This is due to the different treatment of the collinear-photon radiation.

We have also analyzed a scenario with increased slepton
masses, so that $\tilde{\chi}_2^0$ can only undergo genuine three-body
decays. We find that the total $\tilde \chi_2^0$ decay width is enhanced
by $1.2\%$ when one-loop corrections are considered,
while the branching ratios of $\tilde \chi_2^0$ decay into electron and muon pairs are  
enhanced by about $13.6\%$ at the one-loop level.
One also finds that the shape of dilepton invariant mass $M_{l^+l^-}$ distributions are also 
affected by the real photon emission.
Moreover, these distributions have a rather complicated shape, showing the contributions from $Z$
exchange near the upper endpoints. 
In this case the shape of the distribution away from the
endpoint also carries information about slepton masses and neutralino mixing
angles. Fitting tree-level distributions to real data might therefore give
wrong results for these physical parameters.
In this context a careful
analysis of the collinear-photon radiation is also important, since differences in the
energy measurements of electrons and muons could lead to spurious differences
of fitted selectron and smuon masses.
Here the collinear-photon radiations for electrons and muons are treated as discussed beforehand.
One finds that the one-loop shapes of the $M_{e^+e^-}$ and $M_{\mu^+\mu^-}$ distributions
are different, though the selectrons and smuons have equal masses in our calculations.

\begin{appendix}
\chapter{Notations and SM Parameters}\label{Notations}
In this thesis we adopt standard relativistic units, i.e. $\hbar= c = 1$. A general covariant four-vector is
denoted by
 \begin{eqnarray}
a_{\mu} & = & \left(a_0, a_1, a_2, a_3\right ) = \left (a_0, \vec a\right )\,
 \end{eqnarray}
and a contravariant four-vector is
 \begin{eqnarray}
a^{\mu} & = & \left(a^0, a^1, a^2, a^3\right ) = \left (a_0, -\vec a\right )\, .
 \end{eqnarray}
They are connected by the metric tensor
 \begin{eqnarray}
g^{\mu\nu} & = & g_{\mu\nu}  = {\rm diag}\left(1, -1, -1, -1\right )
 \end{eqnarray}
via the relations 
\begin{eqnarray}
a^{\nu} & = & g^{\mu\nu}a_{\mu}\, .
 \end{eqnarray}
The product of the four-vectors are defined as
\begin{eqnarray}
ab & \equiv & a^\mu b_\mu = a_0b_0 - \vec a \cdot
\vec b\, .
 \end{eqnarray}
The four-gradients $\partial_\mu$ and $\partial^\mu$ are defined 
\begin{eqnarray}
\partial_\mu & \equiv & \frac{\partial}{\partial x^\mu} = \left (\frac{\partial}{\partial t}, \vec \nabla\right )\, , \nonumber \\
\partial^\mu & \equiv & \frac{\partial}{\partial x_\mu} = \left (\frac{\partial}{\partial t}, -\vec \nabla\right )\, .
 \end{eqnarray}
We also use the compact "Feynman slash" notation
\begin{eqnarray}
\not\!{a} = \gamma^\mu a_\mu\, ,
 \end{eqnarray}
where $\gamma^\mu$ are Dirac matrices.
\section{Pauli and Dirac Matrices}
The pauli matrices are defined as
\begin{eqnarray}
\sigma^1 =\left( \begin{array}{cc}
0 & 1\\
1 & 0
\end{array}\right )\, , \ \ \sigma^2 =\left( \begin{array}{cc}
0 & -i\\
i & 0
\end{array}\right )\, ,\ \ \sigma^3 =\left( \begin{array}{cc}
1 & 0\\
0 & -1
\end{array}\right )\, .
 \end{eqnarray}
They satisfy the commutator relation
\begin{eqnarray}
[\sigma^i, \sigma^j] & = & 2 i\varepsilon^{ijk}\sigma^k, \ \ \ i, j, k = 1, 2, 3\, .
\end{eqnarray}
The totally antisymmetric tensors in three dimensions are defined
as
\begin{eqnarray}
\varepsilon_{ijk} & = & \left \{ \begin{array}{l}
+1\, , \ \ {\rm for \ even \ permutations \ of }\ 123\\
-1\, , \ \ {\rm for \ odd \ permutations \ of }\  123\\
0\, , \ \ \ \ {\rm otherwise}\, .
\end{array}\right.
\end{eqnarray}
One can arrange the Pauli matrices as  
\begin{eqnarray}
\sigma^\mu & = & \left (\sigma^0, \vec\sigma\right )= \left (\sigma^0, \sigma^1, \sigma^2,  \sigma^3\right )\, ,\\
\bar{\sigma}^\mu & = & \left (\sigma^0, -\vec\sigma\right )\, ,
\end{eqnarray}
where $\sigma^0 = \left( \begin{array}{cc}
1 & 0\\
0 & 1
\end{array}\right ).$
Anti-symmetric matrices $\sigma^{\mu\nu}$ and $\bar{\sigma}^{\mu\nu}$ are defined by
\begin{eqnarray}
\sigma^{\mu\nu} & = & \frac{i}{4}\left (\sigma^\mu\bar{\sigma}^\nu - \sigma^\nu\bar{\sigma}^\mu\right )\, , \\
\bar{\sigma}^{\mu\nu} & = & \frac{i}{4}\left (\bar\sigma^\mu\sigma^\nu - \bar\sigma^\nu\sigma^\mu\right )\, .
\end{eqnarray}

The Dirac $\gamma$-matrices are defined via the anticommutation relations
\begin{eqnarray}
\left \{\gamma^\mu, \gamma^\nu\right \} = 2g^{\mu\nu}\, .
\end{eqnarray}
A fifth $\gamma$-matrix is defined by
\begin{eqnarray}
\gamma^5 \equiv \gamma_5 \equiv i\gamma^0\gamma^1\gamma^2\gamma^3\, . 
\end{eqnarray}
From these definitions one can easily obtain the following properties for the $\gamma$-matrices,
\begin{eqnarray}
\left\{\gamma^5, \gamma^\mu\right \} = 0\, , \ \ 
(\gamma^5)^2 = 0\, .
\end{eqnarray}
In the chiral or Weyl representation the explicit expressions for the Dirac $\gamma$-matrices are 
 \begin{eqnarray}
\gamma^\mu = \left (\begin{array}{cc}
0 & \sigma^\mu \\
\bar{\sigma}^\mu & 0 
\end{array}\right )\, , \ \ \gamma^5 = \left (\begin{array}{cc}
-1 & 0 \\
0 & 1
\end{array}\right )\, .
\end{eqnarray}
The left- and right-handed operators are defined by
\begin{eqnarray}
\omega_L = \frac{1}{2}\left (1-\gamma_5\right )\, , \ \ 
\omega_R = \frac{1}{2}\left (1+\gamma_5\right )\, .
\end{eqnarray}
\section{Spinors}
The components of the two-component (Weyl) spinor are Grassmann numbers, i.e.
 \begin{eqnarray}
\left \{\chi_\alpha, \chi_\beta\right \}&=& \left \{\chi^\alpha, \chi^\beta\right \}
= \left \{\chi_\alpha, \chi^\beta\right \} = 0\, ,\nonumber \\
\left \{\bar\xi_{\dot\alpha}, \bar\xi_{\dot\beta}\right \}&=& \left \{\bar\xi^{\dot\alpha}, \bar\xi^{\dot\beta}\right \}
= \left \{\bar\xi_{\dot\alpha}, \bar\xi^{\dot\beta}\right \} = 0\, ,
\end{eqnarray}
and they also have anticommutation relations with other Grassmann numbers.
Here the indices $\alpha (\dot\alpha)=1,2$ and $\beta (\dot \beta)= 1, 2$.
The scalar product of two-component spinors $\chi$ and $\xi$ is defined as
 \begin{eqnarray}
 \chi\xi &\equiv & \chi^\alpha \xi_\alpha\, , \nonumber \\
 \bar\chi\bar\xi &\equiv & \bar\chi_{\dot \alpha}\bar \xi^{\dot \alpha}\, , \nonumber \\
\chi\sigma^\mu\bar\xi &\equiv & \chi^\alpha \sigma_{\alpha \dot\alpha}^\mu \bar \xi^{\dot \alpha}\, .
\end{eqnarray}

A four-component (Dirac) spinor $\Psi$ in the Weyl representation can be constructed via
 \begin{eqnarray}
\Psi & = & \left (\begin{array}{c}
\xi_\alpha\\
\bar\chi^{\dot\alpha}
\end{array}\right )\, ,
\end{eqnarray}
where $\xi_\alpha$ and $\bar\chi^{\dot\alpha}$ are Weyl spinors.
The Dirac-adjoint spinor $\bar \Psi$ is expressed as 
\begin{eqnarray}
\bar \Psi &= & \Psi^\dagger \gamma_0 =  \left (\begin{array}{cc}
\chi^\alpha &
\bar\xi_{\dot\alpha}
\end{array}\right )\, .
 \end{eqnarray}
The charge conjugation of the Dirac spinor $\Psi$ is defined via 
 \begin{eqnarray}
\Psi^c = C\bar{\Psi}^T = \left (\begin{array}{c}
\chi_\alpha\\
\bar\xi^{\dot\alpha}
\end{array}\right )\, ,
\end{eqnarray}
where the charge conjugation matrix $C$ is expressed as
 \begin{eqnarray}
C & = & i \left (\begin{array}{cc}
(\sigma^2\bar \sigma^0)_\alpha^\beta & 0\\
0 & (\bar\sigma^2\sigma^0)^{\dot\alpha}_{\dot\beta} 
\end{array}\right )\, .
\end{eqnarray}
A Dirac spinor $\Psi$ is also a Majorana spinor
if the relation $\Psi = \Psi^c$ is satisfied.
Hence a Majorana spinor $\lambda$ can be written as
 \begin{eqnarray}
\lambda & = & \left (\begin{array}{c}
\xi_\alpha\\
\bar\xi^{\dot\alpha}
\end{array}\right )\, .
\end{eqnarray}

The left- and right-handed components of a Dirac spinor
can be written as
 \begin{eqnarray}
\Psi_L = \omega_L \Psi = \left (\begin{array}{c}
\xi_\alpha\\
0
\end{array}\right )\, , \\
\Psi_R = \omega_R \Psi =
 \left (\begin{array}{c}
0\\
\bar\chi^{\dot\alpha}
\end{array}\right )\, .
\end{eqnarray}

Some useful relations between the four- and two-component spinors are  
 \begin{eqnarray}
\bar{\Psi}_1\Psi_2 &= &\chi_1\xi_2 + \bar\xi_1\bar\chi_2\, ,\\
\bar{\Psi}_1\gamma^\mu\Psi_2 &= & \bar\xi_1\bar\sigma^\mu\xi_2 - \bar\chi_2\bar \sigma^\mu\chi_1\, ,\\
\bar{\Psi}_1\gamma^5\Psi_2 &= &-\chi_1\xi_2 + \bar\chi_2\bar\xi_1\, ,\\
\bar{\Psi}_1\gamma^\mu\gamma^5\Psi_2 &= & -\bar\xi_1\bar\sigma^\mu\xi_2 - \bar\chi_2\bar \sigma^\mu\chi_1\, ,\\
\bar{\Psi}_1\gamma^\mu \partial_\mu\Psi_2 &= &\chi_1\sigma^\mu\partial_\mu \bar\chi_2 + \bar\xi_1\bar \sigma^\mu\partial_\mu\xi_2
\, ,\\
\bar{\Psi}_1\omega_L\Psi_2 &= &\chi_1\xi_2\, ,\\
\bar{\Psi}_1\omega_R\Psi_2 &= &\bar\xi_1\bar\chi_2\, ,\\
\bar{\Psi}_1\gamma^\mu\omega_L\Psi_2 &= &\bar\xi_1\bar \sigma^\mu \xi_2\, ,\\
\bar{\Psi}_1\gamma^\mu\omega_R\Psi_2 &= &-\bar\chi_2\bar \sigma^\mu \chi_1\, ,\\
\bar{\Psi}_1\gamma^\mu \omega_L\partial_\mu\Psi_2 &= & \bar \xi_1\bar \sigma^\mu \partial_\mu\xi_2\, , \\
\bar{\Psi}_1\gamma^\mu \omega_R\partial_\mu\Psi_2 &= & \chi_1\sigma^\mu \partial_\mu\bar\chi_2 \, .
\end{eqnarray}
\section{SM Parameters}\label{Parameter}
For the numerical evaluation, the following values of the SM parameters are
used:
\begin{eqnarray}
m_e &=& 0.510999 {\rm MeV}\, , \hspace*{3mm}m_\mu = 105.6584 {\rm MeV}\,  , \hspace*{3mm}
m_\tau =   1.777{\rm GeV}\ , \nonumber \\ 
m_u &=& 53.8 {\rm MeV}\, ,  \hspace*{1.1cm}m_c =  1.5 {\rm GeV}\, , \hspace*{14mm}m_t =   175{\rm GeV} \, ,\nonumber \\ 
m_d &= & 53.8 {\rm MeV}\, ,  \hspace*{1.1cm}m_s = 150 {\rm MeV}\, , \hspace*{12mm}m_b = 4.7 {\rm GeV}\, , \nonumber \\
%
 m_W & = & 80.45{\rm GeV} ,  \hspace*{10mm} m_Z = 91.1875{\rm GeV}\, ,\nonumber \\
\alpha & = &1/137.0359895, \hspace*{10mm}G_\mu = 1.1663910\times 10^-5 {\rm GeV}^{-2} \, . \nonumber 
\end{eqnarray}

\chapter{One-loop Integrals}\label{one-loopintegrals}
\section{Definition of the One-loop Integrals}
We define the one-loop integrals in the same notation as in {\em LoopTools}.
As discussed in Chapter~\ref{ReMSSM}, dimensional reduction is used, where only the momenta are calculated in 
$D$ dimensions, while the fields and the Dirac algebra are kept $4$-dimensional. 
The definition for the scalar one-loop integrals are shown in the following, their tensor integrals are defined by adding
the momenta $q_\mu, q_{\mu}q_{\nu}\cdots $ to the numerator.\\
\hspace*{4cm}
\begin{minipage}{0.05\linewidth}
\vspace*{1cm}
\includegraphics[width=\linewidth]{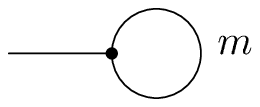}
\end{minipage}
\hspace*{5mm}
\begin{minipage}{0.65\linewidth}
\vspace*{5mm}
\begin{equation}
A_0\left(m^2\right ) = \frac{(2\pi\mu)^{4 - D}}{i \pi^2}\int \frac{d^D q}{q^2 - m^2}\, ,\nonumber
\end{equation}
\end{minipage}\\
\begin{minipage}{0.2\linewidth}
\vspace*{1cm}
\includegraphics[width=\linewidth]{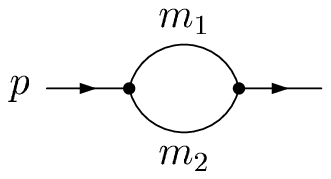}
\end{minipage}
\hspace*{5mm}
\begin{minipage}{0.65\linewidth}
\begin{eqnarray}
&&B_0\left(p^2, m_1^2, m_2^2\right ) \, \nonumber \\
&&= \frac{(2\pi\mu)^{4 - D}}{i \pi^2}\int \frac{d^D q}{\bigl [q^2 - m_1^2\bigr ]
\bigl [(q + p)^2-m_2^2\bigr ]}\, ,\nonumber
\end{eqnarray}
\end{minipage}\\
\begin{minipage}{0.25\linewidth}
\vspace*{1cm}
\includegraphics[width=\linewidth]{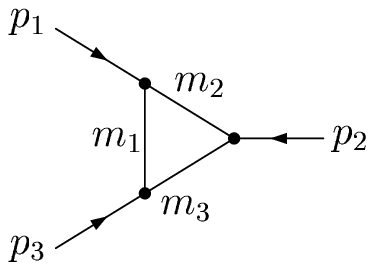}
\end{minipage}
\hspace*{5mm}
\begin{minipage}{0.65\linewidth}
\begin{eqnarray}
C_0\left(p_1^2, p_2^2, (p_1 + p_2)^2, m_1^2, m_2^2, m_3^3\right ) \, \nonumber
\end{eqnarray}
\end{minipage}
\begin{eqnarray}
= \frac{(2\pi\mu)^{4 - D}}{i \pi^2}\int \frac{d^D q}{\bigl [q^2 - m_1^2\bigr ]
\bigl [(q + p_1)^2-m_2^2\bigr ] \bigl [(q + p_1 + p_2)^2-m_3^2\bigr ]}\, ,\nonumber
\end{eqnarray}
\begin{minipage}{0.3\linewidth}
\vspace*{1cm}
\includegraphics[width=\linewidth]{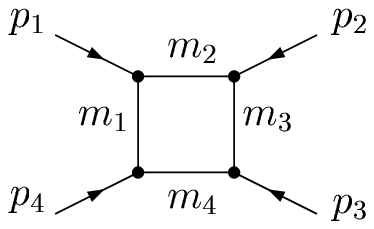}
\end{minipage}
\hspace*{3mm}
\begin{minipage}{0.65\linewidth}
$D_0\left(p_1^2, p_2^2, p_3^2, p_4^2, (p_1 + p_2)^2, (p_2 + p_3)^2, 
m_1^2, m_2^2, m_3^3, m_4^2\right )$ 
\end{minipage}
{\small \begin{eqnarray}
 = \frac{(2\pi\mu)^{4 - D}}{i \pi^2}
\int \frac{d^D q}{\bigl [q^2 - m_1^2\bigr ]
\bigl [(q + p_1)^2-m_2^2\bigr ] \bigl [(q + p_1 + p_2)^2-m_3^2\bigr ]\bigl [(q + p_1 + p_2 + p_3)^2-m_4^2\bigr ]}\, .\nonumber 
\end{eqnarray}}
\section{Scalar One-loop Integrals}
The general formula for the scalar one-, two-, three- and four-point functions were derived in \cite{soft}. 
Here we not only outline their general formula but also give the explicit expressions in some special cases.
\subsection*{Scalar One-point Function}
The scalar one-point function can be written as
\begin{eqnarray}
A_0(m^2) & = & m^2\left (\Delta - \ln\left (\frac{m^2}{\mu^2}\right ) + 1\right )\, ,
\end{eqnarray}
where the UV-divergent part $\Delta$ is defined via
\begin{eqnarray}
\Delta & =& \frac{2}{4-D} - \gamma_E + \ln 4\pi
\end{eqnarray}
with $\gamma_E$ is Euler's constant, and $\mu$ is renormalization scale.
\subsection*{Scalar Two-point Function}
The scalar two-point function can be written as
\begin{equation}
B_0(p^2, m_1^2, m_2^2) = 
\Delta - \int_0^1 dx\ln\left [ \frac{x^2p^2 -x(p^2 -m_2^2 +m_1^2) + m_1^2 -i\epsilon}{\mu^2}\right ]\, ,
\end{equation}
where $\epsilon$ is a infinitesimal real number.
Below are some special cases,
\begin{align}
B_0(m_l^2, m_l^2, 0)& = \Delta - \ln \left (\frac{m_l^2}{\mu^2}\right ) + 2\, ,\\
B_0(0, m_l^2, 0)& = \Delta - \ln(\frac{m_l^2}{\mu^2}) + 1\, ,\\
B_0(p^2, m_l^2, m_l^2)& =  \Delta- \ln\left (\frac{-p^2-i\epsilon}{\mu^2}\right ) +2, \ \ (p^2 \gg m_l^2)\, ,\\
B_0(p^2, 0, m_{\tilde l_s}^2)& = \Delta - \ln \left (\frac{m_{\tilde l_s}^2}{\mu^2}\right ) + 2\, ,\ \ (p^2 \simeq m_{\tilde l_s}^2)\, .
\end{align}
\subsection*{Scalar Three-point Function}
The scalar three-point function can be expressed as
\begin{eqnarray}
C_0\left(p_1^2, p_2^2, (p_1 + p_2)^2, m_1^2, m_2^2, m_3^3\right )  = 
-\int_0^\infty d x_1 d x_2 d x_3 \frac{\delta (1- x_1 - x_2 - x_3)}{(x_1 + x_2 + x_3) g(x_1, x_2, x_3)} \, ,
\label{general3point}
 \end{eqnarray}
where 
\begin{eqnarray}
g(x_1, x_2, x_3) & = & (m_1^2 x_1^2 + m_2^2 x_2^2 + m_3^2 x_3^2)(x_1 + x_2 + x_3) -\, \nonumber \\
&& {} p_1^2 x_1 x_2 - p_2^2 x_2 x_3 -(p_1 + p_2)^2 x_1 x_3 -i\epsilon\, .
\end{eqnarray}
A special case can be expressed as
\begin{eqnarray}
 C_0\left (m_l^2, m_l^2, (p_1 + p_2)^2, 0, m_l^2, m_l^2\right ) &= & \frac{1}{(p_1 + p_2)^2}\Biggl [
\ln\left(\frac{m_l^2}{-(p_1 + p_2)^2- i\epsilon}\right )\ln \left (\frac{\lambda^2}{m_l^2} \right ) + \, \nonumber \\
&&{} \frac{1}{2}
\ln^2\left (\frac{m_l^2}{-(p_1 + p_2)^2- i\epsilon}\right ) - \frac{\pi^2}{6}\Biggr ]\, ,
 \end{eqnarray}
where $\lambda$ is the photon mass regulator, and we have assumed $(p_1 + p_2)^2\gg m_l^2$\, .
This can also be found in Ref.~\cite{loopfunctiona}.

In the calculations for $\tilde \chi_2^0$ decays one has to  
calculate the scalar three-point function
$C_0^a = C_0(m_l^2, m_{\tilde \chi_2^0}^2, (p_1 + p_2)^2, 0, m_l^2, m_{\tilde{l_s}}^2)$~$((p_1 + p_2)^2 \gg m_l^2)$ analytically.
It can be obtained via the calculation of (\ref{general3point}).
Below are the analytical expressions in different cases.
\newpage

For the general $\tilde \chi_2^0$ three-body decays, i.e. $(p_1 + p_2)^2 \neq m_{\tilde{l_s}}^2$, it can be expressed as follows,
\begin{equation}
\begin{split}
C_0^a & = - \frac{1}{(p_1 + p_2)^2 -m_{\tilde \chi_2^0}^2}\left \{ - \ln\left(\frac{m_l^2}{(p_1 + p_2)^2 -m_{\tilde \chi_2^0}^2- i\epsilon}\right )
\ln\left(\frac{m_{\tilde{l_s}}^2 -m_{\tilde \chi_2^0}^2 -i\epsilon}{m_{\tilde{l_s}}^2-(p_1 + p_2)^2 -i \epsilon}\right )+ \right.\\
& \quad \left. \ln\left (\frac{m_{\tilde{l_s}}^2}{m_{\tilde{l_s}}^2-(p_1 + p_2)^2-i\epsilon}\right )
\ln\left (\frac{(p_1 + p_2)^2 -m_{\tilde \chi_2^0}^2- i\epsilon}{(p_1 + p_2)^2}\right )
+ \frac{1}{2}\ln^2\left (\frac{m_{\tilde{l_s}}^2}{m_{\tilde{l_s}}^2-m_{\tilde \chi_2^0}^2 -i\epsilon}\right )
\right.\\
& \quad \left. + \frac{\pi^2}{6} 
- Li_2 \left (\frac{m_{\tilde{l_s}}^2-(p_1 + p_2)^2-i\epsilon}{m_{\tilde{l_s}}^2-m_{\tilde \chi_2^0}^2 -i\epsilon}\right )
+  Li_2 \left (\frac{(- m_{\tilde \chi_2^0}^2 -i\epsilon)(m_{\tilde{l_s}}^2-(p_1 + p_2)^2-i \epsilon)}
{m_{\tilde{l_s}}^2(-m_{\tilde \chi_2^0}^2 + (p_1 + p_2)^2-i\epsilon })\right )\right.\\
& \quad \left. - Li_2\left (\frac{- m_{\tilde \chi_2^0}^2 - i\epsilon}{(p_1 + p_2)^2 -m_{\tilde \chi_2^0}^2- i\epsilon}\right )\right \}\, .
\end{split}
\end{equation}
The dilogarithm $Li_2(x)$ is defined as
\begin{equation}
Li_2(x) = -\int_0^1 dt \frac{\ln (1-xt)}{t}\, . 
\end{equation}

When the sleptons $\tilde{l_s}$ can be on shell, i.e. $(p_1 + p_2)^2$ is close to $m_{\tilde{l_s}}^2$, using complex masses
$ m_{\tilde{l_s}}^2 \to m_{\tilde{l_s}}^2-i\Gamma_{\tilde{l_s}} m_{\tilde{l_s}}$, one obtains
\begin{equation}
\begin{split}
C_0^a & = - \frac{1}{(p_1 + p_2)^2 -m_{\tilde \chi_2^0}^2}\left \{ 
\frac{1}{2}\ln^2 \left (\frac{m_{\tilde{l_s}}^2}{m_{\tilde{l_s}}^2-m_{\tilde \chi_2^0}^2 -i\epsilon}\right ) + \frac{\pi^2}{6}-
 \right.\\
& \quad \left.\ln\left (\frac{m_l^2}{(p_1 + p_2)^2 -m_{\tilde \chi_2^0}^2- i\epsilon}\right )
\ln\left (\frac{m_{\tilde{l_s}}^2 -m_{\tilde \chi_2^0}^2 -i\epsilon}{m_{\tilde{l_s}}^2-(p_1 + p_2)^2 -i\Gamma_{\tilde{l_s}} m_{\tilde{l_s}}-
i \epsilon}\right )+ \right.\\
& \quad \left. \ln \left (\frac{m_{\tilde{l_s}}^2}{m_{\tilde{l_s}}^2-(p_1 + p_2)^2-i\Gamma_{\tilde{l_s}} m_{\tilde{l_s}}
-i \epsilon}\right )\ln \left(\frac{(p_1 + p_2)^2 -m_{\tilde \chi_2^0}^2- i\epsilon}{(p_1 + p_2)^2}\right )-
  \right.\\
& \quad \left. Li_2 \left (\frac{- m_{\tilde \chi_2^0}^2 - i\epsilon}
{(p_1 + p_2)^2 -m_{\tilde \chi_2^0}^2- i\epsilon}\right ) \right \}\, .
\end{split}
\end{equation}
Here we have used $Li_2(0) = 0$. This formula is for the complete calculation.

In this case one can also treat $\tilde \chi_2^0$ decays as production and decay of $\tilde{l_s}$,
i.e. $(p_1 + p_2)^2 = m_{\tilde{l_s}}^2$, the function $C_0^a$ is IR divergent. The explicit expressions are
\begin{eqnarray}
C_0^a &=& \frac{1}{m_{\tilde{l_s}}^2-m_{\tilde \chi_2^0}^2}\left \{- \ln\left(\frac{m_l m_{\tilde l_s}}
{m_{\tilde{l_s}}^2-m_{\tilde \chi_2^0}^2 -i\epsilon}\right ) \ln \left (\frac{\lambda^2}{m_l m_{\tilde{l_s}}}\right )
+ Li_2\left (\frac{-m_{\tilde \chi_2^0}^2 -i\epsilon}{m_{\tilde{l_s}}^2-m_{\tilde \chi_2^0}^2 -i\epsilon}\right )\right. \nonumber \\
&&{}\left. - \frac{1}{2}\ln \left (\frac{m_l^2}{m_{\tilde{l_s}}^2-m_{\tilde \chi_2^0}^2 -i\epsilon}\right )
\ln \left (\frac{m_{\tilde{l_s}}^2}{m_{\tilde{l_s}}^2-m_{\tilde \chi_2^0}^2 -i\epsilon}\right )\right \}\, ,
\end{eqnarray}
which is also given in Ref. \cite{loopfunctiona}.
\subsection*{Scalar Four-point Function}
The scalar four-point function 
$D_0(m_l^2, m_{\tilde \chi_2^0}^2, m_{\tilde \chi_1^0}^2, m_l^2, (p_1 + p_2)^2, (p_2 + p_3)^2, 0, m_l^2, m_{\tilde{l_s}}^2, m_l^2)$ is necessary
for the calculation of the three-body decays of $\tilde \chi_2^0$.
Its analytical expressions can be obtained from Ref. \cite{loopfunctiona},
\begin{eqnarray}
&& D_0(m_l^2, m_{\tilde \chi_2^0}^2, m_{\tilde \chi_1^0}^2, m_l^2, (p_1 + p_2)^2, (p_2 + p_3)^2, 0, m_l^2,
m_{\tilde{l_s}}^2, m_l^2) \, \nonumber\\
&& = \frac{1}{(p_2 + p_3)^2 (m_{\tilde{l_s}}^2-(p_1 + p_2)^2)}\left \{
\ln^2 \left (\frac{m_{\tilde{l_s}}m_l}{m_{\tilde{l_s}}^2 -m_{\tilde \chi_2^0}^2 -i\epsilon}\right )+
\ln^2\left (\frac{m_{\tilde{l_s}}m_l}{m_{\tilde{l_s}}^2 -m_{\tilde \chi_1^0}^2 -i\epsilon}\right )+  \right. \, \nonumber\\
&& \left. \frac{\pi^2}{3}
 - 2 \ln\left (\frac{m_l^2}{-(p_2 + p_3)^2-i\epsilon}\right ) 
\ln \left (\frac{m_{\tilde{l_s}}\lambda}{m_{\tilde{l_s}}^2 -(p_1 + p_2)^2-i m_{\tilde{l_s}}\Gamma_{\tilde{l_s}}-
i \epsilon}\right ) + \right. \, \nonumber\\
&&\left. Li_2\left (1- \frac{(m_{\tilde \chi_2^0}^2- m_{\tilde{l_s}}^2 +i\epsilon)(m_{\tilde \chi_1^0}^2- m_{\tilde{l_s}}^2 +
i\epsilon)}{m_{\tilde{l_s}}^2(-(p_2 + p_3)^2- i\epsilon)}\right )\right \}\, .
\end{eqnarray}
Here we focused on the case where $(p_1 + p_2)^2$ is close to $m_{\tilde{l_s}}^2$
and used the complex masses $ m_{\tilde{l_s}}^2 \to m_{\tilde{l_s}}^2-i\Gamma_{\tilde{l_s}} m_{\tilde{l_s}}$.
\subsection*{Photonic Part of the Fermion (Sfermion) Field \mbox {Renormalization} Constants}\label{photonicfermionZ}
The photonic part of the fermion self-energies are given by
\begin{eqnarray}
\Sigma_{ij}^{f, L}(p^2){\Big |}_{\rm photonic}& = &\Sigma_{ij}^{f, R}(p^2){\Big |}_{\rm photonic}
= -\frac{\alpha}{4 \pi}\delta_{ij}Q_f^2\bigl [2 B_1(p^2, m_{f_i}^2, 0) + 1\bigr ]\, ,\nonumber \\
\Sigma_{ij}^{f, S}(p^2){\Big |}_{\rm photonic}
&=& -\frac{\alpha}{2 \pi}\delta_{ij}Q_f^2 m_{f_i}\bigl [2 B_0(p^2, m_{f_i}^2, 0) - 1\bigr ]\, .
 \end{eqnarray}
The fermion field-renormalization constants have been presented in (\ref{eqn:SMfermion-Reconstantb}) and (\ref{eqn:SMfermion-Reconstantc}) in
Chapter~\ref{ReMSSM}. In order to calculate them in the case of light fermions we need the relations
\begin{eqnarray}
\frac{\partial B_0(p^2, m_f^2, 0)}{\partial p^2}{\Big |}_{p^2 = m_f^2}&= &- \frac{1}{m_f^2}(\ln\frac{\lambda}{m_f} + 1)\, \nonumber \\
B_1(m_f^2, m_f, 0)& = &\frac{1}{2} (\ln(\frac{m_f^2}{\mu^2}) - \Delta -3)\, \nonumber \\
\frac{\partial B_1(p^2, m_f^2, 0)}{\partial p^2}{\Big |}_{p^2 = m_f^2}&=& \frac{1}{m_f^2}(\ln\frac{\lambda}{m_f}+ \frac{3}{2})\, .
 \end{eqnarray}
Here we have used the general relations shown as follows~(see also \cite{SMDenner}),
\begin{eqnarray}
B_1(p^2, m_1, m_2)& =& \frac{m_2^2 - m_1^2}{2p^2}(B_0(p^2, m_1, m_2) - B_0(0, m_1, m_2))-\frac{1}{2}B_0(p^2, m_1, m_2)\, ,\nonumber \\
\frac{\partial B_1(p^2, m_1, m_2)}{\partial p^2} &= &\frac{m_2^2 - m_1^2}{2p^4}(B_0(p^2, m_1, m_2) - B0(0, m_1, m_2))+\, \nonumber \\
&&{}\frac{m_2^2 - m_1^2-p^2}{2p^2}\frac{\partial B_0(p^2, m_1, m_2)}{\partial p^2}\, ,\nonumber \\
\frac{\partial B_0(p^2, m_1, m_2)}{\partial p^2} &= &-\frac{m_1^2 - m_2^2}{p^4}\ln\frac{m_2}{m_1} + 
\frac{m_1m_2}{p^4}\left (\frac{1}{r}-r \right )\ln r\, \nonumber \\
&&{}- \frac{1}{p^2}\left (1 + \frac{r^2 + 1}{r^2 -1}\ln r\right )\, ,
\label{eqn:generalforB}
 \end{eqnarray}
where $r$ and $\frac{1}{r}$ are determined from the equation
\begin{eqnarray}
x^2 + \frac{m_1^2 + m_2^2 -p^2 -i \epsilon}{m_1m_2} x + 1 = (x +r)(x + \frac{1}{r})\, . 
\end{eqnarray}
The photonic part of the field renormalization constants for the light fermions can be expressed as
\begin{eqnarray}
\delta Z_{ii}^{f, L} |_{\rm photonic} = \delta Z_{ii}^{f, R} |_{\rm photonic} =  
\frac{\alpha}{4\pi}Q_f^2 \left (\ln\left (\frac{m_f^2}{\mu^2}\right ) - 4\ln\left (\frac{\lambda}{m_f}\right )-\Delta - 4\right )\, .
\end{eqnarray}

The photonic part of the sfermion field renormalization constants can been written as
 \begin{eqnarray}
\delta Z_{\tilde f_{ii}}{\big |}_{\rm photonic}&=& - \tilde{Re}\frac{\partial\Sigma_{\tilde f_{ii}}(p^2)}{\partial p^2}
{\Big |}_{\rm photonic}^{p^2 = m_{\tilde f_i}^2}
\, ,
\end{eqnarray}
where the sfermion self-energies are expressed as
 \begin{eqnarray}
\Sigma_{\tilde f_{ii}}(p^2) & = &
-\frac{\alpha}{\pi}Q_f^2 p^2\left (B_0(p^2, 0, m_{\tilde f_i}^2) - B_1(p^2, 0, m_{\tilde f_i}^2) \right )\, .
\end{eqnarray}
Using the relations in (\ref{eqn:generalforB}) one obtains the expression for $\delta Z_{\tilde f_{ii}}{\big |}_{\rm photonic}$.
The IR-singular part reads
\begin{eqnarray}
\delta Z_{\tilde f_{ii}}{\big |}_{\rm photonic}^{\rm sing} & = & -\frac{\alpha}{\pi} Q_f^2\ln \left (\frac{\lambda}{ m_{\tilde f_i}}\right )\, .
\end{eqnarray}
\newpage
\chapter{Multi-Channel Monte Carlo Method}\label{mcamcp}
\section{Principles of the Monte Carlo Method}
The Monte Carlo method~(see Ref. \cite{colliderphysics}) is a way to calculate the integrals with a large number 
of integration variables. The integration variables $x$ are mapped to a set of random numbers $r$ via 
\begin{eqnarray}
x = h(r)\, , \ \ 0 \le r \le 1\, .
\end{eqnarray}
An integral can be written as
 \begin{eqnarray}
I = \int f(x)dx =\int_0^1 \frac{f(h(r))}{g(h(r))}dr\, , 
 \end{eqnarray}
where $g(h(r))$ is density, it is defined as 
\begin{eqnarray}
\frac{1}{g(h(r))} =\frac{\partial h(r)}{\partial r}\, .
\end{eqnarray}
This integral can be approximated by sampling the integrand $N$ times and taking the average,
 \begin{eqnarray}
\bar I = \frac{1}{N}\sum_{i =1}^N \frac{f(h(r_i))}{g(h(r_i))}\, .
 \end{eqnarray}
The integration error is defined by
 \begin{eqnarray}
\sigma = \frac{\sqrt{\overline{I^2}- \bar{I}^2}}{N}\, .
 \end{eqnarray}
When the integrand $f$ varies strongly in the phase space, 
the efficiency of the Monte Carlo method is improved by {\em Importance Sampling},
i.e. more events are sampled in the important region where $f$ becomes large. 
This is implemented by choosing the variables $x$, the mappings between these variables and the random numbers $r$
in such a way that the resulting $f/g$ is much smoother than $f$.

In practice, we choose the variables in such a way that the Lorentz invariants corresponding to the propagators are included.
The mapping is chosen such that the density $g$ behaves in a similar way as the propagator.
Below are the mappings belonging to the different propagator types \cite{mutichannelb}.
\begin{itemize}
\item Propagator with vanishing width $\frac{1}{x-m^2}$:\\
The explicit expressions for the mapping $x = h(r)$ and the resulting density are written as follows,  
\begin{align}
h(r, m^2, \nu, x_{\rm min}, x_{\rm max})&=[r(x_{\rm max}-m^2)^{1-\nu} + (1-r)(x_{\rm min}-m^2)^{1-\nu}]^{\frac{1}{1-\nu}}+m^2,
\nonumber \\
g_x(x, m^2, \nu, x_{\rm min}, x_{\rm max})&=\frac{1-\nu}{[(x_{\rm max}-m^2)^{1-\nu}-(x_{\rm min}-m^2)^{1-\nu}](x-m^2)^{\nu}}
\label{eqn:densitya}
\end{align}
for $m^2 < x$ and $\nu \neq 1$,
\begin{align}
h(r, m^2, \nu, x_{\rm min}, x_{\rm max})&=-[r(m^2-x_{\rm max})^{1-\nu} + (1-r)(m^2-x_{\rm min})^{1-\nu}]^{\frac{1}{1-\nu}}+m^2,
\nonumber \\
g_x(x, m^2, \nu, x_{\rm min}, x_{\rm max})&=\frac{-(1-\nu)}{[(m^2-x_{\rm max})^{1-\nu}-(m^2-x_{\rm min})^{1-\nu}](m^2-x)^{\nu}}
\end{align}
for  $m^2 > x$ and $\nu \neq 1$ and
\begin{align}
h(r, m^2, \nu, x_{\rm min}, x_{\rm max})&={\rm exp} \bigl [r\ln (x_{\rm \rm max}-m^2) - (1 -r)\ln (x_{\rm min}-m^2)\bigr ] + m^2 ,
\nonumber \\
g_x(x, m^2, \nu, x_{\rm min}, x_{\rm max})&= \frac{1}{\bigl [\ln (x_{\rm \rm max}-m^2)-\ln (x_{\rm min}-m^2)\bigr ](x - m^2)}
\end{align}
for $\nu = 1$.
\item Breit-Winger propagator $\frac{1}{x-m^2+ i m\Gamma}$:\\
The variable $x$ is mapped to
\begin{equation}
 h(r, m^2-i m \Gamma, 2, x_{\rm min}, x_{\rm max})=m \Gamma \tan[y_1 +(y_2-y_1)r]+m^2\, ,
\label{eqn:mapplingb}
\end{equation}
the resulting density is 
\begin{equation}
g_x(x, m^2-i m\Gamma, 2, x_{\rm min}, x_{\rm max})=\frac{m \Gamma}{(y_2-y_1)[(x-m^2)^2+m^2\Gamma^2]}\, ,
\label{eqn:densityb}
\end{equation}
where
\begin{equation}
y_{1/2}=\arctan(\frac{s_{\rm min/max}-m^2}{m \Gamma})\, .
\end{equation}
\end{itemize}
The parameter $\nu$ can be tuned to optimize the Monte Carlo integration and should be chosen $\nu \ge 1$.
Other variables, i.e. the polar and azimuthal angels $\theta$ and $\phi$, are generated as following,
\begin{equation}
\cos\theta = 2 r-1\, , \ \ \ \phi = 2\pi r\, .
\label{eqn:anglemapping}
\end{equation}
\section{Kinematics}\label{Kinematics}
For a mutiparticle process, i.e. a particle with momentum $k_1$\ ( mass $m_1$) 
decays into $n$ particles with momenta $k_i$ \ (mass $m_i$),
the phase-space element is given by
\begin{eqnarray}
d\Phi_{1 \to n} & = & \Biggl [\prod_{ i = 2}^{n+1}\frac{d^3 k_i}{2 k_{i0}}\Biggr]
\delta^4 (k_1 - \sum_{i=2}^{n+1} k_i)\, .
 \end{eqnarray}
This process can be described by $3 n-4$ independent variables. 
In order to obtain the kinematics we treat the multiparticle process as taking place via cascade decays, 
where the intermediate states are unstable particles which then
decay to others and eventually form the final states particles. 
Hence a mutiparticle process can be composed by isotropic particle decays, which are described as follows.

One particle with momentum $k_1$ decays into two particles with momenta $k_2$ and 
$k_3$, masses $m_2$ and $m_3$. The polar angle $\phi$ and azimuthal angle $\theta$ in the rest frame of the decaying
particle are chosen to be the suitable integration variables. The phase-space element is defined as
\begin{equation}
\begin{split}
\int d\Phi(k_1, m_2^2, m_3^2)&= \int \frac{d^3 k_2}{2k_{20}} \frac{d^3 k_3}{2k_{30}} \delta^{(4)}(k_1 - k_2 -k_3) \\
&= \frac{\lambda^{1/2}(k_1^2, m_2^2, m_3^2)}{8k_1^2}\int_0^{2\pi}d \phi 
\int_{-1}^1 d \cos\theta\, , 
\end{split}
\end{equation}
$\lambda$ is defined as follows,
\begin{equation}
\lambda(x, y, z) = x^2 +y^2 + z^2 -2xy -2yz -2xz\, . 
\end{equation}
Using the Monte Carlo method, the angles have to be mapped to the random numbers as (\ref{eqn:anglemapping}),
hence the density can be written as
\begin{eqnarray}
g_d(k_1^2, m_2^2, m_3^2) = \frac{2k_1^2}{\lambda^{1/2}(k_1^2, m_2^2, m_3^2)\pi}\, .
\label{eqn:twobodydensity}
 \end{eqnarray}

Since the laboratory frame usually does not coincide with the rest frame of the decaying particle,
the Lorentz transformation is introduced~\cite{kinematics}. The Lorentz transformation of momentum $k$
into the rest frame of the particle with momentum $p$ is defined by
\begin{equation}
k^{'} = B(\gamma, \beta)k\, ,
\end{equation}
where $\gamma = \frac{p_0}{m}$, $\gamma\beta =\frac{|\vec{p}|}{m}$, and $m = \sqrt {p^2}$;
the explicit formula for $B(\gamma, \beta)$ can be written as 
\begin{equation}
 B(\gamma, \beta)= \left ( \begin{array}{llcl}
\gamma & 0 & 0 & \gamma\beta\\
0 & 1 & 0 & 0\\
0 & 0 & 1 & 0\\
\gamma\beta & 0 & 0 & \gamma
\end{array} \right )\, .
\end{equation}
The inverse Lorentz transformations is defined by replacing $\vec p$ by $-\vec p$,
\begin{equation}
k = B(\gamma, - \beta)k^{'}\, . 
\end{equation}

The orientation of the coordinate system can be arbitrarily chosen because the decay is isotropic.
The momentum of the outgoing particle can be written as
\begin{equation}
 k = R(\phi, \theta)B(\gamma, - \beta)k^{'}\, ,
\end{equation}
with the explicit rotation
\begin{equation}
\begin{split}
 R(\phi, \theta)&= \left ( \begin{array}{llcl}
1 & 0 & 0 & 0\\
0 & \cos\phi & \sin\phi & 0\\
0 & - \sin\phi & \cos\phi & 0\\
0 & 0 & 0 & 1
\end{array} \right )
\left ( \begin{array}{llcl}
1 & 0 & 0 & 0\\
0 & \cos\theta & 0 & \sin\theta\\
0 & 0 & 1 & 0\\
0 & -\sin\theta & 0 & \cos\theta
\end{array} \right )\\
&= \left ( \begin{array}{llcl}
1 & 0 & 0 & 0\\
0 & \cos\theta\cos\phi & \sin\phi & \sin\theta\cos\phi\\
0 & -\cos\theta\sin\phi & \cos\phi & -\sin\theta\sin\phi\\
0 & -\sin\theta & 0 & \cos\theta
\end{array} \right )\, .
\end{split}
\end{equation}

\begin{figure}[b]
\begin{center}
\begin{tabular}{c}
\includegraphics[width=0.5\linewidth]{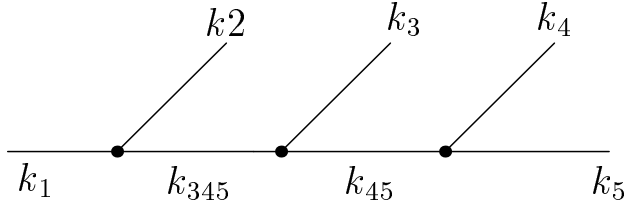}
\end{tabular}
\caption{The $1\to 4$ process expressed as $3$ isotropic decays\label{Figure:kinematics}}
\end{center}
\end{figure}

For example, a $1\to 4$ process can be expressed as $3$ isotropic decays (see Figure \ref{Figure:kinematics}):
\begin{equation}
k_1 \to k_2 + k_{345}\, , \ \ \  \ k_{345} \to k_3 + k_{45}\, , \ \ \ \ k_{45} \to k_4 + k_5\, ,
\end{equation}
where $k_{345} = k_3 + k_4 + k_5$, $k_{45} = k_4 + k_5$\, .
The phase-space element can be written as 
\begin{equation}
\begin{split}
\int d\Phi_{1\to 4}&=\int d\Phi(k_1, k_2^2, k_{345}^2) d\Phi(k_{345}, k_3^2, k_{45}^2)
 d\Phi(k_{45}, k_4^2, k_{5}^2)d k_{45}^2 d k_{345}^2\\
& = \int_{(k_{45}^2)^{min}}^{(k_{45}^2)^{max}}\int_{(k_{345}^2)^{min}}^{(k_{345}^2)^{max}}
\frac{\lambda^{1/2}(k_1^2, k_2^2, k_{345}^2)}{8k_1^2}
\frac{\lambda^{1/2}(k_{345}^2, k_3^2, k_{45}^2)}{8k_{345}^2}
\frac{\lambda^{1/2}(k_{45}^2, k_4^2, k_{5}^2)}{8k_{45}^2}\\
& \quad  d k_{45}^2 d k_{345}^2\int_0^{2\pi}d \phi_1 \int_{-1}^1 d\cos\theta_1 \int_0^{2\pi}d \phi_2 
\int_{-1}^1 d \cos\theta_2 \int_0^{2\pi} d \phi_3 \int_{-1}^1 d \cos\theta_3\, , 
\end{split}
\end{equation}
with 
\begin{eqnarray}
(k_{45}^2)^{min}=(m_4 +m_5)^2\, ,&& \ \ \ (k_{45}^2)^{max}=(m_1-m_2-m_3)^2\, , \nonumber \\
 (k_{345}^2)^{min}=(m_3 + \sqrt{k_{45}^2})^2\, ,&& \ \ \  
(k_{345}^2)^{max}=(m_1-m_2)^2 \, . \nonumber 
\end{eqnarray}
Similarly to the $1 \to 4$ process, the phase-space element of the $1\to 3$ process can be written as,
\begin{equation}
\begin{split}
\int d\Phi_{1\to 3}&= \int d\Phi(k_1, k_2^2, k_{34}^2) d\Phi(k_{34}^2, k_3^2, k_{4}^2)d k_{34}^2\\
& =\int_{(k_{34}^2)^{min}}^{(k_{34}^2)^{max}}
\frac{\lambda^{1/2}(k_1^2, k_2^2, k_{34}^2)}{8k_1^2}
\frac{\lambda^{1/2}(k_{34}^2, k_3^2, k_{4}^2)}{8k_{34}^2}d k_{34}^2\\
 & \quad \int_0^{2\pi}d \phi_1 \int_{-1}^1 d\cos\theta_1 \int_0^{2\pi}d \phi_2 
\int_{-1}^1 d \cos\theta_2\, .
\end{split}
\end{equation}
The decay width of the $1 \to n$ process is defined as
\begin{eqnarray}
d\Gamma_{1\to n} & = & \frac{1}{(2\pi)^{3n-4}}\frac{1}{2 m_1}\int \sum\Bigl |M \Bigr |^2 d\Phi_{1\to n}\, ,
\label{generalwidth}
\end{eqnarray}
where $M$ is the matrix element of all the diagrams for this process, it is squared and averaged over the spin of the
external particles. 
The production process, i.e. the $2 \to n$ process, is treated analogously.
\newpage
\section{Multi-Channel Approach}
The contributions of the real photon bremsstrahlung can be expressed as (\ref{eqn:realGamma}).
The singular part is separated by the phase-space-slicing method and calculated analytically.
The finite part is calculated by the Monte Carlo method. The amplitude of the real photon bremsstrahlung 
has different propagators corresponding to different diagrams. These propagators behave differently in different phase-space regions.
Therefore, in order to obtain a stable numerical result and to reduce the Monte Carlo integration error, 
we use a multi-channel Monte Carlo method~\cite{mutichannela, mutichannelb}.

In the case of the n-body decay, the decay width is expressed in (\ref{generalwidth}).
For each type of propagator we choose a suitable variable set $x_k$, the decay width can be expressed as
\begin{eqnarray}
d\Gamma_{1\to n} & = & \int f(x_k)\rho(x_k)dx_k\, ,
\label{eqn: mcmc1} 
\end{eqnarray}
where
\begin{eqnarray}
f(x_k) = \frac{1}{(2\pi)^{3n-4}}\frac{1}{2 m_1}\sum \bigl |M \bigr |^2 (x_k)
\end{eqnarray}
and $\rho(x_k)$ is the phase-space density. 
Accordingly a mapping $x_k = h_k(r)$ with the random number $0\le r_i \le 1$ is chosen, 
\begin{eqnarray}
d\Gamma_{1\to n} & = & \int_0^1 \frac{f(h_k(r))}{g(h_k(r))}dr\, .
\end{eqnarray}
The resulting density 
\begin{eqnarray}
g(h_k(r)) = \frac{1}{\rho (h_k(r))\frac{\partial h_k(r)}{\partial r}}
\end{eqnarray}
describes the particular behavior of this propagator. 
All densities $g(h_k(r))$ are combined into one total density $g_{\rm tot}$ which is expected to smooth the integrand 
over the whole integration region. 
The phase space integral  (\ref{eqn: mcmc1}) can be written as 
\begin{eqnarray}
d\Gamma_{1\to n} & = & \sum_{k=1}^{M}\int dx_k\rho(x_k)g(x_k)
\frac{f(x_k)}{g_{\rm tot}(x_k)}\, \nonumber \\
& = &  \sum_{k=1}^{M}\int_0^1 dr \frac{f(h_k(r))}{g_{\rm tot}(h_k(r))}\, .
\end{eqnarray}
The total density is defined as
\begin{eqnarray}
g_{\rm tot}(x_k)& = & \sum_{k=1}^{M}g(x_k)\, ,
\end{eqnarray}
where $M$ is the number of the mappings (channels).
\section*{An example}
For the decay $\tilde \chi_2^0 \to \tilde \chi_1^0 l^- l^+ \gamma$, we have 14 channels. Each channel smoothes 
a particular behavior of a propagator. The channels for interference contributions are not included.
\begin{figure}[htb]
\begin{center}
\begin{tabular}{c}
\includegraphics[width=0.5\linewidth]{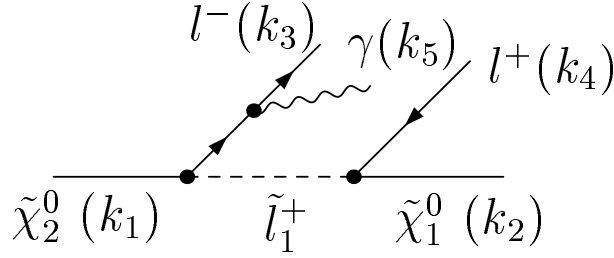}
\end{tabular}
\caption{An example of diagram of the decay $\tilde \chi_2^0 \to \tilde \chi_1^0 l^- l^+ \gamma$\label{exmplefigure}}
\end{center}
\end{figure}
A diagram is shown as an example in Figure \ref{exmplefigure}. The phase space integral can be decomposed as
\begin{eqnarray}
\int d\Phi & =&  \int_{(k_{35}^2)_{\rm min}}^{(k_{35}^2)_{\rm max}} dk_{35}^2 
\int_{(k_{24}^2)_{\rm min}}^{(k_{24}^2)_{\rm max}} dk_{24}^2\int d\Phi (k_1, k_{35}^2, k_{24}^2)\, \nonumber \\
&&{}d\Phi (k_{35}, k_{3}^2, k_{5}^2)\int  d\Phi (k_{24}, k_{2}^2, k_{4}^2)\, ,
 \end{eqnarray}
where $k_{35} = k_3 + k_5$,  $k_{24} = k_2 + k_4$.
The upper and lower limits on the variables $k_{35}^2$ and  $k_{24}^2$ are 
\begin{eqnarray}
(k_{35}^2)_{\rm min}=(k_{35}^2)_{\rm cut}\, ,&& \ \ \ 
(k_{35}^2)_{\rm max}= (m_{\tilde \chi_2^0}-m_{\tilde \chi_1^0}- m_l)^2\, , \nonumber \\
 (k_{24}^2)^{\rm min}=(m_{\tilde \chi_1^0} + m_l)^2\, ,&& \ \ \  
(k_{24}^2)^{\rm max}= (m_{\tilde \chi_2^0}- m_{35})^2\, . \nonumber 
\end{eqnarray}
where $m_{35}= \sqrt {k_{35}^2}$.
The infrared and collinear singularities are excluded by the cut $(k_{35}^2)_{\rm cut}$.
For the SPS1a parameter set where $\tilde l_1$ can be on shell, the variables $k_{35}^2$ and $k_{24}^2$
are mapped via
\begin{eqnarray}
k_{35}^2 &= &h(r_1, m_l^2, \nu, (k_{35}^2)_{\rm min}, (k_{35}^2)_{\rm max})\, ,
\nonumber \\
k_{24}^2 &=& h(r_2, m_{\tilde l_1}^2- i  m_{\tilde l_1}\Gamma_{\tilde l_1}, 2, 
(k_{24}^2)^{\rm min}, (k_{24}^2)^{\rm max})\, ,
\end{eqnarray}
where the function $h$ has been defined in (\ref{eqn:densitya}) and (\ref{eqn:mapplingb}).
The total density for this set of mappings is
\begin{eqnarray}
g &= &g_x(k_{24}^2,  m_{\tilde l_1}^2- i  m_{\tilde l_1}\Gamma_{\tilde l_1}, 2, 
(k_{24}^2)^{\rm min}, (k_{24}^2)^{\rm max})g_x(k_{35}^2, m_l^2, \nu, (k_{35}^2)_{\rm min}, (k_{35}^2)_{\rm max})\, \nonumber \\
&&{}g_d(m_{\tilde \chi_2^0}^2, k_{35}^2, k_{24}^2) g_d(k_{24}^2, k_{2}^2, k_{4}^2)g_d(k_{35}^2, k_{3}^2, k_{5}^2)\, ,
\label{eqn:example-density}
\end{eqnarray}
where the densities $g_x$ and $g_d$ are defined in (\ref{eqn:densitya}), (\ref{eqn:densityb}) and (\ref{eqn:twobodydensity}),
respectively. The density in (\ref{eqn:example-density}) 
describes the propagator of the diagram in Figure \ref{exmplefigure}. Other diagrams for the decay
$\tilde \chi_2^0 \to \tilde \chi_1^0 l^- l^+ \gamma$ have been shown in Figure \ref{realdiagram}. 
Their variables and mappings are chosen similarly to this example.
\chapter{Feynman Diagrams}\label{non-photonicdiagrams}
In this appendix we present the generic Feynman diagrams for the virtual corrections of the decays
$\tilde \chi_2^0 \to \tilde \chi_1^0 l^- l^+ (l = e, \mu, \tau)$.
These Feynman diagrams are classified into vertex diagrams, self-energy diagrams and box diagrams.
The notations are as follows.
\begin{equation}
\begin{array}{ll}
{\rm Leptons} & E = \left \{\nu_e, e, \nu_\mu, \mu, \nu_\tau, \tau \right \}\\
{\rm Sleptons} & \tilde E = \left \{\tilde\nu_e, \tilde e, \tilde\nu_\mu, \tilde\mu, \tilde\nu_\tau, \tilde\tau \right \}\\
{\rm Gauge \ Bosons} & V =  \left \{\gamma, Z, W^\pm\right \}\\
{\rm Neutral \ Higgs \ Bosons \ and \ Goldstone \ Boson} & \phi^0 = \left \{h^0, H^0, A^0, G^0\right \}\\
{\rm Higgs \ Bosons \ and \ Goldstone \ Bosons} & \phi = \left \{\phi^0, H^\pm, G^\pm\right \}\\
{\rm Neutralinos \ and \ Charginos}& \tilde \chi = \left \{\tilde \chi_i^0, \tilde \chi_j^\pm \right \} 
(i = 1, \cdots 4, \ j = 1, 2)\\
{\rm MSSM \ Fermions} & F = \left \{E, u, d, c, s, t, b, \tilde \chi \right \}\\
{\rm MSSM \ Scalars}  & S = \left \{\tilde E, \tilde u, \tilde d, \tilde c, \tilde s, \tilde t, \tilde b, \phi \right \}\\
{\rm Fadeev-}{\rm Popov \ Ghost} & U = \left \{u^Z, u^\pm\right \}\, .
\end{array} 
\nonumber 
\end{equation}
\begin{figure}[htb]
\begin{tabular}{c}
\includegraphics[width=0.95\linewidth]{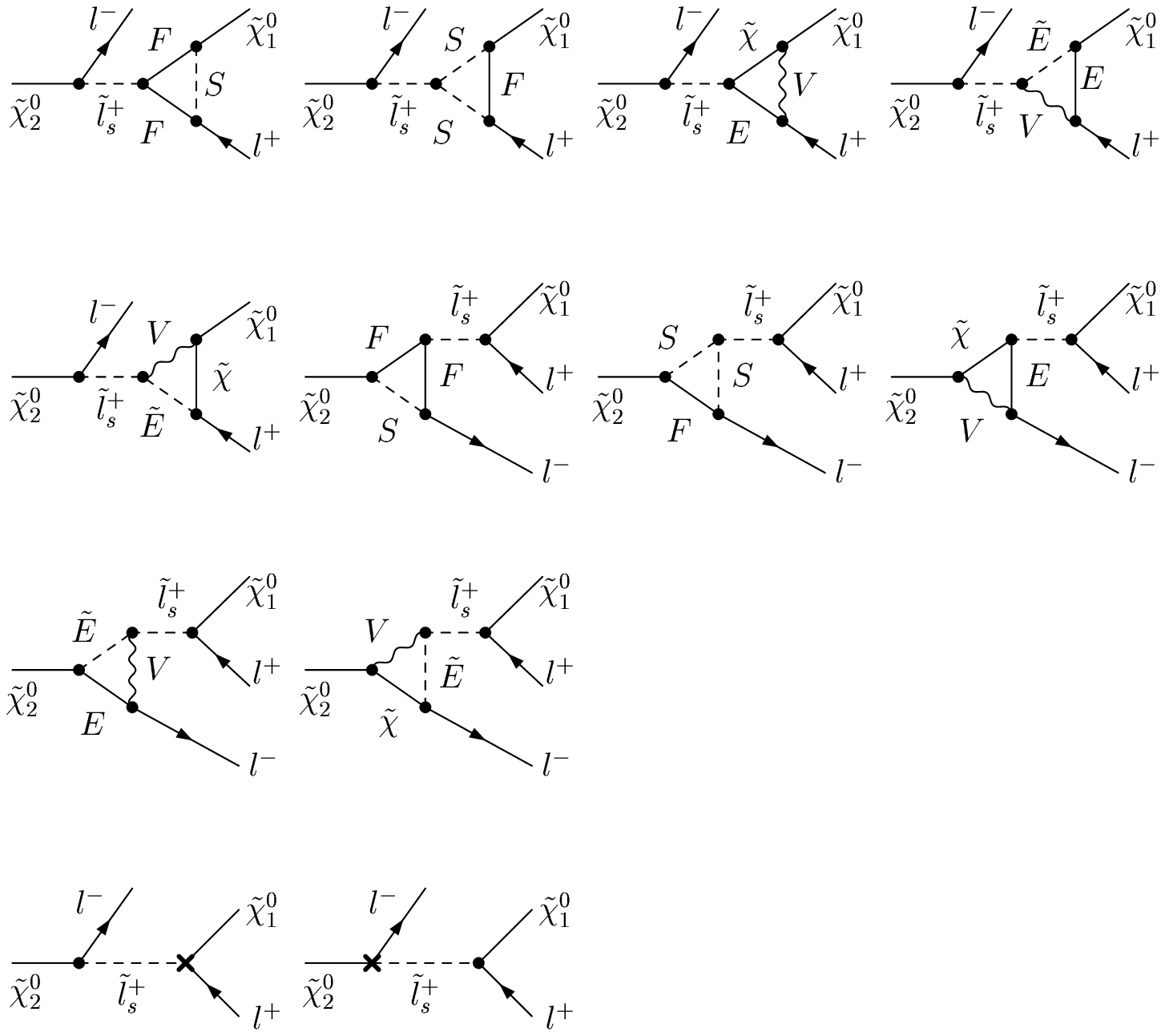} 
\end{tabular}
\caption{Vertex Feynman diagrams for $\tilde \chi_2^0$ decay into $\tilde \chi_1^0$ and two leptons $l^- l^+$
via the the exchange of the sleptons $\tilde l_s^+$, $s = 1, 2$ labels the slepton mass eigenstates.
The diagrams in the last line are the corresponding counterterm diagrams.}
\end{figure}
\begin{figure}[htb]
\begin{tabular}{c}
\includegraphics[width=0.95\linewidth]{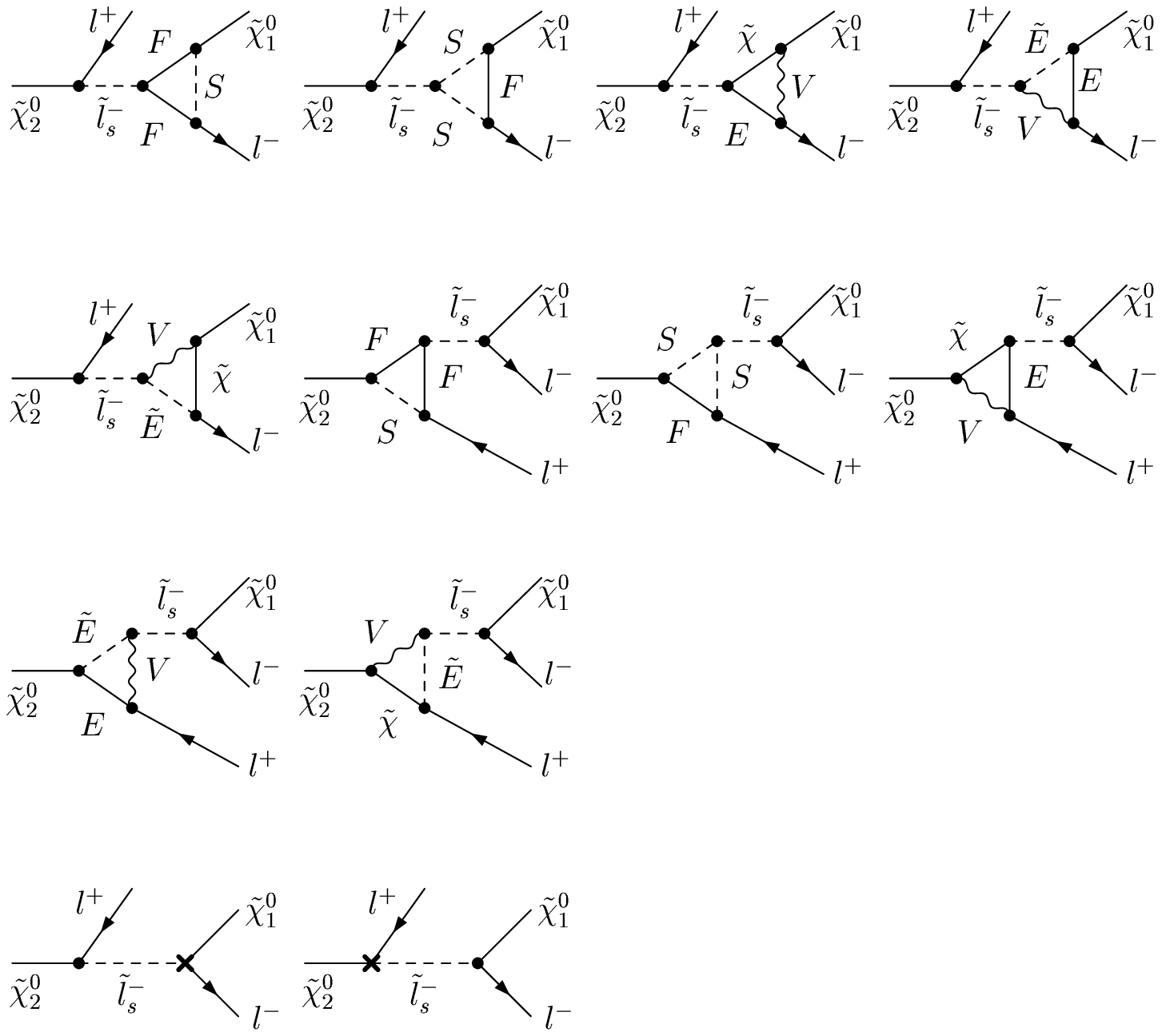} 
\end{tabular}
\caption{Vertex Feynman diagrams for $\tilde \chi_2^0$ decay into $\tilde \chi_1^0$ and two leptons $l^- l^+$
via the exchange of the sleptons $\tilde l_s^-$, $s = 1, 2$ labels the slepton mass eigenstates.
The diagrams in the last line are the corresponding counterterm diagrams.}
\end{figure}

\begin{figure}[htb]
\begin{tabular}{c}
\includegraphics[width=0.95\linewidth]{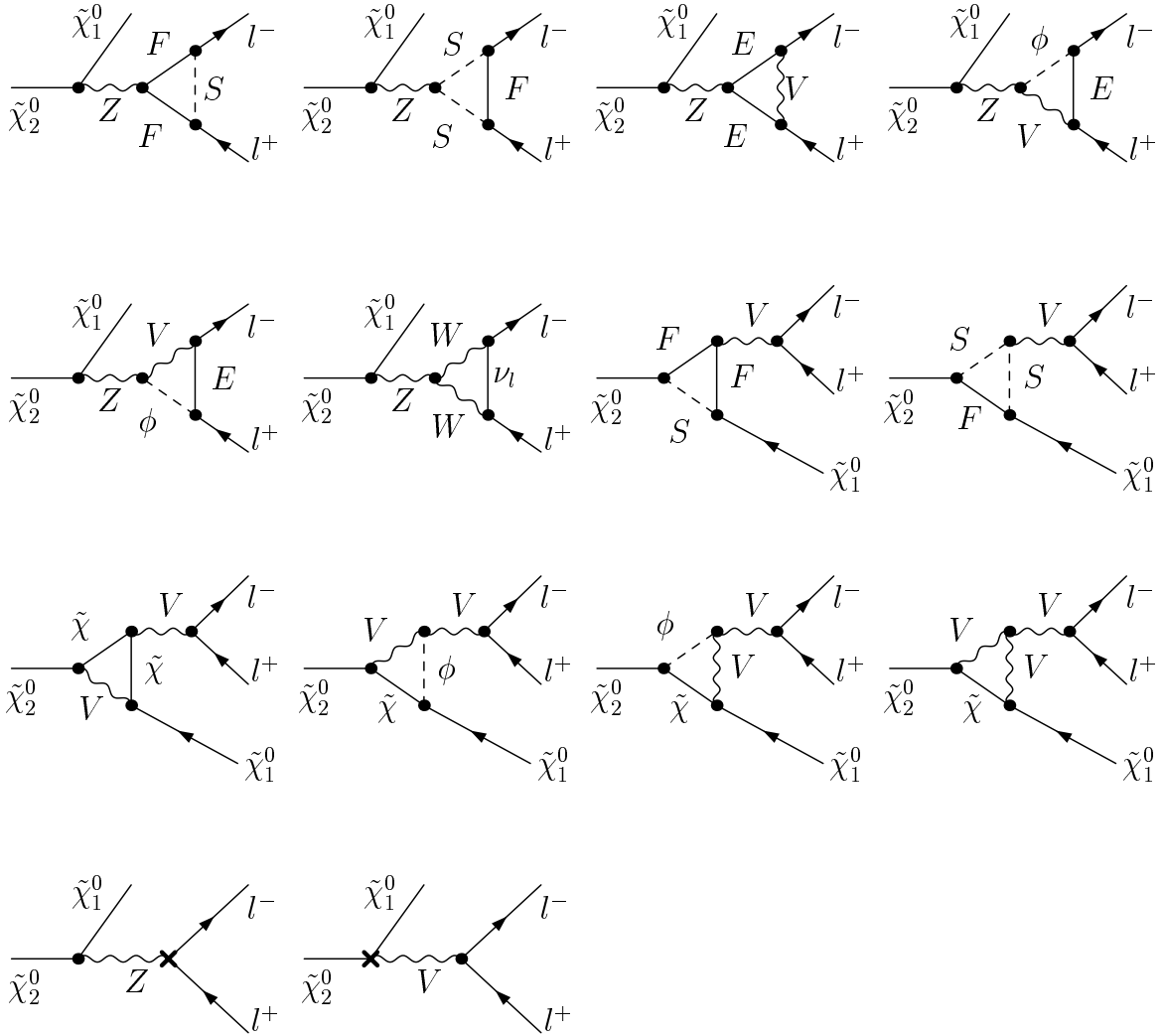} 
\end{tabular}
\caption{Vertex Feynman diagrams for $\tilde \chi_2^0$ decay into $\tilde \chi_1^0$ and two leptons $l^- l^+$
via the the exchange of the gauge bosons $V$.
The diagrams in the last line are the corresponding counterterm diagrams.}
\end{figure}
\begin{figure}[htb]
\begin{tabular}{c}
\includegraphics[width=0.95\linewidth]{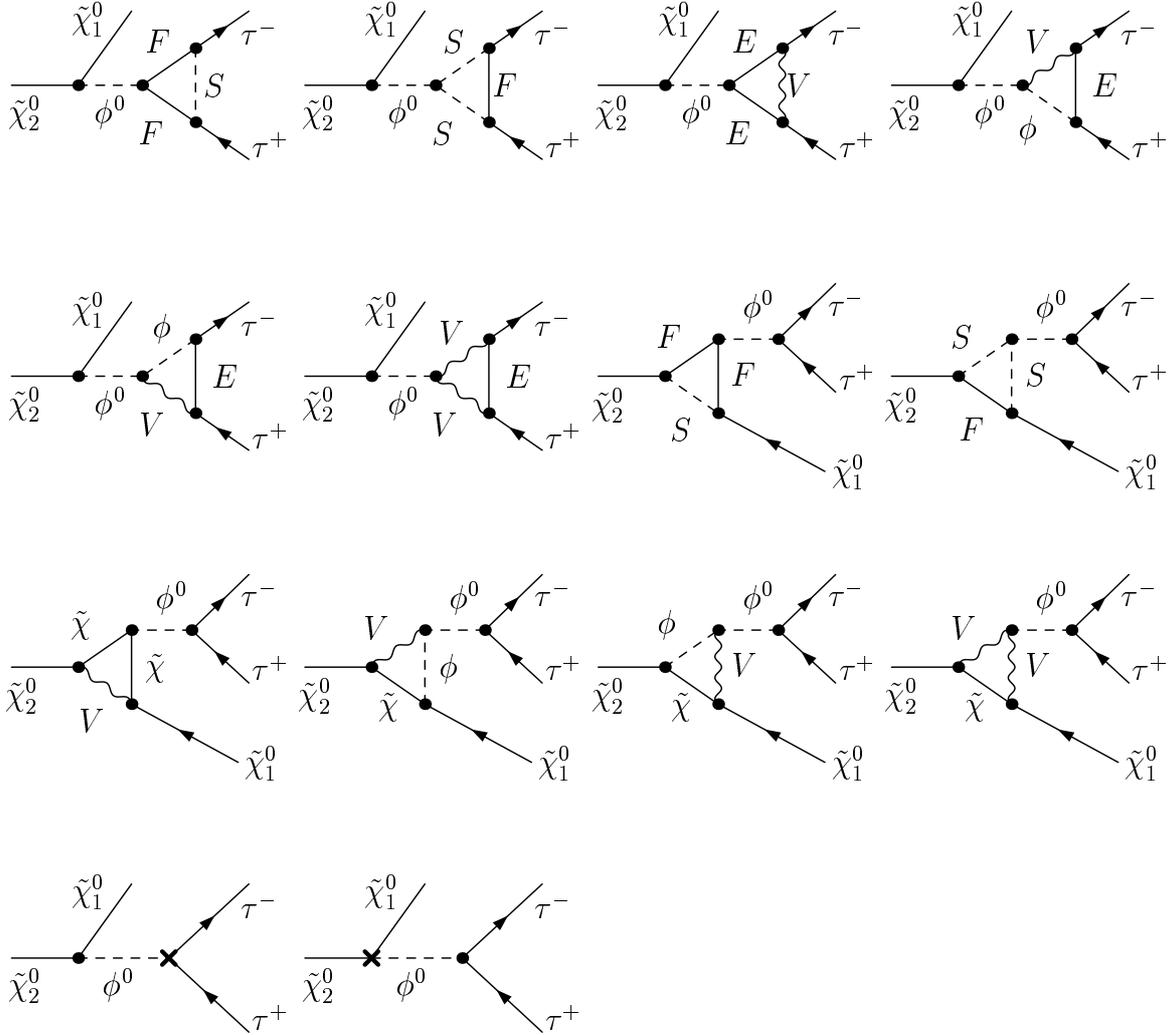} 
\end{tabular}
\caption{Vertex Feynman diagrams for $\tilde \chi_2^0$ decay into $\tilde \chi_1^0$ and $\tau^- \tau^+$
via the exchange of the neutral Higgs bosons and Goldstone boson $\phi^0$.
The diagrams in the last line are the corresponding counterterm diagrams. The Higgs intermediate states are neglectd 
for $l = e, \mu$.}
\end{figure}

\begin{figure}
\begin{tabular}{c}
\includegraphics[width=0.95\linewidth]{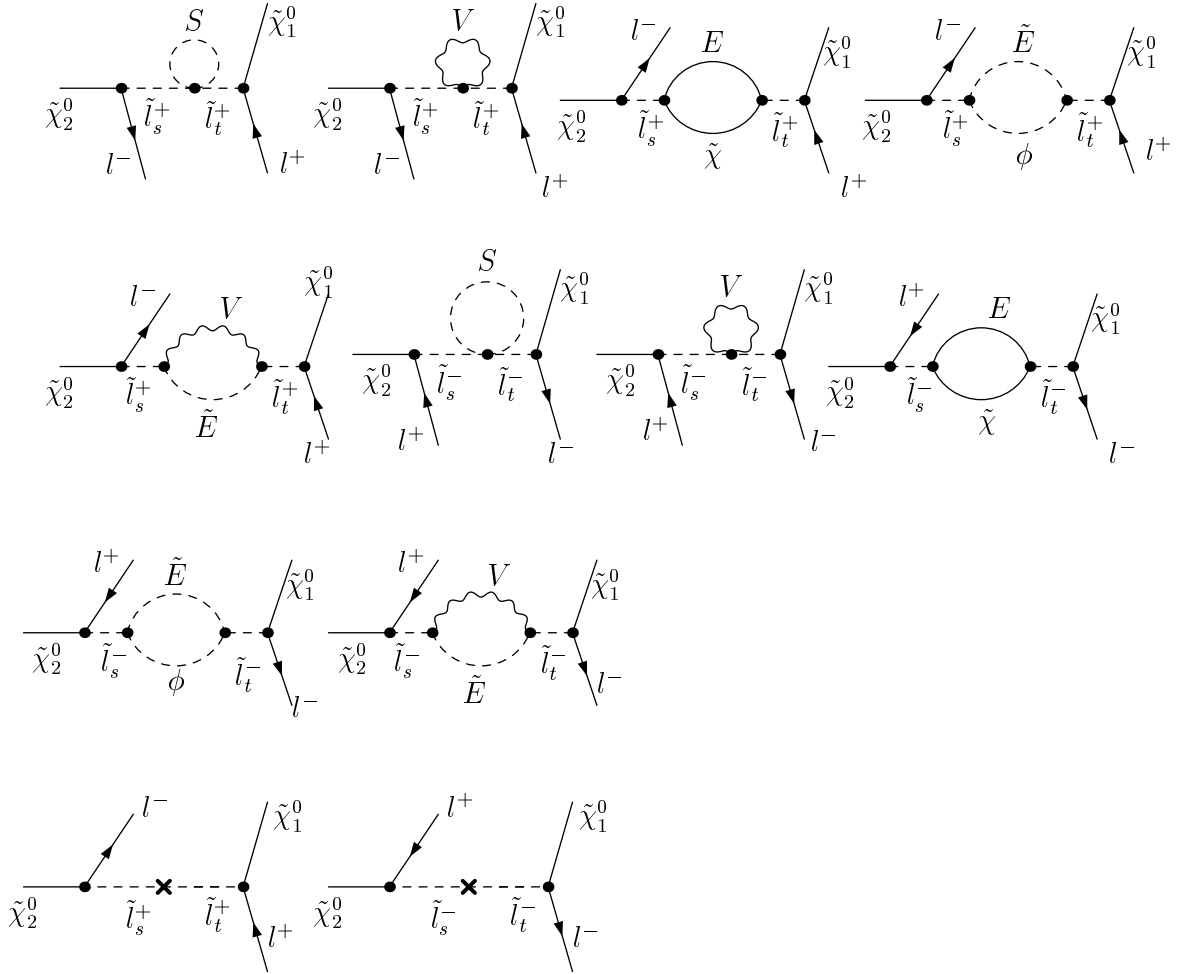} 
\end{tabular}
\caption{Self-enery diagrams for the decays $\tilde \chi_2^0 \to \tilde \chi_1^0 l^- l^+ $ in the case of 
the slepton mixing, $s(t) = 1, 2$ labels the slepton mass eigenstates.
The diagrams in the last line are the corresponding counterterm diagrams.}
\end{figure}
\begin{figure}
\begin{tabular}{c}
\includegraphics[width=0.95\linewidth]{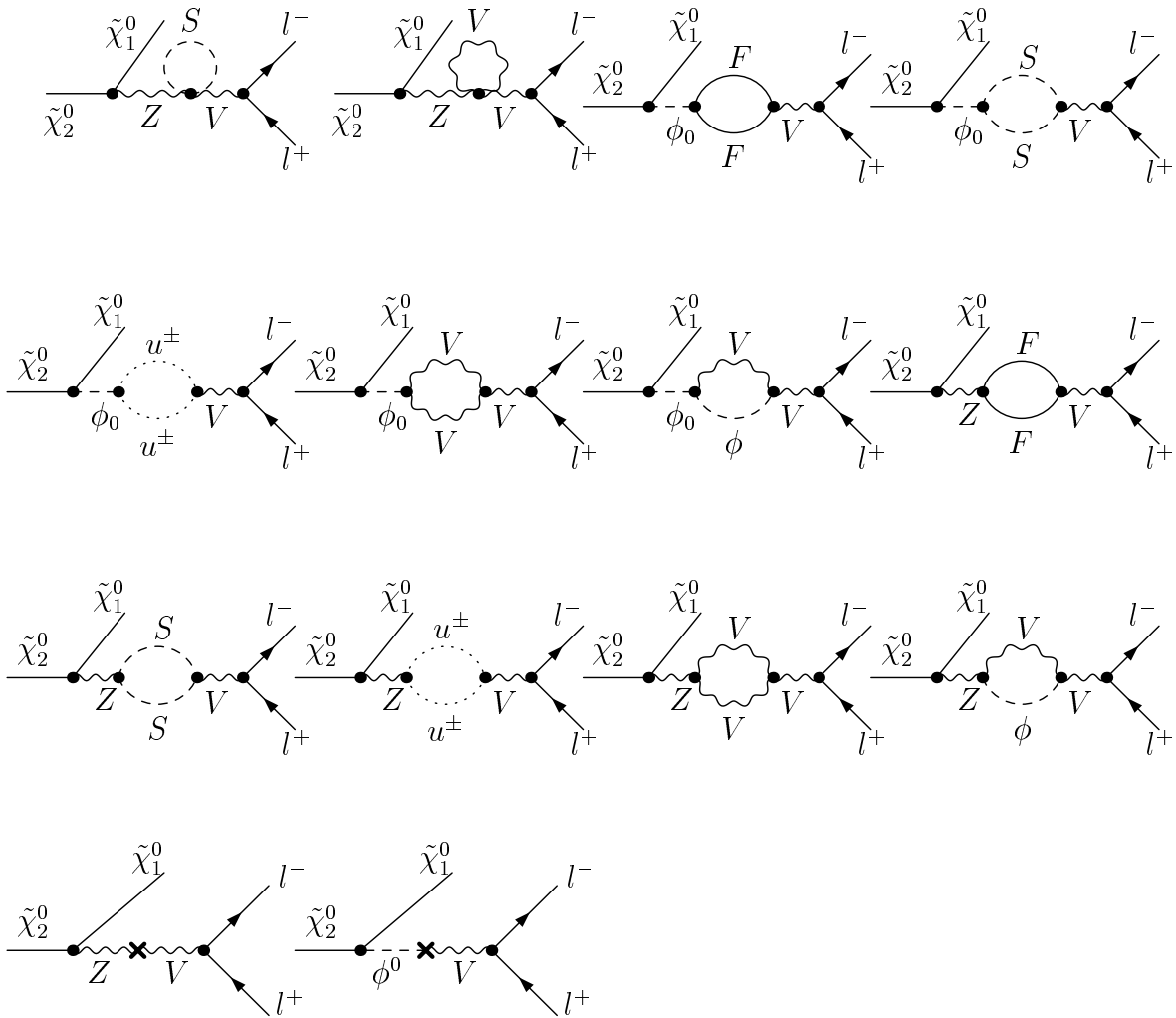} 
\end{tabular}
\caption{Self-enery diagrams for $Z-V$ and $\phi^0-V$ mixing.
The diagrams in the last line are the corresponding counterterm diagrams.} 
\end{figure}
\begin{figure}
\begin{tabular}{c}
\includegraphics[width=0.95\linewidth]{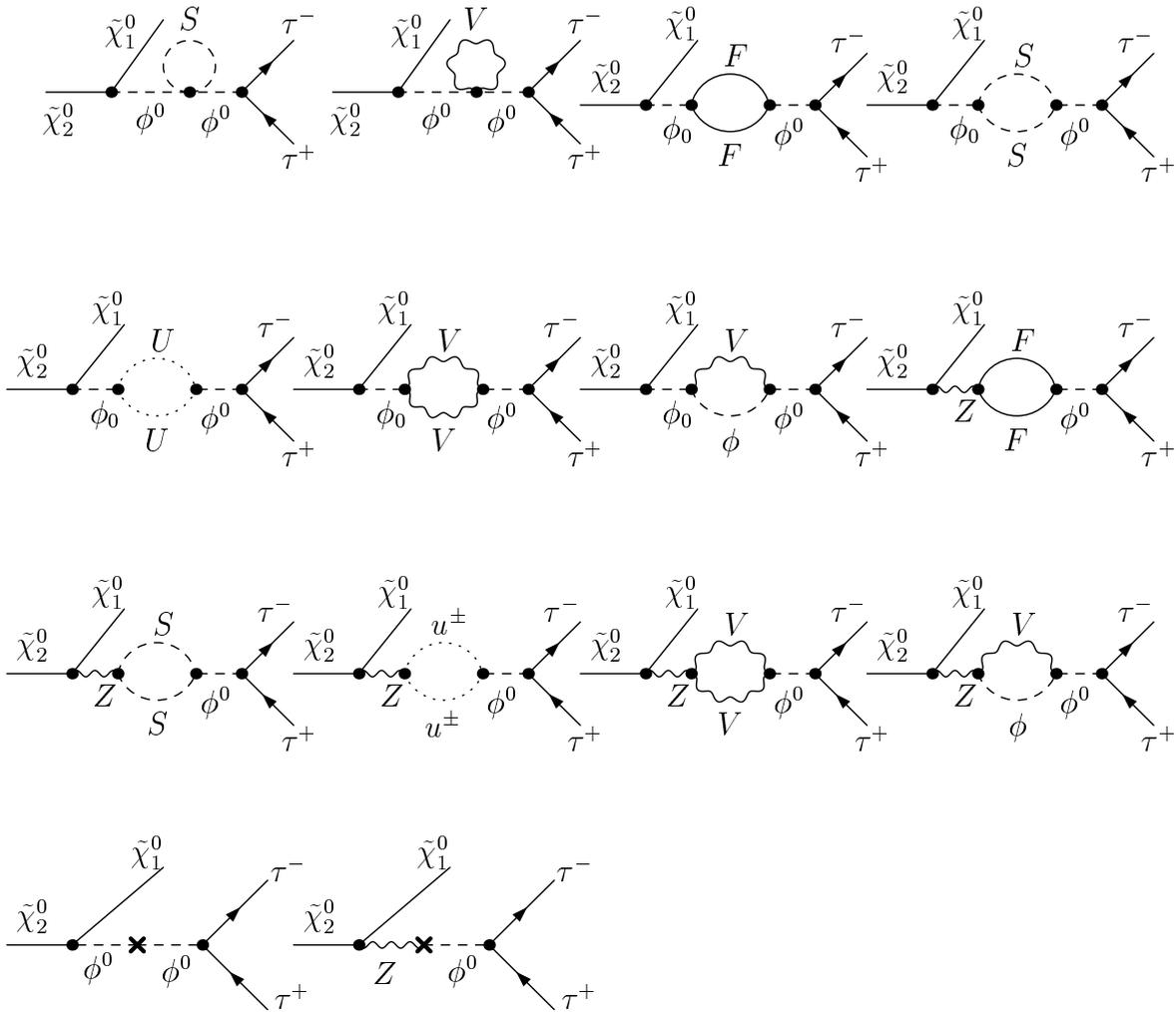} 
\end{tabular}
\caption{Self-enery diagrams for $\phi^0-\phi^0$ and $Z-\phi^0$ mixing, 
where the Higgs intermediate states are neglectd for $l = e, \mu$.
The diagrams in the last line are the corresponding counterterm diagrams.} 
\end{figure}
\begin{figure}[htb]
\begin{tabular}{c}
\includegraphics[width=0.95\linewidth]{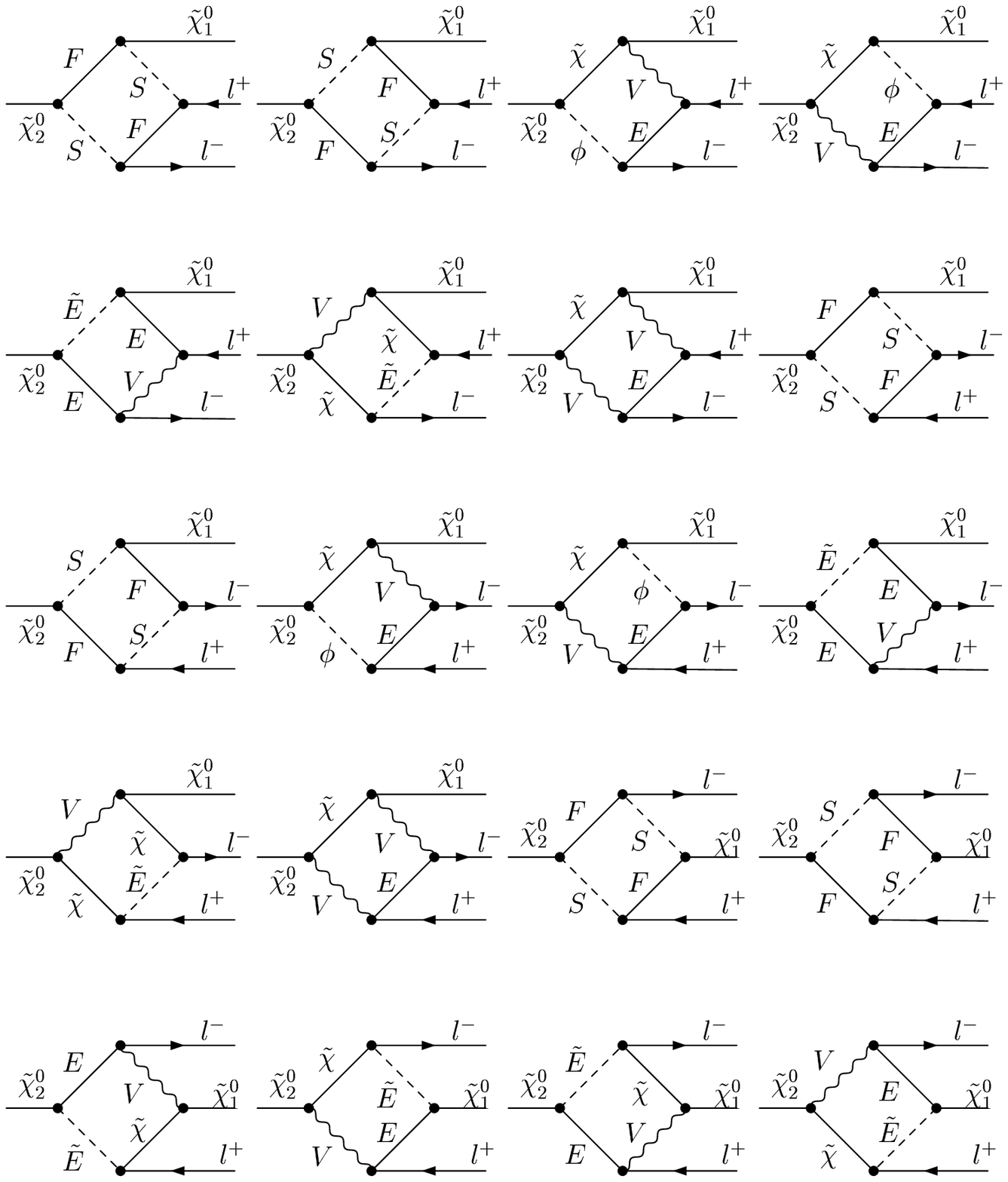} 
\end{tabular}
\caption{Box Feynman diagrams for $\tilde \chi_2^0$ decay into $\tilde \chi_1^0$ and two leptons $l^- l^+$.}
\end{figure}
\end{appendix}

\chapter*{Acknowledgements}
First of all, I am very much grateful to my supervisor Professor W. Hollik for the interesting topic of my thesis and
for supporting me during the work of this thesis.  I benefited a lot from his great experience and good knowledge 
in electroweak physics and SUSY and enjoyed all the days that I have spent with him.

\vspace*{5mm} 
I would like to thank my supervisor Professor M. Drees for the interesting topic of my thesis,
and for his support and encouragement during the work of my thesis. 

\vspace*{5mm}
Many thanks are given to the Max-Planck-Instute for Physics (Werner-Heisenberg-Insitut),
especially to the theory group. It is a great working environment which I enjoyed very much.
I am grateful to T. Fritzsche and T. Hahn for their useful help with the packages: FeynArts, FormCalc and LoooTools, 
and M. Roth for the kind help in calculating the hard photon bremsstrahlung contribution. 
I thank H. Rzehak for the useful discussions about the renormalization,
and for her friendship, advice and encouragement. 
I would like to thank A. Bredenstein for the useful discussions about the collinear photon radiation and
for reading the manuscript of my thesis.  
Thanks go to M. Rauch for his careful reading the manuscript of my thesis and 
for his translating the abstract of this thesis into German.
I am grateful to Mrs. Jurgeleit and Mrs. Reinke for their friendly help during these years.

\vspace*{5mm}
I would like to thank my family, in particular my parents, my parents in law and my husband Yu Wang,  
for their unconditional support and encouragement.
 
\end{document}